\documentclass[a4paper,11pt]{article}
\pdfoutput=1 

\usepackage{jheppub} 

\usepackage[T1]{fontenc} 
\usepackage{setspace}

\usepackage{stmaryrd}

\usepackage{amssymb}
\usepackage{amsmath,amsbsy}

\numberwithin{equation}{section}
\usepackage[titletoc,toc,title]{appendix}
\usepackage{mathtools}
\usepackage{mathrsfs}
\usepackage{braket}
\usepackage{float}
\usepackage{graphicx}

\usepackage{ytableau}

\usepackage{mmacells}

\usepackage[...]{youngtab}
\usepackage{tikz}
\usepackage{tikz-cd}
\usepackage{color}
\usepackage{epic}
\usepackage{pict2e}

\usepackage{hyperref}
\hypersetup{
    colorlinks,
    citecolor=magenta,
    filecolor=black,
    linkcolor=black,
    urlcolor=magenta
}

\newcommand{\GG}{\mathbb{G}}
\newcommand{\JJ}{\mathbb{J}}

\newcommand{\RR}{\mathbb{R}}
\newcommand{\ZZ}{\mathbb{Z}}
\newcommand{\QQ}{\mathbb{Q}}
\newcommand{\CC}{\mathbb{C}}
\newcommand{\lF}{\mathcal{F}}
\newcommand{\lN}{\mathcal{N}}
\newcommand{\lM}{\mathcal{M}}

\newcommand{\lT}{\mathcal{T}}
\newcommand{\lD}{\mathcal{D}}

\newcommand{\lO}{\mathcal{O}}
\newcommand{\lH}{\mathcal{H}}
\newcommand{\lS}{\mathcal{S}}
\newcommand{\lR}{\mathcal{R}}
\newcommand{\lC}{\mathcal{C}}
\newcommand{\lL}{\mathcal{L}}

\newcommand{\lV}{\mathcal{V}}
\newcommand{\lA}{\mathcal{A}}
\newcommand{\Sa}{\mathbb{S}}

\newcommand{\lE}{\mathcal{E}}

\newcommand{\lB}{\mathcal{B}}

\newcommand{\gla}{\mathfrak{gl}}
\newcommand{\sua}{\mathfrak{su}}
\newcommand{\sla}{\mathfrak{sl}}

\newcommand{\qdet}{{\rm qdet}}

\newcommand{\GL}{\mathsf{GL}}
\newcommand{\SL}{\mathsf{SL}}
\newcommand{\GT}{\mathsf{GT}}
\newcommand{\GP}{\mathsf{GP}}

\newcommand{\bT}{\textbf{T}}

\newcommand{\lY}{\mathcal{Y}}
\newcommand{\gl}{\mathfrak{gl}}
\newcommand{\wT}{\mathbb{T}}

\newcommand{\bB}{\textbf{B}}
\newcommand{\bN}{\textbf{N}}
\newcommand{\bb}{\textbf{b}}
\newcommand{\bc}{\textbf{c}}
\newcommand{\bC}{\textbf{C}}
\newcommand{\bD}{\textbf{D}}
\newcommand{\bS}{\textbf{S}}
\newcommand{\T}{\mathbb{T}}

\newcommand{\Q}{\mathbb{Q}}

\newcommand{\CH}{{\mathcal{H}}}

\newcommand{\es}{{\emptyset}}
\newcommand{\svx}{{\mathsf{x}}}
\newcommand{\svy}{{\mathsf{y}}}
\newcommand{\svX}{{\mathsf{X}}}
\newcommand{\svP}{{\mathsf{P}}}

\newcommand{\brax}{\bra{\mathsf{x}}}

\newcommand{\fs}{{\bar\es}}
\newcommand{\bmu}{{\bar\mu}}

\newcommand{\gn}{\mathsf{n}}
\newcommand{\gm}{\mathsf{m}}
\newcommand{\gln}{\gla(\mathsf{n})}
\newcommand{\su}{\mathfrak{su}}

\newcommand{\YT}{{\mathcal{T}}}

\newcommand{\hq}{{\mathbbm{q}}}
\newcommand{\hhq}{{\hat{\mathbbm{q}}}}
\newcommand{\comment}[1]{}
\newcommand{\gloE}{{\mathcal{E}}}
\newcommand{\rhs}{\mbox{r.h.s.} }
\newcommand{\lhs}{\mbox{l.h.s.} }

\newcommand{\psu}{\mathfrak{psu}}
\newcommand{\fkQ}{\mathfrak{Q}}
\newcommand{\fkP}{\mathfrak{P}}
\newcommand{\fkC}{\mathfrak{C}}
\newcommand{\fkU}{\mathfrak{U}}
\newcommand{\ga}{\mathfrak{g}}
\newcommand{\bs}{\textbf{s}}
\newcommand{\be}{\begin{eqnarray}}
\newcommand{\ee}{\end{eqnarray}}
\newcommand{\sfQ}{\mathsf{Q}}
\newcommand{\sfq}{\mathsf{q}}
\newcommand{\sfp}{\mathsf{p}}
\newcommand{\sfT}{\mathsf{T}}
\newcommand{\sft}{\mathsf{t}}
\newcommand{\sfe}{\mathsf{e}}
\newcommand{\sfE}{\mathsf{E}}
\newcommand{\sfi}{\mathsf{i}}

\newcommand{\sfh}{\mathsf{h}}

\newcommand{\bA}{\mathbf{A}}

\newcommand{\detl}{\left\llbracket\,}
\newcommand{\detr}{\,\right\rrbracket}

\title{Integrable systems, separation of variables and the Yang-Baxter equation}

\author{Paul Ryan$^{r}$}

\affiliation[r]{
Mathematics Department, King's College London,
The Strand, London WC2R 2LS, UK
}

\emailAdd{paul.1.ryan$\bullet$kcl.ac.uk}

\abstract{This article, based on the author's PhD thesis, reviews recent advancements in the field of quantum integrability, in particular the separation of variables (SoV) program for high-rank integrable spin chains and the boost mechanism for solving the Yang-Baxter equation. We begin with a general overview of quantum integrable systems with special emphasis on their description in terms of quantum algebras. We then provide a detailed account of the Yangian $\mathcal{Y}(\gl(\gn))$ of $\gl(\gn)$ in particular the Bethe algebra, fusion and T- and Q-systems. We then introduce the notion of separation of variables in integrable systems and build on Sklyanin's work in rank $1$ models and extend to higher rank. By exploiting a novel link between SoV and quantum algebra representation theory we construct the separated variables for high-rank $\gla(\gn)$ bosonic spin chains for arbitrary compact representations of the symmetry algebra and develop various new tools along the way. Next, we build on the previous part and develop a new technique for the computation of scalar products in the SoV framework which we call Functional SoV or FSoV. Unlike the work in the previous part, which was operatorial, this approach is functional and is based on the Baxter TQ equations. After developing this technique we supplement it with a new operator construction providing a unified view of functional and operatorial SoV. Then, we generalise the results of the previous part from compact spin chains to non-compact spin chains.

The final part of this work is based on the development of tools for solving the Yang-Baxter equation. We develop a bottom-up approach for this based on the so-called Boost automorphism and uses the spin chain Hamiltonian as a starting point. Our approach allows us to classify numerous families of solutions in particular a complete classification of $4\times 4$ solutions which preserve fermion number which have applications in the AdS/CFT correspondence.
}

\begin{document} 

\maketitle
\flushbottom
\newpage

\section*{Author's publications}

This review is based on the following published works of the author 
\begin{itemize}
    \item [\cite{Ryan:2018fyo}] P.~Ryan and D.~Volin, ``Separated variables and wave functions for rational gl(N) spin chains in the companion twist frame,''
J. Math. Phys. \textbf{60} (2019) no.3, 032701.

    \item [\cite{deLeeuw:2019zsi}] M.~De Leeuw, A.~Pribytok and P.~Ryan,
``Classifying two-dimensional integrable spin chains,''
J. Phys. A \textbf{52} (2019) no.50, 505201.

    \item [\cite{Gromov:2019wmz}] N.~Gromov, F.~Levkovich-Maslyuk, P.~Ryan and D.~Volin,
``Dual Separated Variables and Scalar Products,''
Phys. Lett. B \textbf{806} (2020), 135494.

    \item [\cite{deLeeuw:2019vdb}] M.~De Leeuw, A.~Pribytok, A.~L.~Retore and P.~Ryan,
``New integrable 1D models of superconductivity,''
J. Phys. A \textbf{53} (2020) no.38, 385201.
    
    \item [\cite{Ryan:2020rfk}] P.~Ryan and D.~Volin,
``Separation of Variables for Rational $\mathfrak {gl}(\mathsf {n})$ Spin Chains in Any Compact Representation, via Fusion, Embedding Morphism and B\"acklund Flow,''
Commun. Math. Phys. \textbf{383} (2021) no.1, 311-343.
    \item [\cite{deLeeuw:2020ahe}] M. de Leeuw, C. Paletta, A. Pribytok, A. L. Retore, and P. Ryan, “Classifying Nearest-Neighbor Interactions and Deformations of AdS,” Phys. Rev. Lett. 125 (2020),
no. 3 031604.

    \item [\cite{Gromov:2020fwh}] N.~Gromov, F.~Levkovich-Maslyuk and P.~Ryan,
``Determinant form of correlators in high rank integrable spin chains via separation of variables,''
JHEP \textbf{05} (2021), 169.

    \item [\cite{deLeeuw:2020xrw}] M.~de Leeuw, C.~Paletta, A.~Pribytok, A.~L.~Retore and P.~Ryan,
``Yang-Baxter and the Boost: splitting the difference'', SciPost Phys. \textbf{11} (2021), 069.
\end{itemize}

\newpage

\section*{Acknowledgements}

This article summarises my work from September 2017 to June 2021 during my PhD studies, and I was somehow fortunate enough to end up with not one but two great supervisors. Thank you to Dima and Marius for all of your hard work and our countless discussions and for supporting me in every possible way. Perhaps this is the best place to share my favourite memories with both of you from my PhD. With Marius, shortly after we started to make some progress on the YBE Anton and I spent a week doing blackboard calculations and running up very excitedly to Marius’ office every few hours with some new insight we had found. With Dima, on one Saturday in Uppsala we arrived at the office early in the morning and left late in the evening and spent the entire day doing calculations in Dima’s office and testing ideas. Plus, Dima brought pizzas for lunch. Again, thanks to you both.  

\medskip

I also want to extend my thanks to my other collaborators during this time -- Ana, Anton, Chiara, Fedor, Kolya and Sébastien -- for drastically improving all of our publications. It has been a great pleasure working with you all. 

\medskip

I have benefited from many discussions with my collaborators as well as countless others, in particular with George Korpas, Juan-Miguel Nieto, Simon Ekhammar (special thanks for taking a look at my thesis and seemingly reading it even more carefully than I did), Dmitry Chernyak, Rob Klabbers, Jules Lamers, Christian Marboe and Alessandro Torrielli. 

\medskip

Thanks to everyone at TCD in particular my rotation of office mates Anne, Anton and Martijn, the other PhD students and postdocs, of which there are too many to name, as well as the TCD admin staff Ciara, Emma, Helen, Karen and Mirela.

\medskip

My trips to Nordita were made painless thanks to the hard work of the admin staff, in particular Hans for helping with any day-to-day issues I encountered (and who always shared interesting stories over lunch) and Elizabeth, Jimmie and Olga for helping to organise my accommodation and flights. 

\medskip

My work was supported in part by a Nordita Visiting PhD Fellowship and by SFI and Royal Society grant UF160578. I am also grateful to Kostya Zarembo and Tristan McLoughlin -- Kostya for financing my numerous trips to Stockholm and Tristan for supporting my trip to IGST in Copenhagen.

\medskip

Special thanks to Cathal and Seleana for coffees, movies and boardgames and to my parents and Tinne's family for their continued support. Finally, thanks to Tinne (and our pets), the person who deserves the most thanks and certainly the most praise for somehow managing to put up with me.

\medskip

I am currently supported by the European Research Council (ERC) under the European Union’s Horizon 2020 research and innovation programme – 60 – (grant agreement No. 865075) EXACTC.

\newpage

\tableofcontents

\section{Introduction}
Quantum integrable systems \cite{Faddeev:1979gh} are one of the cornerstones of theoretical physics. Typically these are models which possess a large number of conserved quantities. They are simple enough to be an important testing ground for new techniques as well as being rich enough to have direct physical applications. Famous integrable models such as the Heisenberg XXX spin chain and one-dimensional Hubbard model have made appearances in statistical mechanics applications and in the context of the AdS/CFT correspondence or gauge/gravity duality. Furthermore their study often leads to new ideas in various areas of pure mathematics such as knot theory.

\paragraph{Spectral problem of $\lN=4$ SYM}

A large amount of motivation for this work comes from maximally supersymmetric Yang-Mills theory in $4d$ ($\lN=4$ SYM) with gauge group $SU(N)$ which is dual under the AdS/CFT correspondence \cite{Maldacena:1997re} to Type IIB superstrings on ${\rm AdS}_5\times S^5$. The theory enjoys $\psu(2,2|4)$ supersymmetry which contains the conformal algebra $\mathfrak{so}(2,4)\simeq \su(2,2)$ and the conformal symmetry remains unbroken \cite{Sohnius:1981sn} at all loop orders. As such the primary objects of interest are its conformal data -- scaling dimensions $\Delta$ of all local operators and three-point structure constants. Once these are determined the theory is considered solved. 

\medskip

Shortly after the turn of the millennium it was discovered that the one-loop spectral problem is integrable in the planar limit $N\rightarrow \infty$. Namely, it was observed \cite{Minahan:2002ve} that the one-loop dilatation operator could be mapped to the Hamiltonian of an integrable spin chain with single-trace local operators corresponding to spin chain states. Around the same time it was discovered that the non-linear sigma model describing classical superstrings on the $AdS_5\times S^5$ background is also classically integrable \cite{Bena:2003wd} and admits an infinite number of Poisson commuting integrals of motion. Since its discovery integrability has also been found at higher loops and appears to hold at all loops and has provided a novel framework for making testable predictions on both sides of the AdS/CFT correspondence, see \cite{Beisert:2010jr} for a review. 
\begin{figure}[htb]
\centering
  \includegraphics[width=100mm,scale=25]{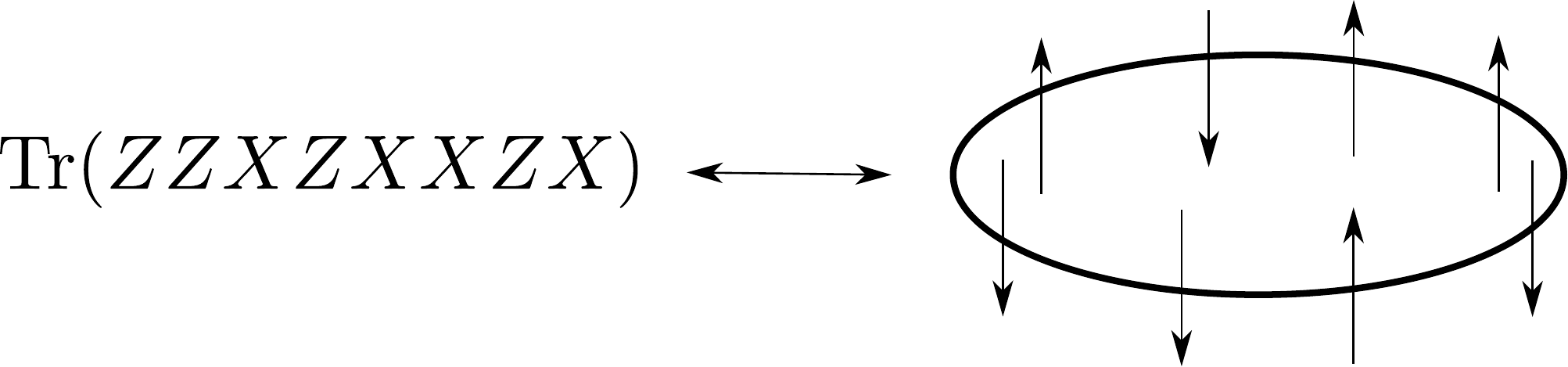}
  \caption{Correspondence between single trace local operators in the $\sua(2)$ sector, comprised of fields $Z$ and $X$, of $\lN=4$ SYM and spin chain states. Fields $Z$ correspond to ``up" states $\uparrow$ and fields $X$ correspond to ``down" states $\downarrow$. Cyclicity of the trace corresponds to periodic boundary conditions on the spin chain.}
  \label{statespinchain}
\end{figure}

\paragraph{The road to the exact spectrum} 

\begin{figure}[htb]
\centering
  \includegraphics[width=80mm,scale=25]{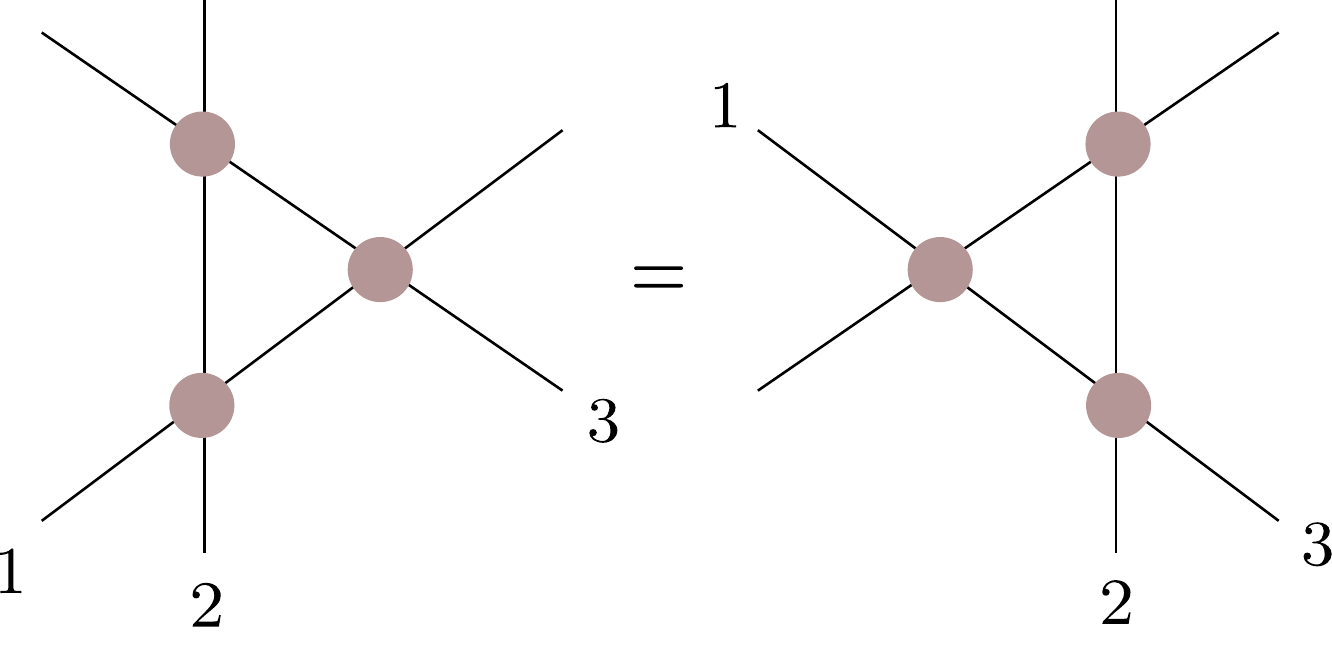}
  \caption{The two ways to factorise a three-particle scattering process into a sequence of two-particle scattering processes. Their equality leads to the Yang-Baxter equation.}
  \label{smatrixfactor}
\end{figure}

Since the discovery of integrability a significant amount of work was put towards the problem of computing the spectrum of anomalous dimensions at finite coupling. Under the AdS/CFT correspondence this is equivalent to finding the energies of string states. The key tool for this is the Thermodynamic Bethe Ansatz (TBA) which was pioneered in the work of Zamolodchikov \cite{zamolodchikov1990thermodynamic} for relativistic theories in $1+1$ dimensions. In essence the TBA allows one to compute the finite-volume spectrum of an integrable quantum field theory using its infinite-volume scattering data. Thankfully, integrability highly constrains this scattering data -- there is no particle production as well as factorised scattering meaning the number of particles before and after the collision is preserved and a multi-particle scattering process factorises into a product of two-particle scattering events. Consistency of this factorisation then leads to the celebrated Yang-Baxter equation for the S-matrix
\begin{equation}
S_{12}S_{13}S_{23} = S_{23}S_{13}S_{12}\,,
\end{equation}
see Figure \ref{smatrixfactor} and \cite{Bombardelli:2016scq} for a review.
In the uniform light-cone gauge the ${\rm AdS}_5\times S^5$ string sigma model is defined on a cylinder of circumference $L$. In the decompactifying limit $L\rightarrow \infty$ the model defines a massive $1+1$-dimensional QFT with elementary excitations transforming in two copies of the defining representation of the algebra $\sua(2|2)_{\rm ce}:=\sua(2|2)\rtimes \RR^3$ \cite{Arutyunov:2009ga}, an enhancement of the superalgebra $\sua(2|2)$ containing additional central charges. This symmetry is constraining enough that it guarantees that the S-matrix satisfies the Yang-Baxter equation \cite{Beisert:2005fw,Beisert:2005tm}.

\medskip

The starting point for the TBA is as follows. We consider a $1+1$-dimensional QFT defined on a cylinder of circumference $L$ with its finite-temperature $\beta^{-1}$ partition function given by $Z=\sum_{n}e^{-\beta E_n(L)}$ where $E_n(L)$ are a complete set of energies of the theory. In the zero-temperature $\beta\rightarrow 0$ limit the partition function is dominated by the ground state energy $E_0(L)$
\begin{equation}
Z(\beta,L)\rightarrow e^{-\beta E_0(L)}\,.
\end{equation}
On the other hand, the theory on the cylinder can be viewed as the $R\rightarrow \infty$ limit of a theory defined on a torus with circumferences $R$ and $L$ and coordinates $(\tau,\sigma)$. One then performs a double Wick-rotation introducing new coordinates $(\tilde{\tau},\tilde{\sigma})$ by $\tau=\sfi\,\tilde{\sigma}$ and $\sigma=\sfi\,\tilde{\tau}$ obtaining a new theory, the so-called \textit{mirror model}. The zero-temperature limit $\beta\rightarrow \infty$ in the original theory corresponds to finite-temperature $\frac{1}{L}$ in the mirror theory but in infinite volume. 

\medskip

For relativistic models the mirror model coincides with the original model. This is not the case for the ${\rm AdS}_5\times S^5$ superstring, in uniform light-cone gauge where it lives on a cylinder of circumference $L$, which lacks worldsheet Lorentz invariance and hence the mirror model deserves a separate investigation. This was carried out in \cite{Arutyunov:2007tc,Bombardelli:2009ns,Arutyunov:2009ur} leading to a detailed account of its finite-temperature thermodynamics and TBA equations, an infinite set of nonlinear integral equations on functions $Y_{a,s}(u)$ of a complex variable $u$ living on a T-shaped lattice of points $(a,s)$ \cite{Gromov:2008gj,Gromov:2009tv}. The TBA equations describe the exact spectrum of the theory.

\paragraph{Y-system and T-system}

The TBA equations are highly complicated but were nevertheless suitable for numerical studies of the spectrum \cite{Gromov:2009zb,Frolov:2010wt}. It was realised that the Y-functions appearing in the TBA equations could be packaged into the Y-system \cite{Gromov:2008gj,Gromov:2009tv}, an infinite set of functional relations on Y-functions $Y_{a,s}(u)$ reading
\begin{equation}\label{Yfunctions}
Y_{a,s}^+ Y_{a,s}^- = \frac{(1+Y_{a,s+1})(1+Y_{a,s-1})}{\left(1+\frac{1}{Y_{a-1,s}}\right)\left(1+\frac{1}{Y_{a+1,s}}\right)}
\end{equation}
where we use the notation $f^\pm = f\left(u\pm\frac{\sfi}{2}\right)$. The functional relations \eqref{Yfunctions} are not completely equivalent to the TBA equations -- one still needs to specify the analytic structure of the Y-functions which have square-root discontinuities \cite{Cavaglia:2010nm}. The Y-system \eqref{Yfunctions} together with the necessary analytic properties became known as an \textit{analytic Y-system}. 

\medskip

The study of the spectrum simplifies even further when one recasts the analytic Y-system as a T-system \cite{Gromov:2009tv} of functions $\sfT_{a,s}$ related to the Y-functions as 
\begin{equation}
Y_{a,s}=\frac{\sfT_{a,s+1}\sfT_{a,s-1}}{\sfT_{a+1,s}\sfT_{a-1,s}}
\end{equation}
and satisfying 
\begin{equation}\label{Tsystem}
\sfT_{a,s}^+ \sfT_{a,s}^- = \sfT_{a+1,s}\sfT_{a-1,s}+\sfT_{a,s+1}\sfT_{a,s-1}
\end{equation}
subject to certain analytic constraints resulting in an \textit{analytic T-system}. The T-system \eqref{Tsystem} is also known as the Hirota bilinear equation \cite{hirota1981discrete} and is one of the key equations in the study of integrable systems both classical and quantum and both discrete and continuous.

\paragraph{Quantum spectral curve}
The ultimate solution of the $\lN=4$ SYM spectral problem takes the form of an \textit{analytic Q-system} dubbed quantum spectral curve (QSC) \cite{Gromov:2013pga,Gromov:2014caa} -- a set of functional relations on a set of $2^8$ Q-functions $\sfQ_{A|I}$ $A,I\subset \{1,2,3,4\}$ with certain analytic properties called QQ-relations reading 
\begin{equation}
\begin{split}
& \sfQ_{A|I}\sfQ_{Aab|I} = \sfQ_{Aa|I}^+ \sfQ_{Ab|I}^- - \sfQ_{Aa|I}^- \sfQ_{Ab|I}^+  \\
& \sfQ_{A|I}\sfQ_{A|Iij} = \sfQ_{A|Ii}^+ \sfQ_{A|Ij}^- - \sfQ_{A|Ii}^- \sfQ_{Ab|Ij}^+  \\
& \sfQ_{Aa|I}\sfQ_{A|Ii} = \sfQ_{Aa|Ii}^+ \sfQ_{A|I}^- -\sfQ_{Aa|Ii}^- \sfQ_{A|I}^+  \\
\end{split}
\end{equation}
forming a Q-system. The precise expressions for the T-functions $\sfT_{a,s}$ is not universal and depends on the specific choice of $a,s$ but always takes the form of simple determinants in Q-functions and for this reason the Q-functions define a \textit{Wronskian solution} of the T-system. Remarkably, the complicated analytic structure first appearing in the TBA equations simplifies drastically when reduced to the analytic Q-system. 

\medskip

The QSC formulation of the $\lN=4$ SYM spectral problem has led to a plethora of remarkable results. It has successfully been applied as a tool for perturbative QFT computations at weak coupling \cite{Marboe:2014gma} enabling the dimension of the $\sla(2)$-sector Konishi operator to be computed to $10$ loops which was subsequently generalised to the full theory \cite{Marboe:2017dmb,Marboe:2018ugv} and $11$ loops. The QSC has also allowed to probe the structure of the theory at strong coupling \cite{Gromov:2014bva} and at finite coupling numerically \cite{Gromov:2015wca} and in particular analyse the theory when continued to non-integer spin $S$, including the $S=-1$ case which is closely related to high-energy QCD scattering amplitudes \cite{Kuraev:1977fs,Balitsky:1978ic}. In addition to these developments it has also been possible to extend the QSC from application to single-trace local operators to cusped Wilson loops \cite{Gromov:2015dfa}, which remarkably only requires a simple modification of the large-$u$ asymptotics of the Q-functions, and was then used to analyse the so-called quark-anti-quark potential \cite{Gromov:2016rrp}. Finally, the QSC has been extended to a range of other theories such as ABJM \cite{Cavaglia:2014exa,Bombardelli:2017vhk} based on the $\mathfrak{osp}(4|6)$ algebra and the $\eta$-deformed ${\rm AdS}_5 \times S^5$ superstring \cite{Klabbers:2017vtw}, a $q$-deformation of the original ${\rm AdS}_5 \times S^5$ superstring based on the $U_q(\mathfrak{psu}(2,2|4)$ algebra. It is the triumph of the integrability-based approach to $\lN=4$ SYM, see \cite{Gromov:2017blm,Kazakov:2018ugh,Levkovich-Maslyuk:2019awk} for reviews. 

\medskip

\paragraph{Towards QSC for correlators - Separation of Variables} 

Despite the tremendous success of using integrability techniques for the calculation of scaling dimensions $\Delta$ of local operators in $\lN=4$ SYM the situation is far less satisfactory when it comes to computing three-point correlation functions. It is tempting to hope that something similar to the TBA, which worked so wonderfully for the spectral problem, can also be carried out for correlation functions but this has not yet been realised, although there has been some progress for other quantities such as the so-called $g$-function \cite{Caetano:2020dyp}. A novel approach for computing higher-point correlation functions using integrability is the so-called Hexagon formalism \cite{Basso:2015zoa} but this approach suffers from only being valid in the asymptotic regime prior to the appearance of so-called wrapping effects. 

\medskip

Since its discovery it has been hoped that the QSC, which works so well for the spectrum, could also be used to develop a non-perturbative finite-size formalism for correlation functions. One of the main intuitions for this comes from the fact that in integrable spin chains the Q-functions are the building blocks of the wave functions $\Psi(\svx)$ of conserved charges in a certain coordinate system $\svx_\alpha$ dubbed Sklyanin's separated variables \cite{10.1007/3-540-15213-X_80,Sklyanin:1991ss,Sklyanin:1992eu,Sklyanin:1992sm,Sklyanin:1995bm}. The special feature of these variables, as the name suggests, is that the wave-function in this coordinate system factorises into a product of one-particle wave functions
\begin{equation}
\Psi(\svx) \sim \prod_{\alpha} \sfQ(\svx_\alpha)
\end{equation}
for some choice of Q-functions $\sfQ$. As a result of this, in these coordinates the matrix elements of an operator $\lA$ can be expressed in this basis as 
\begin{equation}
\bra{\Psi^A}\lA\ket{\Psi^B} =\displaystyle \int {\rm d}\svx\, {\rm d}\svx^\prime \mu(\svx)\mu(\svx^\prime)\left(\prod_{\alpha} \sfQ(\svx_\alpha)\right)\lA(\svx,\svx^\prime)\left(\prod_{\beta} \sfQ(\svx^\prime_\beta)\right)
\end{equation}
for some appropriate measure $\mu$. Such a construction was successfully realised \cite{Cavaglia:2018lxi} in the context of cusped Wilson loops of $\lN=4$ SYM in the so-called \textit{ladders limit} where only a certain family of Feynman diagrams contribute. The result is that a certain three-point structure constant $C_{123}^{\bullet\bullet\circ}$ could be expressed in terms of Q-functions $q_1$ and $q_2$ \footnote{Related to the QSC Q-functions by appropriate symmetry transformations.} as
\begin{equation}\label{SoVstructureconstant}
C_{123}^{\bullet\bullet\circ}=\frac{\langle q_1 q_2 e^{-\phi_3 u}\rangle}{\sqrt{\langle q_1^2\rangle \langle q_2^2\rangle}}
\end{equation}
where for a function $f(u)$ the bracket operation $\langle f\rangle$ is defined by 
\begin{equation}
\langle f\rangle := \left(2 \sin\frac{\beta}{2}\right)^\alpha \displaystyle\int^{c+i\infty}_{c-i\infty}\frac{{\rm d}u}{2\pi i u} f(u),\quad c>0\,.
\end{equation}
Similar expressions have also been found in a different regime \cite{Giombi:2018qox}. These results made clear that a separation of variables type approach to correlation functions along the lines of Sklyanin could be within reach. Unfortunately, while tremendously successful for $\gl(2)$-based models, Sklyanin's separation of variables program remained almost completely undeveloped for higher-rank or supersymmetric models. The result \eqref{SoVstructureconstant} put the need to develop the SoV program for higher-rank supersymmetric systems, in particular those related to $\mathfrak{psu}(2,2|4)$ needed for $\lN=4$ SYM, firmly in the spotlight and was one of the main driving factors in a flurry of research which followed. 

\medskip

\paragraph{$R$-matrix program} 

The Q-functions entering the QSC and the related T-functions are expected to be eigenvalues of some yet-to-be-constructed Q and T-operators as is the case in integrable spin chains. Unfortunately the governing algebraic structure is still not well-understood and it is not known how to write down an algebra $\lA$ with a commutative subalgebra generated by such Q-operators at finite length and finite coupling. At one-loop the corresponding algebra is given by a so-called Yangian algebra, in particular the Yangian of $\psu(2,2|4)$. In the lightcone gauge and asymptotic limit of operators with large length but finite coupling the algebra is known to be related to that of the one-dimensional Hubbard model \cite{Hubbard_1965RSPSA,Beisert:2005tm} which is described by a deformed Yangian of the centrally extended algebra $\sua(2|2)$ \cite{Beisert:2014hya}. 

\medskip

The algebras describing quantum integrable systems generally fall into the realm of quasi-triangular Hopf algebras, see \cite{Chari:1994pz} for an extensive treatment. Given a Hopf algebra $\lA$, which in particular means that it is an algebra equipped with a coproduct $\Delta$, we say that $\lA$ is quasi-triangular if there exists an invertible element $\lR\in\lA\otimes \lA$ with the property that for all $a\in\lA$ we have 
\begin{equation}\label{quasi-triangular}
\Delta^{\rm op}(a) = \lR^{-1} \Delta(a) \lR
\end{equation}
where $\Delta^{\rm op}(a)$ denotes the ``opposite" coproduct on $\lA$ obtained by permuting factors. Together with certain other assumptions this leads to the quantum Yang-Baxter equation
\begin{equation}\label{YBEintro}
\lR_{12}\lR_{13}\lR_{23}=\lR_{23}\lR_{13}\lR_{12}
\end{equation}
on the triple tensor product $\lA\otimes \lA\otimes \lA$ where the indices $\lR_{ij}$ indicate on which of the three factors $\lR$ is acting on.

\medskip

The universal $\lR$-matrix is an extremely powerful tool. In physical applications one is generally interested not in the algebra $\lA$ itself but in certain representations. For example, in the TBA for the ${\rm AdS}_5\times S^5$ superstring one needs to know the scattering matrix for elementary excitations as well as for bound states \cite{Arutyunov:2009zu,Arutyunov:2009mi}. If one had access to the universal $\lR$-matrix these could be simply obtained by evaluating it in the given representation. There are also various other applications of the universal $\lR$-matrix, for example its use \cite{Meneghelli:2015sra} in constructing lattice-discretizations of integrable quantum field theories, a powerful method of dealing with UV divergences in a rigourous way \cite{Faddeev:1985qu,volkov1992quantum,Ridout:2011wx}. 

\medskip

The quantum algebra describing the one-dimensional Hubbard model, the Yangian of centrally extended $\sua(2|2)$, is not quasi-triangular. However, it is possible that the algebra can be extended to a new algebra which does admit a universal $\lR$-matrix. This is known as the quantum double construction \cite{drinfeld1986quantum}. Although this has not yet been carried out for the deformed $\sua(2|2)$ Yangian it has been done for a simpler but related algebra in \cite{Beisert:2016qei} giving hope that the procedure can be extended for the full Hubbard model. Despite the algebra not being quasi-triangular it is however ``almost" quasi-triangular \cite{Beisert:2014hya}. This means that while we cannot construct a universal $\lR$-matrix an operator $R$ satisfying the quantum Yang-Baxter equation can be constructed at the level of representations. For practical applications this is usually enough. Unfortunately one is then tasked with constructing the operator $R$, simply called an $R$-matrix, for every situation at hand. For case of ${\rm AdS}_5\times S^5$ strings scattering elementary excitations it is a $16\times 16$ matrix \cite{Beisert:2005tm}. Hence, one needs an efficient method for solving the Yang-Baxter equation \eqref{YBEintro}. 

\medskip

In this work we aim to make advancements in both of the discussed directions. We will develop the SoV framework for high rank spin chains and develop new efficient techniques for solving the Yang-Baxter equation.

\medskip

\paragraph{Outline}
This article is organised as follows. 

\begin{enumerate}
\item \textbf{Part 1: Quantum algebras and quantum integrability} In this part we review the basic objects which will be used throughout the text. We will begin with a quick review of the XXX spin chain -- the prototypical example of a quantum integrable system. We will then move on to the notion of quantum algebras which are the mathematical framework for discussing quantum integrable systems. The primary object of interest will be the so-called Yangian algebra and we will discuss its representation theory and how the conserved charges of the XXX spin chain fit into a certain commutative subalgebra, the Bethe algebra. We will then present a detailed review of the Bethe algebra including the fusion procedure for transfer matrices, Baxter equations and Q-system. 

\item \textbf{Part 2: Separation of Variables} The second part of this  article focuses on the recent developments of the SoV program for higher-rank integrable systems. After a short review of separation of variables in the classical XXX spin chain we will discuss Sklyanin's quantum separation of variables and the recent progress made for its higher rank generalisation. We will place particular emphasis on the relation between SoV and Yangian representation theory for compact spin chains via Gelfand-Tsetlin patterns. This is based on the author's publications \cite{Ryan:2018fyo} and \cite{Ryan:2020rfk}.

\item \textbf{Part 3: Functional orthogonality and scalar products} Next, we discuss a method for the calculation of scalar products in the SoV framework based on the Baxter TQ equations. We obtain determinant formulas for these scalar products and develop an operatorial construction to supplement the functional approach. This is based on the publications \cite{Gromov:2019wmz} and partly on \cite{Gromov:2020fwh}.

\item \textbf{Part 4: Non-compact spin chains} 
In this Part we switch our attention from compact spin chains to non-compact ones. We start with a brief overview of the corresponding representation theory and explain how the functional scalar products of the previous Part can be generalised to this case and construct a corresponding operatorial framework. We give explicit examples of our constructions in $\sla(2)$ and $\sla(3)$ spin chains of low length and explain how to calculate a number of non-trivial correlation functions, including form-factors of local operators. This is based on \cite{Gromov:2020fwh}.

\item \textbf{Part 5: Solving the Yang-Baxter equation} 
This Part has a different focus. We study the Yang-Baxter equation and develop an efficient approach for obtaining and classifying its solutions via the so-called Boost operator. In particular we classify all $4\times 4$ $R$-matrices which preserve fermion numbers. As an application, we classify all integrable deformations of the ${\rm AdS}_2\times S^2 \times T^6$ S-matrix. This is based on the publications \cite{deLeeuw:2019zsi,deLeeuw:2019vdb,deLeeuw:2020ahe,deLeeuw:2020xrw} of the author.

\end{enumerate}

\part{Quantum algebras and quantum integrability}\label{Part1}
\section{A first look at the XXX spin chain}
Some of the most common methods for solving integrable systems go by the name of the Bethe ansatz and are the Coordinate, Algebraic, Analytic, Functional Bethe ansatz. In essence, all of them consist of proposing a suitable ansatz for the eigenvectors of the conserved charges. Physical requirements such as periodicity of these eigenvectors then leads to a set of quantisation conditions known as the Bethe Ansatz equations\footnote{Actually, in the Analytical Bethe Ansatz an ansatz is instead made for the eigenvalues of the conserved charges. Imposing certain analytical properties then leads to the Bethe ansatz equations.}. The first incarnation, the Coordinate Bethe ansatz, was used by Hans Bethe \cite{Bethe:1931hc} to write down the wave function in a simple model of interacting electrons -- the XXX spin chain. 

\subsection{XXX Hamiltonian, symmetries and higher charges}

\paragraph{Hamiltonian}
The Heisenberg XXX spin chain consists of $L$ spin-$\frac{1}{2}$ particles on a circle with the interaction governed by the following Hamiltonian 
\begin{equation}\label{XXXham}
H=\displaystyle\sum_{\alpha=1}^L \lH_{\alpha,\alpha+1},\quad \lH_{\alpha,\alpha+1}=S^j_\alpha S^j_{\alpha+1}-\frac{1}{4},\quad j=x,y,z,
\end{equation}
where as usual $S^j = \frac{1}{2}\sigma^j$. The interaction is clearly only between nearest-neighbours on the spin chain, manifest from the fact that the Hamiltonian is a sum of nearest-neighbour densities $\lH_{\alpha,\alpha+1}$. Periodic boundary conditions are assumed, that is $L+\alpha=\alpha\, {\rm mod}\,L$, $\alpha=1,2,\dots,L-1$. 

\paragraph{Symmetries}
The Hamiltonian \eqref{XXXham} commutes with the global generators $S_{x,y,z}$ of spin
\begin{equation}
S_{x,y,z}=\sum_{\alpha=1}^L S_{x,y,z}^\alpha\,.
\end{equation}
Hence, the eigenstates of \eqref{XXXham} arrange themselves into irreducible representations of $\sua(2)$.
Momentum can also be shown to be a conserved quantity. The momentum operator $P$ is defined as generating discrete shifts along the spin chain. Denoting by $U=e^{iP}$ the operator 
\begin{equation}
U X_n U^{-1} = X_{n+1}
\end{equation}
which shifts a local operator $X_n$ at site $n$ by one site we have
\begin{equation}
[U,H]=0\,.
\end{equation}

\paragraph{Higher conserved charges} 
Although it is not at all obvious from the definition, the Hamiltonian actually commutes with higher conserved charges. The first of these charges, to be denoted $\JJ_3$ is defined as 
\begin{equation}\label{J3charge}
\JJ_3 = \sum_{\alpha=1}^L [\lH_{\alpha,\alpha+1},\lH_{\alpha+1,\alpha+2}]
\end{equation}
and is a range $3$ operator, meaning it is a sum of densities which act on $3$ neighbouring spin chain sites, in contrast to the Hamiltonian which was a range $2$ operator. In fact, there are further independent conserved charges $\JJ_4,\JJ_5,\dots$ which can be constructed and it is this tower of higher charges which signals the integrability of the model. 

\medskip

Starting from the Hamiltonian it is impossible to guess that these higher charges exist. However, they can actually be constructed in a systematic fashion which involves embedding the Hamiltonian into a commutative subalgebra of some appropriate quantum algebra. It is this embedding which renders a given quantum Hamiltonian integrable. We will see in Part \ref{YBEboost} how this procedure can be turned bottom-up allowing the quantum algebra itself to be obtained from the Hamiltonian and a single higher charge. For now however we will proceed with the direct diagonalisation of the Hamiltonian.

\subsection{Coordinate Bethe ansatz}
We can now try to diagonalise the Hamiltonian \eqref{XXXham}. We start with an appropriate vacuum state $\ket{0}$ where each spin site has spin up along the $z$-axis 
\begin{equation}
\ket{0}=\bigotimes_{\alpha=1}^L \ket{\uparrow}_\alpha
\end{equation}
where $\ket{\uparrow}_\alpha=\left(\begin{array}{c}
1 \\
0
\end{array} \right)$ in the $\alpha$-th copy of $\CC^2$. It can be easily checked that this state is an eigenvector of the Hamiltonian with eigenvalue $0$. 

\medskip

We now look for excited states obtained by flipping some of the spin up states to spin down. A state  $\ket{M}$ with $M$ excitations, dubbed magnons, is constructed as a superposition of states with $M$-flipped spins
\begin{equation}
\ket{M} = \sum_{1\leq n_1 < \dots < n_{M}\leq L} a(n_1,\dots,n_M) S^-_{n_1} \dots S^-_{n_M} \ket{0}\,.
\end{equation}
The coordinate Bethe ansatz then involves making the following ansatz for the coefficients $a(n_1,\dots,n_M)$
\begin{equation}
a(n_1,\dots,n_M) = \sum_{\sigma \in \mathfrak{S}_M} A_\sigma(p_1,\dots,p_M) e^{\sfi\, p_{\sigma_i} n_i}
\end{equation}
where $p_i$ are complex numbers referred to as magnon momenta and the sum is over elements $\sigma$ of the permutation group $\mathfrak{S}_M$ on $M$ objects. The condition that this is an eigenstate of the Hamiltonian, together with the periodic boundary conditions, leads to a quantization condition for the magnon momenta 
\begin{equation}\label{Bethe1}
e^{\sfi p_jL}=\prod_{k\neq j}\lS(p_k,p_j),\quad j=1,\dots,M
\end{equation}
where $\lS(p_1,p_2)$ is the magnon S-matrix
\begin{equation}
\lS(p_1,p_2)=\displaystyle\frac{\cot\frac{p_1}{2}-\cot\frac{p_2}{2}-2\sfi}{\cot\frac{p_1}{2}-\cot\frac{p_2}{2}+2\sfi}
\end{equation}
through which the coefficients $A_\sigma$ can also be expressed. The equations \eqref{Bethe1} are the so-called Bethe Ansatz equations and all variants of the Bethe ansatz eventually lead to these equations. Once these equations are solved various physical quantities can be computed, for example the energy for an $M$-magnon state is given by
\begin{equation}
E=\sum_{k=1}^M E(p_k),\quad E(p)= 4\sin^2\frac{p}{2}\,.
\end{equation} 
Physically, the Coordinate Bethe Ansatz is very reasonable and nothing beyond textbook quantum mechanics is required to solve the model. On the other hand, it masks a very rich and elegant underlying algebraic structure. As well as this, it is not at all obvious how to tell from a given Hamiltonian if there exists higher conserved charges rendering the model integrable. A beautiful reformulation of the problem was constructed by the Leningrad school \cite{Faddeev:1979gh} which puts the notion of quantum integrability into the framework of quantum groups and representation theory. In this language the key object underlying the XXX spin chain is not the Hamiltonian but a certain associative algebra, called Yangian, and integrability, the existence of a large family of commuting operators, is governed by the existence of a maximal commutative subalgebra, dubbed Bethe (sub-)algebra.  Yangian algebras will play a key role in the remainder of this work and we will now start an in-depth analysis of them. 

\section{Quantum algebras}\label{quantumalgebras}

Historically, quantum algebras initially appeared in the work of the Leningrad school relating to the problem of quantizing functions on a Lie group, see \cite{takhtajan1990introduction} for a historical overview and introduction to the subject and \cite{Chari:1994pz} for a textbook treatment which we very closely follow. The phase space $M$ of a classical mechanical system naturally has the structure of a Poisson manifold. The space $\lF(M)$ of differentiable complex-valued functions on $M$ has a Lie bracket 
\begin{equation}
\{-,-\}: \lF(M) \times \lF(M) \rightarrow \lF(M)
\end{equation}
such that for any function $f\in \lF(M)$ its time evolutions is governed by 
\begin{equation}
\frac{{\rm d}}{{\rm d}t}f(\gamma(t)) = \{\lH_{\rm cl},f \}(\gamma(t))
\end{equation}
where $\gamma(t)\in M$ defines the trajectory and $\lH_{\rm cl}$ is the classical Hamiltonian. The problem of quantization roughly speaking involves replacing $\lF(M)$ with operators on some suitable Hilbert space which reduces to $\lF(M)$ in an appropriate classical limit $\hbar \rightarrow 0$. 

\medskip

Naturally, the algebra $\lF(M)$ is commutative. The idea of deformation quantisation is to replace the usual (commutative) product on $\lF(M)$ with a non-commutative one $*_\hbar$ with the resulting non-commutative algebra denoted $\lF_\hbar(M)$ with the property 
\begin{equation}
\lim_{\hbar\rightarrow 0} \frac{f_1*_\hbar f_2 - f_2 *_\hbar f_1}{\hbar} = \{f_1,f_2 \}\,.
\end{equation}
Under some additional technical assumptions the possible deformations are quite restrictive -- these restrictions correspond to so-called ``rigidity theorems" \cite{Chari:1994pz}. The resulting deformed algebra is known as a quantum algebra. Quantum algebras, as we will see, naturally fall into the realm of Hopf algebras, which we will now briefly review. 
 
\subsection{Hopf algebras}

\paragraph{Algebra}
A (unital, associative) algebra over a unital commutative ring $R$ is defined as a triple $(A,\mu,\iota)$ where $A$ is a (left) $R$-module and $\mu:A\otimes A\rightarrow A$ and $\iota:R\rightarrow A$ are linear maps such that the following diagrams commute:
\begin{equation}
\begin{tikzcd}
A\otimes A \otimes A \arrow{r}{\mu\otimes 1} \arrow[swap]{d}{1\otimes \mu} & A\otimes A \arrow{d}{\mu} \\%
A\otimes A \arrow{r}{\mu}& A
\end{tikzcd}
\end{equation}
\begin{equation}
\begin{tikzcd}
A\otimes R \ \arrow{r}{1\otimes \iota} \arrow[swap]{d}{\simeq} & A\otimes A \arrow{d}{\mu} \\%
A \arrow{r}{1}& A
\end{tikzcd}\quad \quad
\begin{tikzcd}
R \otimes A \arrow{r}{\iota\otimes 1} \arrow[swap]{d}{\simeq} & A\otimes A \arrow{d}{\mu} \\%
A \arrow{r}{1}& A
\end{tikzcd}
\end{equation}
Here $1$ is the identity map from $A$ to itself. $\mu$ is called the product and $\iota$ is called unit and the first diagram expresses the associativity of multiplication. In the above diagrams $\simeq$ denotes the natural isomorphism between $R\otimes A$ and $A$. For most of our purposes $R$ will simply be the field $\CC$ of complex numbers, but we will also consider the ring $\CC[[\hbar]]$ of formal power series in an indeterminate $\hbar$. 

\medskip

\paragraph{Coalgebra}
A coalgebra is defined by simply reversing all of the arrows in the above commutative diagrams in the usual manner of obtaining a co-object from an object in category theory. Namely, a coalgebra is a triple $(A,\Delta,\varepsilon)$ where $A$ is an $R$-module and $\Delta:A\rightarrow A\otimes A$ and $\varepsilon:A\rightarrow R$ are linear maps such that the following diagrams commute:
\begin{equation}
\begin{tikzcd}
A \arrow{r}{\Delta} \arrow[swap]{d}{\Delta} & A\otimes A \arrow{d}{1\otimes \Delta} \\%
A\otimes A \arrow{r}{\Delta\otimes 1}& A\otimes A\otimes A
\end{tikzcd}
\end{equation}
\begin{equation}
\begin{tikzcd}
A \arrow{r}{1} \arrow[swap]{d}{\Delta} & A \arrow{d}{\simeq} \\%
A\otimes A \arrow{r}{1\otimes \varepsilon}& A\otimes R
\end{tikzcd}\quad \quad
\begin{tikzcd}
A \arrow{r}{1} \arrow[swap]{d}{\Delta} & A \arrow{d}{\simeq} \\%
A\otimes A \arrow{r}{\varepsilon\otimes 1}& R\otimes A
\end{tikzcd}
\end{equation}
$\Delta$ is referred to as the coproduct and $\varepsilon$ as the counit. The first diagram expresses that the coproduct is coassociative. 

\paragraph{Bialgebra}
A bialgebra is obtained by imposing a coalgebra structure on an algebra (or vice versa) subject to certain compatability conditions. More precisely, a bialgebra is a tuple $(A,\mu,\iota,\Delta,\varepsilon)$ such that $(A,\mu,\iota)$ is an algebra, $(A,\Delta,\varepsilon)$ is a coalgebra and 
\begin{enumerate}
\item $\Delta$ and $\varepsilon$ are algebra homomorphisms
\item $\mu$ and $\iota$ are coalgebra homomorphisms. 
\end{enumerate}

\paragraph{Hopf algebra}
Finally, a Hopf algebra $A$ is a bialgebra equipped with a linear map $S:A \rightarrow A$, called the \textit{antipode}, such that 
\begin{equation}
\mu\circ(1\otimes S)\circ\Delta = \mu \circ(S\otimes 1) \circ \Delta = \iota \otimes \varepsilon\,.
\end{equation}
We end our discussion of Hopf algebras with two examples. 

\paragraph{Functions $\lF(G)$ on a group $G$}
Let $G$ be a group with identity element $1_G$ and consider the space $\lF(G)$ of $\CC$-valued functions on $G$. The vector space and algebra structures on $\lF(G)$ are defined by the usual pointwise addition and multiplication. For the counit $\varepsilon$ and antipode $S$ we define, for $f\in\lF(G)$ and $g\in G$,
\begin{equation}
\varepsilon(f) = f(1_G),\quad S(f)(g) = f(g^{-1})\,.
\end{equation}
For the coproduct we notice that as $\CC$-algebras $\lF(G)\otimes \lF(G) \simeq \lF(G\times G)$ and this isomorphism takes $f_1\otimes f_2$ to the function mapping $(g_1,g_2)\mapsto f_1(g_1)f_2(g_2)$ and so naturally set 
\begin{equation}
\Delta(f)(g_1,g_2)=f(g_1 g_2)\,.
\end{equation}

\paragraph{Universal enveloping algebra $U(\ga)$ of a Lie algebra $\ga$}

If $\ga$ is a $\CC$ Lie algebra then $U(g)$ is a $\CC$-algebra in the usual way. We introduce the Hopf algebra structure on $U(\ga)$ by setting
\begin{equation}
\Delta(x)= x\otimes 1 + 1 \otimes x,\ \forall x\in\ga\,.
\end{equation}
and 
\begin{equation}
S(x) = -x,\quad \varepsilon(x)=0,\ \forall x\in\ga\,.
\end{equation}

\paragraph{Opposite Hopf algebra}

Note that if $(A,\Delta,\varepsilon)$ is a coalgebra then we obtain another coalgebra $(A,\Delta^{\rm op},\varepsilon)$, usually denoted by the shorthand $A^{\rm op}$, by setting 
\begin{equation}
\Delta^{\rm op} = \sigma\circ \Delta
\end{equation}
where $\sigma:A\otimes A \rightarrow A\otimes A$ is the flip operator sending $x\otimes y\mapsto y\otimes x$ for all $x,y\in A$. Note that if $A$ is a Hopf algebra with antipode $S$ then $A^{\rm op}$ becomes a Hopf algebra with antipode $S^{-1}$. 

\paragraph{Cocommutative}
A coalgebra is called cocommutative if $\Delta^{\rm op}=\Delta$. 

\medskip

We see that in the examples in the previous subsection $U(\ga)$ is clearly cocommuative. In general however Hopf algebras are neither commutative nor cocommutative. On the other hand, from the point of view of quantum integrable systems special attention is paid to Hopf algebras which are ``almost" cocommutative, a notion which will now be made precise.

\paragraph{Almost cocommutative}
A Hopf algebra $A$ is called almost cocommutative if there exists an invertible element $\lR\in A\otimes A$ such that for all $x\in A$
\begin{equation}
\Delta^{\rm op}(x) = \lR \Delta(x) \lR^{-1}\,.
\end{equation}
If $A$ is almost cocommutative with such an $\lR$ we denote it with the pair $(A,\lR)$. Since $A^{\rm op}$ must itself be a Hopf algebra this places strong constraints on the form of $\lR$. Indeed, coassociativity of $\Delta^{\rm op}$ is not guaranteed for a generic $\lR$ but a sufficient condition is that
\begin{equation}
\lR_{12}(\Delta\otimes 1)(\lR) = \lR_{23}(1\otimes \Delta)(\lR)
\end{equation}
where $\lR_{12}=\lR\otimes 1$ and $\lR_{23}=1\otimes \lR$. It is convenient to make a stronger assumption -- that $A$ is quasi-triangular.

\paragraph{Quasi-triangular} An almost cocommutative Hopf algebra $(A,\lR)$ is called quasi-triangular if
\begin{equation}
(\Delta\otimes 1)(\lR)=\lR_{13}\lR_{23},\quad (1\otimes \Delta)(\lR)=\lR_{13}\lR_{12}\,.
\end{equation}
If $(A,\lR)$ is quasi-triangular then we call $\lR$ the universal $R$-matrix of $(A,\lR)$. It follows that
\begin{equation}
\lR_{12}\lR_{13}\lR_{23} = \lR_{12}(\Delta\otimes 1)(\lR)=\lR_{23}(1\otimes \Delta)(\lR)=\lR_{23}\lR_{13}\lR_{12}
\end{equation}
and hence $\lR$ satisfies the Yang-Baxter equation
\begin{equation}
\lR_{12}\lR_{13}\lR_{23}=\lR_{23}\lR_{13}\lR_{12}\,.
\end{equation}

\subsection{Quantised function algebras and quantised universal enveloping algebras}

We now return to the question of quantisation of the algebra of functions $\lF(M)$ of a Poisson manifold $M$. In the special case where $M=G$ is a Lie group the algebra of functions $\lF(G)$ naturally acquires the structure of a Hopf algebra with the group multiplication and inverse maps giving rise to the comultiplication and antipode, respectively, as was previously seen. On the other hand, for any Lie group one can associate a second Hopf algebra, namely the universal enveloping algebra $U(\ga)$ of the Lie algebra $\ga$ of $G$. 

\medskip

Quantum algebras, in the sense considered in this work, fall into two categories -- quantized function algebras and quantum universal enveloping algebras. As the name suggests, these appear as quantisations, or deformations, of the Hopf algebra structures on the algebra of functions $\lF(G)$ and the universal enveloping algebra $U(\ga)$, respectively. Again subject to some technical assumptions it can be shown that these two notions are equivalent to each other under an appropriate duality. This duality allows us to discuss the notion of quantizations using two equivalent perspectives -- on either a space of functions or universal enveloping algebras, whichever is most convenient. Indeed, quantization of an algebra of functions coincides with the intuitive notion of quantization of the functions on a phase space in classical mechanics whereas quantised universal enveloping algebras are often easier to deal with. 

\medskip

We will now discuss the notion of deformations of Hopf algebras. Roughly speaking this involves replacing a Hopf algebra $A$ over $\CC$ with a new Hopf algebra $A_{\hbar}$ over $\CC[[\hbar]]$. An important point in this construction is that any algebra over $\CC[[\hbar]]$ comes equipped with a natural topology called the $\hbar$-adic topology in which two elements are ``close" if they only differ by a large power of $\hbar$ and we will refer to a Hopf algebra equipped with this topology as a topological Hopf algebra. 

\paragraph{Deformations of Hopf algebras}

Let $(A,\mu,\iota,\Delta,\varepsilon,S)$ be a Hopf algebra over $\CC$. A deformation of $A$ is a topological Hopf algebra $(A_\hbar,\mu_\hbar,\iota_\hbar,\Delta_\hbar,\varepsilon_\hbar,S_\hbar)$ over $\CC[[\hbar]]$ such that
\begin{enumerate}
\item $A_\hbar \simeq A[[\hbar]]$ as $\CC[[\hbar]]$-modules
\item $\mu_\hbar = \mu\ {\rm mod}\,\hbar$ and $\Delta_\hbar = \Delta\ {\rm mod}\,\hbar$\,.
\end{enumerate}
The first condition formalises the intuitive notion that if we multiply all elements of $A$ by all possible formal power series in $\hbar$ and consider all possible linear combinations we obtain $A_\hbar$. The second condition is the statement that if we send $\hbar \rightarrow 0$ then we obtain the original product and coproduct. The deformed unit $\iota_\hbar$ and counit $\varepsilon_\hbar$ are obtained by simply extending $\CC[[\hbar]]$-linearly those of $A$ and similarly with the antipode. 

\medskip

Having defined deformations of Hopf algebras a technical comment is in order. The notation $A[[\hbar]]$ denotes the algebra of formal power series with coefficients in $A$ which one may intuitively think of as being $A\otimes \CC[[\hbar]]$ but actually the former is ``bigger" and is the completion of the latter in the $\hbar$-adic topology. We will not stress this point and prefer to sweep it under the rug as we will not make much use of it. 

\paragraph{Quantisations of Hopf algebras}

So far we have not yet said anything about a Poisson bracket structure which is of course an important ingredient in quantization. Given a Poisson algebra structure on a Lie group we can ask what is the corresponding structure on its Lie algebra $\ga$ and how this lifts to the universal enveloping algebra $U(\ga)$. The required structure is that of co-Poisson-Hopf algebra and we will discuss how quantisations are formulated in this language. Of course, we can also quantise in the usual sense of ``Poisson bracket becomes commutator", but this former language is more useful for Yangians which we will look at later.

\paragraph{Co-Poisson algebra} A co-Poisson algebra over a commutative ring $R$ is a cocommutative coalgebra $(A,\varepsilon,\Delta)$ with a linear map $\delta: \ga\rightarrow \ga\otimes \ga$ called the Poisson co-bracket which is skew-symmetric\footnote{Let $\sigma$ denote the permutation operator on $\ga\otimes \ga$. Skew-symmetry of $\delta$ means that for all $x\in\ga$ we have $\sigma\,\delta(x)=-\delta(x)$.} and satisfies 
\begin{equation}
{\rm cp} \circ (\delta\otimes 1) \circ \delta =0
\end{equation}
where ${\rm cp}$ denotes summing over cyclic permutations of the factors in the triple tensor product $A\otimes A\otimes A$ (the co-Jacobi-identity)
and 
\begin{equation}
(\Delta\otimes 1)\delta = (1\otimes \delta)\Delta + \sigma_{23} (\delta\otimes 1)\Delta
\end{equation}
where $\sigma_{23}$ permutes the second and third factors in the tensor product. 

\paragraph{Co-Poisson-Hopf algebra}
A co-Poisson-Hopf algebra is a co-Poisson algebra $(A,\varepsilon,\Delta,\delta)$ which is also a Hopf algebra and the two structures are compatible in the sense that
\begin{equation}
\delta(a_1 a_2) = \delta(a_1)\Delta(a_2) + \Delta(a_1) \delta(a_2),\quad \forall a_1,a_2\in A\,.
\end{equation}

\paragraph{Quantisation}
A quantisation of a co-Poisson-Hopf algebra $A$ over $\CC$ is a Hopf-algebra deformation $A_\hbar$ of $A$ such that
\begin{equation}\label{quanthopf}
\delta(x) = \frac{\Delta_\hbar(x)-\Delta_\hbar^{\rm op}(x)}{\hbar}\, {\rm mod}\,\hbar
\end{equation}
where $x\in A$ and $a\in A_\hbar$ is any element such that $x=a\, {\rm mod}\,\hbar$. 

\medskip

\noindent Now we move on to some examples. We will mostly just sketch the details in order to eventually motivate Yangians. 

\medskip

\paragraph{$\lF_{\hbar}(\SL(2))$}

The discussion of the relation between quantum algebras and quantum integrable systems is made clearest in terms of quantised function algebras. 

\medskip

Let us consider the algebra $M_2(\CC)$ of $2\times 2$ complex matrices and the algebra $\lF(M_2(\CC))$ of polynomial functions on $M_2(\CC)$. A general element $T\in M_2(\CC)$ has the form 
\begin{equation}\label{Tmatrix}
T = \left( 
\begin{array}{cc}
a & b \\
c & d
\end{array}
\right)
\end{equation}
and each of the entries $a,b,c,d$ can be viewed as maps $M_2(\CC)\rightarrow \CC$, that is they can be viewed as elements of $\lF(M_2(\CC)) = \CC[a,b,c,d]$. 

\medskip

Matrix multiplication on $M_2(\CC)$ naturally endows $\lF(M_2(\CC))$ with a bialgebra structure and we further set $\det\,T=1$ and obtain $\lF(\SL(2))$ as an appropriate quotient 
\begin{equation}
\lF(\SL(2)) = \lF(M_2(\CC)) / (\det\,T-1)\,.
\end{equation}

\medskip

We now introduce the deformed algebra $\lF_\hbar(\SL(2))$. It is defined by the relations 
\begin{equation}\label{deformedalgebra}
\begin{split}
a c = e^{-\hbar} & \, c a, \ bd = e^{-\hbar}\, db,\ ab = e^{-\hbar}\, ba,\ cd = e^{-\hbar}\, dc \\
& bc=cb,\ ad-da = (e^\hbar - e^{-\hbar})bc
\end{split}
\end{equation}
which define a deformation of $\lF(M_2(\CC))$ together with the requirement $ad-e^{-\hbar}bc=1$. The quantity $ad-e^{-\hbar}bc$ is called the \textit{quantum determinant}. 

\medskip

The resulting deformed algebra can be easily expressed in terms of an $R$-matrix $R\in {\rm End}(\CC^2\otimes \CC^2)$, which should not be confused with the universal $\lR$-matrix introduced above. Let us denote by 
\begin{equation}
T_a = \sum_{i,j=1}^2 e_{ij}\otimes 1 \otimes T,\quad T_b = \sum_{i,j=1}^2 1\otimes e_{ij} \otimes T
\end{equation}
where $T$ is as in \eqref{Tmatrix}. Then the defining relations of the deformed algebra \eqref{deformedalgebra} can be expressed simply as
\begin{equation}
R_{ab}\, T_a\,  T_b = T_b\, T_a\, R_{ab}
\end{equation}
where
\begin{equation}\label{numericRmatrix}
R=\left(
\begin{array}{cccc}
e^\hbar & 0 & 0 & 0 \\
0 & 1 & 0 & 0 \\
0 & e^\hbar - e^{-\hbar} & 1 & 0 \\
0 & 0 & 0 & e^{\hbar}
\end{array}
\right)\,.
\end{equation}
Naturally one can now reverse the logic and, for any invertible operator $R$ on $\CC^{\gn}\otimes \CC^{\gn}$ define an algebra with generators $T=(t_{ij})$, $i,j=1,\dots,\gn$ by the relations 
\begin{equation}
R_{ab}\, T_a\,  T_b = T_b\, T_a\, R_{ab}\,.
\end{equation} 
For any $R$ the resulting algebra has a bialgebra structure given by 
\begin{equation}
\Delta(t_{ij}) = \sum_{k=1}^{\gn} t_{ik}\otimes t_{jk},\quad \varepsilon(T) = 1\,.
\end{equation}
Now we consider the associativity of multiplication in the generated algebra. Since $R$ is invertible, swapping $T_a$ with $T_b$ equivalent to conjugating with $R_{ab}$. Now consider the triple product $T_a T_b T_c$ and note that we can reach $T_c T_b T_a$ in two different ways: 
\begin{equation}
\begin{array}{c}
T_a T_b T_c \rightarrow T_a T_c T_b \rightarrow T_c T_a T_b \rightarrow T_c T_b T_a \\
T_a T_b T_c \rightarrow T_b T_a T_c \rightarrow T_b T_c T_a \rightarrow T_c T_b T_a \\
\end{array}
\end{equation}
which gives 
\begin{equation}
R_{ab}^{-1}R_{ac}^{-1}R_{bc}^{-1} T_c T_b T_a R_{bc} R_{ac} R_{ab} = R_{bc}^{-1}R_{ac}^{-1}R_{ab}^{-1}T_c T_b T_a R_{ab} R_{ac} R_{ab}
\end{equation}
which is obviously satisfied if the (constant) quantum Yang-Baxter equation 
\begin{equation}
R_{ab} R_{ac} R_{bc} = R_{bc} R_{ac}R_{ab} 
\end{equation}
is satisfied. In fact, a huge benefit of this extra assumption from the point of view of integrable systems is that it immediately provides a representation of the quantum algebra where $T = R$. Such representations correspond to integrable spin chains.

\paragraph{$U_q(\sla(2))$} 

We recall that $\sla(2)$ is the Lie algebra generated by elements $\sfe_+,\sfe_-,\sfh$ subject to the commutation relations
\begin{equation}
[\sfe_+,\sfe_-]=\sfh,\quad [\sfh,\sfe_\pm]=\pm2\sfe_{\pm}
\end{equation}
and equip $U(\sla(2))$ with the Poisson co-bracket structure
\begin{equation}
\delta(\sfh)=0,\quad \delta(\sfe_+)=\sfe_+ \wedge \sfh,\quad \delta(\sfe_-)=\sfe_- \wedge \sfh\,.
\end{equation}
The quantised universal enveloping algebra $U_q(\sla(2))$ is the deformation of $U(\sla(2))$ with generators $\sfE_{\pm}, q^{\pm\frac{\sfh}{2}}$ subject to the relations 
\begin{equation}
q^{\frac{\sfh}{2}} \sfE_{\pm} = q^{\pm 1}\, \sfE_{\pm} q^{\frac{\sfh}{2}},\quad [\sfE_+,\sfE_-]= \frac{q^{\frac{\sfh}{2}} -q^{-\frac{\sfh}{2}}}{q-q^{-1}}\,.
\end{equation}
The defining relations of $\sla(2)$ are recovered in the $q\rightarrow 1$ limit if we assume that $q=e^\hbar$ and $\sfE_{\pm} = \sfe_{\pm}+\lO(\hbar)$.

\medskip

The coalgebra structure on $U_q(\sla(2))$ is given by 
\begin{equation}
\begin{split}
& \Delta_\hbar(q^{\pm\frac{\sfh}{2}})=q^{\pm\frac{\sfh}{2}}\otimes q^{\pm\frac{\sfh}{2}} \\
& \Delta_\hbar(\sfE_{+})= \sfE_{+}\otimes q^{\sfh} + 1\otimes \sfE_{+},\quad \Delta_\hbar(\sfE_{-})= \sfE_{-}\otimes  + q^{-\sfh}\otimes \sfE_{-} \\
& \varepsilon_\hbar(\sfE_{\pm})=0,\quad \varepsilon_\hbar(q^{\frac{\sfh}{2}})=1
\end{split}
\end{equation}
and the antipode $S_\hbar$ is given by 
\begin{equation}
S_\hbar(q^{\pm\frac{\sfh}{2}})=q^{\mp\frac{\sfh}{2}},\quad S(\sfE_{\pm})=- q ^{\pm 1} \sfE_{\pm}\,.
\end{equation}
A straightforward calculation easily shows that $U_q(\sla(2))$ is indeed a quantisation of $U(\sla(2))$ in the sense that \eqref{quanthopf} is satisfied. 

\medskip

$U_q(\sla(2))$ is a quasi-triangular Hopf algebra with universal $\lR$-matrix given by 
\begin{equation}
\lR=q^{\frac{1}{2}\sfh \otimes \sfh}\displaystyle \sum_{n\geq 0} \frac{(1-q^{-2})^n}{[n]_q !}q^{\frac{n-n^2}{2}} \left(q^{\frac{n}{2}\sfh} (\sfE_+)^n\right) \otimes \left(q^{-\frac{n}{2}\sfh} (\sfE_-)^n \right)
\end{equation}
where we have used the $q$-factorial $[n]_q!$ defined by
\begin{equation}
[n]_q!:=[n]_q\dots [2]_q [1]_q,\quad [n]_q:=\frac{q^n-q^{-n}}{q-q^{-1}}\,.
\end{equation}

\paragraph{Duality} 
The two examples $\lF_\hbar(\SL(2))$ and $U_q(\sla(2))$ are dual to each other under the previously mentioned duality between quantised function algebras and quantised universal enveloping algebras. We will not concern ourselves with the precise nature of this duality but will comment on the role the $R$-matrix plays in both cases. 

\medskip

In the case of $U_q(\sla(2))$ the universal $R$-matrix $\lR$ is a linear map $U_q(\sla(2))\otimes U_q(\sla(2))\rightarrow U_q(\sla(2))\otimes U_q(\sla(2))$. On the other hand, in the case of $\lF_\hbar(\SL(2))$ the $R$-matrix $R$ is a numeric $4\times 4$ matrix, or equivalently a linear operator on $\CC^2\otimes \CC^2$. In fact the numeric $R$-matrix is simply the image of the universal $\lR$ matrix in the standard representation of $U_q(\sla(2))$ on $\CC^2$. Let $v_\pm$ be a basis of $\CC^2$ with 
\begin{equation}
v_+ = \left(
\begin{array}{c}
1 \\
0
\end{array}
 \right),\quad 
 v_- = \left(
\begin{array}{c}
0 \\
1
\end{array}
 \right)
\end{equation} 
and $\{v_+\otimes v_+,v_-\otimes v_+,v_+\otimes v_-,v_-\otimes v_-\}$ a basis of $\CC^2\otimes \CC^2$ and consider the representation of $U_q(\sla(2))$ defined by
\begin{equation}
\mathsf{h} v_\pm = \pm v_\pm,\quad \sfE_\pm v_\pm=0,\quad \sfE_\pm v_\mp = v_\pm\,.
\end{equation}
In this representation and basis $\lR$ coincides precisely with $R$. 

\subsection{Integrable systems}

Having reviewed the key concepts relating to quantum algebras we are now ready to see how integrable systems fit into the picture. 

\paragraph{Integrable spin chains}
We extend our previous discussion to include quantised function algebras with a $T$-matrix depending on a spectral parameter $u\in \CC$ which naturally appears in the context of quantizing infinite-dimensional algebras such as the current algebra $\ga[u]$ or the loop algebra $\ga[u,u^{-1}]$. We are then lead to the quantum algebra relation 
\begin{equation}\label{generalRTT}
R_{ab}(u,v)\, T_a(u)\,  T_b(v) = T_b(v)\, T_a(v)\, R_{ab}(u,v)
\end{equation}
where $R(u,v)$ is an invertible numeric $n^2\times n^2$ matrix and satisfies the Yang-Baxter equation
\begin{equation}
\boxed{
R_{ab}(u,v) R_{ac}(u,w) R_{bc}(v,w) = R_{bc}(v,w) R_{ac}(u,w) R_{ab}(u,v)}\,.
\end{equation}
One of the main interests in quantum algebras generated in this way is that they have commutative subalgebras which physically can be considered integrals of motion. Indeed, let us denote by $\T(u)={\rm tr}_a T_a(u)$ where the trace is taken over the space $a$ which is $\CC^n$. Then it follows from \eqref{generalRTT} that 
\begin{equation}
\T(u)\T(v) = \T(v)\T(u)\,.
\end{equation}
This is the key relation for integrability and the object $\T(u)$ is called a \textit{transfer matrix}. Under the assumption that $\T(u)$ is analytic around some point (say $u=0$) then we can expand 
\begin{equation}
\T(u) = \sum_{n=0}^\infty I_n u^n
\end{equation}
which implies 
\begin{equation}
[I_n,I_m]=0,\quad n,m=0,1,2,\dots
\end{equation}
and hence $\T(u)$ generates a commutative family of operators $I_n$. One then hopes to construct a representation of the algebra generated by $T$ such that some physical operator of interest, such as a Hamiltonian, belongs to this family of operators. This is indeed the case of the XXX spin chain as we will see later. 

\paragraph{$S$-matrix in $1+1$-dim QFT}
Before discussing the relation to Hopf algebras let us quickly review the properties of the S-matrix in a QFT. 

\medskip

A scattering problem in QFT is naturally formulated using asymptotic “in” and “out” states $\ket{\dots}^{\rm in}$ and $\ket{\dots}^{\rm out}$. If a state with particle content with momenta $p_1, p_2,\dots$ and other quantum numbers $i_1, i_2,\dots$ is prepared at $t\rightarrow -\infty$ then it is an asymptotic 
“in” state $\ket{p_1,p_2}^{\rm in}_{i_1,i_2,\dots}$. Similarly, if a state is found to have particle content with momenta $p_1, p_2,\dots$ and other quantum numbers $i_1, i_2,\dots$  at $t\rightarrow \infty$ then it is the asymptotic “out” state $\ket{p_1,p_2}^{\rm out}_{i_1,i_2,\dots}$.
The S-matrix is the unitary operator which relates the basis of asymptotic “in” and “out” states
\begin{equation}
\lS\ket{\dots}^{\rm in}=\ket{\dots}^{\rm out}\,.
\end{equation}
Asymptotic “in” and “out” states can be constructed by means of creation operators $a^{{\rm in}\,\dagger}_i(p)$ and $a^{{\rm out}\,\dagger}_i(p)$. 
By definition, $a^{{\rm in}\,\dagger}_i(p)$ creates an asymptotic “in” state from the vacuum state corresponding to a particle with momentum $p$ and the index $i$ labels all other quantum numbers. That is 
\begin{equation}
\ket{p}_i^{\rm in}:= a^{{\rm in}\,\dagger}_i(p) \ket{0}\,.
\end{equation}
Similarly a multi-particle state can be constructed as 
\begin{equation}
\ket{p_1,\dots,p_n}_{i_1,\dots,i_n}^{\rm in}:= a^{{\rm in}\,\dagger}_{i_1}(p_1)\dots a^{{\rm in}\,\dagger}_{i_n}(p_n) \ket{0}
\end{equation}
and precisely the same construction goes through for “out” states
\begin{equation}
\ket{p_1,\dots,p_n}_{i_1,\dots,i_n}^{\rm out}:= a^{{\rm out}\,\dagger}_{i_1}(p_1)\dots a^{{\rm out}\,\dagger}_{i_n}(p_n) \ket{0}\,.
\end{equation}
The above discussion is of course valid in any quantum field theory. We will now specialise to an integrable QFT in $1+1$-dimensions. An integrable QFT is characterised by the existence of an infinite number of conserved quantities $\JJ_n$, $n=1,2,\dots$, dubbed higher conserved charges, which are diagonalised in one-particle states 
\begin{equation}
\JJ_n\ket{p}^{\rm in}_i = \omega_i^{(n)}(p)\ket{p}^{\rm in}_i
\end{equation}
and the functions $\omega_i^{(n)}(p)$ are independent. The existence of such conserved quantities places strong constraints on a scattering process \cite{Parke:1980ki}. They are as follows:

\paragraph{Absence of particle production and momentum conservation}
Consider a set of ``in" momenta $\{p_i \}^{\rm in}$ and ``out" momenta $\{q_i \}^{\rm out}$. The eigenvalues $\omega_i^{(n)}(p)$ roughly scale as $p^n$ and since we must have
\begin{equation}
p_1^n + p_2^n + \dots = q_1^n + q_2^n+\dots
\end{equation}
for all $n$ the only way this can be satisfied is if $\{p_i \}^{\rm in} = \{q_i \}^{\rm out}$. 

\paragraph{Factorised scattering}

The most crucial consequence of higher conserved charges is factorised scattering which means that a multi-particle scattering process factorises into a sequence of $2\rightarrow 2$ body scattering processes. This is a consequence of the existence of higher charges and the unique features of scattering in $1+1$-dimensions \cite{Parke:1980ki}.

\paragraph{Zamolodchikov-Faddeev algebra}

The implication of factorised scattering is that the $2\rightarrow 2$ S-matrix simply acts by swapping the two particles. We are then naturally led to the introduction of a new set of creation operators $A_i^\dagger(p)$ such that \cite{Zamolodchikov:1978xm} 
\begin{equation}
\ket{p_1,\dots,p_n}_{i_1,\dots,i_n}^{\rm in}:= A^\dagger_{i_1}(p_1)\dots A^\dagger_{i_n}(p_n) \ket{0},\quad p_1 > \dots > p_n
\end{equation}
\begin{equation}
\ket{p_1,\dots,p_n}_{i_1,\dots,i_n}^{\rm out}:= A^\dagger_{i_1}(p_n)\dots A^\dagger_{i_1}(p_1) \ket{0},\quad p_1 > \dots > p_n\,.
\end{equation}
and for simplicity have assumed the theory only contains bosons to avoid introducing extra sign factors due to fermions.

The $2\rightarrow 2$ S-matrix $\lS(p_1,p_2)$ which relates these in and out states then satisfies
\begin{equation}
\ket{p_1,p_2}^{(\rm in)}_{ij}= \lS(p_1,p_2)\ket{p_1,p_2}^{(\rm out)}_{ij}
\end{equation}
which in component form states
\begin{equation}
\ket{p_1,p_2}^{(\rm in)}_{ij}= \lS^{kl}_{ij}(p_1,p_2)\ket{p_1,p_2}^{(\rm out)}_{kl}
\end{equation}
and we sum over repeated indices. By using the definition of the “in” and “out” states using the creation operators we are then naturally led to the following intertwining relation
\begin{equation}
A^\dagger_i(p_1) A^\dagger_j(p_2) = \lS^{kl}_{ij}(p_1,p_2) A^\dagger_k(p_2)A^\dagger_l(p_1)
\end{equation}
which is just one of the commutation relations of the so-called Zamolodchikov-Faddeev algebra which is obtained by also introducing annihilation operators $A_k(p)$ conjugate to $A^\dagger_k(p)$ and intertwined by the S-matrix, see \cite{Arutyunov:2009ga}. 

\medskip

\begin{figure}[htb]
\centering
  \includegraphics[width=80mm,scale=25]{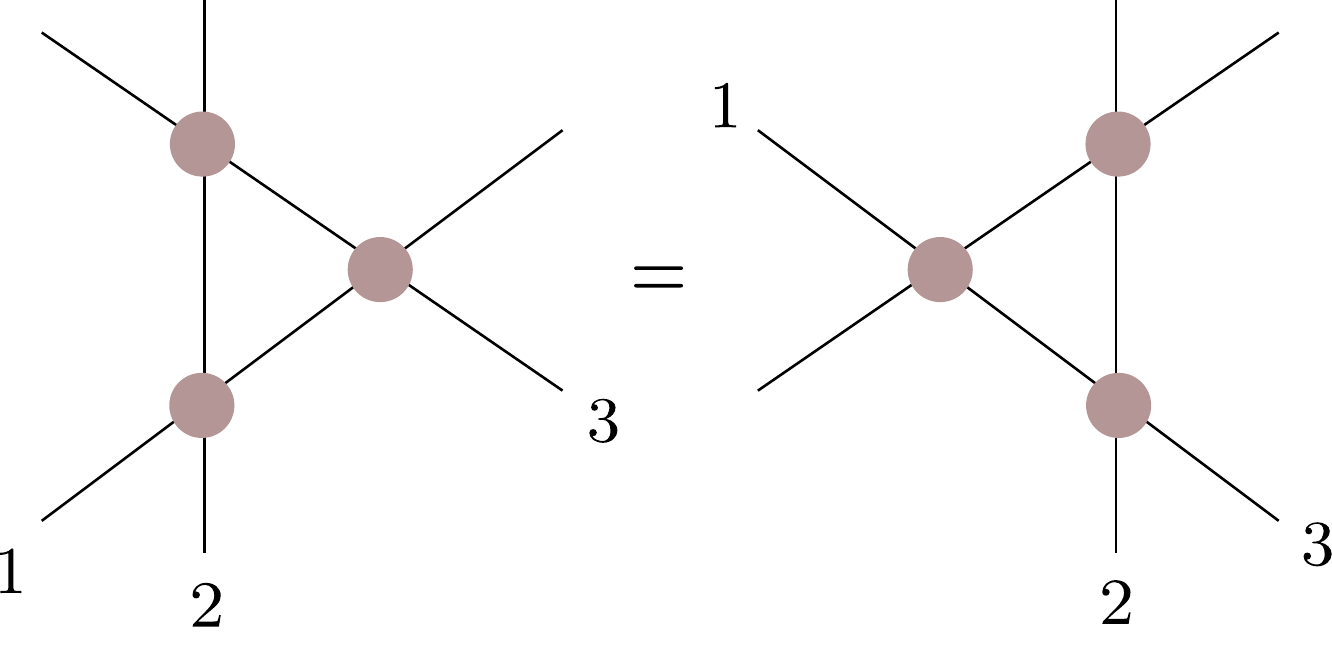}
  \caption{The two ways to factorise a three-particle scattering process into a sequence of two-particle scattering processes. Their equality leads to the Yang-Baxter equation.}
  \label{smatrixfactor2}
\end{figure}

Further relations arise from the consistency of $3\rightarrow 3$ body scattering which can be decomposed into $2\rightarrow 2$ scattering events in two different ways. Let us introduce the matrix $R(p_1,p_2) = \displaystyle\sum R^{i_1 i_2}_{j_1 j_2}(p_1,p_2) \sfe_{i_1 j_1}\otimes \sfe_{i_2 j_2}$ where the matrix elements are given by 
\begin{equation}
R^{i_1 i_2}_{j_1 j_2}(p_1,p_2) = \lS^{i_2 i_1}_{j_1 j_2}\,.
\end{equation}
Imposing the equality of the two different ways to decompose $3\rightarrow 3$ scattering events into $2\rightarrow 2$ scattering events implies \cite{Arutyunov:2009ga} the Yang-Baxter equation
\begin{equation}
R_{12}(p_1,p_2)R_{13}(p_1,p_3)R_{23}(p_2,p_3)=R_{23}(p_2,p_3)R_{13}(p_1,p_3)R_{12}(p_1,p_2),
\end{equation} see Figure \ref{smatrixfactor2}.

\section{Yangian}

We now turn our attention to the primary quantum algebra we will consider in this work -- the Yangian algebra. For any semi-simple complex Lie algebra $\ga$ Drinfeld constructed \cite{drinfeld1986quantum} an algebra $\lY(\ga)$, referred to as the Yangian of $\ga$, as a deformation of the current algebra $\ga[u]$ of $\ga$ and this deformation is unique under appropriate assumptions. 

\medskip

The current algebra $\ga[u]:=\ga \otimes \CC[u]$ is spanned by elements of the form 
\begin{equation}
a\, u^r,\quad  a\in\ga,\ r=0,1,2,\dots 
\end{equation}
where $u$ is some indeterminate. Clearly, $\ga[u]$ is a Lie algebra under the point-wise-defined Lie bracket induced from $\ga$ and can furthermore be identified with the set of polynomial maps $f:\CC \rightarrow \ga$. Hence, a deformation of $U(\ga[u])$ can be viewed as a deformation of a space of functions, in line with the formulation of quantum algebras as deformations of function algebras. The co-bracket $\delta:\ga[u]\rightarrow \ga[u]\otimes \ga[u]\otimes \ga[u]=(\ga\otimes \ga)(u,v)$ is given by 
\begin{equation}
\delta(f)(u,v)=({\rm ad}_{f(u)}\otimes 1 + 1 \otimes {\rm ad}_{f(v)})\left(\frac{t}{u-v} \right)
\end{equation}
where $t\in {\rm sym}^2(\ga\otimes \ga)$ is the Casimir on $\ga$ associated to a fixed bilinear form and equips $U(\ga[u])$ with a co-Poisson-Hopf structure. 

\medskip

It is useful to introduce a basis ${\mathsf J}_i$ of $\ga$ such that the commutation relations read 
\begin{equation}
[\mathsf{J}_i, \mathsf{J}_j]=f^{k}_{ij} \mathsf{J}_k
\end{equation}
where $f^{k}_{ij}$ are the structure constants and summation over $k=1,2,\dots,{\rm dim}\,\ga$ is implied.
Then the deformation $U_{\hbar}(\ga[u])$ is defined by introducing a further set of object $\hat{\mathsf{J}_i}$ with
\begin{equation}
[\mathsf{J}_i, \hat{\mathsf{J}_j}]=f^{k}_{ij} \hat{\mathsf{J}_k}
\end{equation}
which behave as $\hat{\mathsf{J}}_i = u\,\mathsf{J}_i+ \mathcal{O}\left(\hbar\right)$ and so in the limit $\hbar \rightarrow 0$ we have $\hat{\mathsf{J}_i} \rightarrow u\, \mathsf{J}_i$ and the coproduct $\Delta_\hbar$ of the deformed algebra satisfies 
\begin{equation}
\delta(x) = \frac{\Delta(x)-\Delta^{\rm op}(x)}{\hbar}\, {\rm mod}\,\hbar\,.
\end{equation}
Of course there are other technical assumptions but we will not concern ourselves with them. The deformed algebra $U_{\hbar}(\ga[u])$ is then denoted $\lY(\ga)$ and this presentation of $\lY(\ga)$ in terms of $\mathsf{J}_i$ and $\hat{\mathsf{J}_i}$ is called Drinfeld's first realisation. 

\medskip

The algebra we will concern ourselves with is actually not $\lY(\ga)$ for a semi-simple $\ga$ but rather $\lY_{\gn}:=\lY(\gl(\gn))$ which we will define below using a different realisation -- the RTT realisation \cite{kirillov1986yangians}. The algebra $\lY(\sla(\gn))$ can then be obtained from $\lY_{\gn}$ as an appropriate quotient. 

\subsection{Defining relations, symmetries and quantum determinant}
\paragraph{RTT formulation}
We will now begin our discussion of the Yangian algebra $\lY_\gn:=\lY(\gl(\gn))$, see \cite{molev2007yangians} for a comprehensive overview which we closely follow. In the RTT realisation \cite{kirillov1986yangians} it is generated by countably many generators $t_{ij}^{(r)},\ i,j=1,2,\dots,\gn,\ r=1,2,\dots$ subject to the relations
\begin{equation}\label{yangianrel}
[t_{ij}^{(r+1)},t_{kl}^{(s)}]-[t_{ij}^{(r)},t_{kl}^{(s+1)}]=\hbar\left(t_{kj}^{(r)}t_{il}^{(s)}-t_{kj}^{(s)}t_{il}^{(r)}\right)\,.
\end{equation}
As before $\hbar$ is an indeterminate. In fact, changing $\hbar\rightarrow \hbar^\prime$ produces an isomorphic algebra and it is common in mathematics literature to set $\hbar=1$ while in physics the choice $\hbar=\sfi=\sqrt{-1}$ is common. we will not concern ourselves with the value of $\hbar$ and leave it arbitrary for the majority of this work.

\medskip

It is convenient to repackage the generators $t_{ij}^{(r)}$ into formal power series $t_{ij}(u)\in\lY_{\gn}[[u^{-1}]]$ with 
\begin{equation}
t_{ij}(u)= \delta_{ij}1+t_{ij}^{(1)}u^{-1}+t_{ij}^{(2)}u^{-2}+\dots 
\end{equation}
which allows us to write the defining relations \eqref{yangianrel} as 
\begin{equation}\label{RTT1}
(u-v)[t_{ij}(u),t_{kl}(v)]=\hbar \left(t_{kj}(u)t_{il}(v)-t_{kj}(v)t_{il}(u)\right)\,.
\end{equation}
The parameters $u,v$ are called \textit{spectral parameters}. \eqref{RTT1} can be compactly written by introducing the rational $R$-matrix $R \in {\rm End}\left(\CC^\gn\otimes \CC^\gn \right)$ with
\begin{equation}
R(u,v) = 1-\frac{\hbar}{u-v} P
\end{equation}
where $P$ is the permutation operator on $\CC^{\gn}\otimes \CC^{\gn}$ acting as 
\begin{equation}
P(x\otimes y) = y\otimes x
\end{equation}
which is given in the standard basis by 
\begin{equation}
P = \sum_{i,j=1}^\gn \sfe_{ij}\otimes \sfe_{ji}
\end{equation}
where $\sfe_{ij}$ furnish the defining (vector) representation of $\gl(\gn)$ and satisfy 
\begin{equation}
\sfe_{ij}\sfe_{kl}=\delta_{jk}\sfe_{il}\,.
\end{equation}
Then \eqref{RTT1} is equivalent to
\begin{equation}\label{RTT2}
R_{ab}(u-v)t_a(u)t_b(v)=t_b(v)t_a(u)R_{ab}(u-v)
\end{equation}
where the monodromy matrix $t(u)$ has been introduced with 
\begin{equation}
t_a(u) = \sum_{i,j=1}^\gn \sfe_{ij}\otimes 1 \otimes t_{ij}(u),\quad t_b(u) = \sum_{i,j=1}^\gn 1 \otimes \sfe_{ij}\otimes t_{ij}(u)\,.
\end{equation}
The monodromy matrix $t(u)$ is an element of ${\rm End}(\CC^\gn)\otimes \lY_{\gn}[[u^{-1}]]$ with the copy of $\CC^{\gn}$ referred to as the \textit{auxiliary space}, in contrast to the \textit{physical space} where the operators $t_{ij}(u)$ act for some given representation. The notation $R_{ab}$ indicates that the $R$-matrix acts on the two auxiliary spaces. Note that it is trivial to check this $R$-matrix satisfies the Yang-Baxter equation 
\begin{equation}
R_{ab}(u,v)R_{ac}(u,w)R_{bc}(v,w) = R_{bc}(v,w) R_{ac}(u,w)R_{ab}(u,v)\,.
\end{equation}

\paragraph{Hopf algebra}

By definition $\lY_{\gn}$ is a Hopf algebra deformation of $U(\gla(\gn)[u])$. The coproduct is given by
\begin{equation}
\Delta\left(t_{ij}(u)\right)=\displaystyle\sum_{k=1}^\gn t_{ik}(u)\otimes t_{kj}(u)
\end{equation}
while the counit is simply $\varepsilon :t(u) \mapsto 1$ and the antipode is given by 
\begin{equation}
S: t(u) \mapsto t^{-1}(u)\,.
\end{equation}

\bigskip

The Yangian algebra $\lY_{\gn}$ admits a number of transformations which preserve the defining RTT relations and we will make use of several of them throughout the text. We start by considering some automorphisms. 

\paragraph{Rescaling} 

The transformation $t(u)\mapsto f(u) t(u)$ with 
\begin{equation}\label{rescaling}
f(u) = 1 + \mathcal{O}\left(\frac{1}{u} \right)\in \CC[[u^{-1}]]
\end{equation}
clearly preserves the RTT relation.

\paragraph{$\GL(\gn)$ symmetry and change of basis}
Yang's $R$-matrix satisfies an important property -- it is $\GL(\gn)$ invariant. Specifically, for $A\in\GL(\gn)$ we have 
\begin{equation}
[R,A\otimes A]=0\,.
\end{equation}
Hence, if $t(u)$ satisfies the RTT relation then so does $A\, t(u)\,A^{-1}$, where matrix multiplication is performed in the auxiliary space. 

\bigskip

\paragraph{Anti-automorphisms}

$\lY_\gn$ also admits a few anti-automorphisms which we will make use of later in the text. They are given by the following three maps 
\begin{enumerate}
\item $t(u) \mapsto t^{-1}(u) $
\item $t(u) \mapsto t(-u) $
\end{enumerate}
It is trivial to verify that these indeed constitute anti-automorphisms provided one notes that the inverse of the $R$-matrix $R(u,v)$ is simply given, up to an overall factor, by $R(v,u)$
\begin{equation}
R(u,v)R(v,u) = -\left((u-v)^2-\hbar^2\right)\,.
\end{equation}
This property is known as \textit{braiding unitarity}. 

\subsection{Finite-dimensional irreducible representations}
We now begin the study of representation theory of $\lY_\gn$. Since $\lY_\gn$ is a deformation of the universal enveloping algebra of $\gla(\gn)[u]$ it is not so surprising that the representation theory of $\lY_\gn$ and $\gl(\gn)$ is similar.

\paragraph{Highest-weight reps of $\gla(\gn)$}
Recall that $\gl(\gn)$ is the complex Lie algebra with generators $\sfE_{ij}$, $i,j=1,\dots,\gn$ subject to the relations 
\begin{equation}
[\sfE_{ij},\sfE_{kl}]=\delta_{jk}\sfE_{il}-\delta_{li}\sfE_{kj}\,.
\end{equation}
Consider the root space decomposition of $\gl(\gn)$ and write 
\begin{equation}
\gla(\gn) = \mathcal{E}^- \oplus \lH \oplus \mathcal{E}^+
\end{equation}
where the Cartan subalgebra $\lH$ is 
\begin{equation}
\lH=\{\sfE_{jj}\, |\, j=1,2,\dots,\gn \}
\end{equation}
and the raising and lowering operators $\mathcal{E}^\pm$ are given by 
\begin{equation}
\begin{split}
\mathcal{E}^+ & =\{\sfE_{jk}\, |\, 1\leq j < k \leq \gn \} \\
\mathcal{E}^- & =\{\sfE_{kj}\, |\, 1\leq j < k \leq \gn \}\,. 
\end{split}
\end{equation}

\medskip

A representation $\lV$ of $\gla(\gn)$ is called \textit{highest-weight} if there exists a vector $\ket{\Omega}\in\lV$ with the property that 
\begin{equation}\label{glnhw}
\begin{split}
\sfE_{jj}\ket{\Omega} & =\lambda_j \ket{\Omega} \\
\mathcal{E}^+ \ket{\Omega}& = 0\,.
\end{split}
\end{equation}
The vector $\ket{\Omega}$ is called the \textit{highest-weight state} and the numbers $\lambda=[\lambda_1,\lambda_2,\dots,\lambda_\gn]$ are called the highest-weights. The representation is finite-dimensional if and only if the differences 
\begin{equation}\label{integerdiff}
\lambda_j-\lambda_{j+1}\in\ZZ_{\geq 0},\quad j=1,2,\dots,\gn-1\,.
\end{equation}
It is a standard result in the theory of Lie algebras that all finite-dimensional irreducible representations of $\gl(\gn)$ are of highest-weight type \cite{fulton2013representation}. 

\paragraph{Highest-weight reps of Yangian}

In analogy with the case of $\gla(\gn)$ we say a representation $\lV$ of $\lY_\gn$ is highest-weight if there exists a vector $\ket{0}\in\lV$ with the properties 
\begin{equation}\label{hwcond1}
t_{jk}(u)\ket{0}=0,\quad j>k
\end{equation}
and
\begin{equation}\label{YangHW}
t_{jj}(u)\ket{0}=\lambda_j(u)\ket{0},\quad \lambda_j(u)\in\CC[[u^{-1}]],\quad j=1,2,\dots,\gn\,.
\end{equation}
Note the order of indices in \eqref{hwcond1} compared to \eqref{glnhw}.

\medskip

It is a well-known fact that all finite-dimensional irreducible representations of $\lY_\gn$ are of highest-weight type \cite{molev2007yangians}\cite{Chari:1994pz}. In fact one can even classify which irreducible representations of $\lY_\gn$ are finite-dimensional. The analogue of the condition \eqref{integerdiff} is replaced by the existence of so-called Drinfeld polynomials.

\paragraph{Drinfeld polynomials}
An irrep of $\lY_\gn$ is finite-dimensional if and only if \cite{drinfeld1987new}, see also \cite{Chari:1994pz,molev2007yangians}, there exists polynomials $P_j(u)\in \CC[u]$, $j=1,2,\dots,\gn-1$ satisfying 
\begin{equation}\label{drinfeldpoly}
\frac{\lambda_{j+1}(u)}{\lambda_{j}(u)}=\frac{P_j(u+\hbar)}{P_j(u)}\,.
\end{equation}
The polynomials $P_j(u)$ are referred to as Drinfeld polynomials and if they exist they are unique \cite{drinfeld1987new,Chari:1994pz,molev2007yangians}. Notice that the $P_j(u)$ can, without loss of generality, be taken to be monic polynomials where we remind the reader that a polynomial $p(u)$ of degree $n$ is said to be monic if $p(u) = u^n + \dots$. Hence, there is a one-to-one correspondence between finite-dimensional irreps of $\lY_\gn$ and monic polynomials. 

\paragraph{Evaluation representations}
Our next task is to actually construct representations. A particularly simple class of representations are the so-called evaluation representations $ev_\theta^\lambda$, $\theta\in\CC$, which produce $\lY_\gn$ representations from representations of $\gl(\gn)$. They are defined by 
\begin{equation}
ev_\theta^\lambda\left(t_{ij}(u)\right)=\delta_{ij}-\frac{\hbar}{u-\theta} \pi^\lambda\left(\sfE_{ji}\right)
\end{equation}
where $\pi^\lambda\left(\sfE_{ij}\right)$ are the images of the $\gl(\gn)$ generators $\sfE_{ij}$ in our chosen representation $\lambda$. A straightforward calculation allows us to easily demonstrate that this is indeed a representation of $\lY_\gn$. In fact the only requirement for this to produce a representation of $\lY_\gn$ is that $\pi^\lambda$ is a representation of $\gl(\gn)$. Hence evaluation representations can be used to construct infinite-dimensional representations and even non-highest-weight representations of $\lY_\gn$. 

\medskip

\begin{figure}[htb]
\centering
  \includegraphics[width=10mm,scale=10]{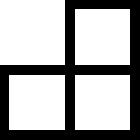}
  \caption{Skew diagram $\lambda/\mu$ obtained by removing the Young diagram $\mu=[1,0,0]$ from the Young diagram $\lambda=[2,2,0]$.}
  \label{simpleskew}
\end{figure}

We can now calculate the Drinfeld polynomials for the evaluation rep $ev_\theta^\lambda$. Clearly, the weight functions $\lambda_j(u)$ are simply given by 
\begin{equation}
\lambda_j(u)=1-\frac{\hbar}{u-\theta}\lambda_j
\end{equation}
and hence
\begin{equation}
\frac{\lambda_{j+1}(u)}{\lambda_j(u)}=1+\frac{\hbar}{u}(\lambda_j-\lambda_{j+1})+\lO\left(u^{-2}\right)\,.
\end{equation}
On the other hand 
\begin{equation}
\frac{P_j(u+\hbar)}{P_j(u)}=1+\frac{\hbar}{u}{\rm deg}P_j +\lO\left(u^{-2}\right)
\end{equation}
and so we see that a polynomials $P_j(u)$, $j=1,\dots,\gn-1$ satisfying \eqref{drinfeldpoly} can exist only if $\lambda_j-\lambda_{j+1}$, $j=1,\dots,\gn-1$, are integers which is precisely the requirement that the $\gl(\gn)$ rep be finite-dimensional. A direct calculation shows that the Drinfeld polynomials are given by 
\begin{equation}\label{Drinfeldpol}
P_j(u) = \prod_{k=0}^{\lambda_j-\lambda_{j+1}-1}(u-\theta-\hbar(\lambda_j-k))\,.
\end{equation}
We need to stress however that not every irrep of $\lY_{\gn}$ is an evaluation representation. We will give a simple example. By using the rescaling symmetry \eqref{rescaling} we can set $\lambda_3(u)=1$ without loss of generality. Then we put 
\begin{equation}
\lambda_1(u) = \frac{u(u-2\hbar)}{(u+\hbar)(u-\hbar)},\quad \lambda_2(u)=\frac{u-2\hbar}{u-\hbar}\,.
\end{equation}
The Drinfeld polynomials for this representation are easily worked out to be 
\begin{equation}
P_1(u) = u,\quad P_2(u) = u-2\hbar
\end{equation}
and it is trivial to check that these do not coincide with \eqref{Drinfeldpol} for any choice of $\theta$, $\lambda_{1,2,3}$. We will not say too much about these types of representations apart from this: finite-dimensional evaluation representations corresponded to finite-dim $\gl(\gn)$ irreps which are labelled by Young diagrams $\lambda$. The non-evaluation representation we have constructed corresponds to a \textit{skew} Young diagram $\lambda/\mu$ obtained by removing a Young diagram $\mu$ from another Young diagram $\lambda$ with their top left corners aligned. Our representation corresponds to the skew diagram in Figure \ref{simpleskew}. We will return to skew Young diagrams in Section \ref{sec:skewdiagrams}.

\paragraph{Spin chain representation}

We now turn to the representations we will focus on -- tensor products of evaluation representations, also known as \textit{spin chain representations} since each tensor factor can be interpreted as the Hilbert space of a particle transforming in that particular $\gl(\gn)$ representation. It is convenient to introduce polynomial Lax operators $\lL^\lambda(u,\theta) = \sum_{i,j}\sfE_{ij}\otimes \lL_{ij}(u,\theta)\in {\rm End}(\CC^{\gn}\otimes \lV^{\lambda})$ with
\begin{equation}
\lL_{ij}^\lambda(u-\theta)=(u-\theta)\,ev_\theta^\lambda(t_{ij}(u))
\end{equation}
and hence 
\begin{equation}\label{generalisedperm}
\lL^\lambda(u-\theta)=(u-\theta)-\hbar\, \mathcal{P}^\lambda,\quad \mathcal{P}^\lambda = \sum_{i,j=1}^\gn \sfe_{ij}\otimes \pi^\lambda(\sfE_{ji})
\end{equation}
where $\mathcal{P}^\lambda$ is referred to as a generalised permutation operator since when $\lambda$ is the defining representation it reduces to the usual permutation operator. 

\medskip

Let us fix a family of Young diagrams $\nu^1,\nu^2,\dots,\nu^L$ and label the corresponding representations spaces $\lV^{\nu^\alpha}$, $\alpha=1,\dots,L$. The number $L$ is the length of the spin chain. By using the Yangian coproduct we can construct a representation of $\lY_{\gn}$ on the $L$-fold tensor product 
\begin{equation}
\lV^{\nu^1} \otimes \dots \otimes \lV^{\nu^L}
\end{equation}
by setting 
\begin{equation}
T(u) = \lL^{\nu^1}(u-\theta_1)\dots \lL^{\nu^L}(u-\theta_L)
\end{equation}
where each $\lL^{\nu^\alpha}$ acts on the same auxiliary space. Note that $T(u)$ satisfies the RTT relation \eqref{RTT2} but unlike $t(u)$ it is a polynomial of degree $L$. The highest-weight representation structure \eqref{YangHW} is left unchanged except now $T_{jj}(u)$ act on the highest-weight state as polynomials $\nu_j(u)$ given by 
\begin{equation}\label{spinchainrep}
\nu_j(u) = \prod_{\alpha=1}^L (u-\theta_\alpha-\hbar\,\nu^\alpha_j)\,.
\end{equation}

\medskip

Let us now consider the expansion of the operators $T_{ij}(u)$ at large $u$. By construction 
\begin{equation}
T_{ij}(u) = u^L \delta_{ij}-u^{L-1} \left(\hbar\,\lE_{ji}+\delta_{ij}\displaystyle \sum_{\alpha=1}^L \theta_\alpha \right) + \mathcal{O}(u^{L-2})
\end{equation}
where $\lE_{ji}$ are the generators of the global $\gl(\gn)$ algebra 
\begin{equation}
\lE_{ij} = \sum_{\alpha=1}^L \pi^{\nu^\alpha}\left(\sfE_{ij} \right)\,.
\end{equation}
By expanding the RTT relation \eqref{RTT2} in powers of $u$ we find the following commutation relation 
\begin{equation}\label{CartanRTT}
[\lE_{ij},T_{kl}(v)]=\delta_{jl}T_{ki}(v) - \delta_{ki}T_{jl}(v)
\end{equation}
which we will make use of later. 

\paragraph{Infinite-dimensional and non-highest-weight representations} 
So far we have focused much of our attention on finite-dimensional irreducible representations of $\lY_\gn$ which, as we established, are of highest-weight type. However, later in this work we will also consider highest-weight representations which are not finite-dimensional. This case corresponds to case where we choose highest-weights $\lambda_j(u)\in\CC[[u^{-1}]]$ for which some or all of the corresponding Drinfeld polynomials do not exist. When we do consider these representations we shall construct them as evaluation representations. 

\medskip

Finally, we note that, while we will not consider them in this work, non-highest-weight representations of Yangian are also of importance and show up in a number of different contexts in physics. We briefly outline a few of these:

\paragraph{Scattering amplitudes in QCD}

It was noticed by Lipatov \cite{Lipatov:1993yb} that certain hadron-hadron scattering amplitudes in high-energy QCD could be described by an integrable system. More precisely, a wave function of $L$ gluons was shown to be an eigenfunctions of certain nearest-neighbour Hamiltonians on a one-dimensional lattice. In \cite{Faddeev:1994zg} it was shown that this integrable system was described by the Yangian $\lY_2$ but in an evaluation representation corresponding to the principal series representations of $\SL(2)$. 

\paragraph{Yangian symmetry in $\lN=4$ SYM}

Scattering amplitudes in planar $\lN=4$ SYM possess Yangian symmetry \cite{Drummond:2009fd} which means for all amplitudes $\mathcal{M}$ we have
\begin{equation}
\JJ\, \mathcal{M}=0
\end{equation}
for all $\JJ\in \lY(\psu(2,2|4))$. The representation is constructed from the infinite dimensional representations of the superconformal algebra $\psu(2,2|4)$ as differential operators with the amplitudes corresponding to certain one-dimensional invariant subspaces called \textit{Yangian invariants} \cite{Frassek:2013xza}. Yangian symmetry is also not just a feature of scattering amplitudes but also of the spectral problem \cite{Dolan:2003uh,Beisert:2010jq} and the one-loop Hamiltonian commutes with the Yangian generators up to boundary terms, and these vanish in the limit $L\rightarrow \infty$. 

\paragraph{Conformal fishnet theory}

$\lN=4$ SYM has a cousin -- $4d$ conformal fishnet theory -- obtained as a certain double scaling limit of $\gamma$-deformed $\lN=4$ SYM \cite{Gurdogan:2015csr}. It maintains the integrability of the former but also manifests it in new ways. The Feynman diagrams contributing to certain two-point functions exhibit a simple iterative structure with each loop order corresponding to action with a certain \textit{graph-building operator} which corresponds to the Hamiltonian of an infinite-dimensional spin chain \cite{Gromov:2017cja,Grabner:2017pgm}. The Yangian symmetry of $\lN=4$ SYM also remains in the fishnet theory \cite{Chicherin:2017cns,Chicherin:2017frs}.

\paragraph{${\rm AdS}_5$ fish chain}

The ${\rm AdS}_5$ fishchain \cite{Gromov:2019bsj,Gromov:2019jfh} corresponds to an $L$-fold tensor product of evaluation representations of the Yangian of $\sla(4) \simeq \mathfrak{so}(1,5)$ with each site carrying a representation defined on functions on ${\rm AdS}_5$ subject to certain other physical constraints. The model is holographically dual to $4d$ conformal fishnet theory \cite{Gromov:2019aku,Gromov:2019bsj,Gromov:2019jfh}, with the Hamiltonian wave functions of the fishchain corresponding to an $L+1$-point correlation function of the fishnet theory. 

\subsection{Algebraic Bethe ansatz}
Having reviewed the basic features of the Yangian $\lY_{\gn}$ and its representations we will now review how to extract the XXX Hamiltonian \eqref{XXXham} and diagonalise it. The procedure for doing this is called the Algebraic Bethe ansatz \cite{Faddeev:1996iy}.

\medskip

We start by considering the Yangian monodromy matrix $T(u)$
\begin{equation}
T(u) = \left( 
\begin{array}{cc}
T_{11}(u) & T_{12}(u) \\ 
T_{21}(u) & T_{22}(u) \\ 
\end{array}
\right)\,.
\end{equation}
For notational simplicity it is common to relabel the algebra generators $T_{ij}(u)$ as operators $A,B,C,D$ as follows
\begin{equation}
T(u) = \left( 
\begin{array}{cc}
A(u) & B(u) \\ 
C(u) & D(u) \\ 
\end{array}
\right)\,.
\end{equation}
As was already mentioned in the previous section the aim when solving a quantum integrable system is to diagonalise the family of commuting operators obtained from the transfer matrix $\T(u)={\rm tr}\, T(u)$. We will start our considerations by examining the Yangian representation constructed of $L$ copies of the evaluation representation with each site carrying the defining representation of $\gl(2)$ before eventually moving on to the general case. In order to make manifest the fact that we are trying to interpret the representation space as a chain of spin-$\frac{1}{2}$ particles we will use $\sua(2)$ generators instead of $\gl(2)$ generators, and so introduce the spin operator $S^z$ along the $z$-axis by 
\begin{equation}
\sfE_{11}=\frac{1}{2}+S^z,\quad \sfE_{22}=\frac{1}{2}-S^z,\quad S^z=\left( 
\begin{array}{cc}
\frac{1}{2} & 0 \\
0 & -\frac{1}{2}
\end{array}
\right)\,.
\end{equation}
In terms of these operators the local Lax operator $\lL_\alpha$ is given by 
\begin{equation}
\lL_\alpha(u) = \left( 
\begin{array}{cc}
u-\theta_\alpha -\frac{\hbar}{2}-\hbar S^z_\alpha & -\hbar S^-_\alpha \\
-\hbar S^+_\alpha & u-\theta_\alpha -\frac{\hbar}{2}+\hbar S^z_\alpha
\end{array}
\right)\,.
\end{equation}

\paragraph{Extracting the Hamiltonian}

The transfer matrix $\T(u) = {\rm tr}\,T_a(u)$ is the key object which allows us to embed the Hamiltonian \eqref{XXXham} into the quantum algebra construction. Indeed, as was already mentioned the transfer matrix generates a commuting family of operators owing to the commutativity condition 
\begin{equation}
[\T(u),\T(v)]=0\,.
\end{equation}
Under an appropriate identification of the parameters $\theta_\alpha$ and $\hbar$ the XXX Hamiltonian belongs to this commuting family of operators. 

\medskip

The first step is to take the homogeneous limit $\theta_\alpha\rightarrow -\frac{\hbar}{2}$. This guarantees that the conserved charges generated by $\T(u)$ are local, which is certainly true of the XXX Hamiltonian. The next is to notice that at the point $u=\frac{\hbar}{2}$ each Lax operator becomes the permutation operator 
\begin{equation}
\lL_\alpha\left(\frac{\hbar}{2}\right)= P_{a\alpha}
\end{equation}
which permutes vectors on the auxiliary space and the $\alpha$-th spin chain site. At this point the transfer matrix can be computed explicitly leading to 
\begin{equation}
U:=\T\left(\frac{\hbar}{2}\right) = {\rm tr}_a \left(\prod_{\alpha=1}^L P_{a\alpha}\right)= P_{L-1,L}\dots P_{23}P_{12}
\end{equation}
which is a shift operator along spin sites -- if $X_\alpha$ is an operator acting non-trivially on site $\alpha$ then 
\begin{equation}
U X_\alpha U^{-1} = X_{\alpha-1}
\end{equation}
subject to periodic boundary conditions. Hence, all of the conserved charges are translationally invariant. 

\medskip

Finally, we compute the first logarithmic derivative of the transfer matrix and evaluate at $u=\frac{\hbar}{2}$ and have
\begin{equation}
\left.\frac{{\rm d}}{{\rm d}u}\log\,\T(u)\right|_{u=\frac{\hbar}{2}}:=\T\left(\frac{\hbar}{2}\right)^{-1}\T'\left(\frac{\hbar}{2}\right)=U^{-1}\T'\left(\frac{\hbar}{2}\right)\,.
\end{equation}
A straightforward calculation then yields that 
\begin{equation}
\left.\frac{{\rm d}}{{\rm d}u}\log\,\T(u)\right|_{u=\frac{\hbar}{2}}=H
\end{equation}
where $H$ is the Hamiltonian \eqref{XXXham} up to an overall rescaling and shift of the energy levels. Furthermore, it can be demonstrated that the second logarithmic derivative of $\T(u)$ yields the higher conserved charge $\JJ_3$ mentioned in \eqref{J3charge}. Finally, it can be demonstrated that all of the conserved charges are Hermitian, guaranteeing their mutual diagonalisability. 

\paragraph{Diagonalising the conserved charges} 

We now proceed to the diagonalisation of the transfer matrix for an arbitrary finite-dim irrep with weight functions $\nu_1(u)$ and $\nu_2(u)$ as in \eqref{spinchainrep}.  Since the transfer matrix commutes with itself at different values of the spectral parameter $u$ it follows that its eigenvectors do not depend on $u$ and hence its eigenvectors are eigenvectors for the full family of conserved charges. 

\medskip

We will denote the highest-weight state by $\ket{\Omega}$. On the highest-weight state we have
\begin{equation}
\begin{split}
A(u)\ket{\uparrow^L} & = \nu_1(u)\ket{\Omega} \\
D(u)\ket{\uparrow^L} & = \nu_2(u)\ket{\Omega} \\
C(u)\ket{\uparrow^L} & = 0 \\
\end{split}
\end{equation}
and both $A$ and $D$ are polynomial and hence so is the transfer matrix $\T(u)$. Starting from the highest-weight state $\ket{\Omega}$ we wish to create new eigenvectors. This can be achieved with the help of the operator $B(u)$ which behaves at large-$u$ as 
\begin{equation}\label{Binfinity}
B(u) = -\hbar\, u^{L-1} \lS^-+\lO\left(u^{L-2}\right)\,.
\end{equation}
Hence, any state of the form 
\begin{equation}\label{Bstate}
\prod_{j=1}^M B(u_j)\ket{\uparrow^L}
\end{equation}
is a linear combination of states with $M$-flipped spins. However, not all values of the spectral parameters $u_j$ will produce an eigenvector of the transfer matrix. The Yangian commutation relations impose strong constraints on what values they take. By repeatedly using the relations between $A$, $B$ and $D$ stemming from the RTT relation \eqref{RTT2} one finds that in order for the state \eqref{Bstate} to be an eigenvector of the transfer matrix the following set of algebraic equations, known as Bethe equations, must be satisfied: 
\begin{equation}
\frac{\nu_1(u_j)}{\nu_2(u_j)}=-\frac{q^{[-2]}(u_j)}{q^{[2]}(u_j)},\quad j=1,2,\dots,M\,.
\end{equation}
Here we have introduced the Baxter polynomial or Baxter Q-function $q(u)=\prod_{j=1}^M(u-u_j)$ together with the following notation for shifts of the spectral parameter 
\begin{equation}
f^{[2n]}(u)=f\left(u+n\,\hbar \right),\quad n\in\ZZ
\end{equation}
for some function $f(u)$. These Bethe equations are precisely those appearing in \eqref{Bethe1} upon choosing the spin $\frac{1}{2}$ evaluation representation with $\theta_\alpha=\frac{\sfi}{2}$, $\hbar=-\sfi$ and 
\begin{equation}
u_j = \frac{1}{2}\cot \frac{p_j}{2}\,.
\end{equation}

\paragraph{Eigenvalues}
When these equations are satisfied the corresponding eigenvalue $\sfT$ of the transfer matrix on the state \eqref{Bstate} can be worked out to be 
\begin{equation}\label{Teigenvalue}
\sfT(u)=\nu_1(u)\frac{q^{[2]}(u)}{q(u)}+\nu_2(u)\frac{q^{[-2]}(u)}{q(u)}
\end{equation}
which can be recast as Baxter's famous TQ equation \cite{baxter2016exactly} which defines a finite-difference equation for the function $q(u)$ 
\begin{equation}\label{TQ1}
\sfT(u)q(u) = \nu_1(u)q^{[2]}(u)+\nu_2(u)q^{[-2]}(u)\,.
\end{equation}
At first glance it may seem like the transfer matrix eigenvalue $\sfT(u)$ has a pole at $u\rightarrow u_j$ which is certainly not consistent with the fact that the transfer matrix and hence its eigenvalues is a polynomial function of $u$. Thankfully, the coefficient of this pole is zero thanks to the Bethe equations. In fact, one can reverse the logic and start from \eqref{Teigenvalue} and impose that it is pole-free at $u\rightarrow u_j$. This then leads immediately to the Bethe equations. Deriving the Bethe equations in this way is known as the \textit{Analytical Bethe ansatz} \cite{Kuniba:1994na}.

\paragraph{Symmetry multiplets}
The transfer matrix commutes with the global $\sua(2)$ symmetry generators and hence every eigenstate constructed as in \eqref{Bstate} is also an eigenvector for the global $\sua(2)$. The full spin chain representation space is clearly reducible as a representation of $\su(2)$ and decomposes into a direct sum of irreps. It can be shown that for each Bethe state $\ket{\Psi}$ we have that 
\begin{equation}
\lS^+ \ket{\Psi}=0
\end{equation}
and hence the Bethe states are highest-weight states of the mentioned irreducible representations. Clearly the highest-weight states do not form a basis of eigenstates by themselves and so we must also construct descendants. These are obtained by acting on the Bethe states with the global lowering operator $\lS^-$ or, equivalently, including Bethe roots at infinity, owing to the relation \eqref{Binfinity}. Since the transfer matrix $\T(u)$ commutes with the global $\sua(2)$ generators its eigenvalue is the same on state in a given irreducible $\mathfrak{su}(2)$ representation.

\medskip

Considering the simple example of $L=2$ with the spin $\frac{1}{2}$ evaluation rep, the representation space $\CC^2\otimes \CC^2$ decomposes as 
\begin{equation}
\CC^2\otimes \CC^2={\rm sym}^2\left(\CC^2\right)\oplus \wedge^2\left(\CC^2\right)\,.
\end{equation}
The symmetric space ${\rm sym}^2\left(\CC^2\right)$ is three-dimensional and is spanned by 
\begin{equation}
\ket{\uparrow^L},\, \lS^-\ket{\uparrow^L}, \, \left(\lS^-\right)^2\ket{\uparrow^L}
\end{equation}
while the antisymmetric space $\wedge^2\left(\CC^2\right)$ is one-dimensional and is spanned by 
\begin{equation}
B(u_1)\ket{\uparrow^L}
\end{equation}
where 
\begin{equation}
u_1=\frac{1}{2}\left(\theta_1+\theta_2+\hbar \right)
\end{equation}
satisfies the Bethe equation
\begin{equation}
\prod_{\alpha=1}^2\left(\frac{u_1-\theta_\alpha-\hbar}{u_1-\theta_\alpha} \right) = 1\,.
\end{equation}

\paragraph{Problems with Bethe equations and completeness}

While the Bethe equations lead to a simple characterisation of the transfer matrix (and hence Hamiltonian) spectrum they are not without their faults. Indeed, one quite easily construct various non-physical solutions for which the transfer matrix eigenvalue is not polynomial. At the level of Bethe equations it is not always clear which solutions are physical and the non-physical ones must be removed by hand. On the other hand it is also not clear that every transfer matrix eigenstate can be constructed using the algebraic Bethe ansatz. This is the problem of \textit{completeness} and an extensive amount of effort has been put towards resolving this issue, see for example \cite{kirrilov1987completeness,kerov1988combinatorics,kirillov1988bethe}. 

\medskip

For spin chains in the defining evaluation representation of $\gl(\gn)$ and more recently $\gl(\gm|\gn)$ this has been positively resolved \cite{mukhin2009bethe,2013arXiv1303.1578M,Chernyak:2020lgw}. The resolution is based on the fact that the transfer matrix eigenvalue equation can be recast as 
\begin{equation}\label{BaxteropTQ}
\left(\nu_1(u) \lD - \sfT(u) + \nu_2(u) \lD^{-1} \right)q(u) = 0
\end{equation}
where we have introduced the shift operator $\lD$ which has the following action on functions $f(u)$ 
\begin{equation}\label{shiftop}
\lD^{\pm 1} f(u) = f(u+\hbar)\,.
\end{equation}
\eqref{BaxteropTQ} defines a finite-difference equation of order $2$ and hence has two linearly independent solutions which we denote as $\sfq_1$ and $\sfq_2$. For the defining representation in the homogeneous limit $\theta_\alpha \rightarrow -\frac{\hbar}{2}$ the two solutions satisfy the Wronskian relation 
\begin{equation}
u^L = \sfq_1^{[1]} \sfq_2^{[-1]} - \sfq_2^{[1]} \sfq_1^{[-1]}\,.
\end{equation}
If $\sfq_1$ corresponds to the Baxter polynomial constructed by the algebraic Bethe ansatz then all non-physical solutions correspond to solutions of the Wronskian relation for which $\sfq_2$ is not a polynomial. By imposing that both $\sfq_1$ and $\sfq_2$ be polynomial one obtains only physical solutions and furthermore all transfer matrix eigenstates can by characterised in this way. 

\medskip

\paragraph{Higher-rank generalisation}

The generalisation of the algebraic Bethe ansatz to higher-rank $\gl(\gn)$ cases is known as the \textit{nested Bethe ansatz}, see \cite{Belliard:2008di,Slavnov:2019hdn} for in-depth reviews. We will only sketch some brief details. 

\medskip

Like in the $\sua(2)$ case the transfer matrix $\T(u)={\rm tr}\,T(u)$ commutes with the global $\gl(\gn)$ symmetry algebra and hence the eigenspaces of $\T(u)$ correspond to irreps of $\gl(\gn)$. The eigenvalue $\sfT(u)$ of the transfer matrix on the highest-weight state is given by 
\begin{equation}
\sfT(u) = \sum_{j=1}^\gn \nu_j(u)\,.
\end{equation}
The nested Bethe ansatz procedure for constructing eigenvectors of $\T(u)$ is based on first diagonalising a family of auxiliary transfer matrices $T^{(k)}(u)$, $k=1,\dots,\gn-1$ where $T^{(k)}$ denotes the trace of the principal $k\times k$ submatrix of the monodromy matrix $T(u)$. The procedure is quite involved and the complexity increases drastically with rank so we will not spell out any further details here. The main point is that an eigenvalue $\sfT(u)$ of $\T(u)$ is parameterised by not just one polynomial $q(u)$ like in the $\sua(2)$ case but by $\gn-1$ polynomials $\sfq_1(u),\sfq_{12}(u),\dots,\sfq_{1\dots\gn-1}(u)$ with
\begin{equation}
\sfq_{1\dots j}(u) = \prod_{k=1}^{M_j} (u-u_k^{(j)})\,.
\end{equation}
A generic transfer matrix eigenvalue $\sfT(u)$ can then be expressed as 
\begin{equation}
\sfT(u) = \sum_{j=1}^\gn \Lambda_j(u)
\end{equation}
where $\Lambda_j(u)$ are functions, known as \textit{quantum eigenvalues} \cite{kulish1982gl_3,Sklyanin:1992sm} (of the monodromy matrix), given by 
\begin{equation}
\Lambda_j(u) = \nu_j(u) \frac{\sfq_{1\dots j-1}^{[-2]}}{\sfq_{1\dots j-1}}\frac{\sfq_{1\dots j}^{[2]}}{\sfq_{1\dots j}}\,.
\end{equation}
The Bethe equations describing the transfer matrix eigenstate are then given by 
\begin{equation}
\frac{\nu_k(u)}{\nu_{k+1}(u)} = - \frac{\sfq_{1\dots k-1}^{[-2]}}{\sfq_{1\dots k-1}}\,\frac{\sfq_{1\dots k}^{[-2]}}{\sfq_{1\dots k}^{[2]}}\, \frac{\sfq_{1\dots k+1}}{\sfq_{1\dots k+1}^{[2]}},\quad k=1,\dots,\gn-1
\end{equation}
with both the \lhs and \rhs evaluated at a root $u_j^{(k)}$ of $\sfq_{1\dots k}$ and $\sfq_{1\dots\gn}:=1$. 

\subsection{Twisting and separation of variables: a first look}\label{firstsov}

Owing to the $\gl(\gn)$ symmetry of the transfer matrix the spectrum is highly degenerate. In many cases it is highly desirable to have a situation where the spectrum of conserved charges is non-degenerate giving us a one-to-one correspondence between transfer matrix eigenstates and eigenvalues. The procedure for doing this is known as twisting. 

\medskip

Twisting is based on the $\GL(\gn)$ symmetry of the $R$-matrix
\begin{equation}
[R(u,v),G\otimes G]=0
\end{equation}
where $G$ is any invertible $\gn\times \gn$ matrix. As a result of this, the RTT commutation relation 
\begin{equation}
R_{ab}(u,v)T_a(u)T_b(v) = T_b(v)T_a(u)R_{ab}(u,v)
\end{equation}
remains satisfied if we replace $T(u) \rightarrow \bT(u)=H\,T(u)\,G$ for any two $H,G\in\GL(\gn)$. If we consider the transfer matrix $\T(u)$ as being obtained from the trace of $H\,T(u)\,G$ instead of $T(u)$ then only the product $GH$ contributes due to the cyclicity of the trace and hence without loss of generality set $H=1$. Furthermore, the eigenvalues of the transfer matrix are only sensitive to the eigenvalues of the twist matrix $G$. To see this, we note that the $\GL(\gn)$ symmetry of the $R$-matrix implies $\gl(\gn)$ symmetry 
\begin{equation}
[R(u,v),\mathsf{J} \otimes 1 + 1 \otimes \mathsf{J}],\quad \mathsf{J}\in\gl(\gn)
\end{equation}
which extends to the Lax operator $\lL^\lambda(u,\theta)$ 
\begin{equation}
[\lL^\lambda(u,\theta),\mathsf{J} \otimes 1 + 1 \otimes \pi^\lambda\left(\mathsf{J}\right)],\quad \mathsf{J}\in\gl(\gn)
\end{equation}
which in turn implies $\GL(\gn)$ symmetry of the Lax operator 
\begin{equation}
[\lL^\lambda(u,\theta),G\otimes \Pi^\lambda\left( G \right)]=0,\quad G\in \GL(\gn)
\end{equation}
where $\Pi^\lambda$ denotes the image of the group element $G$ induced from the representation $\pi^\lambda$ on $\gl(\gn)$. Now consider the transfer matrix $\T$ obtained from the twisted monodromy matrix $\bT = T\,G$ constructed from $L$-copies of evaluation representations and consider the change of basis $\Pi^\lambda(K):=\Pi^{\lambda_1}(K)\otimes \dots \otimes \Pi^{\lambda_L}(K)$ where $K$ is such that $K G K^{-1} = g$. Then we have
\begin{equation}
\begin{split}
\Pi^\lambda(K) \T(u) \Pi^\lambda(K^{-1}) & = {\rm tr}_a\left(\Pi^\lambda(K)\,T_a(u)\Pi^\lambda(K^{-1})G_a \right)\\
& = {\rm tr}_a\left(\Pi^\lambda(K^{-1})_a\,T_a(u)\Pi^\lambda(K)_a G_a \right) \\
& = {\rm tr}\left(T(u)\,g \right) \\
\end{split}
\end{equation}
where in the second equality we used the $\GL(\gn)$-invariance of each Lax operator to move the rotation $\Pi^\lambda(G)$ onto the physical space and in the third equality used the cyclicity of the trace. An immediate consequence of this is that if $\ket{\Psi^G}$ is some eigenvector of $\T(u)$ constructed with $G$ then $\ket{\Psi^g}=\Pi^\lambda(K^{-1})\ket{\Psi^G}$ is an eigenvector for $\T(u)$ constructed with $g$. 

\medskip

The physical consequence of twisting is breaking the global symmetry by deforming the integrals of motion while still preserving integrability. The breaking of the global symmetry can be seen by examining the effect of twisting on the Hamiltonian which can still be extracted from the transfer matrix by taking the logarithmic derivative (assuming the defining representation without inhomogeneities). The deformed Hamiltonians $H^g$ reads, as is easily confirmed by direct calculation, 
\begin{equation}
H^g = \lH_{12}+\dots \lH_{L-2,L-1} + g_L^{-1} \lH_{L,1}g_L
\end{equation}
where $g_L$ denotes that the twist matrix $g$ only acts non-trivially on site $L$. As a result of twisting the deformed Hamiltonian no longer commutes with the full $\sua(2)$ algebra and only the Cartan subalgebra $\mathfrak{u}(1)$ generated by the global $S_z$ remains a symmetry. 

\medskip

Throughout this work we will denote the eigenvalues of the twist matrix $G\in\GL(\gn)$ as $z_1,\dots,z_{\gn}$. As a result of twisting the transfer matrix no longer commutes with the full global $\gl(\gn)$ algebra -- only its Cartan subalgebra remains a symmetry. Furthermore the transfer matrix eigenvalues and Bethe equations get modified. Both of these are conveniently described by replacing the Baxter polynomials $\sfq_{1\dots j}$ with what are referred to as \textit{twisted polynomials} which we define to be functions of the form $\kappa^{\frac{u}{\hbar}}p(u)$ where $\kappa\in\CC$ and $p(u)$ is a polynomial. For the situation at hand we define twisted polynomials $\hat{\sfq}_{1\dots j}$ defined by 
\begin{equation}
\hat{\sfq}_{1\dots j} = \left(z_1\dots z_j\right)^{\frac{u}{\hbar}}\sfq_{1\dots j}(u)
\end{equation}
where $\sfq_{1\dots j}$ now denotes a new polynomial, different from the original Baxter polynomial. The modification of the Bethe equations and transfer matrix eigenvalues is then obtained by making the simple replacement $\sfq_{12\dots j}\rightarrow \hat{\sfq}_{1\dots j}$. As an example, the eigenvalue of the transfer matrix with diagonal twist on highest-weight state of the Yangian representation is given by 
\begin{equation}
\sfT(u) = \sum_{j=1}^{\gn} z_j \nu_j(u)\,.
\end{equation}

\paragraph{Separation of Variables}

Before closing this section we will take a brief look at how a separated variable basis can be constructed for the $\sua(2)$ spin chain in the defining evaluation representation, see \cite{Kazama:2013rya} for an introductory overview. We constructed transfer matrix eigenstates $\ket{\Psi}$ by repeatedly acting with the operator $B$ on the highest-weight state $\ket{\Omega}$ 
\begin{equation}
\ket{\Psi} = \prod_{j=1}^M B(u_j)\ket{\Omega}\,.
\end{equation}
Suppose that $B$ were diagonalisable with a basis of left eigenvectors denoted $\bra{\svx}$. Note that in this work we have not equipped the representation space $\lV$ with any metric and so the bra vectors $\bra{v}$ are simply defined to be elements of the dual space $\lV^*$ and the scalar product $\braket{v|w}$ simply denotes the action of a dual vector $\bra{v}$ on a vector $\ket{w}$. In the basis $\bra{\svx}$ of $\lV^*$ the transfer matrix eigenstates will factorise 
\begin{equation}
\braket{\svx|\Psi} = (-1)^{ML} \prod_{\alpha=1}^L \sfq_1(\svx^\alpha)
\end{equation}
and we normalised $\braket{\svx|\Omega}=1$ and $\svx^\alpha$ denote the eigenvalues of the $L$ roots of the polynomial $B(u)$. Each of the individual factors $\sfq_1(\svx^\alpha)$ can then be interpreted as a one-particle wave function and we have succeeded in separating variables. Furthermore, since each of the one-dimensional wave functions are solutions of the Baxter TQ equation we can view the TQ equation as the one-particle Schr{\"o}dinger equation in separated variables. 

\medskip

As it stands however this construction is moot as $B$ is actually a polynomial of degree $L-1$ and when constructed with a diagonal twist is nilpotent since it behaves as lowering operator at large $u$. Thankfully however both of these problems can be removed allowing us to realise the above construction.

\medskip 

Consider the special twist $T(u)g\rightarrow K T(u)g K^{-1}$. This preserves all commutation relations and further leaves the transfer matrix invariant. Hence the new operator $B\rightarrow \bB$ can also be used to build transfer matrix eigenstates. A nice feature is that we can choose $K$ to make $\bB$ diagonalisable and have simple spectrum allowing us to proceed with the above construction. However, this relies on having a twist $g\neq 1$ in the first place and so the presence of twist is a crucial part of the construction. 

\bigskip

\bigskip

We have now finished our review of the Yangian algebra. In the next section we will examine the structure of the conserved charges arising from the transfer matrix in more detail. 

\section{Bethe algebra}

When discussing the generalisation of the algebraic Bethe ansatz to the higher rank case we briefly discussed the diagonalisation of the transfer matrix $\T(u)$. This transfer matrix provides us with $L$ integrals of motion. However, these $L$ integrals of motion are in general not enough to completely characterise an eigenstate as for certain representations it has degenerate spectrum. This can be seen by considering a $\gl(3)$ spin chain of length $L=1$ with a diagonal twist. The transfer matrix is given by
\begin{equation}
\T(u) = (z_1+z_2+z_3)(u-\theta) -\hbar\left(z_1\, \sfE_{11}+z_2\, \sfE_{22} +z_3\,\sfE_{33} \right)\,.
\end{equation}
In this special case the non-trivial part of the transfer matrix is an element of the Cartan subalgebra of $\gl(\gn)$ and the only requirement for the transfer matrix to have degenerate spectrum is that the Cartan subalgebra has degenerate spectrum. This is the case for the $[2,1,0]$ representation, see Section \ref{GLNGT}. In order to remove these degeneracies it is necessary to construct a larger family of integrals of motion. This family is known as the Bethe subalgebra, coined in \cite{nazarov1996bethe}, and we will now present an in-depth review. 

\subsection{Fusion}\label{Fusionsec}
\begin{figure}[htb]
\centering
  \includegraphics[width=80mm,scale=25]{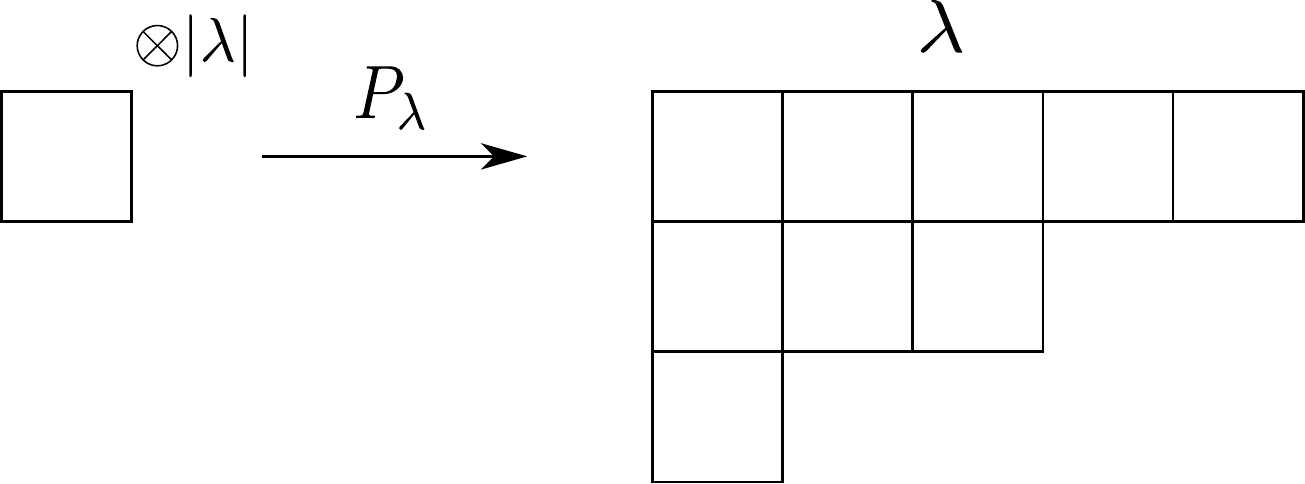}
  \caption{Any finite dimensional irrep $\GL(\gn)$ corresponding to a Young diagram $\lambda$ can be constructed by applying a suitable projection $P_\lambda$ to the tensor product of $|\lambda|$ copies of the defining representation.}
  \label{Youngdiagramfusion}
\end{figure}

Fusion \cite{Kulish:1981gi,kulish1982gl_3,cherednik1982properties,cherednik1986special} is a procedure which allows us to construct new solutions of the Yang-Baxter equation from old ones and is similar to the construction of irreps of $\GL(\gn)$ via Young Symmetrisers \cite{fulton2013representation}, see Figure \ref{Youngdiagramfusion}. See \cite{Zabrodin:1996vm,molev2007yangians} for reviews. 

The rational $R$-matrix $R(u,v)$ acts on two copies of $\CC^\gn$. Viewing $\CC^\gn$ as the defining representation of $\GL(\gn)$ the fusion procedure allows to construct more general $R$-matrices 
\begin{equation}
R^{\lambda\, \mu}\in {\rm End}\left(\lV^\lambda\otimes \lV^\mu\right)
\end{equation}
acting on the tensor product of two finite-dim irreps $\lV^\lambda$ and $\lV^\mu$ of $\GL(\gn)$ and satisfying a more general form of the Yang-Baxter equation 
\begin{equation}
R^{\lambda\, \mu}(u,v)R^{\lambda\, \nu}(u,w)R^{\mu\, \nu}(v)=R^{\mu\, \nu}(v,w)R^{\lambda\, \nu}(u,w)R^{\lambda\ \mu}(u-v)
\end{equation}
for any Young diagrams $\lambda$, $\mu$ and $\nu$. The term ``fusion'' comes from the fact that the defining $R$-matrix can be viewed as the scattering matrix in an integrable field theory and the higher $R$-matrices constructed in this way describe the scattering of bound states obtained by fusing two elementary particles. 

\paragraph{Fusion in the physical space}
We will start with fusion in the physical space and explain how to construct the operator $R^{\Box\ \mu}$ where $\Box$ denotes the defining representation of $\GL(\gn)$. 

\medskip

\begin{figure}[htb]
\centering
  \includegraphics[width=30mm,scale=10]{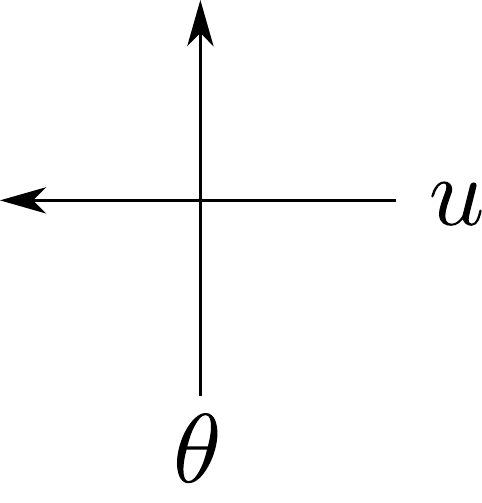}
  \caption{Graphical representation of $R$-matrix $R(u,\theta)$. The horizontal line with label $u$ labels the first space and the vertical line with label $\theta$ labels the second space. The directional arrows can be placed anywhere on a given line.}
  \label{Rmatgraph}
\end{figure}

In order to discuss fusion in simple terms it is convenient to introduce graphical notations for performing calculations. First we need the $R$-matrix $R(u,\theta)$ which we write as in Figure \ref{Rmatgraph}

\begin{figure}[htb]
\centering
  \includegraphics[width=100mm,scale=10]{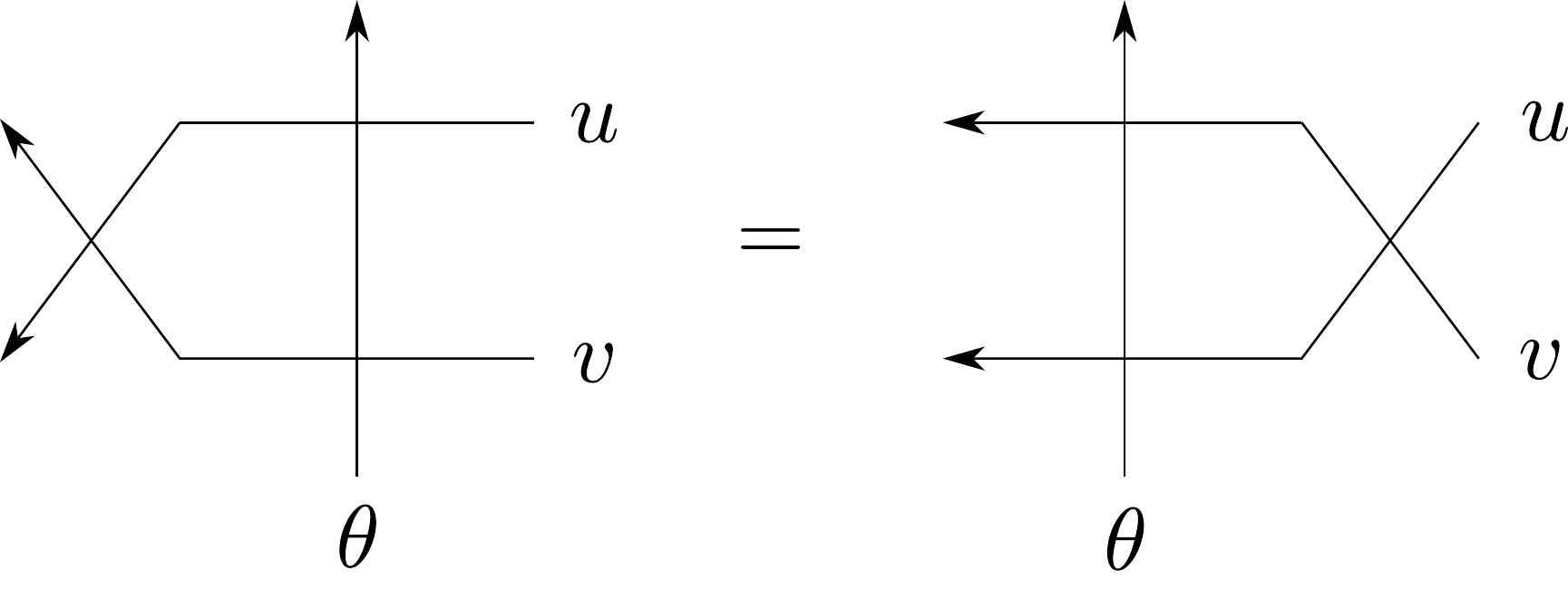}
  \caption{Yang-Baxter equation in graphical notations. By following the directions of arrows from right to left we can distinguish between $R(u,v)$ and $R(v,u)$. Note that objects which appear down and right in graphical notations act first on the Hilbert space.}
  \label{YBEgraph}
\end{figure}

In these graphical notations the Yang-Baxter equation 
\begin{equation}
R_{ab}(u,v)R_{ac}(u,\theta)R_{bc}(v,\theta)=R_{bc}(v,\theta)R_{ac}(u,\theta)R_{ab}(u,v)
\end{equation}
is represented simply by Figure \ref{YBEgraph}.

\medskip

\begin{figure}[htb]
\centering
  \includegraphics[width=15mm,scale=10]{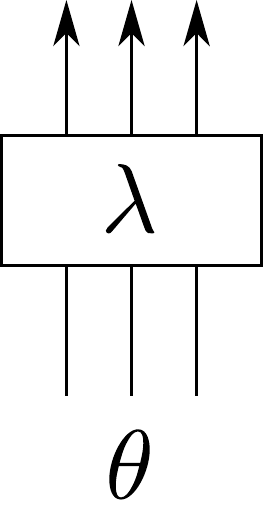}
  \caption{We use this notation to indicate that a collection of incoming particles have been consistently projected onto the irreducible representation $\lambda$. The label $\theta$ indicates that the left-most particle has rapidity $\theta$.}
  \label{projection}
\end{figure}

\begin{figure}[htb]
\centering
  \includegraphics[width=80mm,scale=25]{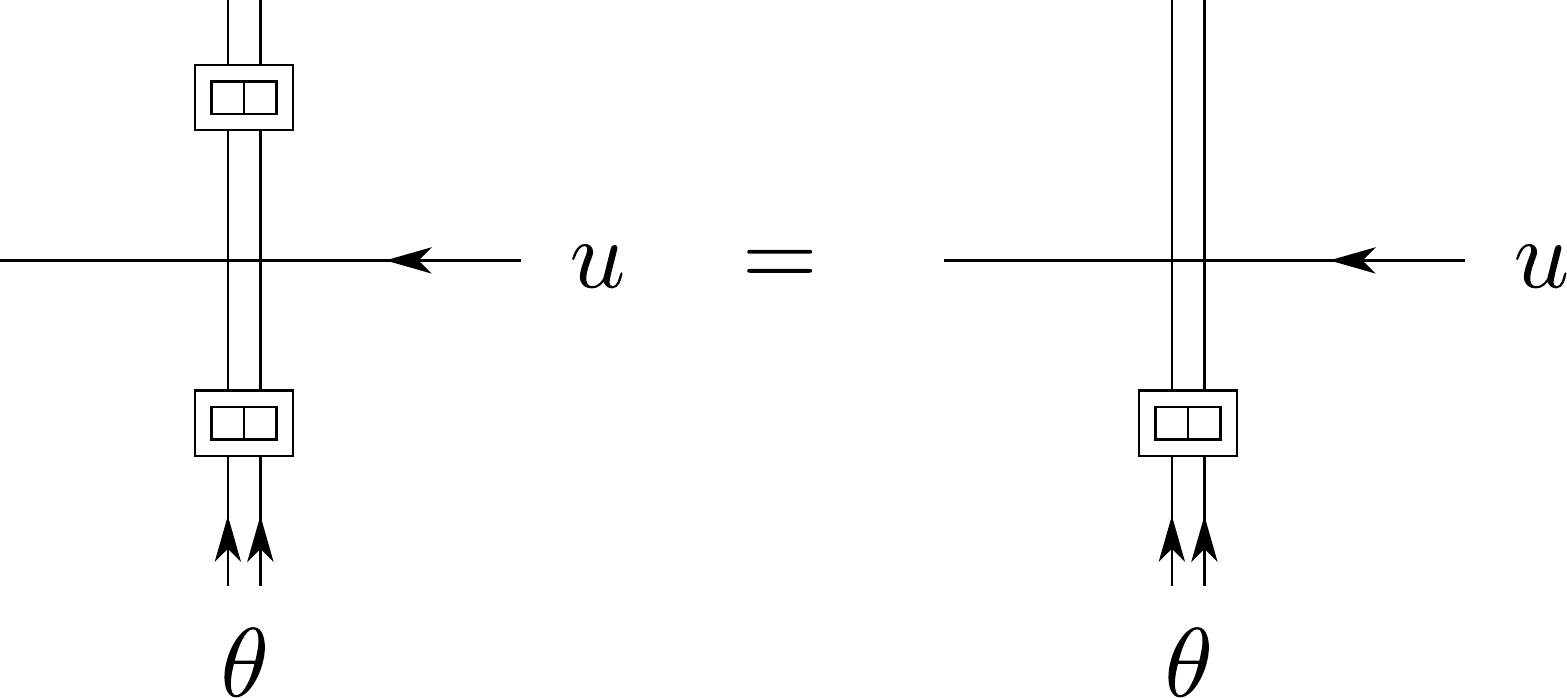}
  \caption{Graphical representation of the fused $R$-matrix $R^{{\raisebox{0.075cm}{\tiny$\Box$}}\  {{\raisebox{0.075cm}{\tiny$\Box$}\hspace{-0.1EM}\raisebox{0.075cm}{\tiny$\Box$}}}}(u,\theta)$ }
  \label{fusedRsymmetric}
\end{figure}

The main point of the fusion procedure is that at certain values of the spectral parameters the $R$-matrix $R(u,v)$ reduces to projectors $P^{{\raisebox{0.075cm}{\tiny$\Box$}\hspace{-0.1EM}\raisebox{0.075cm}{\tiny$\Box$}}}$ and $P^{\, {\raisebox{-0.075cm}{\tiny$\Box$}\hspace{-0.53EM}\raisebox{0.075cm}{\tiny$\Box$}}}$
\begin{equation}
\begin{split}
& R_{12}(v+\hbar,v) = R_{12}(u,u-\hbar)= 2\hbar\, P^{\, {\raisebox{-0.075cm}{\tiny$\Box$}\hspace{-0.54EM}\raisebox{0.075cm}{\tiny$\Box$}}} \\
& R_{12}(v-\hbar,v)=R_{12}(u,u+\hbar) = - 2\hbar\, P^{{\raisebox{0.075cm}{\tiny$\Box$}\hspace{-0.1EM}\raisebox{0.075cm}{\tiny$\Box$}}}\,.
\end{split}
\end{equation}
In the decomposition 
\begin{equation}
\CC^\gn \otimes \CC^\gn={\rm sym}^2(\CC^\gn)\oplus \wedge^2(\CC^\gn)
\end{equation}
of $\GL(\gn)$ irreps the projector $P^{{\raisebox{0.075cm}{\tiny$\Box$}\hspace{-0.1EM}\raisebox{0.075cm}{\tiny$\Box$}}}$ projects onto ${\rm sym}^2(\CC^\gn)$, and similarly $P^{\, {\raisebox{-0.075cm}{\tiny$\Box$}\hspace{-0.54EM}\raisebox{0.075cm}{\tiny$\Box$}}}$ projects onto $\wedge^2(\CC^\gn)$. The projectors are idempotent
\begin{equation}
\begin{split}
P^{{\raisebox{0.075cm}{\tiny$\Box$}\hspace{-0.1EM}\raisebox{0.075cm}{\tiny$\Box$}}}P^{{\raisebox{0.075cm}{\tiny$\Box$}\hspace{-0.1EM}\raisebox{0.075cm}{\tiny$\Box$}}}& =P^{{\raisebox{0.075cm}{\tiny$\Box$}\hspace{-0.1EM}\raisebox{0.075cm}{\tiny$\Box$}}} \\
P^{\, {\raisebox{-0.075cm}{\tiny$\Box$}\hspace{-0.54EM}\raisebox{0.075cm}{\tiny$\Box$}}}P^{\, {\raisebox{-0.075cm}{\tiny$\Box$}\hspace{-0.54EM}\raisebox{0.075cm}{\tiny$\Box$}}} & = P^{\, {\raisebox{-0.075cm}{\tiny$\Box$}\hspace{-0.54EM}\raisebox{0.075cm}{\tiny$\Box$}}}
\end{split}
\end{equation}
and mutually orthogonal
\begin{equation}
P^{{\raisebox{0.075cm}{\tiny$\Box$}\hspace{-0.1EM}\raisebox{0.075cm}{\tiny$\Box$}}}P^{\, {\raisebox{-0.075cm}{\tiny$\Box$}\hspace{-0.54EM}\raisebox{0.075cm}{\tiny$\Box$}}}=0\,.
\end{equation}
Focusing on the symmetric projection $P^{{\raisebox{0.075cm}{\tiny$\Box$}\hspace{-0.1EM}\raisebox{0.075cm}{\tiny$\Box$}}}$, the Yang-Baxter equation implies that \footnote{We have switched from $v$ to $\theta$ to emphasise that the second space corresponds to a physical particle.}
\begin{equation}
R_{a1}(u,\theta)R_{a2}(u,\theta+\hbar)P^{{\raisebox{0.075cm}{\tiny$\Box$}\hspace{-0.1EM}\raisebox{0.075cm}{\tiny$\Box$}}}_{12}=P^{{\raisebox{0.075cm}{\tiny$\Box$}\hspace{-0.1EM}\raisebox{0.075cm}{\tiny$\Box$}}}_{12}R_{a2}(u,\theta+\hbar)R_{a1}(u,\theta)
\end{equation}
which then guarantees that the projection survives scattering with the auxiliary space: 
\begin{equation}
R_{a1}(u,\theta)R_{a2}(u,\theta+\hbar)P^{{\raisebox{0.075cm}{\tiny$\Box$}\hspace{-0.1EM}\raisebox{0.075cm}{\tiny$\Box$}}}_{12}=P^{{\raisebox{0.075cm}{\tiny$\Box$}\hspace{-0.1EM}\raisebox{0.075cm}{\tiny$\Box$}}}_{12}R_{a1}(u,\theta)R_{a2}(u,\theta+\hbar)P^{{\raisebox{0.075cm}{\tiny$\Box$}\hspace{-0.1EM}\raisebox{0.075cm}{\tiny$\Box$}}}_{12}\,.
\end{equation}
Hence, we can view the symmetrised pair of $R$-matrices as a composite -- fused! -- particle transforming in the symmetric representation of $\GL(\gn)$ and which doesn't decompose into its two constituent pieces upon scattering with the particle in the auxiliary space and hence define 
\begin{equation}
R^{{\raisebox{0.075cm}{\tiny$\Box$}}\ {{\raisebox{0.075cm}{\tiny$\Box$}\hspace{-0.1EM}\raisebox{0.075cm}{\tiny$\Box$}}}}(u,\theta)=P^{{\raisebox{0.075cm}{\tiny$\Box$}\hspace{-0.1EM}\raisebox{0.075cm}{\tiny$\Box$}}}_{12}R_{a1}(u,\theta)R_{a2}(u,\theta+\hbar)P^{{\raisebox{0.075cm}{\tiny$\Box$}\hspace{-0.1EM}\raisebox{0.075cm}{\tiny$\Box$}}}_{12}\,,
\end{equation}
see Figures \ref{projection} and \ref{fusedRsymmetric}. 

\medskip

\begin{figure}[htb]
\centering
  \includegraphics[width=90mm,scale=25]{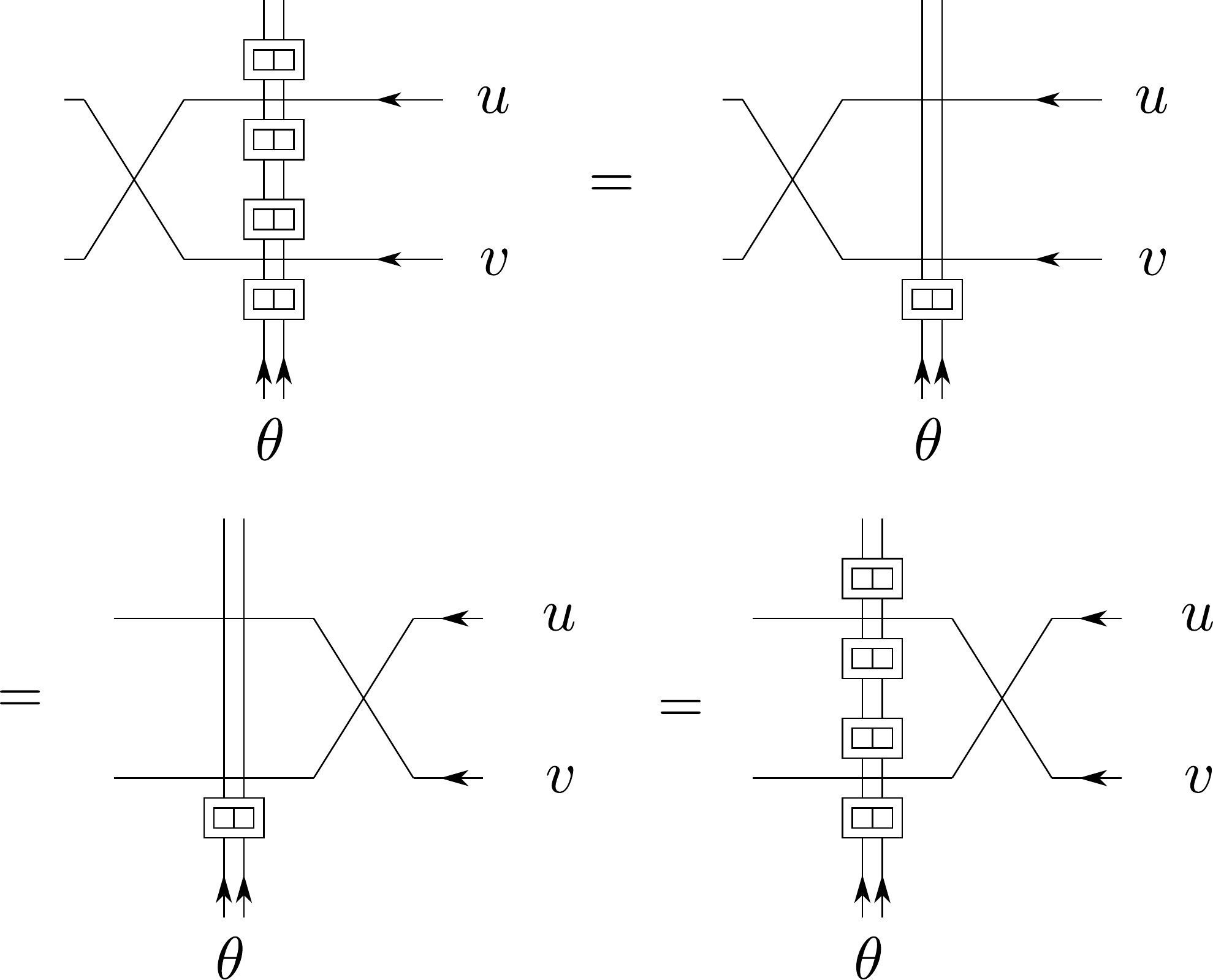}
  \caption{Proof of the Yang-Baxter equation $R^{{\raisebox{0.075cm}{\tiny$\Box$}}\ {\raisebox{0.075cm}{\tiny$\Box$}}}R^{{\raisebox{0.075cm}{\tiny$\Box$}}\ {{\raisebox{0.075cm}{\tiny$\Box$}\hspace{-0.1EM}\raisebox{0.075cm}{\tiny$\Box$}}}}R^{{\raisebox{0.075cm}{\tiny$\Box$}}\ {{\raisebox{0.075cm}{\tiny$\Box$}\hspace{-0.1EM}\raisebox{0.075cm}{\tiny$\Box$}}}}=R^{{\raisebox{0.075cm}{\tiny$\Box$}}\  {{\raisebox{0.075cm}{\tiny$\Box$}\hspace{-0.1EM}\raisebox{0.075cm}{\tiny$\Box$}}}}R^{{\raisebox{0.075cm}{\tiny$\Box$}}\ {{\raisebox{0.075cm}{\tiny$\Box$}\hspace{-0.1EM}\raisebox{0.075cm}{\tiny$\Box$}}}}R^{{\raisebox{0.075cm}{\tiny$\Box$}}\ {\raisebox{0.075cm}{\tiny$\Box$}}}$. The first equality follows from the fact that the symmetric projectors are idempotent and survive scattering with the auxiliary spaces. The second equality follows from using the Yang-Baxter equation to move the physical lines through the auxiliary $R$-matrix. Finally, projectors are restored.}
  \label{fusedYBE}
\end{figure}

A straightforward calculation demonstrates that the YBE 
\begin{equation}
R^{{\raisebox{0.075cm}{\tiny$\Box$}}\ {\raisebox{0.075cm}{\tiny$\Box$}}}(u,v)R^{{\raisebox{0.075cm}{\tiny$\Box$}}\ {{\raisebox{0.075cm}{\tiny$\Box$}\hspace{-0.1EM}\raisebox{0.075cm}{\tiny$\Box$}}}}(u,\theta)R^{{\raisebox{0.075cm}{\tiny$\Box$}}\ {{\raisebox{0.075cm}{\tiny$\Box$}\hspace{-0.1EM}\raisebox{0.075cm}{\tiny$\Box$}}}}(v,\theta)=R^{{\raisebox{0.075cm}{\tiny$\Box$}}\ {{\raisebox{0.075cm}{\tiny$\Box$}\hspace{-0.1EM}\raisebox{0.075cm}{\tiny$\Box$}}}}(v,\theta)R^{{\raisebox{0.075cm}{\tiny$\Box$}}\ {{\raisebox{0.075cm}{\tiny$\Box$}\hspace{-0.1EM}\raisebox{0.075cm}{\tiny$\Box$}}}}(u,\theta)R^{{\raisebox{0.075cm}{\tiny$\Box$}}\ {\raisebox{0.075cm}{\tiny$\Box$}}}(u,v)
\end{equation}
is satisfied on the triple tensor product of representations ${\raisebox{0.0 cm}{$\Box$}}\otimes {\raisebox{0.0 cm}{$\Box$}}\, \otimes {{\raisebox{0.0cm}{$\Box$}\hspace{-0.1EM}\raisebox{0.0cm}{$\Box$}}}$, see Figure \ref{fusedYBE}.

The procedure for constructing the fused $R$-matrix $R^{\Box \  \lambda}(u)$ for any Young diagram $\lambda$ is totally analogous to the construction presented above for $R^{{\raisebox{0.075cm}{\tiny$\Box$}}\  {{\raisebox{0.075cm}{\tiny$\Box$}\hspace{-0.1EM}\raisebox{0.075cm}{\tiny$\Box$}}}}$. 
Namely, we write 
\begin{equation}
R^{\Box \ \lambda}(u,\theta)=\left(\displaystyle \prod_{j=1}^{|\lambda|}R_{aj}(u,\theta+\hbar\, c_j) \right) P^\lambda =  P^\lambda\left(\displaystyle \prod_{j=1}^{|\lambda|}R_{aj}(u,\theta+\hbar\, c_j) \right) P^\lambda 
\end{equation}
where $c_j$ are some appropriate numbers made precise below. The second equality above is the statement that the projection $P^\lambda$ onto the irrep $\lambda$ of $\GL(\gn)$ survives scattering with the auxiliary space. This is achieved by constructing $P^\lambda$ as on appropriate product of fundamental $R$-matrices and repeatedly applying the Yang-Baxter equation, similar to what we did for the symmetric representation above, see for example \cite{Zabrodin:1996vm}. The proof of the Yang-Baxter equation 
\begin{equation}
R^{{\raisebox{0.075cm}{\tiny$\Box$}}\ {\raisebox{0.075cm}{\tiny$\Box$}}}R^{{\raisebox{0.075cm}{\tiny$\Box$}}\ \lambda}R^{{\raisebox{0.075cm}{\tiny$\Box$}}\ \lambda}=R^{{\raisebox{0.075cm}{\tiny$\Box$}}\ \lambda}R^{{\raisebox{0.075cm}{\tiny$\Box$}}\ \lambda }R^{{\raisebox{0.075cm}{\tiny$\Box$}}\ {\raisebox{0.075cm}{\tiny$\Box$}}}
\end{equation}
is then performed in precisely the same way as in Figure \ref{fusedYBE}. 

\medskip

\begin{figure}[htb]
\centering
  \includegraphics[width=80mm,scale=25]{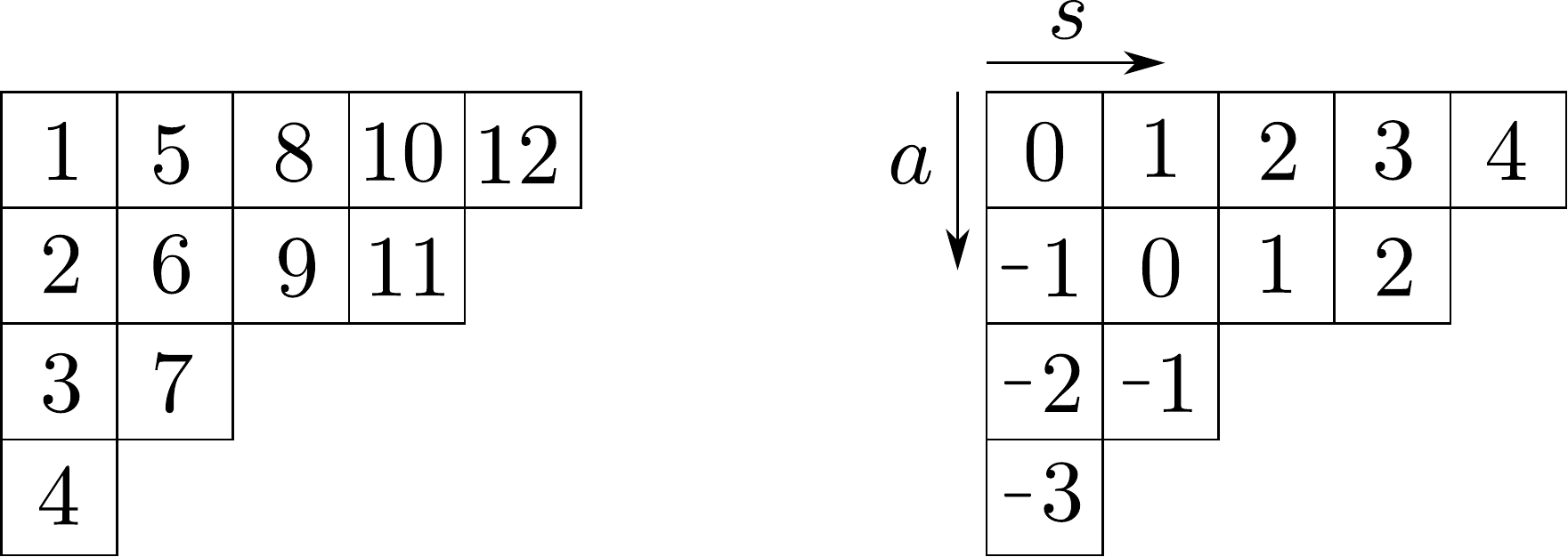}
  \caption{Left: Column-ordering of boxes on the Young diagram $\lambda$. Right: $c_j=s-a$ associated with each box $j$.}
  \label{Youngtableaux}
\end{figure}

The numbers $c_j$ can be read off from the Young diagram $\lambda$. We draw the Young diagram $\lambda$ and label the boxes in column-ordering. Then, to the box numbered $j$ we associate the value $c_j$ with 
\begin{equation}
c_j = s - a 
\end{equation}
if box $j$ has Cartesian coordinates $(a,s)$, see Figure \ref{Youngtableaux}.

\paragraph{Drinfeld Polynomials for fused $R$-matrices}

By constructing the fused $R$-matrix $R^{\Box\ \lambda}(u,\theta)$ we have actually managed to seemingly construct two Yangian representations on the space $\lV^\lambda$. The first is given using the Lax operator 
\begin{equation}
T(u) = \lL^\lambda(u)
\end{equation}
while the second is given using the fused $R$-matrix $R^{\Box\ \lambda}(u)$ 
\begin{equation}
T(u) = R^{\Box\ \lambda}(u)\,.
\end{equation}
Both representations clearly satisfy the RTT relation and, initially, might seem rather different -- the Lax operator is a polynomial of degree $1$, whereas $R^{\Box\ \lambda}(u)$ is a Laurent polynomial.  Thankfully, and perhaps not so surprisingly, these two Yangian representations are isomorphic as can be checked by computing their Drinfeld polynomials.

\paragraph{Fusion in the auxiliary space}

Fusion in the auxiliary space is almost exactly the same as for the physical space. The fused $R$-matrix $R^{\lambda\ \Box}(u,\theta)$ is simply given by 
\begin{equation}
R^{\lambda\ \Box}(u,\theta)=P^\lambda\left( \displaystyle \prod_{j=1}^{|\lambda|}R_{a_j 1}(u+\hbar\, c_j,\theta) \right)  =  P^\lambda\left(\displaystyle \prod_{j=1}^{|\lambda|}R_{a_j 1}(u+\hbar\, c_j,\theta) \right) P^\lambda 
\end{equation}
Note that here we can pull $P^\lambda$ all the way to the left -- without affecting the ordering of $R$-matrices -- while for the physical space we could pull the $R$-matrices all the way to the right. We can still pull $P^\lambda$ all the way to the right in this case, but doing so will change the ordering of $R$-matrices
\begin{equation}
P^\lambda\left( \displaystyle \overrightarrow{\prod_{j}}R_{a_j 1}(u+\hbar\, c_j,\theta) \right) =\left( \displaystyle \overleftarrow{\prod_{j}}R_{a_j 1}(u+\hbar\, c_j,\theta) \right)P^\lambda\,.
\end{equation}

Finally, putting all the pieces together we can construct $R^{\lambda\ \mu}$ as 
\begin{equation}
R^{\lambda \ \mu}(u,\theta)=P^\mu P^\lambda \left( \displaystyle \prod_{j=1}^{|\lambda|}\prod_{k=1}^{|\mu|}R_{a_j k}(u+\hbar\, c^\lambda_j,\theta+\hbar c^\mu_k)\right) P^\mu P^\lambda
\end{equation}
where $c^\lambda_j$ and $c^\mu_k$ denote the content of the Young diagrams $\lambda$ and $\mu$, respectively. Checking the Yang-Baxter equation is now a trivial consequence of the developed techniques, and is most easily performed graphically.

\paragraph{Fusion for monodromy and transfer matrices}

The fusion procedure described above extends immediately to allow us to construct fused monodromy matrices $\bT^\lambda(u)$ which satisfy a generalised version of the RTT relation: 
\begin{equation}\label{fusedRTT}
R^{\lambda,\mu}_{ab}(u,v)\bT_a^\lambda(u)\bT^\mu_b(v)=\bT^\mu_b(v)\bT_a^\lambda(u)R^{\lambda,\mu}_{ab}(u,v)
\end{equation}
The fused monodromy matrix $\bT^\lambda(u)$ is constructed in total analogy with the fused $R$-matrix: 
\begin{equation}
\bT^\lambda(u)=P^\lambda\left( \displaystyle \prod_{j=1}^{|\lambda|}\bT_{a_j}(u+\hbar\, c_j) \right)  =  P^\lambda\left(\displaystyle \prod_{j=1}^{|\lambda|}\bT_{a_j}(u+\hbar\, c_j) \right) P^\lambda \,.
\end{equation}
The proof of the fused RTT relation \eqref{fusedRTT} is identical to that of the fused Yang-Baxter equation. 

\medskip

Since the fundamental $R$-matrices fused to create $R^{\lambda\ \mu}$ are invertible it follows that so is $R^{\lambda\ \mu}$ and hence
\begin{equation}
[\T_\lambda(u),\T_\mu(v)]=0
\end{equation}
where the fused transfer matrix $\T_\lambda(u)$ is given by $\T_\lambda(u) = {\rm tr}\, \bT^\lambda(u)$ and the trace is taken over the fused auxiliary space $\lambda$. The tower of commuting fused transfer matrices $\T_\lambda(u)$, where $\lambda$ ranges over all possible Young diagrams of $\GL(\gn)$, form a commutative subalgebra of $\lY_\gn$ called the Bethe subalgebra. 

\medskip

The fusion procedure presented here relied heavily on the fact that the $R$-matrix degenerated to a projector when the difference of spectral parameters attained certain values. There are numerous integrable systems for which the projectors onto symmetric and anti-symmetric subpsaces can not be so easily extracted such as the one-dimensional Hubbard model. Nevertheless, fusion can still be performed and higher transfer matrices can be obtained -- in \cite{Beisert:2015msa} an analogue of the fusion procedure was developed which only relies on the fact that the rank of the $R$-matrix drops at special points. 

\paragraph{Quantum minors and quantum determinant}

It will be useful for later purposes to consider the matrix elements of the fused monodromy matrices in anti-symmetric representations. In the $\wedge^a(\CC^\gn)$ representation these are given by 
\begin{equation}\label{quantumminor}
\bT\left[^{i_1\dots i_a}_{j_1\dots j_a}\right](u) = \displaystyle\sum_{\sigma\in\mathfrak{S}_a} (-1)^{|\sigma|} \bT_{i_{\sigma(1)}j_1}\bT_{i_{\sigma(2)}j_2}^{[-2]}\dots \bT_{i_{\sigma(a)}j_a}^{[-2(a-1)]}
\end{equation}
and are called \textit{quantum minors} owing to the fact that they are minors of the $\gn\times \gn$ matrix $\bT(u)$ with extra (quantum) shifts included. We have chosen to present the quantum minor with anti-symmetrisation performed over the upper indices. Since anti-symmetrisation commutes with scattering we could just as well have performed the anti-symmetrisation over the lower indices but with the opposite ordering of shifts
\begin{equation}\label{qdet2}
\bT\left[^{i_1\dots i_a}_{j_1\dots j_a}\right](u)=\displaystyle\sum_{\sigma\in\mathfrak{S}_a} (-1)^{|\sigma|} \bT_{i_{1}j_{\sigma(1)}}^{[-2(a-1)]}\bT_{i_{2}j_{\sigma(2)}}^{[-2(a-2)]}\dots \bT_{i_{a}j_{\sigma(a)}}\,.
\end{equation}

A highly useful property of the quantum minors is as follows \cite{molev2007yangians}. Let $I$ and $J$ denote subsets of $\{1,2,\dots,\gn\}$. If $i\in I$ and $j\in J$ then 
\begin{equation}\label{minorcommutativity}
[\bT\left[^I_J\right](u),\bT_{ij}(v)]=0
\end{equation}
for any $u,v$. Of course this immediately implies that $\bT\left[^{12\dots\gn}_{12\dots \gn}\right](u)$, known as the \textit{quantum determinant} $\qdet\,\bT(u)$ \cite{izergin2009lattice,kulish1982quantum}, commutes with all elements of the Yangian $\lY_{\gn}$ and hence is central. In fact, its coefficients in its $u$ expansion generate all central elements. 

\medskip

Since the quantum determinant is central it acts as a scalar multiple of the identity on any irreducible representation. Its value can be easily computed by acting with the presentation \eqref{qdet2} on the highest-weight state leading to 
\begin{equation}
\qdet\,\bT(u) = \det\,G\,\prod_{j=1}^{\gn} \nu_j(u-\hbar(\gn+1-j))\,.
\end{equation}
The Yangian of $\sla(\gn)$ mentioned in the introduction of the previous section is then obtained by the simple quotient \cite{molev2007yangians}
\begin{equation}
\lY(\sla(\gn)) = \lY(\gl(\gn))/(\qdet\,\bT(u)-1)\,.
\end{equation}

\paragraph{Talalaev generating function}

We end this section by presenting an alternative way of constructing the transfer matrices $\T_{a,1}$ corresponding to the representation $\wedge^a(\CC^{\gn})$. $\T_{a,1}$ correspond to traces of the monodromy matrices in anti-symmetric representations and hence can be expressed in terms of quantum minors as 
\begin{equation}
\T_{a,1}(u) = \sum_{1\leq i_1 < \dots < i_a\leq \gn} \bT\left[^{i_1\dots i_a}_{i_1\dots i_a} \right]\,.
\end{equation}
The transfer matrices $\T_{a,1}$ can then be conveniently generated by using Talalaev's formula \cite{Talalaev:2004qi} 
\begin{equation}\label{talalaev}
\det\left(1-\bT(u)\lD^{-1}\right)=\displaystyle \sum_{a=0}^\gn (-1)^a \T_{a,1}(u)\lD^{-a}
\end{equation}
where we have used the shift operator \eqref{shiftop}. This may seem somewhat limited as we have a family of transfer matrices $\T_\lambda$ at our disposal, not just the ones corresponding to antisymmetric representations. In the next section we will see that this is all we need, as all other transfer matrices can be expressed as simple polynomials in $\T_{a,1}$. 

\medskip

It is also worth pointing out that Talalaev's formula has other uses apart from being a tool for generating transfer matrices. Let us define the two finite-difference operators $\overleftarrow{\lO}$ and $\overrightarrow{\lO}$ where $\lO$ is the finite-difference operator
\begin{equation}
\lO = \displaystyle \sum_{a=0}^\gn (-1)^a \T_{a,1}(u)\lD^{-a}
\end{equation}
and the arrows indicate in which direction the shift operators act. These two difference operators define the Baxter equation and dual Baxter equation \cite{Krichever:1996qd}, generalising \eqref{TQ1}. We will return to this at the end of this section. 

\subsection{Transfer matrices and $\mathsf{T}$-system}

Using fusion we managed to construct a large family of integrals of motion -- a transfer matrix for every Young diagram. At first sight it may appear that we have constructed an infinite family of conserved charges. After all, one can write down an infinite number of Young diagrams for every $\GL(\gn)$. On the other hand if we are dealing with finite dimensional representations (and we mostly will be) then clearly the infinite number of transfer matrices $\T_\lambda(u)$ cannot be independent. Another reason can be found by relating transfer matrices to characters of $\GL(\gn)$ group elements.

\paragraph{Quantization of classical characters}
The transfer matrices can be understood as a quantization of classical $\GL(\gn)$ characters. Letting $G\in \GL(\gn)$ denote an invertible matrix with pairwise distinct eigenvalues $z_1,\dots,z_\gn$ then the character $\chi_\lambda$ of $G$ in the representation $\lambda$ can be obtained in the large-$u$ asymptotics of the transfer matrix $\T_\lambda(u)$ constructed with twist $G$:
\begin{equation}
\lim_{|u|\rightarrow \infty}\frac{\T_\lambda(u)}{u^{L|\lambda|}}=\chi_\lambda\,.
\end{equation}
The characters $\chi_\lambda$ are certainly not all independent - they are related \cite{fulton2013representation} by the following formula relating characters $\chi_{a,s}$ corresponding to rectangular Young diagrams with $a$ rows and $s$ columns
\begin{equation}\label{charHirota}
\chi_{a,s}\chi_{a,s}=\chi_{a+1,s}\chi_{a-1,s}+\chi_{a,s+1}\chi_{a,s-1}
\end{equation}
as well as the Jacobi-Trudi formula allowing us to express all characters $\chi_\lambda$ in terms of $\chi_{a,1}$
\begin{equation}\label{JacobiTrudi}
\chi_{\lambda}=\displaystyle\det_{1\leq i,j\leq \lambda_1}\chi_{\lambda^{\prime}_j+i-j,1}
\end{equation}
where $\lambda_j^\prime$ denotes the height of the $j$-th column of the Young diagram $\lambda$. The fact that the transfer matrices are not algebraically independent at large $u$ suggests they are not independent in general. This is indeed the case and both \eqref{JacobiTrudi} and \eqref{charHirota} have analogues for transfer matrices. 

\paragraph{Hirota equation and CBR formula}
The Hirota equation \cite{hirota1981discrete} is the quantum analogue of the relation \eqref{charHirota} and reads
\begin{equation}\label{Hirota}
\T_{a,s}\T_{a,s}^{[2]}=\T^{[2]}_{a+1,s}\T_{a-1,s}+\T_{a,s+1}\T_{a,s-1}^{[2]}
\end{equation}
and initially appeared in the theory of solitions in \textit{classical} systems and describes numerous integrable hierarchies such as those arising from the Korteweg-de Vries (KdV) equation and Kadomtsev-Petviashvili (KP) equation, see \cite{babelon_bernard_talon_2003} for an overview. In our case however the Hirota equation relates \textit{quantum} transfer matrices. As a special case of it we have the following relation between the transfer matrices $\T_{1,1}$, $\T_{2,1}$ and $\T_{1,2}$
\begin{equation}
\T_{1,1}\T_{1,1}^{[2]} = \T_{1,2} + \T_{2,1}^{[2]}
\end{equation}
generalising the familiar character relation following from the decomposition 
\begin{equation}\label{Hirota21}
\CC^\gn \otimes \CC^\gn = {\rm sym}^2\left(\CC^{\gn} \right) \oplus \wedge^2 \left(\CC^{\gn} \right)\,.
\end{equation}

We also have the Cherednik-Bazhanov-Reshetikhin (CBR) \cite{cherednik1987analogue,Bazhanov:1989yk,Kazakov:2007na} formula which is a quantum analog of the Jacobi-Trudi formula which states that for a Young diagram $\lambda$ 
\begin{equation}\label{CBRfla}
\T_{\lambda}(u)=\displaystyle\det_{1\leq i,j\leq \lambda_1}\T_{\lambda^{\prime}_j+i-j,1}(u+\hbar(i-1))\,.
\end{equation}
These formulae should be supplemented with the boundary conditions 
\begin{equation}\label{bdyconds}
\begin{split}
& \T_{\es}(u)=\T_{0,1}(u) = 1 \\
& \T_{a,1}(u)=0,\  a<0
\end{split}
\end{equation}
where $\T_{\es}$ denotes the transfer matrix corresponding to the empty diagram.

\paragraph{Gauge symmetries of Hirota equation} 
The Hirota equation admits a number of symmetries \cite{Saito:1986qx,Zabrodin:1996vm} which we refer to as gauge transformations. Let us introduce a family of functions $g^{\pm,\pm}_{a,s}$ defined by
\begin{equation}
g^{(\pm,\pm)}_{a,s}(u)=f^{(\pm,\pm)}\left(u+\hbar\left(\left(\frac{1}{2}\pm \frac{1}{2} \right)s -\left(\frac{1}{2}\pm \frac{1}{2} \right)a\right)\right)
\end{equation}
Then
\begin{equation}
\T_{a,s}(u) \rightarrow g^{(\pm,\pm)}_{a,s}\T_{a,s}(u)
\end{equation}
is a symmetry of the Hirota equation and so there are four independent gauge transformations which can be performed corresponding to the possible pairs $(\pm,\pm)$. The choice of gauge largely comes down to personal preference. For the most part we choose to work with what we call the \textit{fusion gauge} where all transfer matrices $\T_\lambda$ coincide with those constructed using the fusion procedure of the previous section. Various other gauges are possible and useful. Indeed, one is often interested in the situation where the physical space carries the defining representation and in this case most transfer matrices have a number of overall trivial zeroes resulting from the fusion procedure and it is often convenient to choose a gauge which removes these trivial zeroes, see \cite{Zabrodin:1996vm,Kazakov:2007fy}.

\subsection{Q-system}\label{Qsystem}

We now introduce one of the key concepts in this work -- the $\mathsf{Q}$-system. The $\GL(\gn)$ $\mathsf{Q}$-system is a set \cite{Krichever:1996qd,Tsuboi:2009ud,Bazhanov:2010jq,Kazakov:2010iu} of $2^\gn$ functions $\sfQ_A(u)$ labelled by subsets $A\subset \{1,2,\dots,\gn\}$ subject to relations known as QQ-relations\footnote{The QQ-relations presented here differ from the ones in the Introduction by means of a redefinition of the Q-functions. The convention used here is most convenient for our purposes. The convention in the Introduction is the one most used in AdS/CFT contexts.}
\begin{equation}
\sfQ_{Abc}\sfQ_A^{[-2]} = \sfQ_{Ab}\sfQ_{Ac}^{[-2]}-\sfQ_{Ac}\sfQ_{Ab}^{[-2]}\,.
\end{equation}
It follows from the QQ relations that the $\sfQ$-functions are anti-symmetric in all indices. The Q-functions are related to the (twisted) Baxter polynomials which appeared in the nested Bethe ansatz and we will explain precisely how later.

\paragraph{Geometric interpretation of the $\mathsf{Q}$-system}

Q-functions can be naturally interpreted as Pl{\"u}cker coordinates of certain hyperplanes. First, Q-functions can be packaged into exterior forms and we closely follow \cite{Kazakov:2015efa}. Introduce a basis $\zeta_1,\dots,\zeta_{\gn}$ of $\CC^{\gn}$. Then define $\sfQ_{(k)}\in \wedge^{k}(\CC^\gn)$ by
\begin{equation}
\sfQ_{(k)}=\displaystyle \sum_{|A|=k} \sfQ_A \zeta_A \quad \zeta_{a_1\dots a_k}:=\zeta_{a_1}\wedge \dots \wedge\zeta_{a_k}\,.
\end{equation}
Let $V_{(k)}(u)$ denote a $k$-dimensional linear subspace of $\CC^{\gn}$. Consider the collection 
\begin{equation}
V_{(0)}(u),\ V_{(1)}(u),\ \dots,\ V_{(\gn)}(u)
\end{equation}
with 
\begin{equation}
V_{(k)}(u) = \{\,\textbf{x}\, |\, \textbf{x}\wedge \sfQ_{(k)}(u)=0  \}\,.
\end{equation}
Then the QQ-relations are equivalent to the following union property 
\begin{equation}
V_{(k)}\cup V_{(k)}^{[-2]} = V_{(k+1)}
\end{equation}
for all $u\in\CC$ apart from possibly a discrete set of points.

\medskip

The Q-functions are clearly projective coordinates and so one is free to make the rescaling $\sfQ_A(u)\rightarrow f(u) \sfQ_A(u)$ for any function $f(u)$ without spoiling the Q-system. As a result, one can always use this rescaling freedom to set $\sfQ_\es=1$. 

\paragraph{Hodge dual}

The presentation of the Q-system using exterior forms naturally allows us to introduce a notion of Hodge duality for Q-functions. We define the Hodge dual map 
\begin{equation}
*:\wedge^{k}\left(\CC^\gn \right) \rightarrow \wedge^{\gn-k}\left(\CC^\gn \right)
\end{equation}
which can be used to transform $k$-forms to $\gn-k$-forms and hence define the Hodge dual Q-function $\sfQ^A$ of $\sfQ_A$ by 
\begin{equation}\label{dualQfn}
\sfQ^A = \varepsilon^{\bar{A}A}\sfQ_{\bar{A}}
\end{equation}
where $\bar{A}$ denotes the complement of $A$ in the set $\{1,2,\dots,\gn\}$ and we use the convention $\varepsilon^{12\dots \gn}=1$. Note that there is no sum over $\bar{A}$ in \eqref{dualQfn}. 

\paragraph{Q-system and Baxter equations}

The Q-functions entering the Q-system naturally appear as solutions of the finite-difference Baxter equations \cite{Krichever:1996qd,Kazakov:2007fy,Zabrodin:1996vm}. We have two finite-difference operators $\overleftarrow{\lO}$ and $\overrightarrow{\lO}$ where 
\begin{equation}\label{Baxtereqn}
\lO = \det\left(1-\bT \lD^{-1} \right) = \displaystyle \sum_{a=0}^\gn (-1)^a \T_{a,1}(u)\lD^{-a}\,.
\end{equation}
Since the transfer matrices mutually commute we could just as well consider \eqref{Baxtereqn} with the operators $\T_{a,1}$ replaced with their eigenvalues $\sfT_{a,1}$, also known as T-functions. Let us denote the $\gn$ independent solutions of $\overrightarrow{\lO}f^{[2]}(u)=0$ as $f(u)=\sfQ_1,\dots,\sfQ_\gn$ (the overall shift is for convenience). Since $\overrightarrow{\lO}$ has degree $\gn$ we can formally factorise it as 
\begin{equation}
\overrightarrow{\lO} = \left(1-\Lambda_\gn(u)\lD^{-1} \right)\dots  \left(1-\Lambda_1(u)\lD^{-1} \right)
\end{equation}
where $\Lambda_j$ are some functions to be determined. This is often known as a \textit{quantum Miura transform} \cite{Chervov:2009ck}. We can then fix the functions $\Lambda_j$ uniquely by the property that $\sfQ_1,\dots,\sfQ_j$ satisfy 
\begin{equation}
\left(1-\Lambda_j(u)\lD^{-1} \right)\dots  \left(1-\Lambda_1(u)\lD^{-1} \right)\sfQ^{[2]}_k=0,\quad k=1,\dots,j
\end{equation}
and hence 
\begin{equation}
\Lambda_j(u) = \frac{\sfQ_{1\dots j-1}^{[-2]}}{\sfQ_{1\dots j-1}}\frac{\sfQ_{1\dots j}^{[2]}}{\sfQ_{1\dots j}}.
\end{equation}
The action of the operator $\overrightarrow{\lO}$ on a function $f^{[2]}(u)$ can be conveniently expressed as a determinant in the $\gn$ solutions $\sfQ_1^{[2]},\dots,\sfQ_{\gn}^{[2]}$ as 
\begin{equation}
\overrightarrow{\lO}f^{[2]} = \frac{1}{\sfQ_\fs}\left| 
\begin{array}{cccc}
f^{[2]} & f & \dots & f^{[2(1-\gn)]} \\
\sfQ_1^{[2]} & \sfQ_1 & \dots & \sfQ_1^{[2(1-\gn)]} \\
\vdots & \vdots  & \ddots & \vdots \\
\sfQ_\gn^{[2]} & \sfQ_\gn & \dots & \sfQ_\gn^{[2(1-\gn)]} \\
\end{array}
\right|
\end{equation}
which allows us to conveniently express the T-functions $\sfT_{a,1}$ as 
\begin{equation}\label{antisymtran}
\sfT_{a,1}(u) = \frac{*\left(\sfQ_{(a)}^{[2]}\wedge \sfQ_{(\gn-a)}^{[-2a]} \right)}{\sfQ_{\fs}}
\end{equation}
where Hodge duality has been performed to convert $\gn$-forms to $0$-forms, i.e. functions. 

\medskip

We can now also demonstrate how the solutions $f(u)$ of the dual Baxter equation $f\,\overleftarrow{\lO}$ fit into the picture. In the factorised expression for the Baxter operator $\overrightarrow{\lO}$ the left-most factor is given by $(1-\Lambda_{\gn}\lD^{-1})$ and it is obvious that 
\begin{equation}
\frac{\sfQ^\gn}{\left(\sfQ^\fs \right)^{[2]}}(1-\Lambda_{\gn}\lD^{-1})=0
\end{equation}
and hence can be verified in general that 
\begin{equation}
\frac{\sfQ^i}{\left(\sfQ^\es \right)^{[2]}}\overleftarrow{\lO}=0\,.
\end{equation}
The presence of $\sfQ^\es$ in the denominator may look like there is an asymmetry between solutions for $\overrightarrow{\lO}$ and $\overleftarrow{\lO}$ but this is just a manifestation of that fact that we set $\sfQ_\es=1$ and in general we have
\begin{equation}
\frac{\sfQ^i}{\left(\sfQ^\es \right)^{[2]}}\overleftarrow{\lO}=0,\quad \overrightarrow{\lO}\frac{\sfQ_i^{[2]}}{\sfQ_\es}=0
\end{equation}
which follows from the fact that $\sfQ_{\es}$ can be restored in the QQ-relations by transforming 
\begin{equation}\label{restoreempty}
\sfQ_{A} \rightarrow \frac{\sfQ_A}{\sfQ_{\es}^{[-2|A|]}}\,.
\end{equation}

\paragraph{Q-system and quantum Weyl-character formula}

The simplest possible solution of the Q-system is the \textit{character} solution where all Q-functions are given by 
\begin{equation}
\sfQ_A = \lN_A \prod_{a\in A}z_a^{\frac{u}{\hbar}}
\end{equation}
where $\lN_A$ is a normalisation factor needed to ensure the QQ-relations are satisfied. It is a simple exercise to work out that 
\begin{equation}
\lN_A = \prod_{a< b}(z_a-z_b),\quad a,b\in A
\end{equation}
where we assume that $A$ is ordered in an increasing sequence. The character solution is referred to as such because it is closely related to the Weyl-character formula 
\begin{equation}
\chi_\lambda = \frac{\displaystyle \det_{1\leq i,j,\leq \gn}z_i^{\hat{\lambda}_j}}{\displaystyle\det_{1\leq i,j\leq \gn}z_i^{1-j}},\quad \hat{\lambda}_j = \lambda_j-j+1
\end{equation}
where $z_1,\dots,z_{\gn}$ are the eigenvalues of the $\GL(\gn)$ group element whose character we are computing. By using the character solution $\chi_\lambda$ can be expressed in terms of Q-functions
\begin{equation}
\chi_\lambda = \frac{\displaystyle \det_{1\leq i,j,\leq \gn}\sfQ_i^{[2\hat{\lambda}_j]}}{\sfQ_\fs},\quad \hat{\lambda}_j = \lambda_j-j+1\,.
\end{equation}
Since transfer matrices $\T_\lambda$ are quantisations of the characters $\chi_\lambda$ it is natural to conjecture that in general one also has 
\begin{equation}\label{Wronsk}
\sfT_\lambda(u) = \frac{\displaystyle \det_{1\leq i,j,\leq \gn}\sfQ_i^{[2\hat{\lambda}_j]}}{\sfQ_\fs}\,.
\end{equation}
This is indeed correct, and can be checked explicitly in the case of $\sfT_{a,1}$ where \eqref{Wronsk} reproduces \eqref{antisymtran}. By restoring $\sfQ_\es$ using the transformation \eqref{restoreempty} we have in general that
\begin{equation}\label{Wronsk2}
\sfT_\lambda(u) = \frac{\sfQ_{\es}^{[-2\gn]}}{\sfQ_\fs}\displaystyle \det_{1\leq i,j,\leq \gn}\left(\frac{\sfQ_i}{\sfQ_\es^{[-2]}}\right)^{[2\hat{\lambda}_j]}\,.
\end{equation}
\paragraph{Q-operators}
Since the functions $\sfT_\lambda$ are eigenvalues of the transfer matrices $\T_\lambda(u)$ one can naturally ask if the Baxter Q-functions $\sfQ_A$ are eigenvalues of some yet to be constructed Baxter Q-operators $\QQ_A(u)$. This is indeed the case and these Q-operators have been constructed using various different means, perhaps the most versatile of which corresponds to obtaining them as traces of mondodromy matricies satisfying the RTT relation but carrying infinite-dimensional representations of certain oscillator algebras in the auxiliary space \cite{Bazhanov:1996dr,Bazhanov:2010ts,Bazhanov:2010jq,Frassek:2011aa}. There have also been other constructions, such as defining them as the traces of certain factorised $R$-matrices \cite{Derkachov:2003qb} or through the elegant co-derivative formalism \cite{Kazakov:2010iu}. All of the relations we have written between T-functions and Q-functions continue to hold at the operatorial level, in particular \cite{Bazhanov:2010jq,Bazhanov:2010ts,Frassek:2011aa}
\begin{equation}
\T_\lambda(u) = \frac{\QQ_{\es}^{[-2\gn]}}{\QQ_\fs}\displaystyle \det_{1\leq i,j,\leq \gn}\left(\frac{\QQ_i}{\QQ_\es^{[-2]}}\right)^{[2\hat{\lambda}_j]}\,.
\end{equation}

\paragraph{Analytic structure of Q-functions}

For highest-weight representations of $\gl(\gn)$ it was demonstrated \cite{Frassek:2011aa} from the explicit construction of Baxter Q-operators that all Q-functions have the form 
\begin{equation}
\sfQ_A(u) = \lN_A \times z_A^{\frac{u}{\hbar}}\times \sfq_A(u) \times \prod_{j=1}^{|A|}\Gamma[\lambda_j^{[2(1-j)]}]
\end{equation}
where $\mathsf{q}_A(u)$ is a polynomial and we use the following convention for the $\Gamma$-function 
\begin{equation}
\Gamma[\nu_j(u)]:=\prod_{\alpha=1}^L \displaystyle\Gamma\left[\frac{u-\theta_\alpha-\hbar\, \nu^\alpha_j}{\hbar}\right]
\end{equation}
which in particular implies that if $f(u)$ is a monic polynomial in $u$ then 
\begin{equation}
\Gamma[f(u+\hbar)]:=f(u)\Gamma[f(u)]\,.
\end{equation}

\paragraph{Symmetries}

The Q-system admits a number of symmetries. The first are referred to as gauge transformations and correspond to
\begin{equation}
\sfQ_A(u) \rightarrow f_{|A|}(u) \sfQ_A(u)\,.
\end{equation}
The QQ-relations are preserved for functions $f_{|A|}$ satisfying 
\begin{equation}
f_{|A|+1}f_{|A|+2}^{[-2]} = f_{|A|+2}f^{[-2]}_{|A|}
\end{equation}
which can be solved by introducing two functions $h(u)$ and $g(u)$ with 
\begin{equation}
f_{|A|}(u)=\frac{h(u)}{g^{[-2|A|]}(u)}\,.
\end{equation}
The second class of symmetries are called H-rotations and correspond to 
\begin{equation}
\sfQ_\es(u) \mapsto \sfQ_\es(u),\quad \sfQ_a(u) \mapsto \sum_{b=1}^{\gn}H_{ab}\sfQ_b(u)
\end{equation}
where $H$ is some invertible $\gn\times \gn$ matrix. The transformation properties of all Q-functions then are given by 
\begin{equation}
\sfQ_{a_1 \dots a_j} \mapsto H_{a_1 b_1}\dots H_{a_j b_j} \sfQ_{b_1 \dots b_j}
\end{equation}
and we sum over repeated indices. Note that H-rotations are local -- they do not depend on the spectral parameter $u$.

\part{Separation of Variables}
 
\section{Separation of Variables in the classical XXX spin chain}\label{classical}

We are now ready to turn our attention to the main concept in this work -- separation of variables. 
We will begin with an overview of how separation of variables works in the classical XXX spin chain before constructing a quantum analogue. 

\subsection{Classical XXX spin chain}

The classical XXX spin chain is obtained from the quantum XXX spin chain in an appropriate $\hbar\rightarrow 0$ limit. We expand the $R$-matrix as
\begin{equation}
R(u,v) = 1 - \hbar\, r(u,v) +\mathcal{O}\left(\hbar^2 \right)
\end{equation}
where $r(u,v)$ is called the \textit{classical r-matrix} 
\begin{equation}
r_{ab}(u,v)=\frac{P_{ab}}{u-v}
\end{equation}
and then expand the RTT relation which provides the semi-classical formula 
\begin{equation}
[\bT_a(u),\bT_b(v)] = \hbar\, [r_{ab}(u,v),\bT_a(u)\bT_b(v)]\,.
\end{equation}
In the classical limit $[-,-]\rightarrow \hbar \{-,-\}$ we then obtain the defining relations of the classical XXX spin chain
\begin{equation}
\{ \bT_a(u),\bT_b(v)\} = [r_{ab}(u,v),\bT_a(u)\bT_b(v)]\,.
\end{equation}
which in component-form reads 
\begin{equation}
\{\bT_{ij}(u),\bT_{kl}(v) \}= \frac{1}{u-v}\left(\bT_{kj}(u)\bT_{il}(v)-\bT_{il}(u)\bT_{kj}(v)\right)
\end{equation}
where now $\bT_{ij}(u)$ are functions on an appropriate phase space. 

\bigskip

Representations of this algebra can be constructed in the same way as in the quantum case, namely by taking products of Lax operators. We define 
\begin{equation}
\lL^{(\alpha)}(u) = u\, 1 - \sfE^{(\alpha)}
\end{equation}
where $\sfE^{(\alpha)}$ is the $\gn\times \gn$ matrix whose $(i,j)$-th entry is $\sfE_{ji}^{(\alpha)}$ which are the generators of the classical $\gl(\gn)$ algebra 
\begin{equation}\label{classicalgln}
\{\sfE_{ij}^{(\alpha)},\sfE_{kl}^{(\beta)}\}=\delta^{\alpha\beta}\left(\delta_{jk}\sfE_{il}^{(\alpha)}-\delta_{li}\sfE_{kj}^{(\alpha)}\right)\,.
\end{equation}
The bracket $\{-,-\}$ denotes the Poisson bracket of the classical system and a realisation of it in terms of canonical variables will be given below. We can then construct, as in the quantum case, the classical monodromy matrix $\bT(u)$ as
\begin{equation}
\bT(u) = \lL^{(L)}(u-\theta_L)\dots \lL^{(1)}(u-\theta_1)G
\end{equation}
where $G$ is the twist matrix which is assumed to have distinct eigenvalues $z_1,\dots,z_\gn$. As was the case for the quantum model the twist simply corresponds to a deformation of the classical integrals of motion while still preserving integrability.

\paragraph{Phase space and local coordinates}
While it is possible to stick with the abstract Poisson structure introduced above it can be useful to have a concrete representation in mind. For example in $\gl(2)$ one could use
\begin{equation}\label{localcoords}
\begin{split}
\sfE_{11}& =\lambda_1 - q\,p\\
\sfE_{22}& =\lambda_2 + q\,p\\
\sfE_{12}& =p \\
\sfE_{21}& =(\lambda_1-\lambda_2) q -q^2 p
\end{split}
\end{equation}
where $q$ and $p$ are canonically conjugate coordinates $\{p,q\}=1$. This representation can be obtained from the quantum representation \eqref{gl2diffops1}-\eqref{gl2diffops3} in the classical limit $x\rightarrow q$, $\partial_x \rightarrow p$. By taking $\{p,q\}=1$ it is a simple computation to verify the realisation of $\sfE_{ij}$ in \eqref{localcoords} indeed satisfies the relation \eqref{classicalgln}.

\paragraph{Integrals of motion}
The classical spin chain is Liouville integrable -- we can construct $d$ independent integrals of motion $F_j$, $j=1,\dots,d$ which mutually Poisson commute
\begin{equation}
\{F_j,F_k \}=0\,.
\end{equation}
This is a trivial consequence of the fusion procedure. Indeed, all fused $R$-matrices $R^{\lambda\,\mu}$ can be shown by an easy calculation to have the structure 
\begin{equation}
R^{\lambda\,\mu}(u,v)=1 - \hbar\,r^{\lambda\,\mu}(u,v) + \mathcal{O}(\hbar^2)
\end{equation}
This then implies that 
\begin{equation}
\{ \bT_a^{\lambda}(u),\bT_b^{\mu}(v)\} = [r_{ab}^{\lambda\,\mu}(u,v),\bT_a^{\lambda}(u)\bT_b^{\mu}(v)]
\end{equation}
and as a result 
\begin{equation}
\{\sfT_\lambda(u),\sfT_\mu(v)\}=0,\quad \sfT_\lambda(u) = {\rm tr}_a\, \bT^\lambda_a(u)\,.
\end{equation}
For generic values of the twist eigenvalues all integrals of motion are independent \cite{Sklyanin:1992eu}. Since each $\sfT_{a,1}$ is a polynomial of degree $u^{aL}$ the total number of integrals of motion is $\frac{\gn}{2}(\gn-1)$ which matches half the dimension of the phase space hence the model is integrable. Note that any of the integrals of motion $F_k$ can be considered the Hamiltonian of the model and generate a family of commuting flows where $F_k$ generates shifts in the time $t_k$. If one is to use the local coordinates \eqref{localcoords} then it can be checked explicitly, for example for $\gn=2$ and length $L=2$, that the integrals of motion obtained from the expansion of $\T_{1,1}(u)$ are indeed independent for $z_1\neq z_2$. Writing the expansion of the transfer matrix $\sfT_{1,1}(u)$ as 
\begin{equation}
\sfT_{1,1}(u)={\rm tr}G\,u^2 - F_1\,u+F_2
\end{equation}
then an explicit calculation yields 
\begin{equation}
F_1 = z_1 \mathcal{E}_{11} +z_2 \mathcal{E}_{22},\quad \mathcal{E}_{jj} = \sfE_{jj}^{(1)}+\sfE_{jj}^{(2)}
\end{equation}
and 
\begin{equation}
F_2 = z_1\left(\sfE_{11}^{(2)} \sfE_{11}^{(1)} + \sfE_{21}^{(2)} \sfE_{12}^{(1)} \right)+z_2\left(\sfE_{12}^{(2)} \sfE_{21}^{(1)} + \sfE_{22}^{(2)} \sfE_{22}^{(1)} \right)
\end{equation}
where $\sfE_{ij}^{(\alpha)}$ can be obtained from the coordinate representation \eqref{localcoords} by simply replacing $(p,q) \rightarrow (p^\alpha,q^\alpha)$. In the case that $z_1=z_2$ then the transfer matrix only provides one of the two integrals of motion needed to ensure integrability since in this case $F_1$ becomes proportional to one of the Casimir operators generated by $\det\bT(u)$. The global Cartan charge $\mathcal{E}_{11}$ provides the second needed integral of motion to ensure integrability. The independence of the two integrals of motion can be directly checked in the coordinate representation by verifying that the one-forms ${\rm d}F_1$ (or ${\rm d}\lE_{11}$) are linearly independent.

\subsection{Spectral curve and action-angle variables} 

A powerful technique for solving classical integrable systems is based the spectral equation 
\begin{equation}\label{speceqn}
\Gamma:\ \det\left(\lambda(u)-\bT(u)\right)=0
\end{equation}
of the monodromy matrix.  For simplicity we will restrict our attention to the case $\gn=2$. In this case the spectral curve $\Gamma$ defines a 2-sheeted covering of the complex plane -- a point $P$ on this curve is a pair $(u,\lambda(u))$ and to each point $u\in\CC$ there exists $2$ points $(\lambda_+(u),\lambda_-(u))$ corresponding to the two eigenvalues $\lambda_\pm(u)$ of $\bT(u)$. 

\medskip

The eigenvalues $\lambda_{\pm}(u)$ are completely fixed in terms of the integrals of motion -- by expanding \eqref{speceqn} we obtain
\begin{equation}
\lambda^2(u) - \sfT_{1,1}(u)\lambda(u) + \sfT_{2,1}(u)=0,\quad \lambda(u) = \lambda_\pm(u)
\end{equation}
and
\begin{equation}
\lambda_{\pm}(u)=\frac{1}{2}\left(\sfT_{1,1}(u)\pm \Delta(u)\right)
\end{equation}
where we have introduced $\Delta(u)=\sqrt{\sfT_{1,1}(u)^2-4\, \sfT_{2,1}(u)}$. The spectral equation encodes all kinematical information -- the values of the conserved charges. All dynamical information is contained in the eigenvectors of the monodromy matrix as we will now see.

\paragraph{Baker-Akhiezer function}
The Baker-Akhiezer function $\Omega(u)$ which can be viewed as a function on the spectral curve $\Gamma$ is eigenvector of monodromy matrix 
\begin{equation}\label{eigenveceqn}
\bT(u)\Omega^\pm(u) = \lambda_\pm(u)\Omega^{\pm}(u),\quad \Omega^\pm(u) = \left( \begin{array}{c}
\Omega_1^\pm(u) \\
\Omega_2^\pm(u)
\end{array}\right)\,.
\end{equation}
The eigenvector is not yet completely fixed and we need to impose a normalisation. For simplicity we impose $\Omega_1^\pm(u)=1$ and so the eigenvalue equation \eqref{eigenveceqn} implies
\begin{equation}
\Omega_2^\pm(u)= \frac{\lambda_\pm(u)-\bA(u)}{\bB(u)} = \frac{\bC(u)}{\lambda_\pm(u)-\bD(u)}\,.
\end{equation}
Hence, the poles of the Baker-Akhiezer function may be given by the zeroes of the function $\bB$ and the points where $\lambda(u)-\bD(u)=0$. Let $\svx^\alpha$ be such that $\bB(\svx^\alpha)=0$. At this point the monodromy matrix $\bT(\svx^\alpha)$ becomes upper triangular and hence its eigenvalues are given by $\bA(\svx^\alpha)$ and $\bD(\svx^\alpha)$. Hence, at one of the two points above $\svx^\alpha$ on the spectral curve $\lambda(u)-\bA(u)$ must vanish and so $\Omega$ must have only one pole at the point $P_\alpha$ with coordinates $(\svx^\alpha,\sfp^\alpha=\bD(\svx^\alpha))$. The set of points $P_\alpha$, $\alpha=1,\dots,L$ label the so-called \textit{dynamical divisor} \cite{babelon_bernard_talon_2003}.

\medskip

A straightforward calculation with the RTT relation allows one to easily deduce that the coordinates $(\svx^\alpha,\log\,\sfp^\alpha)$ are canonically conjugate \cite{babelon_bernard_talon_2003,10.1007/3-540-15213-X_80}
\begin{equation}
\{\svx^\alpha,\svx^\beta\}=0=\{\sfp^\alpha,\sfp^\beta\},\quad \{\svx^\alpha, \log\, \sfp^\beta \}=\delta^{\alpha\beta}
\end{equation}
and hence the coordinates $(\svx^\alpha,\log\,\sfp^\alpha)$ form a set of separated variables. We now demonstrate how the equations of motion for the separated variables linearise. We have $L$ independent Hamiltonians at our disposal contained in $\sfT_{1,1}(u)$ each generating independent flows. For a function $F$ on the phase space let $\dot{F}$ denote the evolution along a generic flow 
\begin{equation}
\dot{F} = \{\sfT_{1,1}(u),F \}\,. 
\end{equation}
By using the RTT relations we can calculate the time evolution of the separated coordinates $\svx^\alpha$ to be given by
\begin{equation}
\dot{\svx}^\alpha = \left(\bA(\svx^\alpha)-\bD(\svx^\alpha) \right)\frac{\bB(u)}{(u-\svx^\alpha)\bB'(\svx^\alpha)}\,.
\end{equation}
Note that at the points $\svx^\alpha$ we have $\bA(\svx^\alpha)\bD(\svx^\alpha)=\sfT_{2,1}(\svx^\alpha)$ and hence $\bA(\svx^\alpha)-\bD(\svx^\alpha)=-\Delta(\svx^\alpha)$ and so the evolution equation for $\svx^\alpha$ reads
\begin{equation}\label{SoVtime}
\dot{\svx}^\alpha = -\Delta(\svx^\alpha)\frac{\bB(u)}{(u-\svx^\alpha)\bB'(\svx^\alpha)}\,.
\end{equation}
Clearly $\dot{\svx}^\alpha$ is a polynomial of degree $L-1$ with zeroes at $\svx^\beta$, $\beta\neq \alpha$ and $\dot{\svx}^\alpha(\svx^\alpha)=-\Delta(\svx^\alpha)$.  

\medskip

Linearisation of the equations of motion \eqref{SoVtime} is then achieved by introducing the \textit{Abel maps} $\lA_\alpha$ \cite{babelon_bernard_talon_2003} defined by
\begin{equation}\label{abelmap}
\lA_\alpha = \displaystyle \sum_{\beta=1}^L \int_{P_0}^{P_\beta} \omega_\alpha
\end{equation}
where $P_0$ is some base-point and $\omega_\alpha$ are Abelian differentials 
\begin{equation}
\omega_\alpha = \frac{u^\alpha}{\Delta(u)}{\rm d}u,\quad \alpha=1,\dots,L\,.
\end{equation}
It can then be easily worked out \cite{babelon_bernard_talon_2003} that the evolution equations \eqref{SoVtime} are equivalent to the linear evolution equations 
\begin{equation}\label{linearisation}
\dot{\lA}_\alpha = u^\alpha\,.
\end{equation}
Indeed, the time evolution of $\lA_\beta$ is worked out from \eqref{abelmap} to be 
\begin{equation}
\dot{\lA}_\beta(u) = \sum_{\beta=1}^L \frac{(\svx^\alpha)^\beta \dot{\svx}^\alpha}{\Delta(\svx^\alpha)}
\end{equation}
which when combined with \eqref{SoVtime} shows that $\dot{\lA}_\alpha$ is a polynomial in $u$ of degree $L-1$. It can then be fully reconstructed using Lagrange interpolation by evaluating it at the points $\svx^\beta$, $\beta\neq \alpha$ which leads to \eqref{linearisation}.

\subsection{Quantisation}

Having discussed separation of variables in the classical XXX spin chain we now turn to the quantum model. We saw in Section \ref{firstsov} that the Bethe algebra wave functions, in the basis diagonalising the $\bB$ operator, were given by a simple product of Q-functions
\begin{equation}
\Psi(\svx) = \prod_{\alpha=1}^L \sfq_1(\svx^\alpha)\,.
\end{equation}
We will now motivate this from a different perspective, namely as a natural consequence of canonically quantizing the classical XXX spin chain in the separated variables, see also \cite{Babelon:2002wt}. The canonical quantisation prescription is given by 
\begin{equation}
\svx\mapsto \svX,\quad \mathsf{p}\mapsto\mathsf{P}
\end{equation}
subject to the commutation relations
\begin{equation}
[\mathsf{P},\svX]=\hbar\, \mathsf{P}
\end{equation}
and so $f(\svx,\sfp)\mapsto f(\svX,\mathsf{P})$ and we implicitly assume some normal ordering prescription, for example we choose to place all $\svX$'s to the left of all $\mathsf{P}$'s. Hence, in the coordinate representation of the separated variables $\svx$ we have
\begin{equation}
\mathsf{P} \psi(\svx) = \Delta(\svx)e^{-\hbar \partial_\svx}\,\psi(\svx)
\end{equation}
where $\Delta(\svx)$ is an expected cocycle factor and the classical equation $H(\svx,\sfp)=E$ is then replaced with the eigenvector equation
\begin{equation}
\left(H(\svx,e^{-\hbar\partial_{\svx}})-E\right)\psi(\svx)=0\,.
\end{equation}
The cocycle factor is present due to the fact that for finite-dimensional models the allowed range of values of $\svx$ must be finite and $\Delta(\svx)$ is present to ensure that the action of $\mathsf{P}$ on the wave function will eventually be zero ensuring the space of states is finite dimensional. 

For the $\gl(2)$ spin chain the separated equations of motion had the form 
\begin{equation}
1 - \sfT_{1,1}(\svx^\alpha)\mathsf{P}^\alpha+\sfT_{2,1}(\svx^\alpha)\mathsf{P}^{\alpha\,2}=0,\quad \alpha=1,2,\dots,L\,.
\end{equation}
Under the canonical quantisation prescription this classical equation is then replaced by the Schrodinger equation
\begin{equation}
\left(1 - \sfT_{1,1}(\svx^\alpha)\Delta(\svx^\alpha)e^{-\hbar\partial_{\svx^\alpha}}+\sfT_{2,1}(\svx^\alpha)\Delta(\svx^\alpha)\Delta^{[-2]}(\svx^\alpha)e^{-2\hbar\partial_{\svx^\alpha}}\right)\psi(\svx^\alpha)=0
\end{equation}
where now $\sfT_{a,1}$ denote the eigenvalues of integrals of motion on the state $\psi$. By choosing the cocycle factor appropriately this Schrodinger equation is none other than the Baxter equation \eqref{TQ1} with $\psi(\svx^\alpha)=\sfq^{[2]}(\svx^\alpha)$. 

\medskip

All information about the eigenvalues of the integrals of motion can be reconstructed from the $L$ separated equations. Since $\T_{2,1}(u)$ takes the same value for every state it is non-dynamical and so the only dynamical information comes from $\sfT_{1,1}(u)$. Since $\sfT_{1,1}(u)$ is a polynomial of degree $L$ with asymptotics $\sfT_{1,1}(u)\sim \chi_{1,1}u^L$ it is fixed by its value at $L$ distinct points, say $\svx^\alpha$, and can be fully reconstructed by Lagrange interpolation. The $L$ separated equations provide the values $\sfT_{1,1}(\svx^\alpha)$. The full wave function $\Psi(\svx)=\prod_{\alpha=1}^L \psi_\alpha(\svx^\alpha)$ then satisfies 
\begin{equation}
\T_{1,1}(u)\Psi(\svx) = \sfT_{1,1}(u)\Psi(\svx),\quad \Psi(\svx)=\prod_{\alpha=1}^L \sfq_1(\svx^\alpha)
\end{equation}
where $\T_{1,1}$ is now considered as a normal-ordered operator in $\svX$ and $\svP$. 

\bigskip

We have just seen that wave functions of the XXX spin chain are given by Baxter Q-functions in a natural way arising from the canonical quantisation of classical separated variables. This quantisation however assumes some square-integrability properties on the wave functions $\psi(\svx)$ is not justified for finite-dimensional spin chains. On the other hand, we saw by direct construction in the $\gl(2)$ case that the wave functions were still given by Q-functions and we will now proceed with developing this further. 

\section{Towards separation of variables for higher rank}

The problem of generalising the previously outlined classical construction for $\gl(\gn)$ spin chains was initiated by Sklyanin in \cite{Sklyanin:1992eu} for $\gl(3)$ spin chains. Sklyanin constructed the classical $\bB$ function given by 
\begin{equation}\label{ClassicalB}
\bB(u) = \bT_{23}\bT\left[^{12}_{23}\right]+\bT_{13}\bT\left[^{12}_{13}\right]
\end{equation}
where the minors \eqref{quantumminor} are now \textit{classical} minors which commute with all other functions and do not possess shifts in $\hbar$. By constructing a suitable generalisation of the $D(u)$ function used in the $\gl(2)$ case it was demonstrated that these operators provide a set of canonically conjugate variables precisely as in the $\gl(2)$ case. The generalisation to classical $\gl(\gn)$ spin chains was later carried out by Scott \cite{Scott:1994dz} and Gekhtman \cite{gekhtman1995}.

\medskip

The quantisation of the classical separated variables for $\gl(3)$, resulting in a quantum $\bB$ operator, was carried out by Sklyanin in \cite{Sklyanin:1992sm} at a formal operatorial level without appealing to a specific representation and the separated Baxter equations were derived. A generalisation of this construction was subsequently performed by Smirnov in \cite{2001math.ph...9013S} for the case of $U_q(\widehat{\sla(\gn)})$. Sklyanin's expression for the $\bB$ operator of $\lY_3$  is given explicitly by
\begin{equation}\label{Bgl3}
\bB(u) = \bT_{23}\bT^{[2]}\left[^{12}_{23}\right]+\bT_{13}\bT^{[2]}\left[^{12}_{13}\right]
\end{equation}
which coincides with the classical expression \eqref{ClassicalB} up to quantum corrections, that is shifts in $\hbar$. It is worth noticing that expression \eqref{Bgl3} is \textit{cubic} in Yangian generators $\bT_{ij}$ in contrast to the $\lY_2$ case where $\bB=\bT_{12}$ is only linear. 

\medskip

There are however some unresolved questions relating to these higher-rank construction. The most pressing is how to actually construct the factorised wave functions and the SoV basis for some concrete representations.  

\subsection{Initial advancements}

A crucial observation was made about $\bB$ in \cite{Gromov:2016itr} which shed a lot of light on this issue. It was shown that for spin chains carrying the defining representation of $\gl(3)$ the eigenvectors $\ket{\Psi}$ of the transfer matrix could be constructed as 
\begin{equation}\label{Bgeneratesstates}
\ket{\Psi}=\prod_{j=1}^M \bB(u_j)\ket{\Omega}
\end{equation}
where $u_j$ are the roots of the Q-function $\sfQ_1$ of the $\gl(3)$ Q-system. This formula is remarkable -- it shows that one only needs a \textit{single} operator $\bB$ to generate transfer matrix eigenstates in stark contrast to the nested Bethe ansatz approach \cite{Slavnov:2019hdn} where one needs to use a family of operators in a complicated nesting procedure. The formula \eqref{Bgeneratesstates} was proven in \cite{Gromov:2016itr} for $\gl(3)$ for spin chains of any length $L$ for states with $1$ and $2$ magnons (excitations above the vacuum), and further checks were carried out numerically for larger numbers of magnons. A full proof for any number of magnons was subsequently obtained in \cite{Liashyk:2018qfc} in the framework of the nested Bethe ansatz. 

\medskip

It may seem surprising however, and perhaps even somewhat contradictory, that the transfer matrix eigenstates can be generated entirely using the Bethe roots of $\sfq_1$ despite the fact that the nested Bethe ansatz also requires one to use the roots of other Q-functions explicitly which also appear in the transfer matrix eigenvalues 
\begin{equation}\label{Tsu3eigenvalue}
\sfT_{1,1}(u) = z_1 Q_\theta^{[-2]} \frac{\sfq_1^{[2]}}{\sfq_1}+z_2 Q_\theta \frac{\sfq_1^{[-2]}}{\sfq_1}\frac{\sfq_{12}^{[2]}}{\sfq_{12}}+z_3 Q_\theta \frac{\sfq_{12}^{[-2]}}{\sfq_{12}}
\end{equation}
as the roots of the polynomial $\sfq_{12}$. Here we have introduced the polynomial $Q_\theta(u) = \prod_{\alpha=1}^L (u-\theta_\alpha)$ as for the defining representation we have $\nu_1(u)=Q_\theta^{[-2]}(u)$ and $\nu_{2,3}(u)=Q_\theta(u)$. This was argued to be consistent in \cite{Liashyk:2018qfc} at the level of Bethe equations where it was shown that the full set of nested Bethe ansatz equations could be expressed solely as equations on the roots of $\sfq_1$. Here we will present an alternative proof of this fact. Since the transfer matrix eigenvalue is a polynomial of degree $L$ with asymptotics 
\begin{equation}
\sfT_{1,1}(u)\sim (z_1+z_2+z_3)u^L
\end{equation}
it is fixed by its value at $L$ distinct points, say $\theta_\alpha$, $\alpha=1,\dots,L$, using Lagrange interpolation
\begin{equation}
\sfT_{1,1}(u) =(z_1+z_2+z_3)\prod_{\alpha=1}^L(u-\theta_\alpha)+\displaystyle\sum_{\alpha=1}^L\prod_{\beta\neq \alpha} \frac{u-\theta_\beta}{\theta_\alpha-\theta_\beta}\sfT_{1,1}(\theta_\alpha)\,.
\end{equation}
At these points the eigenvalue \eqref{Tsu3eigenvalue} reduces to 
\begin{equation}
\sfT_{1,1}(\theta_\alpha) = z_1 Q_\theta^{[-2]}(\theta_\alpha) \frac{\sfq_1^{[2]}(\theta_\alpha)}{\sfq_1(\theta_\alpha)}
\end{equation}
which only contains the roots of $\sfq_1$.

\medskip

The eigenvalues of $\bB$ were also explicitly computed in \cite{Gromov:2016itr} and yield, for $\gl(3)$ in the defining representation,
\begin{equation}\label{Beigenvalues1}
\kappa\prod_{\alpha=1}^L(u-\svx^\alpha_{1})(u-\svx^\alpha_{2})(u-\theta_\alpha+\hbar),\quad \svx^\alpha_{j} = \theta_\alpha+\hbar\,\lambda^\alpha_j
\end{equation}
where $\lambda^\alpha_j=0,1$ with $\lambda^\alpha_1\leq \lambda^\alpha_2$ labels all possible eigenvectors and $\kappa$ is some normalisation. We see that $\bB$ contains an overall trivial factor $\displaystyle\prod_{\alpha=1}^L(u-\theta_\alpha+\hbar)$ which can be factored out allowing us to define $\bb(u)$ by 
\begin{equation}
\bB(u) = \kappa\,\bb(u) \displaystyle\prod_{\alpha=1}^L(u-\theta_\alpha+\hbar)\,.
\end{equation}
The Bethe algebra eigenstates $\ket{\Psi}$ can then be constructed as 
\begin{equation}
\ket{\Psi} = \prod_{j=1}^M \bb(u_j) \ket{\Omega}
\end{equation}
and hence in the basis $\bra{\svx}$ diagonalising $\bB$ we have
\begin{equation}
\braket{\svx|\Psi} = \prod_{\alpha=1}^L \sfq_1(\svx^\alpha_1)\sfq_1(\svx^\alpha_2)
\end{equation}
demonstrating separation of variables. 

\medskip

The preceding remarks were also generalised to $\gl(\gn)$ in \cite{Gromov:2016itr} where the following $\bB$ operator was proposed 
\begin{equation}\label{Bgln}
\bB(u)=\sum_{J_1,\dots,J_{\gn-1}}T\left[^{J_1}_\gn\right]T^{[2]}\left[^{J_2}_{J_1\, \gn}\right]T^{[4]}\left[^{J_3}_{J_2\, \gn}\right]\dots T^{[2n-4]}\left[^{J_{\gn-1}}_{J_{n-2}\,\gn}\right]
\end{equation}
where $J_k$ is a multi-index with 
\begin{equation}
J_k=(j_k^1,\dots,j_k^k)
\end{equation}
and we sum over configurations with $1\leq j_k^1 < j_2^2 <\dots <j_k^k\leq \gn$. 

\medskip

The form of this operator was based on several requirements 
\begin{enumerate}
\item It is constructed as a sum of products of quantum minors of increasing size.
\item In the classical $\hbar \rightarrow 0$ limit the known classical expressions \cite{Scott:1994dz,gekhtman1995} should be reproduced.
\item $[\bB(u),\bB(v)]=0$.
\item The transfer matrix eigenstates can be generated by repeated action of $\bB(u)$ on the transfer matrix vacuum state. 
\end{enumerate}

\medskip

Further developments were made in \cite{Maillet:2018bim} where an explicit construction of states $\bra{\svx}$ was proposed and is based on the following observation. Suppose for some generic enough covector $\bra{S}$ we can generate a basis of the representation space by repeatedly acting with conserved charges, for example by repeatedly acting with transfer matrices evaluated at some specific points. For example, for $\gl(2)$ spin chains carrying $L$ copies of the defining representation one could take 
\begin{equation}\label{SoVbasis1}
\bra{n_1,\dots,n_L}:=\bra{S}\prod_{\alpha=1}^L \T_{1,1}(\theta_\alpha)^{n_\alpha},\quad n_\alpha=0,1\,.
\end{equation}
The main feature of this is that if \eqref{SoVbasis1} forms a basis then the wave functions $\ket{\Psi}$ of the transfer matrix $\T_{1,1}$ immediately factorise
\begin{equation}
\T_{1,1}(u)\ket{\Psi}=\sfT_{1,1}(u)\ket{\Psi} \ \rightarrow \ \braket{n_1,\dots,n_L|\Psi} =\prod_{\alpha=1}^L \sfT_{1,1}(\theta_\alpha)^{n_\alpha}
\end{equation}
where we have chosen to normalise $\braket{0,\dots,0|\Psi}=1$ for convenience. 

\medskip

This construction is extremely powerful. It doesn't just provides a convenient mechanism for constructing a basis of factorised wave functions but can also be used to completely solve the integrable system. An important ingredient in the construction comes from the so-called closure relation
\begin{equation}
\T_{1,1}(\theta_\alpha)\T_{1,1}(\theta_\alpha+\hbar)=\T_{2,1}(\theta_\alpha+\hbar)=z_1 z_2 Q_\theta^{[-2]}(\theta_\alpha)Q_\theta^{[2]}(\theta_\alpha)
\end{equation}
which is a consequence of the Hirota equation. The closure relation allows one to completely characterise the action of the transfer matrix $\T_{1,1}(u)$ on the basis \eqref{SoVbasis1} since
\begin{equation}
\bra{\dots n_\alpha=0\dots}\T_{1,1}(\theta_\alpha) = \bra{\dots n_\alpha=1\dots}
\end{equation}
and hence
\begin{equation}
\bra{\dots n_\alpha=1\dots}\T_{1,1}(\theta_\alpha+\hbar) = z_1 z_2 Q_\theta^{[-2]}(\theta_\alpha)Q_\theta^{[2]}(\theta_\alpha)\bra{\dots n_\alpha=0\dots}\,.
\end{equation}
Hence, by using Lagrange interpolation we can determine the action of $\T_{1,1}(u)$ on any state of the form \eqref{SoVbasis1}. From here, one can completely characterise the spectrum of the conserved charges and re-derive various known tools of integrability such as the Baxter TQ equation. 

\medskip

It was proven in \cite{Maillet:2018bim} that \eqref{SoVbasis1} indeed forms a basis for a generic enough $\bra{S}$. Furthermore, by choosing $\bra{S}$ appropriately the basis \eqref{SoVbasis1} even diagonalises Sklyanin's $\bB$ operator. This was proven in \cite{Maillet:2018bim} for $\gl(2)$ models and demonstrated for chains of small length for $\gl(3)$. Hence, this approach seems to naturally complement that of \cite{Gromov:2016itr} where it is simple to build transfer matrix eigenstates but not clear how to build eigenstates of $\bB$ and the interplay between these two approaches is one of the key focuses of this work. It is important to stress however that the construction proposed by Maillet and Niccoli is independent of the existence of any form of $\bB$ operator and works in cases where the construction of a $\bB$ operator is less clear, such as in supersymmetric spin chains \cite{Gromov:2018cvh,Maillet:2019ayx}, or in the long-standing case of the XXZ spin chain with diagonal twist \cite{Maillet:2018bim} which has previously only been amenable to SoV techniques with so-called anti-periodic boundary conditions \cite{Niccoli:2012ci,Niccoli:2014sfa,Levy-Bencheton:2015mia,pei2020scalar}. On the other hand, in the cases where Sklyanin's $\bB$ operator \textit{is} available the two approaches are complementary. Since the basis \eqref{SoVbasis1} is not unique -- since one can change the reference vector $\bra{S}$ and even the conserved charges used -- there is a question of what a natural choice for the SoV basis is. The simplicity of the closure relations provides one notion of naturalness but even this does not single out the SoV basis uniquely as was demonstrated in \cite{Maillet:2019nsy} for higher-spin representations of $\gl(2)$. Having an explicit operator, the $\bB$ operator, easily constructable within the given quantum algebra provides another notion of naturalness. 

\medskip

Having discussed some initial facts about SoV for higher rank we will now begin a more in-depth analysis, focusing on the interplay between Sklyanin's approach using the operator $\bB$ and the basis construction proposed in \cite{Maillet:2018bim}. 

\subsection{Designing a good twist}

As was already mentioned, the explicit expression for $\bB$ in terms of the bare monodromy matrix elements $T_{ij}$ depends on the specific choice of twist and the resulting operator can be more or less complicated depending on the twist used. In this section we will design a twist with the objective to make $\bB$ as simple as possible. For the moment we will focus our attention on the defining representation in the physical space. For a generic choice of twist the $\bB$ operator is rather complicated. On the other hand, it was demonstrated to be diagonalisable in \cite{Gromov:2016itr} with very simple eigenvalues \eqref{Beigenvalues1}. Hence, there exists a basis in the representation where, after normalising $\bB$ to be a monic polynomial in $u$, we have 
\begin{equation}\label{Bdiagonal}
\bB(u) = (u-\theta+\hbar)
\left( 
\begin{array}{ccc}
(u-\theta-\hbar)^2 & 0 & 0 \\
0 & (u-\theta-\hbar)(u-\theta) & 0 \\
0 & 0 & (u-\theta)^2
\end{array}
\right)\,.
\end{equation}
One may naturally wonder what the significance of the property $\bB(\theta-\hbar)=0$ is. After the next section its meaning will be clear and so we postpone discussion of this fact until then, see immediately below \eqref{eq:variablesx}.

\medskip

We now examine the transfer matrix $\T_{1,1}(u)$ for length $L=1$ which is given by 
\begin{equation}
\T_{1,1}(u)= {\rm tr}_a \left(\lL(u-\theta)G\right)
\end{equation}
and by explicitly performing the trace we end up with
\begin{equation}
\T_{1,1}(u)=(u-\theta){\rm tr}G\times 1-\hbar\, G\,.
\end{equation}
Our aim is to find a twist $G$ such that $\bB$ is given by \eqref{Bdiagonal}. We now demand that the transfer matrix has eigenvectors 
\begin{equation}
\ket{\Omega},\quad \bB(u_1)\ket{\Omega},\quad \bB(u_1)\bB(u_2)\ket{\Omega}
\end{equation}
where $\ket{\Omega}$ is a generic column vector corresponding to the $\T_{1,1}$ vacuum eigenvalue 
\begin{equation}
z_1 Q_\theta^{[-2]}(u) + (z_2+z_3)Q_\theta(u)
\end{equation}
and $\{u_1 \}$ and $\{u_1,u_2\}$ are roots of $\sfq_1$ for two different states and $\bB$ is given by its diagonal representation \eqref{Bdiagonal}. Since the transfer matrix eigenvalues are also known, by explicit computation or by plugging the Bethe roots into Baxter TQ relations the transfer matrix and hence the twist $G$ can be totally reconstructed which follows from the simple fact that a diagonalisable matrix is completely determined by its eigenvectors and eigenvalues. We point out that this procedure is in fact more straightforward than the naive approach of simply computing $\bB$ with generic twist and solving for the entries of the twist to reproduce \eqref{Bdiagonal} due to the fact that $\bB$ is a polynomial in twist matrix entries of degree $3$ and so solving the resulting equations is very difficult. Instead, with the approach outlined above one only needs to solve linear equations for the transfer matrix entries. 

\medskip

The conclusion is that the twist is fixed to be 
\begin{equation}
G = \left( 
\begin{array}{ccc}
\chi_1 & -\frac{\chi_2}{w_1} & \frac{\chi_3}{w_1 w_2} \\
w_1 & 0 & 0 \\
0 & w_2 & 0 \\
\end{array}
\right)
\end{equation}
where $\chi_1$, $\chi_2$ and $\chi_3$ are the elementary symmetric polynomials in the twist eigenvalues $z_1,z_2,z_3$
\begin{equation}
\begin{split}
& \chi_1 = z_1+z_2 +z_3 \\
& \chi_2 = z_1 z_2 + z_1 z_3 + z_2 z_3 \\
& \chi_3 = z_1 z_2 z_3
\end{split}
\end{equation}
and the parameters $w_1,w_2$ are arbitrary and can take any non-zero value. Furthermore the eigenvalues of $G$ do not depend on their value and they can always be set to $1$ by means of a diagonal change of basis. Hence, we refer to them as \textit{auxiliary twist parameters}. Instead of setting them to $1$ we will keep them arbitrary for now as they will be useful later.

\medskip

The procedure for determining $G$ can be repeated for $\gn=2$ and $\gn=4$ leading to the following forms of $G$ in the cases $\gn=2,3,4$:

\begin{equation}
\label{eq:growth}
\left(\begin{array}{cc}
\chi_1 & -\frac{\chi_2}{w_1} \\
w_1 & 0
\end{array} \right),\quad
\left(\begin{array}{ccc}
\chi_1 & -\frac{\chi_2}{w_1} & \frac{\chi_3}{w_1 w_2} \\
w_1 & 0 & 0 \\
0 & w_2 & 0
\end{array} \right),\quad
\left(\begin{array}{cccc}
\chi_1 & -\frac{\chi_2}{w_1} & \frac{\chi_3}{w_1 w_2} & -\frac{\chi_4}{w_1 w_2 w_3} \\
w_1 & 0 & 0 & 0\\
0 & w_2 & 0 & 0\\
0 & 0 & w_3 & 0
\end{array} \right)\,.
\end{equation}

Hence, we are naturally led to use the following matrix $G$ for $\gl(\gn)$ with components $G_{ij}$ given by:
\begin{equation}\label{MCTcomponent}
G_{ij}=\frac{\chi_j \delta_{i1}}{w_{|j-1|}}+\delta_{i,j+1}w_j,\quad w_{|j|}:=(-1)^{j}\prod_{k=1}^j w_k\,.
\end{equation}
For the special case $w_j=1$, $j=1,2,\dots,\gn$ the matrix \eqref{MCTcomponent} is known as the \textit{companion matrix} $C$ for the eigenvalues $z_1,\dots,z_\gn$. A rather nice feature of the companion matrix is that it generates the standard basis of (the dual space of) $\CC^\gn$. Indeed, letting $\bra{\sfe_1},\dots,\bra{\sfe_\gn}$ denote the standard basis such that $\bra{\sfe_j}$ has $1$ in position $j$ and $0$ everywhere else then 
\begin{equation}
\bra{\sfe_{j-1}}=\bra{\sfe_j}C,\quad j=2,3,\dots,\gn\,.
\end{equation}
An immediate corollary of this is that any matrix with pairwise-distinct eigenvalues can generate a basis by repeatedly acting on some generic enough reference vector and it is this which forms the basis, excusing the pun, for the construction of Maillet and Nicolli \cite{Maillet:2018bim}. 

\medskip

Since the matrix \eqref{MCTcomponent} only differs from the companion matrix by the presence of the additional parameters $w_1,\dots,w_{\gn-1}$ we naturally refer to it as the \textit{modified companion} twist matrix, or MCT matrix.

\paragraph{Computing $\bB$}
Having identified a twist which seems to result in a rather simple form of $\bB$ we will now compute it directly. The main property we will use is the transformation law for quantum minors. That is if $\bT(u)=T(u)G$ then 
\begin{equation}
\bT\left[^{i_1\dots i_a}_{j_1\dots j_a} \right]=T\left[^{i_1\dots i_a}_{k_1\dots k_a} \right]G_{k_1 j_1}\dots G_{k_a j_a}
\end{equation}
and we sum over all $k_i\in\{1,2,\dots,\gn\}$. Examining the quantum minors $\bT\left[^{J_{a+1}}_{J_{a}\,\gn}\right]$ which make up $\bB$ we have
\begin{equation}\label{minorsum}
\bT\left[^{J_{a+1}}_{J_{a}\,\gn}\right] = T\left[^{J_{a+1}}_{k_1\dots k_{a+1}}\right]G_{k_1 j_a^1}\dots G_{k_a j_a^a}G_{k_{a+1} \gn}\,.
\end{equation}
We now exploit the properties of the MCT matrix. Most importantly, we have 
\begin{equation}
G_{k_{a+1}\gn}=(-1)^{\gn-1}\delta_{k_{a+1} 1}\displaystyle\frac{\chi_\gn}{w_1\dots w_{\gn-1}}
\end{equation}
and so the only terms which can survive in \eqref{minorsum} are those with $k_{a+1}=1$. Then by the antisymmetry of quantum minors we must have $k_i>1$, $i=1,2,\dots,1$ in which case $G_{k_i l}=w_{l}\,\delta_{k_i l+1}$ which results in
\begin{equation}
\bT\left[^{J_{a+1}}_{J_{a}\,\gn}\right] =(-1)^{\gn+a-1}\chi_\gn T\left[^{J_{a+1}}_{1\, J_{a}+1}\right]\, \frac{w_{j_1}\dots w_{j_{a}}}{w_1\dots w_{\gn-1}}
\end{equation}
where $J_{a}+1:=\{j_a^1+1,\dots,j_a^a+1\}$. Hence we arrive at
\begin{equation}\label{Bincomp}
\boxed{
\bB(u)=\sum_{J_1,\dots,J_{\gn-1}}T\left[^{J_1}_1\right]T^{[2]}\left[^{J_2}_{1\ J_1+1}\right]\dots T^{[2n-4]}\left[^{J_{\gn-1}}_{1\ J_{n-2}+1}\right]\frac{w_{J_1}w_{J_2}\dots w_{J_{\gn-1}}}{w_1 w_{12}\dots w_{1\dots\gn-1}}}\,,
\end{equation}
where $w_{J_k}:=\prod\limits_{i=1}^k w_{j_{ki}}$ and we have removed an overall sign arising from the permutation of indices in quantum minors and normalised the expression to be monic in $u$. As before we sum over configurations with $1\leq j_k^1 < j_2^2 <\dots <j_k^k\leq \gn$. 

\medskip

At first glance this expression may not appear to be much of an improvement over \eqref{Bgln}. However the first term in the sum with $J_k=\{1,2,\dots,k\}$, which we denote as $\bB^{\rm GT}$ is special and is given by
\begin{equation}
\bB^{\rm GT}(u)=T\left[^1_1\right]T^{[2]}\left[^{12}_{12}\right]T^{[4]}\left[^{123}_{123}\right]\dots T^{[2n-4]}\left[^{12\dots \gn-1}_{12\dots \gn-1}\right]\,.
\end{equation}
The objects $T\left[^{12\dots \gn-1}_{12\dots \gn-1}\right]$ are the generators of the so-called \textit{Gelfand-Tsetlin subalgebra} of $\lY_\gn$ which are diagonalised in the \textit{Gelfand-Tsetlin basis} of $\lY_\gn$. Since the Gelfand-Tsetlin subalgebra is one of the key concepts in this work we will now begin an in-depth review of it. A complete account can be found in \cite{molevgelfand,nla.cat-vn1878494} for $\gl(\gn)$ and in \cite{Molev1994,molev2007yangians} for the Yangian case and we closely follow these works.

\section{Gelfand-Tsetlin algebra}

\subsection{GT basis for $\gl(\gn)$}\label{GLNGT}

Let us consider some finite-dim irrep $\lV^\lambda$ of $\gl(\gn)$. By definition, we have 
\begin{equation}
\sfE_{jj}\ket{\Omega}=\lambda_j \ket{\Omega},\quad \sfE_{jk}\ket{\Omega}=0,\ j<k
\end{equation}
where $\ket{\Omega}$ is the highest-weight vector. $\gl(\gn)$ has a subalgebra naturally identified with $\gl(\gn-1)$, with 
\begin{equation}
\gl(\gn-1)=\{\sfE_{jk}:1\leq j,k\leq \gn-1\}
\end{equation}
and we can consider the action of $\gl(\gn-1)$ on the $\gl(\gn)$ representation $\lV^\lambda$. In general this action will be reducible and so the restriction $\lV^\lambda\rvert_{\gl(\gn-1)}$ of $\lV^\lambda$ to $\gl(\gn-1)$ decomposes into a direct sum of irreducible $\gl(\gn-1)$ modules $\lV^\mu$ of highest-weight $\mu$:
\begin{equation}\label{restrictionrep}
\lV^\lambda\rvert_{\gl(\gn-1)}=\bigoplus_{\mu}\lV^{\mu}\,.
\end{equation}
It is a central result of representation theory \cite{molevgelfand,nla.cat-vn1878494} that the decomposition of \eqref{restrictionrep} is multiplicity-free, meaning each possible weight $\mu$ appears at most once, and that the only weights $\mu$ which can appear are those which satisfy the \textit{branching-rule} 
\begin{equation}
\lambda_j \geq \mu_j\geq \lambda_{j+1}\,.
\end{equation}
We can then consider the restriction of each of the $\gl(\gn-1)$ irreps $\lV^\mu$ to $\gl(\gn-2)$ and so on. In total, we consider the chain of subalgebras 
\begin{equation}
\gl(1)\subset \gl(2)\subset \dots \subset \gl(\gn-1)\subset \gl(\gn)
\end{equation}
and let $[\lambda_{k1},\dots,\lambda_{kk}]$ denote the highest-weight of the $\gl(k)$ subalgebra appearing in the above chain. Owing to the fact that all $\gl(1)$ irreps have dimension $1$ it follows that there exists a basis in the original $\lV^\lambda$ irrep which is parameterised by the following array 
\begin{equation}
\begin{array}{ccccccccc}
\lambda_{\gn 1} &  & \lambda_{\gn 2} &  & \dots & \ & \lambda_{\gn ,\gn-1} &  & \lambda_{\gn\gn} \\
 & \lambda_{\gn-1,1} &  & \lambda_{\gn-1,2} & \dots & \lambda_{\gn-1,\gn-2}&  \ & \lambda_{\gn-1,\gn-1} & \ \\
 & \ & \dots & \ & \dots & \ & \dots &  \\
 & \ & \ & \lambda_{21} & \ & \lambda_{22} & \ & & \\
  & \ & \ & \  & \lambda_{11} & \ & \ & &
\end{array}
\label{GTintroduced}
\end{equation}
where $\lambda_{\gn j}=\lambda_j$ are nothing other than the original $\gl(\gn)$ highest weights and all nodes on the array satisfy the branching rule $\lambda_{jk}\geq \lambda_{j-1,k-1}\geq \lambda_{j+1,k}$. Such an array is known as a \textit{Gelfand-Tsetlin pattern}, and the corresponding basis is known as the Gelfand-Tsetlin basis \cite{gelfand1950finite}.
\paragraph{Examples} As an example we will construct the GT patterns for two representations of $\gl(3)$, the defining representation $\lambda=[1,0,0]$ and the adjoint representation (of $\sla(3)$) $\lambda=[2,1,0]$. For $[1,0,0]$ there are $3$ such patterns 
\begin{equation}
\begin{array}{ccccc}
1 & & 0 & & 0 \\
 & 1 & & 0 & \\
 & & 1 & & 
\end{array}, \quad 
\begin{array}{ccccc}
1 & & 0 & & 0 \\
 & 1 & & 0 & \\
 & & 0 & & 
\end{array},\quad
\begin{array}{ccccc}
1 & & 0 & & 0 \\
 & 0 & & 0 & \\
 & & 0 & & 
\end{array}
\end{equation}
and for $[2,1,0]$ there are $8$ patterns 
\begin{equation}
\begin{split}
& \begin{array}{ccccc}
2 & & 1 & & 0 \\
 & 2 & & 1 & \\
 & & 2 & & 
\end{array},\quad 
\begin{array}{ccccc}
2 & & 1 & & 0 \\
 & 2 & & 1 & \\
 & & 1 & & 
\end{array},\quad 
\begin{array}{ccccc}
2 & & 1 & & 0 \\
 & 1 & & 1 & \\
 & & 1 & & 
\end{array},\quad 
\begin{array}{ccccc}
2 & & 1 & & 0 \\
 & 2 & & 0 & \\
 & & 2 & & 
\end{array}, \\
& \begin{array}{ccccc}
2 & & 1 & & 0 \\
 & 2 & & 0 & \\
 & & 1 & & 
\end{array},\quad 
\begin{array}{ccccc}
2 & & 1 & & 0 \\
 & 2 & & 0 & \\
 & & 0 & & 
\end{array},\quad 
\begin{array}{ccccc}
2 & & 1 & & 0 \\
 & 1 & & 0 & \\
 & & 0 & & 
\end{array},\quad 
\begin{array}{ccccc}
2 & & 1 & & 0 \\
 & 0 & & 0 & \\
 & & 0 & & 
\end{array}\,.
\end{split}
\end{equation}
It is worthwhile to note that in the example of $[1,0,0]$ we see that the node corresponding to $\lambda_{22}$ is always fixed to $0$, which is a consequence of the branching rule and the fact that $\lambda_2=\lambda_3=0$. This is an example of a rather general feature which is that coinciding weights results in nodes on the pattern which do not change between states -- they are \textit{non-dynamical}. We will return to this point later when discussing separation of variables.

\paragraph{Gelfand-Tsetlin algebra}
We now turn to the question of constructing the operators which are diagonalised in the GT basis. To this end, we recall that the values of the Casimir operators of $U(\gl(\gn))$ on any finite-dimensional irrep allow us to determine the representation uniquely. In other words, knowing the values of the Casimirs is equivalent to knowing the weights of the representation. 

\medskip

The Casimir operators $\lC_k$, $k=1,\dots,\gn$ can be generated by the following row-ordered determinant (Capelli determinant)
\begin{equation}
\GG_\gn(u)=\det_{1\leq j,k\leq \gn}\left[\left(\left(u-\hbar(\gn-1)\right)\delta_{jk}-\hbar \sfE_{kj}\right)e^{\hbar \partial_u}\right]e^{-n\hbar\partial_u}
\end{equation}
and we have 
\begin{equation}
\GG_\gn(u)=u^\gn + u^{\gn-1} \lC_1 + \dots + \lC_\gn\,.
\end{equation}
Since $\GG_\gn(u)$ is central it takes a constant value on any irreducible representation which can be found by evaluating it on the highest-weight state, yielding
\begin{equation}
\GG_\gn(u) = \prod_{j=1}^{\gn}(u-\hbar( \hat{\lambda}_{\gn j}+\gn-1))
\end{equation}
where $\hat{\lambda}_{\gn j}:=\lambda_{\gn j}-j+1$ are the so-called shifted weights. The fact that the centre is generated by symmetric polynomials in the shifted weights is the Harish-Chandra isomorphism \cite{Harish}.

\medskip

Similarly, for $k=1,2,\dots,\gn-1$ we can compute $\GG_k(u)$. On each irreducible $\gl(k)$ representation $\GG_k(u)$ takes the value 
\begin{equation}
\GG_k(u) = \prod_{j=1}^{k}(u-\hbar( \hat{\lambda}_{k j}+k-1))\,.
\end{equation}
Since each $\GG_k(u)$ is an element of the centre $Z(U(\gl(k)))$ of $U(\gl(k))$ and we have the chain of subalgebras 
\begin{equation}
U(\gl(1)) \subset U(\gl(2))\subset \dots \subset U(\gl(\gn))
\end{equation} 
the set of $\GG_k(u)$, $k=1,2,\dots,\gn$ form a commutative subalgebra of $U(\gl(\gn))$, called the \textit{Gelfand-Tsetlin subalgebra}. By construction, each $\GG_k(u)$ acts diagonally on the Gelfand-Tsetlin basis. Letting $\ket{\Lambda}$ denote the GT basis element corresponding to the GT pattern $\Lambda$ we have 
\begin{equation}\label{Gdeteigen}
\GG_k(u)\ket{\Lambda}=\prod_{j=1}^{k}(u-\hbar( \hat{\lambda}_{k j}+k-1)) \ket{\Lambda}\,.
\end{equation}

\paragraph{Non-degeneracy}
For our purposes, the most crucial property of the Gelfand-Tsetlin algebra is that it has non-degenerate spectrum for any finite-dim irrep $\lambda$. Indeed, the set of all possible eigenvalues of the GT subalgebra correspond to all possible GT patterns which in turn label a basis of the representation $\lV^\lambda$. 

\medskip

This non-degeneracy is in stark contrast to the Cartan subalgebra of $\gl(\gn)$ which is only non-degenerate for special classes of representations. Indeed, the Cartan subalgebra is contained in the Gelfand-Tsetlin subalgebra and so is also diagonalised in the GT basis. This follows from the fact that 
\begin{equation}
\GG_k(u) = u^k -\hbar u^{k-1}\left(\sum_{j=1}^k k-j+\sfE_{jj}\right)+\lO\left(u^2\right)
\end{equation} 
and so comparing with \eqref{Gdeteigen} we find
\begin{equation}
\left(\sum_{j=1}^k\sfE_{jj}\right)\ket{\Lambda} = \left(\sum_{j=1}^k\lambda_{kj}\right)\ket{\Lambda}\,.
\end{equation}
Hence, for a given GT pattern $\Lambda$ the eigenvalue of $\sfE_{jj}$ is simply given by 
\begin{equation}
\displaystyle \sum_{k=1}^j \lambda_{jk} - \sum_{k=1}^{j-1}\lambda_{j-1,k}
\end{equation}
i.e. the sum total of the $j$-th row minus that of the $(j-1)$-th row as read from the bottom upwards. 

\medskip

The degeneracy of the Cartan subalgebra can be readily seen by examining the $[2,1,0]$ representation, where it is clear that both of the GT patterns
\begin{equation}
\begin{array}{ccccc}
2 & & 1 & & 0 \\
 & 2 & & 0 & \\
 & & 1 & & 
\end{array}, \quad 
\begin{array}{ccccc}
2 & & 1 & & 0 \\
 & 1 & & 1 & \\
 & & 1 & & 
\end{array}
\end{equation}
correspond to states with Cartan weights $[1,1,1]$ and hence are indistinguishable using the Cartan subalgebra alone. Thankfully, they are distinguishable when we extend from the Cartan subalgebra to the Gelfand-Tsetlin algebra. 

\medskip

The only representations where the Cartan subalgebra can distinguish all states in the GT basis are special cases of so-called \textit{rectangular representations} $(S^A)$ corresponding to Young diagrams of rectangular shape with $S$ columns and $A$ rows. Of these, the ones for which the Cartan subalgebra is non-degenerate are the symmetric powers of the defining representation $(S^1)$ and their conjugates $(S^{\gn-1})$ as well as the anti-symmetric powers of the defining representation $(1^A)$. For example, for $\gl(5)$ the family of representations for which the GT algebra has non-degenerate spectrum correspond to GT patterns of the form 
\begin{equation}
\begin{split}
& \begin{array}{cccccccccccc}
S & \ & 0 & \ & 0 & \ & 0 & \ & 0 \\
\ & * & \ & 0 & \ & 0 & \ & 0 \\
\ & \ & * & \ & 0 & \ & 0 \\
\ & \ & \ & * & \ & 0 \\
\ & \ & \ & \ & * 
\end{array}\quad \begin{array}{cccccccccccc}
1 & \ & 1 & \ & 0 & \ & 0 & \ & 0 \\
\ & 1 & \ & * & \ & 0 & \ & 0 \\
\ & \ & * & \ & * & \ & 0 \\
\ & \ & \ & * & \ & * \\
\ & \ & \ & \ & * 
\end{array} \\
& \begin{array}{cccccccccccc}
1 & \ & 1 & \ & 1 & \ & 0 & \ & 0 \\
\ & 1 & \ & 1 & \ & * & \ & 0 \\
\ & \ & 1 & \ & * & \ & * \\
\ & \ & \ & * & \ & * \\
\ & \ & \ & \ & * 
\end{array}\quad \begin{array}{cccccccccccc}
S & \ & S & \ & S & \ & S & \ & 0 \\
\ & S & \ & S & \ & S & \ & * \\
\ & \ & S & \ & S & \ & * \\
\ & \ & \ & S & \ & * \\
\ & \ & \ & \ & * 
\end{array} 
\end{split}
\end{equation}
where $S$ denotes some element of $\ZZ_{\geq 0}$ and $*$ denotes dynamical nodes. 

\subsection{Gelfand-Tsetlin algebra for Yangian}\label{sec:GTYang}
We now turn to the construction of the Gelfand-Tsetlin basis for Yangian \cite{Molev1994,molev2007yangians}. The idea is similar to the case of $\gl(\gn)$ -- namely, we simultaneously diagonalise the centres 
\begin{equation}
Z(\lY_1),Z(\lY_2),\dots,Z(\lY_\gn)
\end{equation}
and call the resulting commutative subalgebra the Gelfand-Tsetlin subalgebra of $\lY_\gn$. 

\medskip

Recall that the centre of $\lY_\gn$ is generated by the quantum determinant ${\rm qdet}\,T(u)$ which coincides with the transfer matrix $\T_{\gn,1}(u)$ in the totally anti-symmetric representation which was written in terms of quantum minors as 
\begin{equation}
\T_{\gn,1}(u)=T\left[^{12\dots \gn}_{12\dots \gn}\right](u)\,.
\end{equation}
We define the objects $\GT_a(u)$ by 
\begin{equation}
\GT_a(u)=T\left[^{12\dots a}_{12\dots a}\right](u)
\end{equation}
which manifestly form a commutative algebra by the property \eqref{minorcommutativity}.

\paragraph{GT algebra from Bethe algebra}
Note, that we can interpret $\GT_a(u)$ as transfer matrices in a special singular-twist limit. Indeed, the transfer matrices in anti-symmetric representations $\T_{a,1}$, with a diagonal twist $g={\rm diag}(z_1,\dots,z_\gn)$, are given explicitly by 
\begin{equation}
\T_{a,1} = \displaystyle \sum_{J} z_J T\left[^J_J \right]
\end{equation}
where $z_J = \prod_{j\in J} z_j$. Then in the limit
\begin{equation}
z_1\gg z_2 \gg\dots \gg z_{\gn} \gg 1
\end{equation}
we obtain 
\begin{equation}
\T_{a,1}(u) \rightarrow \GT_a(u)
\end{equation}
after appropriate normalisation. The fact that transfer matrices commute for different values of the spectral parameter then guarantees the same property for the Gelfand-Tsetlin generators.

\medskip

We can learn a lot about the Gelfand-Tsetlin algebra by considering the $\gn=2$ case. In this case the
GT algebra is generated by $T_{11}(u)$. We can compute its eigenvalue on the lowest-weight state $\bra{0}$ of the representation leading to 
\begin{equation}
\bra{0}T_{11}(u) = \nu_2(u) \bra{0}\,.
\end{equation}
We also have access to raising and lowering operators generated by $T_{12}(u)$ and $T_{21}(u)$. Let $\bra{\Lambda}$ be some eigenvector of $T_{11}(u)$ with eigenvalue $\lambda_{11}(u)$ with roots $\theta_\alpha+\hbar\lambda_{11}^\alpha$. It then follows from the RTT relation that 
\begin{equation}
\bra{\Lambda}T_{12}(\theta_\alpha+\hbar \lambda^\alpha_{11}),\quad \bra{\Lambda}T_{21}(\theta_\alpha+\hbar \lambda^\alpha_{11})
\end{equation}
are both eigenvectors of $T_{11}(u)$ with new eigenvalue where $\lambda_{11}^\alpha$ has been replaced with $\lambda_{11}^\alpha+1$ and $\lambda_{11}^\alpha-1$ respectively. By starting from the lowest-weight state $\bra{0}$ we can then repeatedly act with $T_{12}$ creating more and more eigenvectors of $T_{11}(u)$ and eventually spanning the space -- this is the Gelfand-Tsetlin basis of the Yangian $\lY_2$. The constructed states are all non-zero provided the numbers $\lambda_{11}^\alpha$ label a Gelfand-Tsetlin pattern. 

\medskip

We will now outline the general case. All $\GT_a$ are diagonalisable and their eigenstates $\bra{\Lambda^{\rm GT}}$ are labelled as follows \cite{molev2007yangians}. Each $\Lambda$ is an $L$-tuple
\begin{equation}
\Lambda=\left( \Lambda^1,\Lambda^2,\dots,\Lambda^L\right)\,,
\end{equation}
where each $\Lambda^\alpha$ is a GT pattern. Namely, it is an array 
\begin{equation}
\begin{array}{ccccccccccc}
\nu_{1}^\alpha & \ & \nu_{2}^\alpha & \ & \dots & \ & \nu_{\gn}^\alpha \\
\ & \lambda_{\gn-1,1}^\alpha & \ & \dots & \ & \lambda_{\gn-1,\gn-1}^\alpha \\
\ & \ & \dots & \ & \dots \\
\ & \ & \lambda_{21}^\alpha & \ & \lambda_{22}^\alpha \\
\ & \ & \ & \lambda_{11}^\alpha
\end{array}
\end{equation}
in which the nodes $\lambda^\alpha_{aj}\in\ZZ$ are subject to the \textit{branching rules}
\begin{equation}\label{compactbranch}
\lambda^\alpha_{a+1,j}\geq \lambda_{aj}^\alpha\geq \lambda_{a+1,j+1}^\alpha,\quad a=1,2,\dots,\gn-1,\quad j=1,2,\dots,a\,,
\end{equation}
and $\nu_{j}^{\alpha}\equiv\lambda_{\gn,j}^{\alpha}$ are fixed numbers defined by the chosen representation $\nu^{\alpha}=(\nu_{1}^{\alpha},\ldots,\nu_{\gn}^{\alpha})$ at $\alpha$-th site of the spin chain.

\medskip

The eigenvalues of $\GT_a$ are
\begin{equation}\label{GTspectrum}
\bra{\Lambda^{\rm GT}}\GT_a(u)=\displaystyle\prod_{\alpha=1}^L\prod_{j=1}^a(u-\theta_\alpha-\hbar(\lambda_{aj}^\alpha+a-j))\bra{\Lambda^{\rm GT}}\,.
\end{equation}
We see that $\GT_a(u)$ measures the value of the $a$-th rows of the GT patterns which make up  $\bra{\Lambda^{\rm GT}}$. This hierarchical organisation comes from the original procedure to build up GT patterns: one considers the tautological homomorphism $\phi^{\rm GT}:T_{ij}\to T_{ij}$ which, for $i,j$ being restricted to range $1,2,\ldots, a$, can be considered as an injection  of  $\lY_a$ into {\it e.g.} $\lY_{a+1}$. One then builds the ascending chain
\begin{equation}
\lY_1\xrightarrow{\phi^{\rm GT}}\ldots \lY_a\xrightarrow{\phi^{\rm GT}} \lY_{a+1}\ldots \xrightarrow{\phi^{\rm GT}} \lY_{\gn}
\end{equation}
for which $\GT_a$ are precisely the central elements (quantum determinants) of $\lY_a$. The center of $\lY_a$ acts as
\begin{equation}
\bra{\Lambda^{\rm GT}}\GT_{\gn}(u)=\displaystyle\prod_{j=1}^{\gn}\nu_j(u-\hbar(\gn-j))\bra{\Lambda^{\rm GT}}\,,\quad \nu_j(u):=\prod_{\alpha=1}^{L}(u-\theta_\alpha-\hbar\ \nu^\alpha_j)\,.
\end{equation}
For each $\GT_a$ there is also a corresponding raising operator $\GP^+_a$ and a lowering operator $\GP^-_a$ which act on the GT basis as \cite{molev2007yangians}
\begin{equation}
\bra{\Lambda^{\rm GT}}\GP^\pm_a(\theta_\alpha+\hbar(\lambda_{aj}^\alpha+a-j))\propto \bra{\Lambda\pm\delta^\alpha_{aj}{}^{\rm GT}}\,.
\end{equation}
Here $\Lambda\pm\delta^\alpha_{aj}$ denotes a GT pattern where the node $(a,j)$ of the $\alpha$-th pattern has been changed by $\pm 1$. The coefficient of proportionality is non-zero provided that the pattern $\Lambda\pm\hbar\,\delta^\alpha_{aj}$ satisfies the branching rules, {\it i.e}. corresponds to a consistent GT pattern. Each $\GP^\pm_a(u)$ can be written explicitly in terms of quantum minors. Specifically,
\begin{equation}
\GP^+_a(u)=T\left[^{12\dots a-1\ a}_{12\dots a-1 \ a+1} \right](u),\quad \GP^-_a(u)=T\left[^{12\dots a-1\ a+1}_{12\dots a-1 \ a} \right](u)\,.
\end{equation}

\paragraph{Dual diagonals} We will find it convenient to introduce an alternative labelling of the GT pattern entries, by $\mu^\alpha_{kj}$, where $\mu^\alpha_{kj}=\lambda^\alpha_{\gn-k+j-1,j}$. For example, for $\gl(4)$ we have
\begin{equation}\label{mupattern}
\begin{array}{ccccccccccc}
\nu_{1}^\alpha & \ & \nu_{2}^\alpha & \ & \nu_{3}^\alpha & \ & \nu_{4}^\alpha \\
\ & \mu_{11}^\alpha & \ & \mu_{22}^\alpha & \ & \mu_{33}^\alpha \\
\ & \ & \mu_{21}^\alpha & \ & \mu_{32}^\alpha \\
\ & \ & \ & \mu_{31}^\alpha
\end{array}\,.
\end{equation}
This new labelling naturally suggests to parameterise GT patterns by what we refer to as \textit{dual diagonals} $\mu^\alpha_k$ where we define
\begin{equation}\label{dualdiagonals}
\mu^\alpha_k=(\mu^\alpha_{k1},\mu^\alpha_{k2},\dots,\mu^\alpha_{kk}),\quad k=1,\dots,\gn-1\,.
\end{equation}
Since the minimum value of each $\mu^\alpha_{kj}$ allowed by the branching rules is $\mu^\alpha_{kj}=\nu^\alpha_{k+1}$, it is also convenient to introduce the parameters 
\begin{equation}
\label{eq:defmbar}
\bar\mu^\alpha_{kj}=\mu^\alpha_{kj}-\nu^\alpha_{k+1}
\end{equation}
which measure how much a given dual diagonal has been excited above its minimum value. Clearly, $\bar\mu^\alpha_k$ corresponds to a $\gl(k)$ Young diagram. As we will see, dual diagonals turn out to be a natural labelling of GT patterns in the context of separation of variables.

\section{Spectral problem for $\bB$}

\subsection{$\bB$ eigenvalues}

We will now begin using our knowledge of the Gelfand-Tsetlin algebra and the associated basis to learn some things about $\bB$. We will begin with its spectrum. 

\medskip

Consider again the expression for $\bB$ in the MCT frame 
\begin{equation}\label{Bdefn2}
\bB(u)=\sum_{J_1,\dots,J_{\gn-1}}T\left[^{J_1}_1\right]T^{[2]}\left[^{J_2}_{1\ J_1+1}\right]\dots T^{[2n-4]}\left[^{J_{\gn-1}}_{1\ J_{n-2}+1}\right]\frac{w_{J_1}w_{J_2}\dots w_{J_{\gn-1}}}{w_1 w_{12}\dots w_{1\dots \gn-1}}
\end{equation}
and the leading term $\bB^{\rm GT}(u)$ given by
\begin{equation}
\bB^{\rm GT}(u)=T\left[^1_1\right]T^{[2]}\left[^{12}_{12}\right]T^{[4]}\left[^{123}_{123}\right]\dots T^{[2n-4]}\left[^{12\dots \gn-1}_{12\dots \gn-1}\right]\,.
\end{equation}
We are going to show that 
\begin{equation}
\bB(u) = \bB^{\rm GT}(u) + {\rm Nil}
\end{equation}
where ${\rm Nil}$ refers to a nilpotent term which is strictly upper triangular in an appropriate ordering of the Gelfand-Tsetlin basis which diagonalises $\bB^{\rm GT}$. To do this we introduce a partial ordering on the Gelfand-Tsetlin basis vectors $\ket{\Lambda}$ with respect to their global Cartan weight $\lambda=[\lambda_1,\dots,\lambda_{\gn}]$ with 
\begin{equation}
\lE_{ii}\ket{\Lambda} = \lambda_i \ket{\Lambda}\,.
\end{equation}
We define $\lambda \succ \lambda^\prime$ if and only if $\lambda_i > \lambda_i^\prime$ for the smallest $i$ such that $\lambda_i \neq \lambda^\prime_i$ and hence say $\ket{\Lambda}\succ \ket{\Lambda^\prime}$ if and only if $\lambda \succ \lambda^\prime$.

\medskip

Next we recall the commutation relation \eqref{CartanRTT} between the global $\gl(\gn)$ generators $\lE_{ij}$ and monodromy matrix elements $T_{kl}(v)$ which we repeat here for convenience
\begin{equation}
[\mathcal{E}_{ij},T_{kl}(v)]=\delta_{jl}T_{ki}(v)-\delta_{ki}T_{jl}(v)
\end{equation}
which implies
\begin{equation}\label{globalcartancom}
[\mathcal{E}_{jj},T_{kl}(v)]=(\delta_{jl}-\delta_{kj})T_{kl}(v)\,.
\end{equation}
Let $\lO$ be an operator appearing as an individual summand of \eqref{Bdefn2}
\begin{equation}
\lO = T\left[^{J_1}_1\right]T^{[2]}\left[^{J_2}_{1\ J_1+1}\right]\dots T^{[2n-4]}\left[^{J_{\gn-1}}_{1\ J_{n-2}+1}\right]
\end{equation}
with no sum over $J$'s. We are going to show that the action of $\lO\neq \bB^{\rm GT}$ on a Gelfand-Tsetlin basis vector $\ket{\Lambda}$ is either zero or strictly positive in the sense of the partial order $\succ$. More precisely, we have
\begin{equation}
\lO \ket{\Lambda} = \displaystyle\sum_{r} c_r \ket{\Lambda_r}
\end{equation}
where the sum $r$ ranges over all elements $\Lambda_r$ of the GT basis. We will show that if $c_r\neq 0$ then $\ket{\Lambda_r}\succ \ket{\Lambda}$ and hence $\lO$ is strictly upper triangular in the GT basis with our ordering. 

\medskip

Let $A=\left[^{\alpha}_{\beta}\right]$ where $\alpha$ denotes the upper indices appearing in $\lO$ and $\beta$ denotes the lower indices, that is 
\begin{equation}
\alpha= \{J_1, J_2, \dots\},\quad \beta = \{1,1,J_1+1,1,J_2+1,\dots \}
\end{equation}
and we have, from \eqref{globalcartancom},
\begin{equation}\label{Ocomm}
[\lE_{ii},\lO]=\left(\sum_{b\in\beta}\delta_{ib}-\sum_{a\in\alpha}\delta_{ia} \right)\lO\,.
\end{equation}
We construct $A^{\rm reg}=\left[^{\alpha_{\rm reg}}_{\beta_{\rm reg}}\right]$ from $A$ by repeatedly removing pairs $(\alpha_i,\beta_j)$ with $\alpha_i=\beta_j$ until no such pairs are left. Clearly this removal does not affect \eqref{Ocomm}. The only possibility that $A^{\rm reg}=\left[^\es_\es \right]$ is $\lO = {\rm B}^{\rm GT}$. For all other cases it is easy to see that $\min[\beta_{\rm reg}]< \min[\alpha_{\rm reg}]$ which clearly implies that action of $\lO$ on the elements of the GT basis, if non-zero, is strictly positive in the above-defined sense, completing the proof. 

\bigskip

Now we are ready to introduce the operators $\svX^\alpha_{kj}$ which will turn out to be the separated variables. One can follow the following logic: Whereas $\bB$ and $\bB^{\rm GT}$ do not coincide, their spectrum is nevertheless equal, since ${\rm Nil}$ is upper-triangular. Therefore we will label the eigenvectors of $\bB$ by $\bra{\Lambda^{\bB}}$, where $\Lambda$ is an $L$-tuple of GT patterns, but note that $\bra{\Lambda^{\bB}}\neq \bra{\Lambda^{\rm GT}}$ in general. We introduce a labelling of the operatorial zeros $\svX^\alpha_{kj}$ of $\bB$ by
\begin{equation}
\label{eq:BX}
\bB(u)=\prod_{\alpha=1}^L\prod_{k=1}^{\gn-1}\prod_{j=1}^{k}(u-\svX_{kj}^\alpha)\,
\end{equation}
with their eigenvalues $\svx^\alpha_{kj}$ given by
\begin{equation}\label{eq:variablesx}
\svx^\alpha_{kj}=\theta_\alpha+\hbar\,(\mu^\alpha_{kj}-j+1)
\end{equation}
where $\mu^\alpha_{kj}$ label the dual diagonals \eqref{dualdiagonals}. Note that the overall trivial factor of $(u-\theta+\hbar)$ present in \eqref{Bdiagonal} is now understood as simply being a consequence of the fact that for the defining representation we have $\mu^\alpha_{22}=0$ for all states.

\medskip

The presented logic has however some subtleties and weak points. By subtlety we mean a statement that requires further clarification and by weak point we mean a statement that requires further arguments to prove being correct.

\medskip

One technical subtlety is that $\bB$ is a symmetric polynomial in $\svX$'s, so in \eqref{eq:variablesx} we actually agree on a way to define {\it e.g.} $\svX_1,\svX_2$ from their known combinations $\svX_1+\svX_2,\svX_1\svX_2$.

\medskip

The second subtlety is that $\bB^{\rm GT}$ has a degenerate spectrum, even assuming that $\theta_{\alpha}$ are distinct. Indeed, $\bB^{\rm GT}$ is only a product of operators generating the GT algebra, so it bears less information. The special cases when $\bB^{\rm GT}$ is still non-degenerate are the rectangular representations with $A=1$, $A=\gn-1$, or $S=1$, that is same ones when the Cartan subalgebra is non-degenerate. Similarly to $\bB^{\rm GT}$, $\bB$  turns out to be degenerate as well, and hence not all $\bra{\Lambda}$ are uniquely defined from the fact that they are $\bB$-eigenvectors so we should provide a separate prescription for which basis diagonalising $\bB$ we would like to choose.

\medskip

The first weak point is that we never showed or even mentioned that $[\bB(u),\bB(u')]=0$ for arbitrary $u,u'$. In fact, there is already an example of a super-symmetric analogue of $\bB$ \cite{Gromov:2018cvh} which is non-commuting. In our case, it turns out that $\bB(u)$ and $\bB(u')$ commute indeed. This was demonstrated in \cite{Sklyanin:1992sm} for the $\gl(3)$ case by Sklyanin via direct computation. We will confirm this fact in a more pedestrian way by computing the eigenvectors of $\bB$ in a $u$-independent way.
Note that it was shown in \cite{2001math.ph...9013S} by Smirnov for the $U_q\left(\widehat{\sla(\gn)} \right)$ case that $[\bB(u),\bB(u')]=0$. It is expected that the $\bB$ operator there reproduces our $\bB$ in an appropriate $q\rightarrow 1$ limit which would confirm commutativity in our case but such a computation has yet to be performed.

\medskip

The second weak point is that $\bB$ was not proven to be diagonalisable, and in principle it might be not the case as $\bB$ is equal to the degenerate diagonal matrix $\bB^{\rm GT}$ added with an upper-triangular matrix. Again, we shall not prove diagonalisability of $\bB$ directly, but this will follow after we construct enough of linearly independent eigenvectors $\bra{\Lambda^{\bB}}$.

\medskip

We therefore see that it is not enough to consider \eqref{eq:variablesx} simply as a consequence of \eqref{eq:BX}. For the above-outlined reasons, we need to construct the basis of $\bra{\Lambda^{\bB}}$'s in an independent way. We provide such a construction in the next section. Furthermore we will obtain results about factorisation of the wave functions that allow one to indeed consider $\svX$'s as separated variables.

\subsection{SoV basis for $\gn=2$}
In this section we will outline a procedure for constructing an SoV basis for $\lY_2$, in a way that reproduces the proposal of \cite{Maillet:2018bim} for the defining representation and also makes it precise for higher-spin representations. This will serve as a precursor for our study of the higher-rank cases. We omit some of the proofs as they are either available in the literature or follow naturally as specialisations of the forthcoming more general discussion.

\medskip

For the modified companion twist we have, after putting $w_1=1$ for convenience,
\begin{equation}
\bB=T_{11}(u),\quad \T_{1,1}(u)=\chi_1 T_{11}(u)+T_{12}(u)-\chi_2 T_{21}(u)\,.
\end{equation}
The state $\bra{0}$ corresponds to the $L$-tuple of GT patterns with all $\lambda_{11}^{\alpha}=\nu_2^\alpha$, and we set $\nu_2^\alpha=0$ for simplicity -- this can always be achieved by a redefinition of inhomogeneities $\theta_\alpha$. This state satisfies 
\begin{equation}
\bra{0}T_{j1}(u)=\delta_{j1}\nu_2(u)\bra{0}\,.
\end{equation}
We then obtain that the action of $\T_{1,1}(\theta_\alpha)$ on $\bra{0}$ simplifies to 
\begin{equation}
\bra{0}\T_{1,1}(\theta_\alpha)=\bra{0}T_{12}(\theta_\alpha)\,.
\end{equation}
By using the RTT relation, one shows that $\bra{0}$ is an eigenvector of $\bB$ with the eigenvalue 
\begin{equation}
(u-\theta_\alpha-\hbar)\prod_{\beta\neq\alpha}(u-\theta_\beta)\,.
\end{equation}
As a result, this state is annihilated by the subsequent action of $T_{11}(\theta_\beta)$, $\beta\neq\alpha$.

\medskip

This state is also annihilated by $T_{21}(\theta_\beta)$, which can be easily checked using RTT. Hence
\begin{equation}
\label{indepinhom}
\bra{0}\T_{1,1}(\theta_\alpha)\T_{1,1}(\theta_\beta)=\bra{0}T_{12}(\theta_\alpha)T_{12}(\theta_\beta)\,,\quad \alpha\neq \beta\,
\end{equation}
which can be shown to be an eigenvector of $\bB$ with the eigenvalue $
(u-\theta_\alpha-\hbar)(u-\theta_\beta-\hbar)\prod_{\gamma\neq \alpha,\beta}(u-\theta_\gamma)\,.$
Then, by induction, it follows that for any subset $I\subset \{1,2,\dots,L\}$ we have that 
\begin{eqnarray}
&&\bra{0}\prod_{\alpha\in I}\T_{1,1}(\theta_\alpha)=\bra{0}\prod_{\alpha\in I}T_{12}(\theta_\alpha)\,,\quad {\rm and}\\ \nonumber
&&\bra{0}\prod_{\alpha\in I}\T_{1,1}(\theta_\alpha)\,\bB(u)=\prod_{\alpha\in I}(u-\theta_\alpha-\hbar)\prod_{\beta\notin I}(u-\theta_\beta)\bra{0}\prod_{\alpha\in I}\wT_{1,1}(\theta_\alpha)\,.
\end{eqnarray}
Furthermore, all of these states are non-zero, since $\T_{1,1}(u)$ has no vanishing eigenvalues at $u=\theta_\alpha$. In this manner we can construct $2^L$ states. This precisely matches the dimension of the Hilbert space if $\lV$ is the defining representation $\nu^\alpha=1$. Hence, for the case of the defining representation, the constructed states form a basis, since each corresponds to a different eigenvalue of $\bB$. 

\medskip

An important point very useful for generalisations is that the constructed states are independent of the twist eigenvalues, as they should be since $\bB$ is independent of these and so naturally its eigenvectors are as well. We will routinely make use of this fact.

\medskip

For a more general case of symmetric power representation $\nu^\alpha=[S,0]$ (we assume all sites carry the same representation for convenience), the constructed states are not sufficient to span the Hilbert space, and we should look for more. As noted in \cite{Maillet:2018bim}, it is natural to conjecture that the basis is not constructed just with $\T_{1,1}(\theta)$, but also with $\T_{1,1}(\theta+n \hbar)$, $n\in\ZZ$. Indeed, this is analogous to the way the GT basis is constructed \cite{molev2007yangians} -- a generic eigenvector of $T_{11}(u)$  can be obtained by acting on $\bra{0}$ with $T_{12}(\theta)T_{12}(\theta+\hbar)T_{12}(\theta+2\hbar)\dots$. To put it more in the perspective of a physicist, one can introduce operators $\svX^{\alpha}$ as operatorial zeros of $T_{11}(u)$ whose spectrum was described in Section \ref{sec:GTYang}. One then finds, using RTT, that  the ladder operators  are
\begin{equation}
\label{ladderoperators}
\svP^+_\alpha= T_{12}(X_{\alpha})\,,\quad \svP^-_{\alpha}= T_{21}(X_{\alpha})\,,
\end{equation}
so that 
\begin{equation}
\bra{s_{\alpha}}\equiv \bra{0}(\svP^+_{\alpha})^s =\bra{0}T_{12}(\theta_{\alpha}+\hbar)\ldots T_{12}(\theta_{\alpha}+\hbar\,(s-1)\hbar)\,
\label{ladderrep}
\end{equation}
is a $\bB$-eigenstate. Normal ordering is used in the above expressions, that is $\svX$'s are placed to the left of other operators. However, the representation \eqref{ladderrep} of the $\bB$-eigenstates is not fully satisfactory as it does not suggest yet that the wave functions would factorise in this basis, so we would like to replace $T_{12}$ with transfer matrices in order to make such conclusions. 

\medskip

The action of $\T_{1,1}(\theta_\alpha)$ (once) is equivalent to the action of $\svP^+_{\alpha}$ (once) on $\bra{0}$, as we learned above. However, the action of $\T_{1,1}(\theta_\alpha)\T_{1,1}(\theta_\alpha+\hbar)$ on $\bra{0}$ does not yield $\bra{0}(\svP_\alpha^+)^2$ as we would like. Instead, it yields
\begin{equation}
\bra{0}\T_{1,1}(\theta_\alpha)\T_{1,1}(\theta_\alpha+\hbar)=\bra{0}T_{12}(\theta_\alpha)T_{12}(\theta_\alpha+\hbar)-\chi_2\bra{0}T_{12}(\theta_\alpha)T_{21}(\theta_\alpha+\hbar)\,.
\end{equation}
The second term is non-vanishing, as can be checked using RTT. Instead, it can be rewritten in a useful form 
\begin{equation}
-\chi_2\bra{0}T_{12}(\theta_\alpha)T_{21}(\theta_\alpha+\hbar)=\chi_2 \bra{0}\left(T_{11}(\theta)T_{22}(\theta_\alpha+\hbar)-T_{12}(\theta_\alpha)T_{21}(\theta_\alpha+\hbar) \right)
\end{equation}
since the first term on the \rhs vanishes. We can recognise the transfer matrix $\T_{2,1}$ in the expression on the \rhs and so we can see that the eigenstate $\bra{2_\alpha}$ is actually given by 
\begin{equation}
\bra{2_\alpha}=\bra{0}(\T_{1,1}(\theta_\alpha)\T_{1,1}(\theta_\alpha+\hbar)-\T_{2,1}(\theta_\alpha+\hbar))=\bra{0}\T_{1,2}(\theta_\alpha)\,,
\end{equation}
where  the Hirota equation \eqref{Hirota21} enjoyed by the transfer matrices was used on the last step. 

\medskip

The $\bB$-eigenvalue of $\bra{2_{\alpha}}$ is $(u-\theta_\alpha-2\hbar)\prod_{\beta\neq \alpha}(u-\theta_\beta)\,.$ It is then natural to guess that, in order to construct the eigenstate $\bra{s_{\alpha}}$ of $\bB$ with eigenvalue $(u-\theta_\alpha-s\ \hbar)\prod_{\beta\neq \alpha}(u-\theta_\beta)$,
we should act on $\bra{0}$ with $\T_{1,s}(\theta_\alpha)$, and therefore
\begin{equation}
\label{T12T11}
\bra{0}\wT_{1,s}(\theta_\alpha)=\bra{0}T_{12}(\theta_\alpha)\dots T_{12}(\theta_\alpha+\hbar(s-1))\,.
\end{equation} 
This can be  seen for instance by recursively using the relation \cite{Zabrodin:1996vm}
\begin{equation}
\label{Zrec}
\T_{1,s+1}(u)=\T_{1,s}(u)\T_{1,1}(u+\hbar\ s)-\T_{1,s-1}(u)\T_{2,1}(u+\hbar\ s)\,.
\end{equation}
which is a special case of the Hirota equation. We won't present this computation here, but instead give a quick argument supporting \eqref{T12T11}. As the result is not expected to depend on twist, let us set $\chi_1=\chi_2=0$ which sets the second term in \eqref{Zrec} to zero and also simplifies $\wT_{1,1}(u)$ to  $\wT_{1,1}(u)=T_{12}(u)$. Then the desired property \eqref{T12T11} is demonstrated immediately by recursion. 

We can also produce formulae of the type \eqref{indepinhom} and finally conclude that all eigenstates of $\bB$, which we label by $\bra{\Lambda}=\bra{s_1,\ldots, s_L}$ for $s_\alpha\in \{0,1,\dots,S\}$, can be constructed as
\begin{equation}
\label{eq:Lambdadir}
\bra{\Lambda}=\bra{0}\prod_{\alpha=1}^L \T_{1,s_\alpha}(\theta_\alpha)\,.
\end{equation}
Their $\bB$-eigenvalues are
\begin{equation}
\bra{\Lambda}\bB(u)=\prod_{\alpha=1}^L (u-\theta_{\alpha}-\hbar\,s_\alpha)\bra{\Lambda}\,.
\label{eq:egvaluesf}
\end{equation}
In the case of $\gn=2$, we were quite lucky to know the ladder operators \eqref{ladderoperators}, so deriving \eqref{T12T11} was sufficient for demonstration of \eqref{eq:egvaluesf}. For higher-rank cases, we won't be able to get a straightforward generalisation of \eqref{ladderoperators}. Instead we will develop a related but more flexible approach to show that the generalisation of \eqref{eq:Lambdadir} are eigenvectors of $\bB$.

\subsection{B-T commutation relation}

In the above the crucial feature was that $\bra{0}$ was annihilated by $T_{j1}(\theta)$, $j=1,2$. This has an obvious extension to higher rank. Hence, in order to diagonalise $\bB$ using transfer matrices we would like a relation of the following form
\begin{equation}\label{BTcom}
\T_\lambda(v)\bB(u) = f_\lambda(u,v) \bB(u) \T_\lambda(v) + \displaystyle \sum_{j=1}^{\gn} T_{j1}(v) \times \dots
\end{equation}
where $f_\lambda(u,v)$ is some function to be determined. 

\paragraph{Null twist}

Since $\bB$ is independent of twist so are its eigenvalues. Hence, if we create eigenvectors of $\bB$ by acting on an appropriate vacuum state $\bra{0}$ by action of transfer matrices this action should coincide with that of transfer matrices constructed with the null twist $\lN$ obtained from the companion twist after setting all $z_j \rightarrow 0$. That is, if $\bra{0}$ is the SoV vacuum state we should have 
\begin{equation}
\bra{0}\T_\lambda(v) = \bra{0}\T_\lambda^{\lN}(v)
\end{equation}
for appropriate values of $v$. Of course, in principle twist can appear in $\bB$ eigenvectors through normalisation but in the $\gn=2$ case in the previous section we indeed needed to set the twist eigenvalues to $0$ and it is reasonable to suspect the same is true in the higher rank case. Hence in what follows we will consider transfer matrices constructed with the null twist. A straightforward calculation yields that all such transfer matrices have the structure 
\begin{equation}
\T_\lambda(v) = \T_\lambda^{\lN}(v) + \sum_{j=1}^\gn T_{j1}(v)\times 
\end{equation}
and we will make use of this fact in what follows. 

\paragraph{Deriving relation}

Our starting point is the fused RTT relation 
\begin{equation}
R^{\mu\lambda}_{ab}(u,v)T_a^\mu(u) T_b^\mu(v) = T_b^\lambda(v) T_a^\mu(u) R^{\mu\lambda}_{ab}(u,v)
\end{equation}
We have included the subscripts $a$ and $b$ to make the ordering of spaces clear -- the fused RTT relation acts on the tensor product $\lV^\mu \otimes \lV^\lambda$, and we can consider each $a$ and $b$ as a multi-index 
\begin{equation}
a:= (a_1,a_2,\dots a_{|\mu|})
\end{equation}
where each $a_j$ labels a copy of $\CC^\gn$ and similarly with $b$.

\medskip

By acting with the inverse of $R^{\mu\lambda}$ we obtain 
\begin{equation}
T_a^\mu(u) T_b^\mu(v)\left(R^{\mu\lambda}_{ab}(u,v) \right)^{-1} = \left(R^{\mu\lambda}_{ab}(u,v) \right)^{-1} T_b^\lambda(v) T_a^\mu(u)
\end{equation}
which will turn out to be much more useful for deriving the relation we desire. On one hand it may seem like taking the inverse\footnote{It may seem like a fused $R$-matrix is not invertible due to the presence of projection operators. However they are invertible on appropriately considered irreducible subspaces which we restrict to.} of a generic fused $R$-matrix is a difficult task, but it is simplified by the following observation. Let us denote by $\bar{R}(u,v)$ the operator
\begin{equation}
\bar{R}_{ab}(u,v) = (u-v) +\hbar P_{ab}\,.
\end{equation} 
Then
\begin{equation}
R_{ab}(u,v)\bar{R}_{ab}(u,v) = (u-v-\hbar)(u-v+\hbar)\,.
\end{equation}
In order to take the inverse of $R^{\mu\lambda}_{ab}(u,v)$ we should reverse the order of all factors and replace each factor with its inverse. Instead of doing this we can reverse the order of all factors and replace each factor of $R$ with $\bar{R}$. The two resulting operators will differ by an overall factor, but this will drop out from the fused RTT relation. Since $R^{\mu\lambda}_{ab}(u,v)$ is defined by
\begin{equation}
R^{\mu\lambda}_{ab}(u,v)=P^\mu \left(\prod_{j=1}^{\rightarrow} \prod_{k=1}^{\rightarrow} R_{a_j b_k}(u+\hbar c^\mu_j,\theta+\hbar c^\lambda_k) \right)P^\lambda
\end{equation}
we are then led to define 
\begin{equation}
\bar{R}^{\mu\lambda}_{ab}(u,v)=P^\lambda \left(\prod_{j=1}^{\leftarrow} \prod_{k=1}^{\leftarrow} \bar{R}_{a_j b_k}(u+\hbar c^\mu_j,\theta+\hbar c^\lambda_k) \right)P^\mu
\end{equation}
which satisfies the fused RTT relation 
\begin{equation}
T_a^\mu(u) T_b^\mu(v)\bar{R}^{\mu\lambda}_{ab}(u,v) = \bar{R}^{\mu\lambda}_{ab}(u,v) T_b^\lambda(v) T_a^\mu(u)\,.
\end{equation}

Let us now return to $\bB$ but first we need to warn the reader of a conventional change. In \eqref{Bdefn2} $\bB$ is defined with the minors increasing in size from left to right. For the calculation we are about to perform it will make our lives easier if we redefine $\bB$ to have the same form but now with the size of minors increasing from right to left. The calculation we present below for this new convention was originally presented in author's publication \cite{Ryan:2018fyo} for the former case, but the proof is more streamlined if we update our conventions. Our main output if this calculation is a commutation relation between $\bB$ and $\T_\lambda$. It is important to understand that all results we will present later in the text depend only on this commutation relation, which is invariant under the two choices of ordering, which will actually imply that $\bB$ itself is invariant under this reordering. Hence, we will allow ourselves to freely (and consistently!) switch between the two orderings depending on what is more convenient.

\medskip

Consider the standard unit vectors $\sfe_j$ in $\CC^\gn$ with components $(\sfe_j)_i = \delta_{ij}$. Clearly, $\bB$, with our updated choice of ordering, is an entry of the following product of fused monodromy matrices
\begin{equation}
\bB(u)\in \bT_{a_1a_2\dots a_\gn}^{\wedge^{\gn}}(u+\hbar(\gn-1))\bT_{a_2\dots a_\gn}^{\wedge^{\gn-1}}(u+\hbar(\gn-2))\dots \bT_{a_\gn}^{\wedge^{1}}(u)
\end{equation}
and $\bB$ is obtained by acting with this on $\sfe_1^{\otimes \gn}$ and projecting onto the physical space, provided we agree that contraction of indices is done using the null twist, that is we define 
\begin{equation}
\bT^i_j \bT^j_k := T^i_{j+1}T^j_k\,.
\end{equation}
For example, using this rule and projecting onto the physical space
\begin{equation}
\bT_{a_1 a_2}^{\wedge^{2}}(u+\hbar)\bT_{a_2}^{\wedge^{1}}(u)\left(\sfe_1 \otimes \sfe_1\right)
\end{equation}
evaluates to 
\begin{equation}
\displaystyle \sum_{j} T\left[^{1 2}_{1\, j+1}\right](u+\hbar)T\left[^j_1\right](u) = \qdet\,T(u+\hbar)\times \bB(u)
\end{equation}
which coincides with $\bB$ in the case $\gn=2$ up to the irrelevant quantum determinant factor. 

\medskip

\begin{figure}[htb]
\centering
  \includegraphics[width=80mm,scale=15]{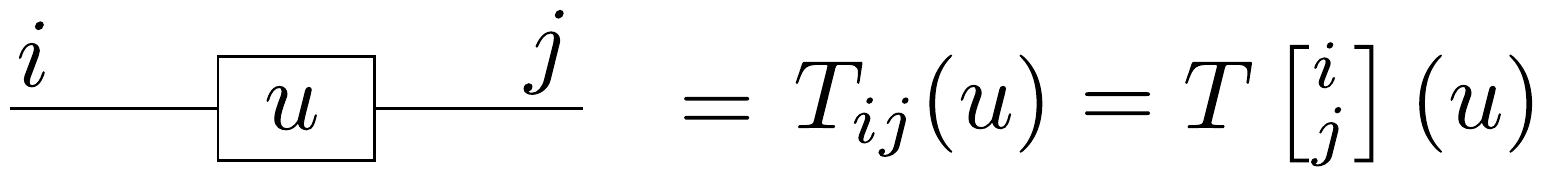}
  \caption{.}
  \label{tijgraphics}
\end{figure}

We will now derive the relation \eqref{BTcom}. We will introduce a graphical framework to make the calculation as simple as possible and avoid unnecessary clutter with indices. We start by defining $T_{ij}(u)$ as in Figure \ref{tijgraphics}.

We will use the convention that objects which appear up and left in graphical notations act last on the representation space. Now consider the following objects:
\begin{eqnarray}
\raisebox{-0.4\height}{
\begin{picture}(20,30)(0,-15)
\put(0,10){
\put(0,0){\line(1,0){3}}
\put(17,0){\line(1,0){3}}
}
\put(0,0){\line(1,0){3}}
\put(17,0){\line(1,0){3}}
\put(0,-10){
\put(0,0){\line(1,0){3}}
\put(17,0){\line(1,0){3}}
}
\polygon(3,14)(17,14)(17,-14)(3,-14)
\put(7,-2){$u$}
\end{picture}
}
\equiv
\raisebox{-0.4\height}{
\begin{picture}(44,30)(-3,-15)
\put(-3,0){\line(1,0){3}}
\put(-3,10){\line(1,0){3}}
\put(-3,-10){\line(1,0){3}}
\put(5,0){\oval(10,30)}
\put(2,-2){${\scriptstyle\wedge}$}
\put(10,0){
\put(0,10){
\put(0,0){\line(1,0){3}}
\put(31,0){\line(1,0){3}}
\polygon(3,4)(31,4)(31,-4)(3,-4)
\put(7,-2){${\scriptstyle u}$}
}
\put(0,0){
\put(0,0){\line(1,0){3}}
\put(31,0){\line(1,0){3}}
\polygon(3,4)(31,4)(31,-4)(3,-4)
\put(7,-2){${\scriptstyle u-\hbar}$}
}
\put(0,-10){
\put(0,0){\line(1,0){3}}
\put(31,0){\line(1,0){3}}
\polygon(3,4)(31,4)(31,-4)(3,-4)
\put(7,-2){${\scriptstyle\cdots}$}
}
}
\end{picture}
}
\,,\quad
\raisebox{-0.4\height}{
\begin{picture}(38,8)(-5,-4)
\put(-5,1){\line(1,0){8}}
\put(-5,-1){\line(1,0){8}}
\put(25,1){\line(1,0){8}}
\put(25,-1){\line(1,0){8}}
\polygon(3,4.5)(25,4.5)(25,-4.5)(3,-4.5)
\put(11,-2.5){$\scriptstyle \lambda$}
\end{picture}
}
\equiv
\raisebox{-0.4\height}{
\begin{picture}(44,30)(-3,-15)
\put(-3,0){\line(1,0){3}}
\put(-3,10){\line(1,0){3}}
\put(-3,-10){\line(1,0){3}}
\put(5,0){\oval(10,30)}
\put(2,-2){${\scriptstyle \Pi}$}
\put(10,0){
\put(0,10){
\put(0,0){\line(1,0){3}}
\put(31,0){\line(1,0){3}}
\polygon(3,4)(31,4)(31,-4)(3,-4)
\put(7,-2){${\scriptstyle v}$}
}
\put(0,0){
\put(0,0){\line(1,0){3}}
\put(31,0){\line(1,0){3}}
\polygon(3,4)(31,4)(31,-4)(3,-4)
\put(7,-2){${\scriptstyle v+\hbar}$}
}
\put(0,-10){
\put(0,0){\line(1,0){3}}
\put(31,0){\line(1,0){3}}
\polygon(3,4)(31,4)(31,-4)(3,-4)
\put(7,-2){${\scriptstyle \cdots}$}
}
}
\end{picture}
}\,.
\label{projdef}
\end{eqnarray}
The first one is the minor defined by \eqref{quantumminor} and here $\bigwedge$ stands for antisymmetrisation. The second object is the generalisation of the minor to the case of arbitrary representations with $\Pi$ denoting the projection onto this irrep. We will call it $\lambda$-minor or, equivalently, (a component of) the fused monodromy matrix.

\medskip

To introduce $\bB$ in graphical notations we first recursively define $\bB_k$, for $k=1,\ldots,\gn$, by
\begin{eqnarray}
\bB_1\equiv 
\raisebox{-0.4\height}{
\begin{picture}(40,20)(0,-10)
\put(0,0){\line(1,0){13}}
\put(27,0){\line(1,0){13}}
\polygon(13,4)(27,4)(27,-4)(13,-4)
\put(17,-2){$u$}
\put(33,3){$\scriptstyle 1$}
\end{picture}
}\,,
\bB_2\equiv \raisebox{-0.4\height}{
\begin{picture}(60,30)(0,-10)
\put(0,0){\line(1,0){3}}
\put(0,10){\line(1,0){3}}
\put(21,0){\line(1,0){10}}
\polygon(3,14)(21,14)(21,-4)(3,-4)
\put(4,4){$\scriptstyle u+\hbar$}
\put(18,0){
\put(27,0){\line(1,0){13}}
\polygon(13,4)(27,4)(27,-4)(13,-4)
\put(17,-2){$\scriptstyle u$}
\put(33,3){$\scriptstyle 1$}
}
\put(0,10){
\put(21,0){\line(1,0){13}}
\put(27,3){$\scriptstyle 1$}
}
\end{picture}
}\,,
\ldots\,,
\bB_{k+1}\equiv \raisebox{-0.4\height}{
\begin{picture}(80,40)(0,-10)
\put(0,0){\line(1,0){3}}
\put(0,20){\line(1,0){3}}
\put(0,12){\line(1,0){3}}
\put(29,0){\line(1,0){8}}
\put(29,12){\line(1,0){8}}
\polygon(3,24)(29,24)(29,-4)(3,-4)
\put(6,8){$\scriptstyle u+k\hbar$}
\put(8,20){
\put(21,0){\line(1,0){13}}
\put(27,3){$\scriptstyle 1$}
}
\put(-4,0){
\polygon(41,-4)(41,16)(80,-4)
\put(48,0){$\scriptstyle \bB_{k}$}
}
\end{picture}
}
\end{eqnarray}
and then $\bB=\frac{1}{\qdet\,T(u+\hbar(\gn-1))}\bB_{\gn}$.

\medskip

We will need scattering of the defining representation through the $\lambda$-minor:
\begin{eqnarray}
{
\raisebox{-0.4\height}{
\begin{picture}(50,30)(-3,-15)
\put(-3,0){\line(1,0){3}}
\put(-3,10){\line(1,0){3}}
\put(-3,-10){\line(1,0){3}}
\put(5,0){\oval(10,30)}
\put(2,-2){${\scriptstyle\Pi}$}
\put(10,0){
\put(0,10){
\put(0,0){\line(1,0){40}}
\put(4,2){${\scriptstyle v}$}
}
\put(0,0){
\put(0,0){\line(1,0){40}}
\put(4,2){${\scriptstyle v+\hbar}$}
}
\put(0,-10){
\put(0,0){\line(1,0){40}}
\put(4,2){${\scriptstyle \cdots}$}
}
}
\put(30,-20){\vector(1,3){14}}
\put(24,-18){$\scriptstyle u$}
\end{picture}
}
}
=\prod_{a,s}(v_{a,s}-u)\left(1+\frac{\hbar}{v-u}\mathcal{P}^{\lambda}\right)\,,
\label{scatter1}
\end{eqnarray}
where $\prod\limits_{a,s}f(v_{a,s})=\prod\limits_{a=1}^{h_\lambda}\prod\limits_{s=1}^{\lambda_a} f(v+\hbar(s-a))\,,$ and where $\mathcal{P}^{\lambda}$ is the generalised permutation \eqref{generalisedperm}.

We will also need scattering in the opposite direction through an antisymmetric representation:
\begin{eqnarray}
{
\raisebox{-0.4\height}{
\begin{picture}(50,30)(-3,-15)
\put(-3,0){\line(1,0){3}}
\put(-3,10){\line(1,0){3}}
\put(-3,-10){\line(1,0){3}}
\put(5,0){\oval(10,30)}
\put(2,-2){${\scriptstyle\wedge}$}
\put(10,0){
\put(0,10){
\put(0,0){\line(1,0){40}}
\put(4,2){${\scriptstyle u}$}
}
\put(0,0){
\put(0,0){\line(1,0){40}}
\put(4,2){${\scriptstyle u-\hbar}$}
}
\put(0,-10){
\put(0,0){\line(1,0){40}}
\put(4,2){${\scriptstyle \cdots}$}
}
}
\put(30,20){\vector(1,-3){14}}
\put(32,16){$\scriptstyle v$}
\end{picture}
}
}
=\prod_{a=0}^{m-1}(v-u+\hbar a)\left(1+\frac{\hbar}{v-u+\hbar (m-1)\hbar}\mathcal{P}\right)\,.
\label{scatter2}
\end{eqnarray}

\medskip

Consider  the following chain of equalities
\begin{equation}
\begin{split}
\raisebox{-0.4\height}{
\begin{picture}(80,40)(0,-10)
\put(-10,0){\line(1,0){13}}
\put(-10,20){\line(1,0){13}}
\put(-10,12){\line(1,0){13}}
\put(29,0){\line(1,0){8}}
\put(29,12){\line(1,0){8}}
\polygon(3,24)(29,24)(29,-4)(3,-4)
\put(6,8){$\scriptstyle u+k\hbar$}
\put(8,20){
\put(21,0){\line(1,0){13}}
\put(27,3){$\scriptstyle 1$}
}
\put(-4,0){
\polygon(41,-4)(41,16)(80,-4)
\put(48,0){$\scriptstyle \bB_{k}$}
}
\put(0,-10){
\drawline(3,-1)(-5,-1)(-5,42)(85,42)(85,-1)(25,-1)
\drawline(3,1)(-3,1)(-3,40)(83,40)(83,1)(25,1)
\put(50,3){\line(5,-2){7.2}}
\put(50,-3){\line(5,2){7.2}}
\polygon(3,4.5)(25,4.5)(25,-4.5)(3,-4.5)
\put(11,-2.5){$\scriptstyle \lambda$}
}
\end{picture}
}
& =
\raisebox{-0.4\height}{
\begin{picture}(95,50)(-10,-10)
\put(-10,0){\line(1,0){13}}
\put(-10,20){\line(1,0){13}}
\put(-10,12){\line(1,0){13}}
\put(29,0){\line(1,0){8}}
\put(29,12){\line(1,0){8}}
\polygon(3,24)(29,24)(29,-4)(3,-4)
\put(6,8){$\scriptstyle u+k\hbar$}
\put(8,20){
\put(21,0){\line(1,0){13}}
\put(27,3){$\scriptstyle 1$}
}
\put(-4,0){
\polygon(41,-4)(41,16)(80,-4)
\put(48,0){$\scriptstyle \bB_{k}$}
}
\put(0,31){
\polygon(3,4.5)(25,4.5)(25,-4.5)(3,-4.5)
\put(11,-2.5){$\scriptstyle \lambda$}
}
\put(0,-10){
\drawline(3,40)(-5,40)(-5,49)(85,49)(85,-1)(32,-1)(32,40)(25,40)
\drawline(3,42)(-3,42)(-3,47)(83,47)(83,1)(34,1)(34,42)(25,42)
\put(50,3){\line(5,-2){7.2}}
\put(50,-3){\line(5,2){7.2}}
}
\end{picture}
}\\
& =
\prod_{a,s}(v_{a,s}-u-k\hbar)
\raisebox{-0.4\height}{
\begin{picture}(95,50)(-10,-10)
\put(-10,0){\line(1,0){13}}
\put(-10,26){\line(1,0){13}}
\put(-10,12){\line(1,0){13}}
\put(29,0){\line(1,0){8}}
\put(29,12){\line(1,0){8}}
\polygon(3,30)(29,30)(29,-4)(3,-4)
\put(6,8){$\scriptstyle u+k\hbar$}
\put(8,26){
\put(21,0){\line(1,0){13}}
\put(27,3){$\scriptstyle 1$}
}
\put(-4,0){
\polygon(41,-4)(41,16)(80,-4)
\put(48,0){$\scriptstyle \bB_{k}$}
}
\put(34,-10){
\polygon(3,4.5)(25,4.5)(25,-4.5)(3,-4.5)
\put(11,-2.5){$\scriptstyle \lambda$}
}
\put(0,-10){
\drawline(37,-1)(32,-1)(32,32)(82,32)(82,-1)(59,-1)
\drawline(37,1)(34,1)(34,30)(80,30)(80,1)(59,1)
\put(65,3){\line(5,-2){7.2}}
\put(65,-3){\line(5,2){7.2}}
}
\end{picture}
}
+\mathcal{R}_k(u,v)\,,\nonumber\\
\end{split}
\label{TBchain}
\end{equation}
where $\mathcal{R}_k=\sum_{j=1}^{\gn} \bB_{k+1}(u)T_{j1}(v)\times\ldots\,,$ with dots standing for expressions whose explicit form is not relevant for further computations and the closed loop indicates that the trace has been taken over the $\lambda$ representation. The expression on the \lhs is the following product 
\begin{equation}
\begin{split}
&{\rm tr}_\lambda\left(\bar{R}^{\wedge^{k+1}\lambda}(u+k\,\hbar,v) \bT^\lambda_b(v)\right)\bB_{k+1}(u) \\ 
&={\rm tr}_\lambda\left(\bar{R}^{\wedge^{k+1}\lambda}(u+k\,\hbar,v)\bT^\lambda_b(v) \bT^{\wedge^{k+1}}(u+k\,\hbar)\bB_k(u)\right) \\
\end{split}
\end{equation}
where we used that the trace is only over the space $\lV^\lambda$. The first equality was obtained by applying the RTT relation between fused monodromy matrices and results in 
\begin{equation}\label{secondscatter}
\bT^{\wedge^{k+1}}(u+k\,\hbar){\rm tr}_\lambda\left(\bar{R}^{\wedge^{k+1}\lambda}(u+k\,\hbar,v)\bT^\lambda_b(v)\right)\bB_k(u)
\end{equation}
where we also used the cyclicity of the trace over $\lV^\lambda$. 

\medskip

The second equality was obtained from the following scattering, {\it cf.} \eqref{scatter1}, 
\begin{eqnarray}
\raisebox{-0.5\height}{
\begin{picture}(45,10)(-5,-8)
\put(-5,1){\line(1,0){8}}
\put(-5,-1){\line(1,0){8}}
\put(25,1){\line(1,0){18}}
\put(25,-1){\line(1,0){18}}
\polygon(3,4.5)(25,4.5)(25,-4.5)(3,-4.5)
\put(11,-2.5){$\scriptstyle \lambda$}
\put(30,-8){\vector(2,5){7}}
\put(37,9){$\scriptstyle 1$}
\end{picture}
}
=
\prod_{\alpha}(v_{\alpha}-u-k\hbar)
\raisebox{-0.5\height}{
\begin{picture}(45,10)(-5,-8)
\put(-5,1){\line(1,0){8}}
\put(-5,-1){\line(1,0){8}}
\put(25,1){\line(1,0){6}}
\put(25,-1){\line(1,0){6}}
\put(35,1){\line(1,0){8}}
\put(35,-1){\line(1,0){8}}
\polygon(3,4.5)(25,4.5)(25,-4.5)(3,-4.5)
\put(11,-2.5){$\scriptstyle \lambda$}
\put(30,-8){\vector(2,5){7}}
\put(37,9){$\scriptstyle 1$}
\end{picture}
}
+
\raisebox{-0.4\height}{
\begin{picture}(44,30)(-3,-15)
\put(-3,0){\line(1,0){3}}
\put(-3,10){\line(1,0){3}}
\put(-3,-10){\line(1,0){3}}
\put(5,0){\oval(10,30)}
\put(2,-2){${\scriptstyle\Pi}$}
\put(10,0){
\put(0,10){
\put(0,0){\line(1,0){3}}
\put(31,0){\line(1,0){5}}
\put(36,-2){$\scriptstyle 1$}
\polygon(3,4)(31,4)(31,-4)(3,-4)
\put(7,-2){${\scriptstyle v}$}
}
\put(0,0){
\put(0,0){\line(1,0){3}}
\put(31,0){\line(1,0){3}}
\polygon(3,4)(31,4)(31,-4)(3,-4)
\put(7,-2){${\scriptstyle v+\hbar}$}
}
\put(0,-10){
\put(0,0){\line(1,0){3}}
\put(31,0){\line(1,0){3}}
\polygon(3,4)(31,4)(31,-4)(3,-4)
\put(7,-2){${\scriptstyle \cdots}$}
}
}
\end{picture}
}
\times\ldots\,.
\label{eq332}
\end{eqnarray}
For the second term in the \rhs., we need only that it is always of the form $\sum_{\lB}T[^\lA_{\lB}]\times\cdots$, where $1\in \lB$. By using the symmetry imposed by the symmetrisation, one can prove that for any $\lB$ which contains $1$, one can represent the $\lambda$-minor as linear combination $T[^\lA_{\lB}]=\sum_{\lB'}\# T[^\lA_{\lB'}]$, where $\#$ stand for numerical coefficients irrelevant for us and all $\lB'$ are such that $\lB_{11}'=1$. Then it follows that the second term in \eqref{eq332} is always of the form $\sum_j T[^j_1](v)\times\ldots$.

\medskip

We use relatons \eqref{TBchain} to pull the trace over the $\lambda$-minors through the $\bB$-operator. At the right-most step one gets $\T_{\lambda}\bB_\gn$ plus $\mathcal{R}$-terms. At the left-most step, one uses the scattering with the fully-antisymmetric representation
\begin{eqnarray}
{
\raisebox{-0.4\height}{
\begin{picture}(125,70)(0,-50)
\put(0,0){
\put(-3,0){\line(1,0){3}}
\put(-3,10){\line(1,0){3}}
\put(-3,-10){\line(1,0){3}}
\put(5,0){\oval(10,30)}
\put(2,-2){${\scriptstyle\Pi}$}
\put(10,0){
\put(0,10){
\put(0,0){\line(1,0){45}}
\put(4,2){${\scriptstyle v}$}
}
\put(0,0){
\put(0,0){\line(1,0){45}}
\put(4,2){${\scriptstyle v+\hbar}$}
}
\put(0,-10){
\put(0,0){\line(1,0){45}}
\put(4,2){${\scriptstyle \cdots}$}
}
}
}
\put(0,-35){
\put(-3,0){\line(1,0){3}}
\put(-3,10){\line(1,0){3}}
\put(-3,-10){\line(1,0){3}}
\put(5,0){\oval(10,30)}
\put(2,-2){${\scriptstyle\wedge}$}
\put(10,0){
\put(0,10){
\put(0,0){\line(1,0){45}}
\put(4,2){${\scriptstyle u+(N-1)\hbar}$}
}
\put(0,0){
\put(0,0){\line(1,0){45}}
\put(4,2){${\scriptstyle \cdots}$}
}
\put(0,-10){
\put(0,0){\line(1,0){45}}
\put(4,2){${\scriptstyle u}$}
}
}
}
\put(55,-10){\line(1,-1){35}}
\put(55,0){\line(1,-1){45}}
\put(55,10){\line(1,-1){55}}
\put(0,-35){
\put(55,10){\line(1,1){35}}
\put(55,0){\line(1,1){45}}
\put(55,-10){\line(1,1){55}}
}
\end{picture}
}
}
=\prod_{a,s}\frac{v_{a,s}-u+\hbar}{v_{a,s}-u}\prod_{k=0}^{N-1}(v_{a,s}-u-k\hbar)\times 1\,
\end{eqnarray}
to take the trace cycle off the chain of $\bB_k$'s hence producing an operator proportional to $\bB\,\T_{\lambda}$.

\medskip

In summary, one gets the following relation
\begin{eqnarray}
\displaystyle \T_{\lambda}^{\lN}(v)\bB(u)=\prod\limits_{a,s}\frac{u-v_{a,s}-\hbar}{u-v_{a,s}}\,\bB(u)\T_{\lambda}^{\lN}(v)+\mathcal{R}(u,v)\,,
\label{TB}
\end{eqnarray}
where $\mathcal{R}(u,v)=\sum\limits_{k=0}^{\gn-1}\sum\limits_{j=1}^{\gn}\bB_k(u)T_{j1}(v)\times\ldots\,,$ and the product over the Young tableau boxes reduces to the following explicit expression
\begin{eqnarray}
\prod\limits_{a,s}\frac{u-v_{a,s}-\hbar}{u-v_{a,s}}=\prod_{a=1}^{h_{\lambda}}\frac{u-v+\hbar\,(a-1-\lambda_a)}{u-v+\hbar\,(a-1)}\,.
\label{TB2}
\end{eqnarray}

\medskip

Next, we use the following commutation relation between a quantum minor and $T_{ij}(u)$ \cite{molev2007yangians}. We have
\begin{equation}
(u-v)[T_{kl}(u),T\left[^A_B\right](v)]=\hbar \left(\displaystyle \sum_{i=1}^m T_{a_i l}(u)T\left[^{A_{i,k}}_B\right](v)- T\left[^{A}_{B_{i,l}}\right](v)T_{k,b_i}(u)\right)
\end{equation}
where the notation $A_{i,k}$ means that $A_i$ has been replaced with $k$ and similarly with $B$. We now restrict to the case where $k=j,l=1$ and $b_1=1$. As a result of the anti-symmetry of quantum minors the only term in the second sum on the \rhs which can contribute is that with $i=1$. Simplifying, we find 
\begin{equation}
T\left[^A_B\right](u)T_{j1}(v) = T_{j1}(v) T\left[^A_B\right](u)\times \dots
\end{equation}
By applying this recursively we obtain that the final form of the required commutation relation
\begin{equation}
\boxed{\T_\lambda(v) \bB(u) = f_\lambda(u,v) \bB(u) \T_\lambda(v) + \displaystyle\sum_{j=1}^{\gn} T_{j1}(v) \times \dots}
\end{equation}
where the null-twist transfer matrices $\T^\lN_\lambda$ have been upgraded to the full transfer matrices $\T_\lambda$ and the function $f_\lambda(u,v)$ is given by 
\begin{equation}
f_\lambda(u,v)=\prod_{a=1}^{h_{\lambda}}\frac{u-v+\hbar\,(a-1-\lambda_a)}{u-v+\hbar\,(a-1)}\,.
\end{equation}

\subsection{Using the commutation relation}

\paragraph{Rectangular representations}
We will now demonstrate how to use this commutation relation to diagonalise $\bB$ in a simple set-up. We consider the length $L=1$ case and the representation of $\gl(3)$ corresponding to the representation $\nu=[2,2,0]$. 

\medskip

Starting from $\bra{0}$ we have $\bra{0}T_{j1}(\theta)=0$ for $j=1,2,3$. Hence, by virtue of the commutation relation \eqref{BTcomm} we have that, for any $\lambda$, the state $\bra{0}\T_\lambda(\theta)$ is an eigenvector of $\bB$. Of course this representation is $6$-dimensional and so the these states are not linearly independent for all $\lambda$. For $\gl(3)$ $\bB$ has non-degenerate spectrum so, assuming the states are non-zero, we just need to construct $6$ states with different $\bB$ eigenvalue. 

\medskip

We will not prove the details here, but if suffices to say that for any subdiagram $\lambda\subset \nu$ the resulting set of states are non-zero and correspond to different eigenvalues of $\bB$, the latter point which can be checked from our commutation relation. These states also form a basis since there are precisely $6$ subdiagrams of $[2,2,0]$ (including the empty diagram). The most remarkable feature however is the precise relation between the Young diagram $\lambda$ and the corresponding eigenvalue of $\bB$. Indeed, eigenvalues of $\bB$ are labelled by Gelfand-Tsetlin patterns, so there should be some relation between $\lambda$ and GT patterns. The precise relation is incredibly simple. 

\medskip

A generic GT pattern for the representation we are considering has the form 
\begin{equation}
\begin{array}{cccccc}
2 & \ & 2 & \ & 0 \\
\ & 2 & \ & \lambda_{22} \\
\ & \ & \lambda_{21} 
\end{array}
\end{equation}
It can be checked that if $\lambda = [\lambda_{21},\lambda_{22},0]\subset [2,2,0]$ is a subdiagram then the state $\bra{0}\T_\lambda(\theta)$ corresponds to precisely this GT pattern! 

\medskip

Owing to the incredible simplicity of this situation we can now conjecture what happens in general. Let us consider a more involved setting with $\gl(6)$ and the representation $[7,7,7,0,0,0]$. The relation between Young diagrams and nodes on GT patterns is illustrated in Figure \ref{SoVaction}. 

\begin{figure}[htb]
\centering
  \includegraphics[width=150mm,scale=10]{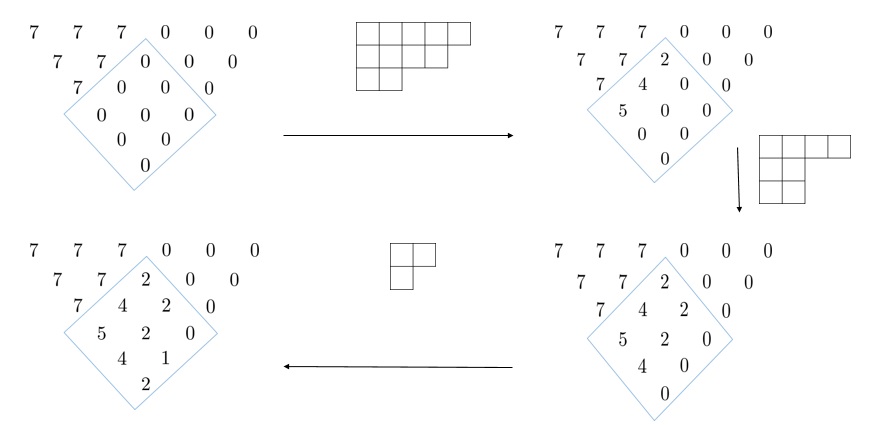}
    \caption{Successive action of transfer matrices corresponding to different Young diagrams on the vacuum state $\bra{0}$. The area in the rectangle corresponds to dynamical nodes which we aim to excite. The transfer matrices act by filling up the dual diagonals with the numbers corresponding to their Young diagram.}
\end{figure}\label{SoVaction}

\paragraph{Degenerate spectrum}

While the above procedure for building eigenvectors of $\bB$ is tremendously simple there is a caveat however and is associated with settings where the spectrum of $\bB$ is degenerate. When we have a representation where $\bB$ has non-degenerate spectrum we can immediately associate to each eigenvector a GT pattern and allows us to easily count linearly independent vectors. On the other hand, for degenerate cases there can be two GT patterns corresponding to the same eigenvalue. The simplest example where this arises is the representation $\nu=[2,2,0,0]$ of $\gl(4)$. Indeed, the two GT patterns 
\begin{equation}
\begin{array}{ccccccccc}
2 & \ & 2 & \ & 0 & \ & 0 \\
\ & 2 & \ & 1 & \ & 0 &  \\
\ & \ & 2 & \ & 0 & \ &  \\
\ & \ & \ & 1 & \ & \ &  \\
\end{array},\quad 
\begin{array}{ccccccccc}
2 & \ & 2 & \ & 0 & \ & 0 \\
\ & 2 & \ & 2 & \ & 0 &  \\
\ & \ & 2 & \ & 0 & \ &  \\
\ & \ & \ & 0 & \ & \ &  \\
\end{array}
\end{equation}
correspond to the same eigenvalue of $\bB^{\rm GT}$. The corresponding eigenvectors of $\bB$ are given by 
\begin{equation}
\bra{0}\T_{[2,1,0]}(\theta) \T_{[1,0,0]}(\theta),\quad \bra{0}\T_{[2,2,0]}(\theta)\,.
\end{equation}
At present we have no way of telling if both of these states are linearly independent. In the next Section we will develop a systematic approach for demonstrating linear independence.

\paragraph{Non-rectangular representations}

Our prescription for constructing eigenvectors for $\bB$ works very well for rectangular representations (where the Young diagram describing the physical space is a rectangle) but obviously not every Young diagram is rectangular. The simplest non-rectangular representation is $\nu=[2,1,0]$ for $\gl(3)$. The issue is that the method described above is not sufficient to generate all eigenstates of $\bB$. As before, we start from the SoV vacuum state $\bra{0}$ corresponding to the GT pattern
\begin{equation}
\bra{0}\ \leftrightarrow \ 
\begin{array}{cccccc}
2 & \ & 1 & \ & 0 \\
\ & 1 & \ & 0 & \\
\ & \ & 0 & \ & 
\end{array}
\end{equation}
We can then proceed to generate the following two states 
\begin{equation}
\bra{0}\T_{1,1}(\theta)\ \leftrightarrow \ 
\begin{array}{cccccc}
2 & \ & 1 & \ & 0 \\
\ & 1 & \ & 0 & \\
\ & \ & 1 & \ & 
\end{array},\quad\quad  \bra{0}\T_{2,1}(\theta)\ \leftrightarrow \ 
\begin{array}{cccccc}
2 & \ & 1 & \ & 0 \\
\ & 1 & \ & 1 & \\
\ & \ & 1 & \ & \,.
\end{array}
\end{equation}
On the other hand, there seems to be no way we can create the state corresponding to the GT pattern
\begin{equation}
\begin{array}{cccccc}
2 & \ & 1 & \ & 0 \\
\ & 2 & \ & 0 & \\
\ & \ & 0 & \ & \,.
\end{array}
\end{equation}
A natural guess would be to act with a transfer matrix $\T_\lambda(\theta+\hbar)$ but then the remainder term $\lR$ in our commutation relation will not vanish. However, the situation is not hopeless. By direct comparison of eigenvalues one can check that the state $\bra{0}\T_{1,2}(\theta)$ corresponds to the GT pattern 
\begin{equation}
\bra{0}\T_{1,2}(\theta)\ \leftrightarrow \ \begin{array}{cccccc}
2 & \ & 1 & \ & 0 \\
\ & 2 & \ & 0 & \\
\ & \ & 1 & \ & 
\end{array}
\end{equation}
Next, consider the state $\bra{0'}$ which is the eigenvector of the Gelfand-Tsetlin algebra corresponding to the pattern
\begin{equation}
\begin{array}{cccccc}
2 & \ & 1 & \ & 0 \\
\ & 2 & \ & 0 & \\
\ & \ & 0 & \ & 
\end{array}
\end{equation}
Remarkably, a direct calculation shows that this is also an eigenstate of $\bB$! Indeed, we have 
\begin{equation}
\bB(u) = \bB^{\rm GT}(u) + T_{21}(u) T^{[2]}\left[^{12}_{13} \right](u)
\end{equation}
and $\bra{0'}T_{21}(u)=0$. Furthermore, we also see that $\lR(u,\theta)$ annihilates $\bra{0'}$ meaning we can use it as an alternate SoV vacuum to generate states from. Indeed, we can immediately see that $\bra{0'}\T_{1,1}(\theta)$ corresponds to the state 
\begin{equation}
\begin{array}{cccccc}
2 & \ & 1 & \ & 0 \\
\ & 2 & \ & 0 & \\
\ & \ & 1 & \ & 
\end{array}
\end{equation}
precisely matching the GT pattern corresponding to $\bra{0}\T_{1,2}(\theta)$! Hence, after appropriate normalisation of $\bra{0'}$ we have 
\begin{equation}
\bra{0'}\T_{1,1}(\theta) = \bra{0}\T_{1,2}(\theta)
\end{equation}
and hence 
\begin{equation}
\bra{0'} = \bra{0}\left(\T_{1,1}(\theta)\right)^{-1}\T_{1,2}(\theta)
\end{equation}
Hence, in lieu of our discussion of rectangular representations and how transfer matrices act on GT patterns we can interpret this as the following sequence of steps
\begin{equation}
\begin{array}{cccccc}
2 & \ & 1 & \ & 0 \\
\ & 1 & \ & 0 & \\
\ & \ & 0 & \ & 
\end{array} 
\ \underrightarrow{\left(\T_{1,1}(\theta)\right)^{-1}}\ 
\begin{array}{cccccc}
2 & \ & 1 & \ & 0 \\
\ & 0 & \ & 0 & \\
\ & \ & 0 & \ & 
\end{array} \ \underrightarrow{\T_{1,2}(\theta)}\ 
\begin{array}{cccccc}
2 & \ & 1 & \ & 0 \\
\ & 2 & \ & 0 & \\
\ & \ & 0 & \ & 
\end{array}
\end{equation}
It is important to point out that the middle array is not a GT pattern as the branching rules are not satisfied. Nevertheless allowing ourselves to imagine the existence of such a pattern makes the procedure for constructing states very intuitive and indeed this intuition extends to all representations and any rank of $\gl(\gn)$. 

\medskip

In the next section we will make this precise. The main tool we will use for this is the so-called \textit{embedding morphism}. 
\section{Embedding Morphism}

\subsection{Embedding morphism and Gelfand-Tsetlin basis}\label{sec:embeddingmorphism}
As was described above, the Gelfand-Tsetlin algebra is constructed by considering the tautological injection $T_{ij}\mapsto T_{ij}$ of $\lY_k$ into $\lY_{k+1}$. Now consider a different (nearly) tautological injection of $\lY_k$ into $\lY_{k+1}$ defined by
\begin{equation}
\label{embtwo}
\phi:T_{ij}(u)\mapsto T_{1+i,1+j}(u)\,.
\end{equation} 
We use it for a different purpose: to construct a special embedding of a $\gl(k)$ spin chain into a $\gl(k+1)$ chain that shall be called \textit{embedding morphism}. Formally the embedding morphism is an induced map $\phi:\CH_{k}\to \CH_{k+1}$, where $\CH_{k}$ is the Hilbert space of the $\gl(k)$ spin chain of length $L$ with spin chain sites in irreps $(\nu_1^{\alpha},\ldots,\nu_k^{\alpha})$, fully defined by the following property
\begin{equation}\label{eq:embphi}
 \phi:\bra{0_k}\mathcal{J}\mapsto \bra{0_{k+1}}\phi(\mathcal{J})\,,
\end{equation}
where $\mathcal{J}$ is any element of $\lY_k$, and $\bra{0_k}$ is the lowest-weight vector of the $\gl(k)$ chain -- the state whose GT pattern has the lowest possible entries $\mu_{ij}^{\alpha}=\nu_{i+1}^{\alpha}$ for $i=1,2,\ldots,k-1$, $j=1,2,\ldots, i$.

\medskip

Define $\lV_{(k)}:=\phi(\CH_{k})$.  By abuse of notation we may also use $\lV_{(k)}=\phi^m(\CH_{k})$, for $m=2,3,\ldots,\gn-k$ and so in particular we think about $\lV_{(k)}$ as a subspace in the full $\gln$ spin chain which represents a smaller $\gl(k)$ chain. Remarkably, the embedding morphism has a simple coordinatisation using GT patterns:
\begin{equation}
\label{eq:im1}
\phi\left(
\mbox{
\begin{picture}(72,26)(8,23)
\put(4,40){
$
\begin{array}{ccccccccccc}
\nu_{1}^\alpha & \ldots & \  & \nu_{k}^\alpha  
\end{array}
$
}
\put(33,20){
$\mu_{ij}^{\alpha}$
}
\drawline(16,32)(70,32)(43,5)(16,32)
\end{picture}
}
\right)
\propto
\mbox{
\begin{picture}(100,30)(8,16)
\put(5,40){
$
\begin{array}{cccccccccccc}
\nu_{1}^\alpha & \ldots & \ & \nu_{k}^\alpha & \nu_{k+1}^{\alpha}
\end{array}
$
}
\put(33,20){
$\mu_{ij}^{\alpha}$
}
\put(53,-2){$\nu_{k+1}^{\alpha}$}
\put(78,23){$\nu_{k+1}^{\alpha}$}
\put(68,9){\rotatebox{45}{$\ldots$}}
\drawline(16,32)(70,32)(43,5)(16,32)
\end{picture}
}
\,,
\end{equation}
{\it i.e.} the image of a state with the GT pattern $\Lambda'$ for the $\gl(k)$ spin chain is the state for the $\gl(k+1)$ chain with the GT pattern which has the right-most dual diagonal at the lowest possible value and the remaining triangular block coinciding with $\Lambda'$. 

\medskip

The above implies the following property of $\lH_{k+1}$ which we will frequently use. If $\bra{\Lambda}\in\lH_{k+1}$ is obtained from a vector in $\lH_{k}$ by action of $\phi$ then $T_{11}(u)=\GT_1(u)\in\lY_{k+1}$ has the eigenvalue $\nu_{k+1}(u)$ on $\bra{\Lambda}$. Since the eigenvalue of $T_{11}$, and hence of the global Cartan generator $\gloE_{11}$, is at its lowest possible value and the eigenvalue of $\gloE_{11}$ is lowered by $T_{j1},\ j>1$ it follows that
\begin{equation}\label{lowering}
\bra{\Lambda}T_{j1}(u)=\delta_{j1}\nu_{k+1}(u)\bra{\Lambda},\quad j=1,\dots,k+1.
\end{equation}
To see why the property \eqref{eq:im1} indeed holds it is enough to check that the raising operators $\GP_a^{+}$ act accordingly because their action generates the whole Hilbert space starting from the lowest-weight state. To this end consider yet another family of homomorphisms \cite{molev2007yangians} $\psi_m:\lY_k\longrightarrow \lY_{k+m}$ for $m=1,2,\ldots$ defined by
\begin{equation}\label{psidefn}
\psi_m:T_{ij}(u)\mapsto \left(\GT_m(u+m\hbar) \right)^{-1}T\left[^{1\ldots m\ m+i}_{1\ldots m\ m+j} \right](u+m\hbar)\,.
\end{equation}
One can show that, for any quantum minor $T\left[^\lA_\lB\right](u)$,
\begin{equation}
\psi_m:T\left[^\lA_\lB\right](u)\mapsto \left(\GT_m(u+m\hbar) \right)^{-1}T\left[^{1\ldots m\ \lA+m}_{1\ldots m\ \lB+m} \right](u+m\hbar)\,,
\end{equation}
and that $\psi_m=(\psi_1)^m$. Then
\begin{equation}
\label{eq:im2}
\psi_1(\GP_a^{\pm}(u))=\left(\GT_1(u+\hbar) \right)^{-1}\GP_{a+1}^{\pm}(u+\hbar)\,.
\end{equation}
Define an embedding morphism of spin chains $\psi_1:\CH_{k}\to\CH_{k+1}$ by \eqref{eq:embphi} with $\phi$ replaced by $\psi_1$. Given \eqref{eq:im2}, relation \eqref{eq:im1} with $\phi$ replaced by $\psi_1$ is obvious: on one hand, \eqref{eq:im2} states that action of raising and lowering operators commutes, up to normalisation, with $\psi_1$. On the other hand, one gets in the image of $\psi_1$ precisely the states of $\CH_{k+1}$ that are generated by $\GP_2^+,\GP_3^+,\ldots,\GP_k^+$ acting on $\bra{0_{k+1}}$. Finally, one notes that the last dual diagonal cannot be excited by these operators if the node $\mu_{k1}^{\alpha}$ attains its lowest value $\mu_{k1}^{\alpha}=\nu_{k+1}^{\alpha}$. But $\mu_{k1}^{\alpha}$ can only change by action of $\GP_1^+$ which cannot be represented as $\psi_1(\GP_a^+)$.

\medskip

Now we remark that the embeddings $\psi_1$ and $\phi$ coincide. Indeed, for any $\bra{\Lambda}$ of the $\gl(k+1)$ chain with $\mu_{k1}^{\alpha}=\nu_{k+1}^{\alpha}$ one has $\bra{\Lambda}T_{j1}(u)=\delta_{j1}\nu_{k+1}(u)\bra{\Lambda}$ as was established above, and so one computes
\begin{equation}
\bra{\Lambda}\psi_1(T_{ij}(u))=(\nu_{k+1}(u+\hbar))^{-1}\bra{\Lambda}T\left[^{1\ 1+i}_{1\ 1+j}\right](u+\hbar)
=\bra{\Lambda}\phi(T_{ij}(u))\,.
\end{equation}
Hence $\psi_1(T_{ij}(u))=\phi(T_{ij}(u))$ when restricted to $\lV_{(k)}$, and so \eqref{eq:im1} holds.

\paragraph{A roadmap to the GT basis}
Now we present a special generation of states in the GT basis based on the embedding morphism. The idea is to consider a recursive procedure
\begin{equation}
\label{eq:CHLV}
 \cdots \to\CH_{k}\lhook\joinrel\xrightarrow{\ \phi\ } \lV_{(k)} \xrightarrow{\lS} \CH_{k+1} \lhook\joinrel\xrightarrow{\ \phi\ }\cdots\,, 
 \end{equation}
where $\lS$ is the introduced-below composite raising operator that excites the largest dual diagonal from its lowest to the desired value. The recursion starts from the lowest weight state of the $\gl(2)$ spin chain which spans $\lV_{(1)}$ and terminates with the full Hilbert space $\CH_{\gn}$.

\medskip

We start by considering a state $\bra{\Lambda}\in\lH_{k+1}$ obtained from a state in $\lH_k$ by action of the embedding morphism. By definition, $\Lambda$ is an $L$-tuple of patterns $\Lambda=(\Lambda^1,\dots,\Lambda^L)$ and each $\Lambda^\alpha$ has $\mu^\alpha_{kj}=\nu^\alpha_{k+1}$, $j=1,\dots,k$. From here we will construct a state where $\mu_{kj}^\alpha=\nu^\alpha_{k+1}+1$, $j=1,\dots,a$, $\mu^\alpha_{kj}=\nu^\alpha_{k+1}$ for $j>a$, for some $1\leq a\leq k$. By the properties of the GT raising operators we know that we can obtain such a state by acting on $\bra{\Lambda}$ with the operators which raise those particular nodes, obtaining 
\begin{equation}\label{creating}
\bra{\Lambda}\GP^+_1\GP^+_2\dots\GP^+_a\,,
\end{equation} 
where each $\GP^+$ is evaluated at $\theta_\alpha+\hbar\,\nu^\alpha_{k+1}$. This can be written explicitly in terms of minors as
\begin{equation}\label{dualraising1}
\bra{\Lambda}T\left[^{1}_{2}\right]T\left[^{12}_{13}\right]\dots T\left[^{12\dots a-1\ a}_{12\dots a-1\ a+1}\right]\,.
\end{equation}
By straightforward application of the quantum column expansion of minors \cite{molev2007yangians} one can show that \eqref{dualraising1} coincides, up to a non-zero coefficient, with
\begin{equation}\label{contraction}
\bra{\Lambda}T\left[^{12\dots a}_{23\dots a+1}\right](\theta_\alpha+\hbar\,\nu^\alpha_{k+1})\,.
\end{equation}
From here, one can further excite the excited nodes, filling up a certain number of nodes successively by $1$ until the full dual diagonal has reached the desired value. In summary, we have the following.
For a Young diagram $\bmu_k$ of height $h_{\bmu_k}\leq k$, let us define a composite operator $\lS_{\bmu_k}(u)$ by
\begin{equation}\label{compositeraising}
\lS_{\bmu_k}(u)=\prod_{j\in {\rm col}(\bmu_k)}^\rightarrow \lS_{\bmu_k,j}(u+\hbar(j-1))\,,
\end{equation}
where the product is over the number of columns ${\rm col}(\bmu_k)$ of $\bmu_k$; and $\lS_{\bmu_{k,j}}$ is the raising operator associated to the $j$-th column of $\bmu_k$. Specifically, if we let $h_{\bmu_k}^j$ denote the number of boxes in the $j$-th column of $\bmu_k$ then 
\begin{equation}
S_{\bmu_k,j}(u)=T\left[^{1 \ 2\ \dots\ h^j_{\bmu_k}}_{2\ 3\ \dots\ h^j_{\bmu_k}+1}\right](u)\,.
\end{equation}
Then $\bra{\Lambda}\prod\limits_{\alpha=1}^L S_{\bmu_k^{\alpha}}(\theta_{\alpha}+\hbar\nu_{k+1}^{\alpha})$ is a state in $\CH_{k+1}$ whose $k$-th dual diagonals are excited to values $\mu_k^1,\mu_k^2,\ldots,\mu_k^L\,.$ Finally, by running the recursion \eqref{eq:CHLV}, we can write any element of the GT basis as 
\begin{equation}
\bra{\Lambda^{\rm GT}}=\bra{0}\displaystyle\prod_{k}^{\leftarrow}\prod_{\alpha=1}^L \phi^{\gn-k-1}\left(\lS_{\bmu^\alpha_{k}}(\theta_\alpha+\hbar\,\nu^\alpha_{k+1})\right)\,,
\end{equation}
where the first product ranges over $k=1,\dots,\gn-1$.

\subsection{Embedding morphism and the $\bB$ operator}\label{bsection}
We will explain how the $\bB$ operator relates to the embedding morphism. The idea is to construct the eigenvectors of $\bB$ by ascending through the spin chains of increasing rank
\begin{equation}
\label{eq:CHLV2}
 \cdots \to\CH_{k}\lhook\joinrel\xrightarrow{\ \phi\ } \lV_{(k)} \xrightarrow{\T_{\bar\mu_k}} \CH_{k+1} \lhook\joinrel\xrightarrow{\ \phi\ }\cdots\,. 
\end{equation}
The procedure is rooted in the following two observations. Firstly,
\begin{equation}
\label{eq:bBk}
\bB^{(k+1)}{|}_{\lV_{(k)}}\sim \phi\left(\bB^{(k)}\right){|}_{\lV_{(k)}}\,,
\end{equation}
where $\bB^{(k)}$ denotes the $\bB$-operator for the $\gl(k)$ spin chain, and $\sim$ means equality up to multiplication by an operator which is proportional to the identity when restricted to $\lV_{(k)}$. This property allows one to build all eigenstates of $\bB^{(k+1)}$ for which the last dual diagonal is not excited, simply by applying the embedding morphism to smaller-rank chains.

\medskip

Secondly, we excite the last dual diagonal of $\gl(k+1)$ patterns by action of transfer matrices  $\T_{\bar\mu_k}$, where the choice of representation $\bar\mu_k$ dictates how the diagonal should be excited. This step closely follows the procedure outlined in the previous section. 

\medskip

Let us now understand how the crucial property \eqref{eq:bBk} comes about. The \rhs of \eqref{eq:bBk} is the image of $\bB^{(k)}$, and $\bB^{(k)}$ is defined by  \eqref{Bincomp} with $\gn$ being replaced with $k$. It is an operator acting on $\CH_{k}$. The \lhs of \eqref{eq:bBk} contains the operator $\bB^{(k+1)}$ acting on $\CH_{k+1}$. We illustrate its restriction to the subspace $\lV_{(k)}$ for the case $k+1=\gn$. From  \eqref{lowering} and the definition of minors \eqref{quantumminor} it follows that $T^{[2r]}\left[^{J_{r+1}}_{1\ J_r+1}\right]$ is only non-zero if $J_{r+1}$ contains $1$. Denote then $J_{r+1}=(1\ J'_{r+1}+1)$ and then simplify, using \eqref{lowering}, $T^{[2r]}\left[^{1\ J'_{r+1}+1}_{1\ J_r+1}\right]=\nu_{\gn}(u+\hbar r)\phi\left(T^{[2(r-1)]}\left[^{J'_{r+1}}_{J_r}\right]\right)$.  Overall, one gets
\begin{equation}
\label{eq:bBk2}
\bB^{(\gn)}{|}_{\lV_{(\gn-1)}}=\prod_{r=0}^{\gn-2}\nu_{\gn}(u+\hbar\,r)\, \phi\left(\bB^{(\gn-1)}\right){|}_{\lV_{(\gn-1)}}\,.
\end{equation}
Obviously, the above conclusion holds when we replace $\gn$ with $k+1$ which confirms \eqref{eq:bBk}.

\medskip

As already outlined, \eqref{eq:bBk} ensures that eigenvectors of $\bB^{(k)}$ become eigenvectors of $\bB^{(k+1)}$ upon using the embedding morphism. Moreover, one guarantees that $\bra{\Lambda^{\bB}}\in\lV_{(k)}\subset\CH_{\gn}$ if and only if at most the first $k-1$ dual diagonals  are excited above their minimal values (for each $\Lambda^\alpha$ of the pattern $\Lambda=(\Lambda^1,\ldots,\Lambda^L)$). This is not a trivial conclusion as $\bra{\Lambda^{\bB}}$ deforms $\bra{\Lambda^{\rm GT}}$ and so its relation to the subspaces $\lV_{(k)}$ could become obscured.  It allows us to consider $\svX_{k'j}^{\alpha}$ as operators defined for any $\gl(k)$ chain with $\svX_{k'j}^{\alpha}=\phi^*(\svX_{k'j}^{\alpha})$, where $\phi^*$ is a pullback of the embedding morphism. For $k>k'$, these operators, for generic representations, are dynamical having all possible eigenvalues permitted by branching rules. For $k\leq k'$,  $X_{k'j}^{\alpha}$ are non-dynamical and they attain only their lowest values.

\paragraph{Diagonalising the $\bB$-operator}
In the previous subsection we clarified how the embedding $\CH_{k}\lhook\joinrel\xrightarrow{\ \phi\ } \lV_{(k)}\subset\CH_{k+1}$ works. This subsection focuses mostly on the excitation step $ \lV_{(k)} \xrightarrow{\T_{\bar\mu_k}} \CH_{k+1}$. We understand by now that one should focus on exciting the longest dual diagonal as all the other diagonals should have been excited to the desired values at lower-rank stages of the recursion.

\medskip

The $\bB$-operator is independent of the twist matrix eigenvalues $z_1,\dots,z_{\gn}$ and hence so are its eigenvectors. Since we expect to construct eigenvectors of $\bB$ with transfer matricies $\T_{\lambda}$, it is natural then to check the case of the null twist first, where the null twist is defined as the MCT with  $z_j=0$. In the previous section we derived the following commutation relation \eqref{BTcom} between $\bB$ and transfer matricies $\T_{\lambda}$:
\begin{equation}\label{BTcomm}
\T_{\lambda}(v)\bB(u)=f_{\lambda}(u,v)\bB(u)\T_{\lambda}(v)+\lR(u,v)\,,
\end{equation}
where $f_{\lambda}(u,v)$ is a function given explicitly by 
\begin{equation}
\label{fvalue}
f_{\lambda}(u,v)=\displaystyle\prod_{a=1}^{h_\lambda}\frac{u-v+\hbar(a-1-\lambda_a)}{u-v+\hbar(a-1)}\,,
\end{equation}
and $\lR(u,v)=\sum_{j=1}^\gn T_{j1}(v)\times \dots$. 

\medskip

Our goal is to engineer a situation when the remainder $\lR(u,v)$ vanishes. Then we can use \eqref{BTcomm} to intertwine between eigenstates of $\bB$. We say that $\bra{\Lambda}$ is an {\label{pos:admissible}admissible} vector at point $v$ if it is an eigenstate of $\bB$ and it satisfies $\bra{\Lambda}T_{j1}(v)=0$ for all $j$ and the given value of $v$. 

\medskip

From \eqref{BTcomm}, it is clear that if $\bra{\Lambda}$ is admissible at point $v$ then $\bra{\Lambda}\T_{\lambda}(v)$ is  an eigenstate of $\bB$ provided that the action of $\T_{\lambda}(v)$ on $\bra{\Lambda}$ is non-zero. We briefly discuss the relevant properties of transfer matrices $\T_{\lambda}$ which we will use.

\medskip

Transfer matrices $\T_{\lambda}(u)$ can be obtained as the trace of the fused monodromy matrix $\bT_{\lambda}$, see Section \ref{Fusionsec}. The elements of $\bT_{\lambda}(u)$ are what we refer to as $\lambda$-minors $\bT_{\lambda}\left[^\lA_\lB\right](u)$. For a $\gl(k+1)$ spin chain, $\lA$ and $\lB$ are sets of indices taking values $1,2,\dots,k+1$ that are in correspondence with semi-standard Young tableaux of shape $\lambda$
\begin{equation}
\lA=\raisebox{-0.4\height}{
\begin{picture}(90,70)(-10,0)
\put(-8,68){\vector(0,-1){30}}
\put(-8,68){\vector(1,0){40}}
\put(-16,42){$a$}
\put(25,72){$s$}
\drawline(0,0)(0,60)
\drawline(20,0)(20,60)
\drawline(40,20)(40,60)
\drawline(60,40)(60,60)
\drawline(80,40)(80,60)
\drawline(0,0)(20,0)
\drawline(0,20)(40,20)
\drawline(0,40)(80,40)
\drawline(0,60)(80,60)
\put(2,48){$\scriptstyle {\mathcal{A}}_{1,1}$}
\put(22,48){$\scriptstyle {\mathcal{A}}_{1,2}$}
\put(44,48){$\ldots$}
\put(60.5,48){$\scriptstyle {\mathcal{A}}_{1\!,\lambda_1}$}
\put(2,28){$\scriptstyle {\mathcal{A}}_{2,1}$}
\put(24,28){$\ldots$}
\put(4,8){$\ldots$}
\end{picture}
}\,,\quad
\lB=\raisebox{-0.4\height}{
\begin{picture}(90,70)(-10,0)
\drawline(0,0)(0,60)
\drawline(20,0)(20,60)
\drawline(40,20)(40,60)
\drawline(60,40)(60,60)
\drawline(80,40)(80,60)
\drawline(0,0)(20,0)
\drawline(0,20)(40,20)
\drawline(0,40)(80,40)
\drawline(0,60)(80,60)
\put(2,48){$\scriptstyle {\mathcal{B}}_{1,1}$}
\put(22,48){$\scriptstyle {\mathcal{B}}_{1,2}$}
\put(44,48){$\ldots$}
\put(60.5,48){$\scriptstyle {\mathcal{B}}_{1\!,\lambda_1}$}
\put(2,28){$\scriptstyle {\mathcal{B}}_{2,1}$}
\put(24,28){$\ldots$}
\put(4,8){$\ldots$}
\end{picture}
}\,.
\end{equation}
$\bT_{\lambda}\left[^\lA_\lB\right](u)$ are constructed by applying appropriate symmetrization of the indices in the ordered product $\overrightarrow{\prod\limits_{a=1}^{h_{\lambda}}\prod\limits_{s=1}^{\lambda_a}}\bT\left[^{\lA_{a,s}}_{\lB_{a,s}}\right](u+\hbar(s-a))$, of which \eqref{quantumminor} is an example for $\lambda=(1^a)$. The transfer matrix $\T_\lambda$ is then defined as $\T_\lambda(u)=\sum_{\lA}\bT_{\lambda}\left[^\lA_\lA\right](u)\,,$ where the sum is over all admissible tableaux $\lA$. It is then a straightforward computation to demonstrate
\begin{equation}
\label{eq:Txi}
\T_{\lambda}(v)=\sum_{\lA}w_\lA T_{\lambda}\left[^\lA_{\lA+1}\right](v)+\sum_{j}T_{j1}(v)\times \lO(z_1,\dots,z_{k+1})\,,
\end{equation}
where  $w_\lA:=\prod_{a\in \lA}w_a$.

\medskip

The first term in \eqref{eq:Txi} coincides with $\T_{\lambda}^\lN$ and we clearly see that the second term vanishes when acting on an admissible vector at point $v$ and thus indeed $\bra{\Lambda}\T_{\lambda}^\lN(v)=\bra{\Lambda}\T_{\lambda}(v)$. One may ask how $ z_1,\ldots  z_{k+1}$ -- the eigenvalues of the MCT of the $\gl(k+1)$ spin chain are related to $z_1,\ldots z_{\gn}$ -- the original MCT eigenvalues. The point here is that none of the constructed states depend on $z_{i}$ and so this relation is immaterial. The auxiliary parameters $w_i$ should however be compatible with the injection \eqref{embtwo} used in the embedding procedure: If $w_i^{(k)}$ denote the auxiliary parameters used for transfer matrices of $\lY(\gl(k))$ then $w_{i+1}^{(k+1)}=w^{(k)}_{i}$, $i=1,\dots,k$. 

\medskip

Let $\bra{\Lambda'}$ be an eigenvector of $\bB^{(k)}$. Then we use \eqref{lowering} to readily see that $\bra{\Lambda}=\phi(\bra{\Lambda'})$ is an admissible vector at points $\theta_{\alpha}+\hbar\,\nu^{\alpha}_{k+1}$. Hence, to excite the $k$-th dual diagonals $\mu^\alpha_{kj}$ of  patterns $\Lambda^{\alpha}$, $\alpha=1,\ldots,L$ we should consider the following product 
\begin{equation}
\label{actiononadmissible}
\bra{\Lambda}\prod_{\alpha=1}^L \T_{\bar\mu_{k}^{\alpha}}(\theta_{\alpha}+\hbar\,\nu^{\alpha}_{k+1})\,
\end{equation}
as one can confirm from the explicit value of $f_\lambda(u,v)$ \eqref{fvalue} for $\lambda=\bar\mu_k^{\alpha}$. The only thing to check is that the action of $\T_{\bar\mu_{k}^{\alpha}}$ at the point $(\theta_{\alpha}+\hbar\,\nu^{\alpha}_{k+1})$ on $\bra{\Lambda}$ results in a vector which is still admissible at points $(\theta_{\beta}+\hbar\,\nu^{\beta}_{k+1})$ for $\beta\neq\alpha$. This is verified by considering the following fused RTT relation \cite{molev2007yangians}
\begin{equation}\label{fusedRTT}
(v-v')[T_{j1}(v),T_{\bar\mu_k}\left[^\lA_\lB\right](v')]=\sum_{a\in \lA}T_{a1}(v)\times\dots - \sum_{a\in \lA}T_{a1}(v')\times\dots\,.
\end{equation}
Taking $v=(\theta_{\beta}+\hbar\,\nu^{\beta}_{k+1})$, $v'=(\theta_{\alpha}+\hbar\,\nu^{\alpha}_{k+1})$ and using \eqref{lowering} and \eqref{fusedRTT} we conclude that if $\bra{\Lambda}$ is admissible at points $v,v'$ then $\bra{\Lambda}\T_{\bar\mu_k}(v)$ is admissible at the point $v'$.

\medskip

Summarising, the recursion \eqref{eq:CHLV2} yields the following recipe for an explicit build up of the eigenstates of the operator $\bB$ with pattern $\Lambda$
\begin{equation}
\label{eq:generating}
\fbox{
$\displaystyle
\bra{\Lambda^{\bB}}=\bra{0}\prod_{\alpha=1}^L\prod_{k=1}^{\gn-1}\phi^{\gn-k-1}\left(\T_{\bar\mu_k^{\alpha}}(\theta_\alpha+\hbar\,\nu_{k+1}^{\alpha})\right)\,.
$
}
\end{equation}
Here $\bra{0}$ is the lowest weight state (the GT vacuum) of the $\gln$ spin chain, and terms in the product with lower values of $k$ should be left of those with higher values of $k$. We remind the reader that $\phi^{r}$ amounts to the simple replacement of all $T_{ij}$ with $T_{i+r,j+r}$.

\medskip

We should still demonstrate that the constructed states are linearly independent. To this end choose null-twist transfer matrices in \eqref{eq:generating} and use the CBR formula \eqref{CBRfla} to rewrite them as a sum over products of transfer matricies in anti-symmetric representations. We then take the auxiliary singular twist limit (ASTL) $w_1 \gg w_2 \gg \dots \gg w_{\gn-1}$ of \eqref{eq:generating}. The leading contribution comes from the term in the CBR expansion with the most number of products\footnote{after using the constraint that the transfer matrix corresponding to the empty diagram $\T_{\es}$ is simply the identity operator}, and it exactly coincides  with the composite raising operator \eqref{compositeraising}. Hence the ASTL of $\bra{\Lambda^{\bB}}$ exists and coincides with $\bra{\Lambda^{\rm GT}}$. So $\bra{\Lambda^{\bB}}$ must be non-zero and moreover all $\bra{\Lambda^{\bB}}$ must be linearly independent for generic enough $w_i$ because $\bra{\Lambda^{\rm GT}}$ are linearly independent. Hence $\bra{\Lambda^{\bB}}$ form a basis (for generic $w_i$) and thus $\bB$ is diagonalisable.

\medskip

One may ask what would happen if $\bar\mu_k^{\alpha}$ in \eqref{actiononadmissible} are chosen to be some arbitrary integer partitions that do not satisfy the branching rules of the GT patterns and hence cannot be interpreted as dual diagonals. Then, if \eqref{actiononadmissible} is non-zero it would be an eigenvector of $\bB$ that is, in general, a linear combination of $\bra{\Lambda^{\bB}}$. Hence the outlined construction \eqref{eq:generating} and generated eigenvectors  $\bra{\Lambda^{\bB}}$ are not unique. However, obvious advantages of the proposed algorithm are that it has clear regular structure and that we can demonstrate that it indeed produces a basis. How one can use this basis is discussed in the next section.

\section{Factorised wave functions}

In this section we show that the basis \eqref{eq:generating}  leads to separation of variables for the Bethe algebra eigenstates.

\medskip

If a basis is generated by action of transfer matrices on some reference state then factorisation of wave functions is immediately obvious \cite{Maillet:2018bim}. One can also use other objects in the Bethe algebra such as Q-operators\footnote{While Q-operators do not belong to the Yangian as an abstract algebra, they do when we descend to representations discussed in this paper. Also note that ``other objects" does not mean new conserved charges but rather their repackaging using {\it e.g.} Q-operators instead of transfer matrices.} to reach the same conclusion. However, this is not how the basis \eqref{eq:generating} is constructed currently because lower rank transfer matrices embedded into $\lY_\gn$ using $\phi$ are typically not elements of the Bethe algebra.

\medskip
 
One of the main results to be demonstrated is that we can generate states \eqref{eq:generating} using auxiliary transfer matricies $\T^{(k)}_{\bar\mu^\alpha_k}$, $k=1,\dots,\gn-1$ who are B{\"a}cklund transforms of the original transfer matrices and who also belong to the Bethe algebra. Namely, we can demonstrate the following equality for any $\bra{\Lambda}\in\lV_{(k)}$
\begin{equation}\label{sov1}
\displaystyle
\bra{\Lambda}\prod_{\alpha=1}^L \phi^{\gn-k-1}\left(\T_{\bar\mu_k^{\alpha}}(\theta_\alpha+\hbar\,\nu_{k+1}^{\alpha})\right)=\bra{\Lambda}\prod_{\alpha=1}^L\T^{(k)}_{\bar{\mu}^\alpha_k}(\theta_\alpha+\hbar\,\nu^\alpha_{k+1})\,.
\end{equation}

We first review the basic properties of the B{\"a}cklund flow in section~\ref{sec:BF} and then focus  on derivation of \eqref{sov1} in section~\ref{sec:ATM}, with some technicalities delegated to appendix \ref{transferaction}. After \eqref{sov1} is established, it is straightforward to use standard Wronskian formulae to obtain factorised wave functions as is demonstrated in sections~\ref{sec:WF} and~\ref{sec:CM}.

\subsection{Quantum Eigenvalues, $Q$-system and B{\"a}cklund Flow}
\label{sec:BF}
Given a Young diagram $\lambda$ and a group element $g\in\GL(\gn)$ with eigenvalues $z_1,z_2,\dots,z_{\gn}$, its character $\chi_\lambda(g)$ in the representation $\lambda$ can be obtained from a summation over semi-standard Young tableaux.  A semi-standard Young tableau $\YT$ of shape $\lambda$ is obtained by filling up each box in the Young diagram $\lambda$ with elements of the set $\{1,2,\dots,\gn\}$ subject to the condition that the numbers weakly decrease in every row and strictly decrease in every column\footnote{Note that our convention is the opposite to the widely used one where the numbers in a tableau strictly increase in each column and weakly increase in each row. The resulting classical character is not sensitive to this difference, however it becomes important for the construction of transfer matrices.}. The character can then be computed as 
\begin{equation}
\chi_\lambda(g)=\sum_{\YT}\prod_{(a,s)\subset \lambda} z_{\#(a,s)}\,,
\end{equation}
where $\#(a,s)$ denotes the number in position $(a,s)$ of the tableau $\YT$ and the product is over all boxes $(a,s)$ of the diagram $\lambda$. 

\medskip

A similar formula exists for transfer matrices \cite{Kuniba:1994na,Tsuboi:1997iq,Tsuboi:1998ne}:
\begin{equation}\label{cbrsoln}
\T_\lambda(u)=\sum_{\YT}\prod_{(a,s)\subset \lambda} \Lambda_{\#(a,s)}(u+\hbar(s-a))\,,
\end{equation}
where the functions $\Lambda_j(u),\ j=1,2,\dots,\gn$ are referred to as quantum eigenvalues of the $\lY_\gn$ monodromy matrix and satisfy 
\begin{equation}
[\Lambda_i(u),\Lambda_j(v)]=0,\ i,j=1,2,\dots,\gn
\end{equation}
and were defined in Section \ref{Qsystem}. In order to avoid needless looking back and forth we will recall their construction here for the convenience of the reader.

\medskip

Recall the generating function \eqref{talalaev} for the transfer matricies $\T_{a,1}$: $\det(1-\bT(u)e^{-\hbar \partial_u})=\sum_{a=0}^\gn (-1)^a \T_{a,1}(u)e^{-a\hbar \partial_u}\,.$ It then follows from \eqref{cbrsoln} that we can write
\begin{equation}\label{qeigen}
\det(1-\bT(u)e^{-\hbar \partial_u})=\left(1-\Lambda_\gn(u)e^{-\hbar \partial_u} \right)\dots \left(1-\Lambda_1(u)e^{-\hbar \partial_u} \right)
\end{equation}
which can easily be seen by expanding the \rhs and comparing coefficients of $e^{-a\hbar \partial_u}$. The $Q$-operators $\QQ_i(u)$, $i=1,\dots,\gn$ are annihilated by the above finite-difference operator
\begin{equation}\label{qdefn}
\det(1-\bT(u)e^{-\hbar \partial_u})\QQ_i^{[2]}(u)=0,\ i=1,2,\dots,\gn\,.
\end{equation}
The complete family of Q-operators comprises operators $\Q_I$, $I\subset \{1,2,\dots,\gn\}$ that are related to $\Q_i$ by means of the $QQ$ relations 
\begin{equation}\label{QQ}
\Q_{Iij}\Q_I^{[-2]}=\Q_{Ii}\Q_{Ij}^{[-2]}-\Q_{Ij}\Q_{Ii}^{[-2]}
\end{equation}
supplemented with $\Q_\es(u)=1$. The analytic structure of $Q$-operators for spin chains in arbitrary representation is known \cite{Frassek:2011aa} to have the following form, see Section \ref{Qsystem},
\begin{equation}\label{qsoln}
\Q_I(u)=N_I\hhq_I(u)\prod_{j=1}^{|I|}\Gamma\left[\hat{\nu}_j^{[2(1-|I|)]}(u)\right]\,,\quad \hhq_I(u):=\hq_I \prod_{j\in I}z_j^{\frac u\hbar}\,,
\end{equation}
where $\hat{\nu}_j(u):=\prod_{\alpha=1}^L(u-\theta_\alpha-\hbar\, \hat{\nu}_j^\alpha)$ with $\hat{\nu}_j^\alpha$ being the shifted weights $\hat{\nu}^\alpha_j:=\nu^\alpha_j-j+1$, $\hq_I(u)$ is an operator-valued monic polynomial, and $\hq_{12\ldots\gn}=1$. Finally $N_I$ is normalisation which is well-defined with $N_I=\prod_{j<k}\frac{z_{i_j}-z_{i_k}}{z_{i_j}z_{i_k}}$ for $I=\{i_1,\dots,i_{|I|}\}$ but is not relevant for our discussion, and $\Gamma[F(u)]$ has the property $\Gamma[F(u+\hbar)]=F(u)\Gamma[F(u)]$.

\medskip

If $I$ is a single index $i$, \eqref{qsoln} becomes
\be\label{gaugetr}
\QQ_i(u)=\hhq_i(u)\Gamma\left[{\nu}_1(u)\right]
\ee
which should be considered as a gauge transformation between two ways to parameterise Baxter Q-operators.

\medskip

By using \eqref{qdefn} together with \eqref{qeigen} it easy to see that a solution for $\Lambda_k(u)$ is given by 
\begin{equation}
\label{eq:QEV1}
\Lambda_k(u)=\frac{\Q_{\sigma(I_{k-1})}^{[-2]}}{\Q_{\sigma(I_{k-1})}}\frac{\Q_{\sigma(I_{k})}^{[2]}}{\Q_{\sigma(I_{k})}},\quad k=1,\dots,\gn\,,
\end{equation}
where $I_k:=\{1,2,\dots,k\}$, while $\sigma$ denotes some element of the permutation group $\mathfrak{S}_\gn$. Clearly, the quantum eigenvalues $\Lambda_k$ are not invariant under choice of $\sigma$ as they are sensitive to the order of terms in the factorisation \eqref{qeigen}. However their (quantum) symmetric combinations, transfer matrices, are invariant under this choice.

\medskip

We will now introduce the notion of the B{\"a}cklund transform. It traces its origins to the solutions of the Hirota bilinear equation on the $\gl(\gn)$ strip \cite{doi:10.1143/JPSJ.45.321,Zabrodin:1996vm,Krichever:1996qd} but we shall define it in more compact terms. Consider the so-called Wronskian solution of the CBR formula \cite{Bazhanov:1996dr,Krichever:1996qd}
\begin{equation}\label{wronsk}
\T_\lambda(u)=\frac{\displaystyle\det_{1\leq i,j\leq n}\Q_{\sigma(i)}^{[2\hat{\lambda}_{\sigma(j)}]}(u)}{\Q_{\sigma(I_\gn)}(u)}\,,
\end{equation}
where $\hat{\lambda}_j=\lambda-j+1$ are the shifted weights and whose equivalence with \eqref{cbrsoln} follows as a result of the QQ-relations. The $(\gn-k)$-th B{\"a}cklund transform of the transfer matrix $\T_\lambda(u)$ that shall be denoted as $\T_\lambda^{(k)}(u)$ is obtained by restricting the range of the determinant in \eqref{wronsk} to $k$ components:
\begin{equation}\label{wronskian2}
\T_\lambda^{(k)}(u)=\frac{\displaystyle\det_{1\leq i,j\leq k}\Q_{\sigma(i)}^{[2\hat{\lambda}_{\sigma(j)}]}(u)}{\Q_{\sigma(I_k)}(u)}\,.
\end{equation}
From \eqref{eq:QEV1}, it is easy to deduce that  $\T_\lambda^{(k)}$ are expressed in terms of quantum eigenvalues as 
\begin{equation}
\label{cbrsolk}
\T_\lambda^{(k)}(u)=\sum_{\YT}\prod_{(a,s)\subset \lambda} \Lambda_{\#(a,s)}(u+\hbar(s-a))\,,
\end{equation}
where the only difference with \eqref{cbrsoln} is that the tableaux $\YT$ are filled with the numbers $\{1,2,\dots,k\}$, instead of the full set $\{1,2,\dots,\gn\}$. Notice that we have the property 
\begin{equation}\label{bcklndprp}
\T_\lambda^{(k)}(\theta_\alpha+\hbar\,\nu^\alpha_k)=\T_\lambda^{(k-1)}(\theta_\alpha+\hbar\,\nu^\alpha_k)
\end{equation}
which follows as a simple consequence of the arguments in the next Section.

\subsection{Action of transfer matrices}
\label{sec:ATM}
We prove \eqref{sov1} in two steps. First, we prove that
\begin{equation}\label{sov2}
\frac{\T_{F^\alpha_k+\bar\mu^\alpha_k}(\theta_\alpha+\hbar\,\nu^\alpha_n)}{\T_{F^\alpha_k}(\theta_\alpha+\hbar\,\nu^\alpha_n)}=\T^{(k)}_{\bar\mu^\alpha_k}(\theta_\alpha+\hbar\,\nu^\alpha_{k+1})\,,
\end{equation}
and then we prove the equality between the \lhs of \eqref{sov2} acting on $\bra{\Lambda}\in\lV_{(k)}$ and the \lhs  \eqref{sov2}. The second step is more technical and we leave it to appendix~\ref{transferaction}, and we also prove in appendix~\ref{invertability} that the ratio of transfer matricies in the \lhs of \eqref{sov2} is well-defined. This subsection deals with \eqref{sov2}.

\medskip

In our proofs we assume that inhomogeneities assume some generic value (that is we avoid a certain subset of measure zero where the invoked arguments could fail). But since the \lhs of \eqref{sov1} is polynomial in inhomogeneities, the final result should be correct for any $\theta_\alpha$. It is however only useful if \eqref{eq:generating} form a basis for which sake a sufficient condition $\theta_\alpha-\theta_\beta\notin\hbar \ZZ$ for pairwise distinct $\alpha,\beta$ is imposed \cite{molev2007yangians}.

\medskip

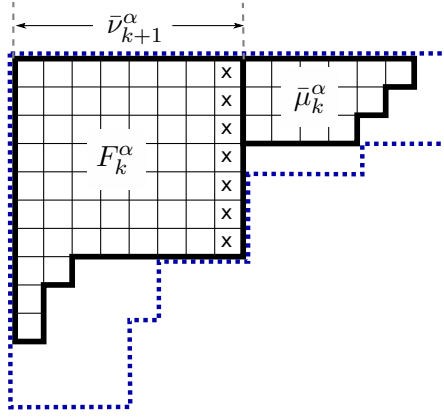
\begin{figure}
\begin{center}
\begin{picture}(170,140)(0,0)
\put(0,0){\includegraphics[width=6cm]{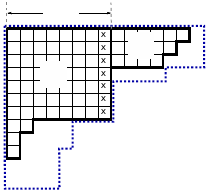}}
\put(36,93){$F^\alpha_k$}
\put(110,116){$\bar{\mu}^\alpha_k$}
\put(40,143){$\bar\nu^\alpha_{k+1}$}
\end{picture}
\end{center}
\caption{\label{fig:Tmu}Gluing of diagrams $F^\alpha_k$ and $\bar{\mu}^\alpha_k$. The dotted line is the boundary of the diagram $\bar\nu^{\alpha}$. Crossed squares depict the column which should be of the same height for $F^{\alpha}_k$ and $\bar\nu^{\alpha}$.}
\end{figure}

In \eqref{sov2}, $\T_{F^\alpha_k+\bar{\mu}^\alpha_k}$ and $\T_{F^\alpha_k}$ are usual $\lY(\gl(\gn))$ transfer matricies and $"+"$ means gluing of Young diagram shapes aligned on top.  Denote by  $\bar\nu^\alpha$ the reduced Young diagram with $\bar\nu^\alpha_j=\nu^\alpha_j-\nu^\alpha_\gn$. Then $F^\alpha_k$ is any Young diagram satisfying the following constraints:  its width (value of the first component $F_{k1}^{\alpha}$) is equal to $\bar\nu^\alpha_{k+1}$, the height of its last column is equal to the height of the $\bar\nu^\alpha_{k+1}$-th column of $\bar\nu^{\alpha}$, and it must be that $F^\alpha_k+\mu^\alpha_k\subset \bar\nu^\alpha$, see Fig~\ref{fig:Tmu}. 

The key feature we need is vanishing of quantum eigenvalues at specific points:
\begin{equation}\label{vanishingeigen}
\Lambda(\theta_\alpha+\hbar\,\nu_r^\alpha)=0,\ \alpha=1,2,\dots,L,\ r=1,\dots,\gn\,.
\end{equation}
It follows from 
\be
\label{QEVq}
\Lambda_r(u)=z_{\sigma(r)}\nu_r(u)\frac{\hq_{\sigma(I_{r-1})}^{[-2]}}{\hq_{\sigma(I_{r-1})}}\frac{\hq_{\sigma(I_{r})}^{[2]}}{\hq_{\sigma(I_{r})}}
\ee 
which can be derived from \eqref{qsoln}, and we assume to avoid situations when the denominator of $\frac{\hq_{\sigma(I_{r-1})}^{[-2]}}{\hq_{\sigma(I_{r-1})}}\frac{\hq_{\sigma(I_{r})}^{[2]}}{\hq_{\sigma(I_{r})}}$ has a pole at $\theta_\alpha+\hbar\,\nu_r^\alpha$.

\medskip

Consider $\T_\lambda(\theta_\alpha+\hbar\,\nu^\alpha_\gn)$ -- the transfer matrix in the representation $\lambda$ evaluated at the point $\theta_\alpha+\hbar\,\nu^\alpha_\gn$, and consider its expansion in quantum eigenvalues \eqref{cbrsoln}. For this special point, only a limited subset of tableaux $\YT$ contribute to this expansion. Indeed, let $\YT$ be a tableau that provides a non-zero contribution to the sum. Then it cannot contain $\gn$ at position $a=1,s=1$ because $\Lambda_{\gn}(\theta_{\alpha}+\hbar\,\nu_\gn^{\alpha})=0$. But since the numbers in a tableau should weakly decrease to the right and strictly decrease down, $\YT$ cannot contain $\gn$ at all. This tableau cannot also contain $\gn-1$ at position $a=1,s=1+\bar\nu^{\alpha}_{\gn-1}$, due to \eqref{vanishingeigen} for $r=\gn-1$. Then any boxes to the right of the column $s=\bar\nu^{\alpha}_{\gn-1}$ cannot contain $\gn-1$. By repeating the argument we get that boxes of $\YT$ to the right of the column $s=\bar\nu^{\alpha}_{k+1}$ can be populated at most by the indices $1,2,\ldots,k$.

\medskip

Now we turn to the case when $\lambda=F^\alpha_k+\bar{\mu}^\alpha_k$. Let $R$ be the maximal number for which $\bar\nu_{R}^{\alpha}=\bar\nu_{k+1}^{\alpha}$, and $r+1$ be the minimal number for which $\bar\nu_{r+1}^{\alpha}=\bar\nu_{k+1}^{\alpha}$. Then we observe two features. Firstly, entries in the $\bar{\mu}^{\alpha}_k$ part of the tableau $\YT$ can be only populated by indices $1,2,\ldots,r$. Secondly, the height of the last column of $F^\alpha_k$ (denoted by crosses in Fig~\ref{fig:Tmu}) is $R$ and, since $\bar\nu_{R+1}^{\alpha}$ is strictly smaller than $\bar\nu_R^{\alpha}$, this last column can be only populated by indices $1,2,\ldots, R$. Hence it is fixed uniquely. Note that an immediate corollary of this discussion is that $\T_\lambda(\theta_\alpha+\hbar\,\nu^\alpha_\gn)=0$ if $\lambda$ is any shape not contained in $\bar\nu^\alpha$, in contrast to the fact that the transfer matrix is invertible otherwise as is shown in appendix \ref{invertability}. 

\medskip

Because for any non-vanishing $\YT$ the last column of the $F^\alpha_k$ part is fixed uniquely, values in other boxes of the $F^\alpha_k$ part do not affect possible values in the boxes of the $\bar{\mu}^{\alpha}_k$ part and vice versa, and so the sum \eqref{cbrsoln} factorises:
\be
\T_{F^\alpha_k+\bar{\mu}^\alpha_k}(\theta_\alpha+\hbar\,\nu^\alpha_\gn)&=&\left(\sum_{\YT_{F}}\prod_{(a,s)\subset F^\alpha_k} \Lambda_{\#(a,s)}(\theta_\alpha+\hbar\,\nu^\alpha_\gn+\hbar(s-a))\right)
\nonumber\\
&\times&\left(\sum_{\YT_{\bar{\mu}}}\prod_{(a,s)\subset \bar{\mu}^\alpha_k} \Lambda_{\#(a,s)}(\theta_\alpha+\hbar\,\nu^\alpha_{k+1}+\hbar(s-a))\right)\,.
\ee
The first factor obviously evaluates to $\T_{F^\alpha_k}(\theta_\alpha+\hbar\,\nu^\alpha_\gn)$. For the second one, recall that the possible entries in the tableaux $\YT_{\bar\mu}$ are constrained to be from the set $\{1,2,\ldots,r\}$, but then this term is precisely $\T_{\bar\mu_k^{\alpha}}^{(r)}(\theta_\alpha+\hbar\,\nu^\alpha_{k+1})$, {\it cf.} \eqref{cbrsolk}. By using the same arguments as we invoked after \eqref{vanishingeigen} we show that all $\T_{\bar\mu_k^{\alpha}}^{(k)}(\theta_\alpha+\hbar\,\nu^\alpha_{k+1})$ for $R-1\geq k\geq r$ are in fact equal to one another and hence \eqref{sov2} indeed holds.

\medskip

We supplement this conclusion with the result of appendix \ref{transferaction} and conclude the remarkable equality \eqref{sov1}. An immediate consequence of \eqref{sov1} is that the basis \eqref{eq:generating} can now be constructed as 
\begin{equation}\label{sov4}
\fbox{
$
\displaystyle
\bra{\Lambda^{\bB}}=\bra{0}\prod_{\alpha=1}^L \prod_{k=1}^{\gn-1}\T^{(k)}_{\bar{\mu}^\alpha_k}(\theta_\alpha+\hbar\,\nu^\alpha_{k+1})
$
}\,.
\end{equation}
We are now one step away from writing concise expressions for wave functions in the SoV basis which is our next goal.

\subsection{Wave functions and separated variables}
\label{sec:WF}
Expressing the basis \eqref{sov4} using the Wronskian solution \eqref{wronskian2} gives
\begin{equation}\label{sov3}
\bra{\Lambda^{\bf B}}=\bra{0}\prod_{\alpha=1}^L\prod_{k=1}^{\gn-1}\frac{\displaystyle\det_{1\leq i,j\leq k}\Q_{\sigma(i)}^{[2\hat{\bar\mu}_j]}(\theta_\alpha+\hbar\,\nu^\alpha_{k+1})}{\displaystyle \Q_{\sigma(I_k)}(\theta_\alpha+\hbar\,\nu_{k+1}^\alpha)}\,.
\end{equation}
It is convenient to introduce a new reference vector $\bra{\Omega_\sigma}:=\bra{0}\prod\limits_{\alpha=1}^L\prod\limits_{k=1}^{\gn-1}\left( \Q_{\sigma(I_k)}(\theta_\alpha+\hbar\,\nu_{k+1}^\alpha)\right)^{-1} $ for which
\begin{equation}\label{normbasis}
\bra{\Lambda^{\bf B}}=\bra{\Omega_\sigma}\prod_{\alpha=1}^L\prod_{k=1}^{\gn-1}\displaystyle\det_{1\leq i,j\leq k}\Q_{\sigma(i)}(x^\alpha_{kj})\,,
\end{equation}
where we have used that $\svx^\alpha_{kj}=\theta_\alpha+\hbar(\mu^\alpha_{kj}-j+1)$, see \eqref{eq:variablesx}. The Gamma-function contribution to the Q-operators \eqref{gaugetr} nicely factorises from the determinants and we accordingly introduce $\brax$ as rescaled basis vectors $\bra{\Lambda^{\bf B}}$:
\be\label{normbasis2}
\brax:=\prod_{\alpha=1}^L\prod_{k=1}^{\gn-1}\frac 1{\Gamma\left[{\nu}_1(x_{kj}^{\alpha})\right]}\bra{\Lambda^{\bf B}}=\bra{\Omega_\sigma}\prod_{\alpha=1}^L\prod_{k=1}^{\gn-1}\displaystyle\det_{1\leq i,j\leq k}\hhq_{\sigma(i)}(x^\alpha_{kj})\,.
\ee
Let us choose the normalisation $\braket{\Omega_\sigma|\Psi}=1$ for all the Bethe algebra eigenvectors $\ket{\Psi}$. Then their wave functions $\Psi(\svx)$ in the constructed basis are 
\begin{equation}
\label{eq:wav}
\displaystyle
\Psi(\svx)=\braket{\svx|\Psi}=\prod_{\alpha=1}^L\prod_{k=1}^{\gn-1}\displaystyle\det_{1\leq i,j\leq k}\hat \sfq_{\sigma(i)}(x^\alpha_{kj})
\end{equation}
where $\hat \sfq_i(u)$ is the eigenvalue of $\hhq_i(u)$ on the state $\ket{\Psi}$. By choosing $\sigma$ to be the identity permutation we immediately obtain
\begin{equation}
\boxed{\Psi(\svx)=\braket{\svx|\Psi}=\prod_{\alpha=1}^L\prod_{k=1}^{\gn-1}\displaystyle\det_{1\leq i,j\leq k}\hat \sfq_{i}(x^\alpha_{kj})}\,.
\end{equation}

\medskip

With the last formula we achieved our goal of wave function factorisation, and its explicit form justifies why the operators $\svX_{kj}^{\alpha}$ --  zeros of $\bB(u)$ whose eigenvalues on  $\brax$ are $x_{kj}^{\alpha}$ should be considered as separated variables. 

\medskip

Define $\ket{\Omega}$ by the property $\braket{\svx|\Omega}=1$ for all $\brax$. Then \eqref{eq:wav} implies that all $\ket{\Psi}$ can be constructed as 
\begin{equation}
\ket{\Psi}=\prod_{\alpha=1}^L\prod_{k=1}^{\gn-1}\displaystyle\det_{1\leq i,j\leq k}\hat \sfq_{\sigma(i)}(\svX^\alpha_{kj})\ket{\Omega}\,.
\end{equation}
We note that $\ket{\Omega}$ is not itself an eigenvector of the Bethe algebra. In some situations it could be beneficial to select a certain Bethe eigenstate $\ket{\omega}$ as a reference and build excitations as
\be
\label{ratio}
\ket{\Psi}=\frac{\prod\limits_{\alpha=1}^L\prod\limits_{k=1}^{\gn-1}\displaystyle\det_{1\leq i,j\leq k}\hat \sfq_{\sigma(i)}(\svX^\alpha_{kj})}{\prod\limits_{\alpha=1}^L\prod\limits_{k=1}^{\gn-1}\displaystyle\det_{1\leq i,j\leq k}\hat \sfq_{\sigma(i)}^{(0)}(\svX^\alpha_{kj})}\ket{\omega}\,,
\ee
where $\hat \sfq_{\sigma(i)}^{(0)}$ is the eigenvalue of $\hhq_{\sigma(i)}$ on $\ket{\omega}$. The most natural candidate for $\ket{\omega}$  is one of the ferromagnetic vacua of the spin chain. It is distinguished by the property  $q_{\sigma(12\ldots k)}^{(0)}=1$, $k=1,\ldots,\gn$. In the reference frame where the twist is diagonal it is the highest-weight vector with respect to an appropriate choice of the Borel subalgebra:
\begin{equation}
T_{ij}(u)\ket{\omega}=0,\quad \sigma^{-1}(i)>\sigma^{-1}(j), \quad T_{jj}(u)\ket{\omega}=\nu_{\sigma^{-1}(j)}(u) \ket{\omega}\,,
\end{equation}
and it should be rotated to the modified companion twist frame which we are using in this paper. 

\medskip

The most drastic simplification of \eqref{ratio} happens  when we consider spin chains in symmetric powers of the fundamental representation. In this case $\nu_j^{\alpha}=0$ for $j>1$ and so, by analysis of section~\ref{sec:ATM}, we can replace $\T^{(k)}_{\bar{\mu}^\alpha_k}$ with $\T^{(1)}_{\bar{\mu}^\alpha_k}$ in  \eqref{sov4}. In particular, $\bar{\mu}^\alpha_k$ consists of a single row. Consequently, \eqref{ratio} becomes
\be
\ket{\Psi}=\frac{\prod\limits_{\alpha=1}^L\prod\limits_{k=1}^{\gn-1}\hat \sfq_{\sigma(1)}(\svX^\alpha_{k1})}{\prod\limits_{\alpha=1}^L\prod\limits_{k=1}^{\gn-1}\displaystyle\hat \sfq_{\sigma(1)}^{(0)}(\svX^\alpha_{kj})}\ket{0}=\prod\limits_{\alpha=1}^L\prod\limits_{k=1}^{\gn-1} \sfq_{\sigma(1)}(\svX^\alpha_{k1})\ket{0}\propto \prod_{r} \bB(u_r)\ket{\omega}\,,
\ee
where $u_r$ are zeros of $\sfq_{\sigma(1)}$. We see that, in this special case, $\prod\limits_{r} \bB(u_r)$ acting on the ferromagnetic vacuum creates all the Bethe states. This result was conjectured based on numerical evidence and analytical tests for low numbers of magnons in \cite{Gromov:2016itr} and then proven for $\gl(3)$ \cite{Liashyk:2018qfc} and $\gl(\gn)$ cases \cite{Ryan:2018fyo}.

\medskip

Finally, we make a few comments about the Bethe equations. To simplify our exposition, we will consider all spin chain sites to have the same representation, that is $\nu^\alpha=\nu$ for all $\alpha=1,\dots,L$. In this case it is convenient to introduce the polynomial $Q_\theta(u)=\prod_{\alpha=1}^L(u-\theta_\alpha)$. We also normalise the twist matrix to $\det G=1$.

\medskip

Originally, the Bethe equations for spin chains in arbitrary representation were written down in \cite{Kulish:1983rd}. These were the equations on zeros of $\sfq_{\sigma(12\ldots)}(u)$ (nested Bethe roots). Instead of such type of Bethe equations, one can write polynomial conditions that should be obeyed by (twisted) polynomials $\hat \sfq_i$. As a consequence of \eqref{QQ} and $\sfq_{\es}=1$ one derives $\det\limits_{1\leq i,j\leq \gn} \sfQ_i(u-\hbar(j-1))=\sfQ_{12\ldots\gn}$. Then the requirement that $\sfq_{12\ldots\gn}=1$ in \eqref{qsoln} provides a quantisation condition on possible values of $\hat \sfq_i$:
\be
\label{quantcond}
\det\limits_{1\leq i,j\leq \gn}\hat \sfq_i(u-\hbar\,(j-1))\propto\prod_{j=2}^\gn \prod^{\nu_1}_{k=\nu_j+1}Q_\theta(u-\hbar(k+\gn-j))\,,
\ee
where $\propto$ means equality up to a constant multiplication. This quantisation condition is the same as the demand that the Wronskian solution \eqref{wronsk} for transfer matrices $\T_{\lambda}$ yields identity if we take $\lambda$ to be the empty Young diagram.

\medskip

There exists also a dual description, in terms of Q-functions $\sfQ^I$ defined by $\sfQ^I:=\varepsilon^{\bar{I}I}Q_{\bar{I}}$, where $\varepsilon$ is the Levi-Civita symbol in $\gn$ dimensions and $\bar{I}$ means the complimentary set to $I$ (no summation over $\bar{I}$ is performed). Again, we can exploit \eqref{QQ} to conclude that $\det\limits_{1\leq i,j\leq\gn}\sfQ^i(u-\hbar(j-1))=\prod\limits_{k=1}^{\gn-1}\sfQ_{12\ldots\gn}(u-\hbar(k-1))$ which, in terms of $\hat \sfq^i:=\varepsilon^{\bar{i}i}\hat \sfq_{\bar{i}}$ becomes
\begin{equation}
\label{hodgequantcond}
\det_{1\leq i,j\leq\gn}\hat \sfq^i(u+\hbar(j-1))\propto\prod_{j=1}^{\gn-1}\prod^{\nu_j}_{k=\nu_\gn+1}Q_\theta(u+\hbar(j-k))\,.
\end{equation}
Note that fixing either $\sfq_i$ or $\sfq^i$ would be sufficient to compute any element of the Bethe algebra.

\medskip

Finally, let us point out that we do not rely on any statements about completeness of Bethe equations. In fact, the situation is quite the opposite one --  an important ingredient of completeness theorems follows immediately from the proposed construction. Namely we showed that the Bethe algebra is a maximal commutative subalgebra of the algebra of the endomorphisms of the spin chain's Hilbert space. Indeed, the SoV basis is generated by action of transfer matrices, but it would be impossible to generate a basis if there was an extra independent operator that commutes with the transfer matrices.

\medskip

Maximality of the Bethe algebra implies that the eigenstates in the Hilbert space can be unambiguously labelled by eigenvalues of Bethe algebra generators. As we can take Q-operators as generators and zeros of the Q-operators satisfy Bethe equations, we conclude that all physical states of the spin chain are labelled, and can be distinguished, by solutions of the Bethe equations.

\medskip

What is not guaranteed by the above argument is that each solution of the Bethe equations labels some physical state. This question can be resolved by explicit counting but this requires certain care, especially for spin chains in arbitrary representations that we consider, as is discussed after \eqref{quantcond}. For the case of the fundamental representation the question was resolved in various ways in the literature. We mention \cite{Maillet:2019ayx} where it was discussed for the supersymmetric $\gl(2|1)$ case and also \cite{Niccoli:2009jq,Niccoli:2011nj} in the SoV framework of the same type as considered in this paper; and \cite{2013arXiv1303.1578M} where completeness is proven for $\gl(\gn)$ spin chains with and without twist, and for any value of inhomogeneities and  for similar results. The results of \cite{2013arXiv1303.1578M} also generalise to the supersymmetric $\gl(\mathsf{m}|\gn)$ case \cite{Chernyak:2020lgw}. 

\medskip

In the case of a spin chain in the defining representation, $\nu=(1,0,\ldots,0)$, the condition \eqref{hodgequantcond} reads $\det\limits_{1\leq i,j\leq \gn}\hat \sfq^i(u+\hbar\,(j-1))\propto Q_\theta(u)$. It  contains only the physical solutions for arbitrary values of inhomogeneities \cite{2013arXiv1303.1578M} and hence can be used alone to fully characterise the spectrum of the model. Similarly, for the conjugate representation $\nu=(1,1,\ldots,1,0)$, the condition \eqref{quantcond} reads $\det\limits_{1\leq i,j\leq \gn}\hat \sfq_i(u-\hbar\,(j-1))\propto Q_\theta(u-\hbar)$ and also is enough to characterise the spectrum.

\medskip

For more complicated representations than the mentioned two, there are more solutions to \eqref{quantcond} or \eqref{hodgequantcond} than the dimension of the Hilbert space. We should then impose extra restrictions. This can be done by the requirement that $\wT_{\lambda}(u)$ should be polynomials in $u$ for any $\lambda$ and that $\hq_I(u)$ computed from $\hq_i(u)$ via \eqref{qsoln} and \eqref{QQ} are also polynomials in $u$ for any $I$.  By generalising the ideas of \cite{Marboe:2016yyn} it is possible to repackage these requirements in a structurally simple manner that allows one simple explicit counting of the physical solutions of \eqref{quantcond} and to confirm that their number coincides with the dimension of the Hilbert space. This result will be presented in some upcoming work. 

\paragraph{Wave function examples}

For certain classes of representations the presented wave functions simplify quite a bit. The largest simplifications occur for symmetric powers of the defining (as was already demonstrated above) or anti-fundamental representations. In these cases, by appropriate changes in normalisation of the SoV bases the wave functions simply become 
\begin{equation}\label{wavefnexamples}
\Psi(\svx) = \prod_{\alpha=1}^L\prod_{j=1}^\gn \sfq_1(\svx^\alpha_{j1}),\quad \Psi(\svx) =\prod_{\alpha=1}^L\det_{1\leq i,j\leq\gn-1}\sfq_i(\svx^\alpha_{\gn-1,j})\,.
\end{equation}
We have singled out these examples in order to easily refer to these formulae later. 

\subsection{Conjugate momenta}
\label{sec:CM}
This work realises to a large extent Sklyanin's SoV program for compact rational $\gl(n)$ spin chains. Indeed, the operators $X_{kj}^{\alpha}$ are naturally a quantisation of zeros $x_\sigma$ of the classical $B(u)$, and wave functions in the proposed SoV basis are products of determinants of Baxter Q-functions who solve \eqref{qdefn} -- a quantisation of the classical spectral curve. 

\medskip

To accomplish the program, we should also quantise $D(u)$ (see section \ref{classical}) to get the conjugate momenta $P^{\alpha}_{kj}$ and then identify the spin chain with a representation of the algebra generated by $P^{\alpha}_{kj}$ and $\svX^\alpha_{kj}$. Quantisation of $D(u)$ (also referred to as $A(u)$ in some literature depending on conventions) was formally suggested in \cite{Sklyanin:1992sm,2001math.ph...9013S}, however the procedure proposed there becomes singular when explicitly applied to highest-weight spin chains, see for example the discussion in \cite{Maillet:2018bim}. Here we shall introduce conjugate momenta by different means and it would be interesting to explore whether our proposal matches a regularised way to quantise $D(u)$.

\medskip

The canonically conjugate momenta 
$P^{\pm\alpha}_{kj}$ associated to the separated coordinates $\svX^\alpha_{kj}$  satisfy the commutation relation 
\begin{equation}
[P^{\pm\alpha}_{kj},\svX^\beta_{k'j'}]=\pm \hbar\, \delta^{\alpha\beta}\delta_{kk'}\delta_{jj'} P^{\pm\alpha}_{kj}.
\end{equation}
We propose their following realisation
\begin{equation}\label{raiselower}
P^{\pm\alpha}_{kj}=c^{\pm\alpha}_{kj} :\frac{\displaystyle\det_{1\leq i,l\leq k} \Q_{\sigma(i)}(\svX^\alpha_{kl}\pm\hbar\delta_{jl})}{\displaystyle\det_{1\leq i,l\leq k} \Q_{\sigma(i)}(\svX^\alpha_{kl})}:\,,
\end{equation}
where $c^{\pm\alpha}_{kj}$ is some simple function of the separated variables to be fixed in a moment. We use a normal ordering prescription $:\ :$ where $\svX$'s are placed to the left of all the coefficients of Baxter $Q$-operators. 
To see that the prescription \eqref{raiselower} works, we utilise \eqref{sov3} and act on $\brax$ with $P^{\pm\alpha}$ as defined above. By using that $\brax \svX^\alpha_{kj}=\svx^\alpha_{kj}\brax$, we immediately obtain (up to normalisation) the state where $\mu^\alpha_{kj}$ has been replaced with $\mu^\alpha_{kj}\pm 1$. In particular the action of $P^{\pm\alpha}_{kj}$ on $\brax$ is well-defined. 
\newline
\newline
The coefficient $c^{\pm\alpha}_{kj}$  in \eqref{raiselower} is required in order to respect the branching rules of GT patterns. Namelly, we have the constraints $\mu^\alpha_{k-1,j}\geq\mu^\alpha_{kj}\geq \mu^\alpha_{k,j+1}$ and $\mu^\alpha_{k,j-1}\geq\mu^\alpha_{kj}\geq \mu^\alpha_{k+1,j}$ on a given GT pattern $\Lambda^\alpha$  and so $P^{+\alpha}_{kj}$ should vanish when we act on a state with $\mu^\alpha_{kj}=\mu^\alpha_{k,j-1}$ or $\mu^\alpha_{kj}=\mu^\alpha_{k-1,j}$, and similarly for $P^{-\alpha}_{kj}$. Using the fact that $\mu^\alpha_{kj}$ is related to $\svx^\alpha_{kj}$ as $\svx^\alpha_{kj}=\theta_\alpha+\hbar(\mu^\alpha_{kj}-j+1)$ we see that we should take \begin{equation}
c^{+\alpha}_{kj}=(\svX^\alpha_{k-1,j}-\svX^\alpha_{kj})(\svX^\alpha_{k,j-1}-\svX^\alpha_{kj}-\hbar)
\end{equation}
and similarly
\begin{equation}
c^{-\alpha}_{kj}=(\svX^\alpha_{kj}-\svX^\alpha_{k+1,j})(\svX^\alpha_{kj}-\svX^\alpha_{k,j+1}-\hbar)\,.
\end{equation}
The separated variables $\svX^\alpha_{kj}$ are defined for indices in the range $1\leq k\leq \gn-1$ and $1\leq j\leq k$, but $c^{\pm\alpha}_{kj}$ can contain factors with $\svX^\alpha_{kj}$ outside of this range. In order to get around this we define operators $\svX^\alpha_{j,j+1}$, $j=0,\dots,\gn-1$ to be scalar multiples of the identity operator with eigenvalue $\theta_\alpha+\hbar(\nu^\alpha_{j+1}-j)$. Furthermore, if $c^{\pm\alpha}_{kj}$ should contain a factor with $\svX^\alpha_{kj}$ outside of this newly established set of operators, we simply declare that factor to be absent.

\part{Scalar products, functional orthogonality relations and dual separated variables}
\section{Baxter equation and functional scalar product}\label{functionalsp}

\subsection{Scalar products and overlaps in integrable systems}

In the previous sections we developed the separation of variables program for compact $\gl(\gn)$ spin chains. 
We would now like to use the developed techniques to compute some quantities of interest. A standard physical quantity of interest is the expectation value $\lO_{A}$ of some operator $\lO$ given by
\begin{equation}\label{formfactor}
\lO_{A}=\frac{\bra{\Psi_A}\lO\ket{\Psi_A}}{\braket{\Psi_A|\Psi_A}}
\end{equation}
where $\ket{\Psi_{A}}$ is a Hamiltonian eigenstate. 

\medskip

At this point it is important that we make our notation clear and we stress that $\bra{\Psi_A}$ is \textit{not} the Hermitian conjugate of $\ket{\Psi_A}$ -- $\bra{\Psi_A}$ is simply a left eigenstate of the Bethe algebra with the same eigenvalue as $\ket{\Psi_A}$ and there is a bijection between such left and right states owing to the non-degeneracy of the Bethe algebra. On the other hand, in many physically reasonable scenarios the Bethe algebra is closed under Hermitian conjugation, for example in the $\sua(2)$ spin chain the various parameters can be chosen so that $\T_{1,1}(u)^\dagger = \T_{1,1}(\bar{u})$ and hence the left eigenstates of the transfer matrix are simply scalar multiples of any left eigenvector with the same eigenvalue. Hence, if we introduce the corresponding inner product $\left( -,-\right)$ on the representation space turning it into a Hilbert space then the following ratios
\begin{equation}
\frac{\bra{\Psi}\lO\ket{\Psi}}{\braket{\Psi|\Psi}} = \frac{\left(\Psi,\lO\Psi\right)}{\left(\Psi,\Psi\right)}
\end{equation}
are equal. Hence we can either work with left eigenvectors directly or the Hermitian conjugate of right eigenvectors. Introducing Hermitian conjugation has several drawbacks however. The transformation properties of the transfer matrices are not as transparent in the higher rank case and it requires some work to prove for a given representation that the Bethe algebra is closed under it. Hence, it will be more convenient to work with left eigenvectors. 

\medskip

The calculation of such expressions have received an extensive amount of attention in the literature in part due to their relation various three-point functions in $\lN=4$ SYM. It was discovered in \cite{Roiban:2004va} that three-point functions could be expressed as in the form \eqref{formfactor}. This direction has been extensively developed \cite{Escobedo:2010xs,Gromov:2012vu,Gromov:2012uv,Foda:2012wf} and has culminated in the elegant Hexagon formalism for correlation functions \cite{Basso:2015zoa,Jiang:2015lda,Fleury:2016ykk}. 

\medskip

The Quantum Inverse Scattering Method (QISM) is one of the main tools for calculating these objects in integrable systems \cite{izergin1984quantum}, see \cite{Korepin:1993kvr} for an extensive treatment. It is based on the fact that the any local operator $\sfE_{ij}^{(\alpha)}$ acting on the $\alpha$-th spin chain site can be expressed as 
\begin{equation}
\sfE_{ij}^{(\alpha)} = \left(\prod_{\beta=1}^{\alpha-1}\T_{1,1}(\theta_\beta)\right) T_{ij}(\theta_\alpha)\left(\prod_{\beta=1}^{\alpha}\T_{1,1}(\theta_\beta)\right)^{-1}
\end{equation}
and hence the calculation of the expectation value $\bra{\Psi}\sfE_{ij}^{(\alpha)} \ket{\Psi}$ amounts to calulating the action of $T_{ij}(u)$ on Bethe vectors. This has been achieved for rank $1$ (i.e. $\gl(2)$-based) spin chains in the SoV framework \cite{Niccoli:2012vq,Niccoli:2012vq,Levy-Bencheton:2015mia,Kitanine:2015jna} which involves exploiting the simple action of the $T_{ij}$ operators on the SoV bases in both finite volume and in the thermodynamic limit \cite{Niccoli:2020zla}. 

\medskip

Of course one still needs to compute the norm of the state or the overlap of a right eigenvector with a left eigenvector $\braket{\Psi|\Psi}$ which has been computed for numerous models \cite{Korepin:1982gg}. In the framework of the algebraic Bethe ansatz these states are constructed as 
\begin{equation}\label{Bethevecs}
\ket{\Psi}=\prod_{j=1}^M B(u_j) \ket{\Omega},\quad \bra{\Psi}=\bra{\Omega}\prod_{j=1}^M C(v_j)
\end{equation}
where $u_j$ and $v_j$ are solutions of the Bethe equations. Overlaps of this type constitute a general class of overlaps called \textit{on-shell/on-shell} owing to the fact that $u_j$ and $v_j$ indeed satisfy Bethe equations. A more general class of overlaps, dubbed \textit{off-shell/on-shell} are obtained when one set of parameters, say $\{v_1,\dots,v_M\}$, do not satisfy the Bethe equations, and similarly one can consider \textit{off-shell/off-shell} overlaps where neither set of rapidities satisfy Bethe equations. For computing the overlap in these cases one can use the celebrated Slavnov determinant formula \cite{slavnov1989calculation}. In fact a determinant representation of these overlaps is a rather universal feature and the reason for this was recently clarified in \cite{Belliard:2019bfz} by noting that the scalar product satisfies a homogeneous system of linear equations and hence has a determinant representation. Such determinant formulas have also been obtained in higher rank models \cite{Belliard:2012pr,Pakuliak:2014fra} and $q$-deformed \cite{slavnov2015scalar} and supersymmetric models \cite{Hutsalyuk:2016yii}. The Bethe ansatz framework is not just limited to scalar products and has also allowed the computation of form factors for higher rank models \cite{Belliard:2012av,pakuliak2014form,Pakuliak:2014ela}. It has also been possible to express the scalar product as a multiple integral formula, see \cite{Kazama:2013rya} for the $\sua(2)$ case using the SoV approach and for higher rank models in the Bethe ansatz formalism \cite{Wheeler:2013zja}.

\medskip

There are also other overlaps which are of direct interest in QFT calculations. One such family of overlaps is the overlap of a Bethe state $\bra{\Psi}$ with a \textit{boundary state} $\bra{B}$ given by
\begin{equation}
\frac{\braket{B|\Psi}}{\braket{\Psi|\Psi}}
\end{equation}
whose interest stems from the fact that they are related to one-point functions in defect conformal field theory \cite{deLeeuw:2015hxa,deLeeuw:2016umh,deLeeuw:2019ebw}, see \cite{deLeeuw:2019usb} for a review. They are also related to the computation of the so-called $g$-function \cite{Caetano:2020dyp} which is one of the simplest quantities one can compute in an integrable QFT beyond its spectrum. These boundary states have also recently been explored in the SoV framework \cite{Gombor:2021uxz}.

\subsection{Functional orthogonality relations}

We now turn to the task of computing such quantities in the SoV framework, aided by the tools we have developed so far. So far we have only discussed the construction of the right eigenstates $\ket{\Psi}$. We also need a way to construct left eigenstates $\bra{\Psi}$ and then we need to calculate their overlap. This should be achieved by constructing the measure in the SoV basis. In principal this should be analogous to calculations which have been done in the rank $1$ sector. However, Martin and Smirnov \cite{Martin:2015eea} suggest that things are not so clear cut in the higher rank case based on a semi-classical calculation. They demonstrate that the expectation values of certain operators do not factorise in the SoV representation and we should expect to run into similar difficulties. 

\medskip

Nevertheless let us proceed. We will now discuss a method for computing the scalar product in the SoV framework without a need to explicitly compute matrix elements of the measure or construct states. This method is based on the use of the Baxter TQ equation and first appeared \cite{Cavaglia:2018lxi} in the computation of three-point structure constants in $\lN=4$ SYM and was subsequently developed for spin chains, first for non-compact spin $-\frac{1}{2}$ representations of $\sla(\gn)$ in \cite{Cavaglia:2019pow} and then for compact $\sua(3)$ spin chains in \cite{Gromov:2019wmz} and compact and non-compact $\sla(\gn)$ spin chains with generic spin in \cite{Gromov:2020fwh}. We will start our analysis with $\sua(2)$ spin chains. 

\paragraph{$\sua(2)$ functional orthogonality relations}

Our starting point is the two Baxter equations
\begin{equation}
\overrightarrow{\lO}\ \frac{\sfQ^{[2]}_i}{\sfQ_\es}=0,\quad \ \frac{\sfQ^i}{\left(\sfQ^\es\right)^{[2]}}\overleftarrow{\lO}=0\,.
\end{equation}
where $\lO$ is the finite-difference operator 
\begin{equation}
\lO = 1 - \sfT_{1,1}(u)\lD^{-1} + \sfT_{2,1}(u)\lD^{-2}
\end{equation}
where we remind the reader that the shift operator is $\lD:=e^{\hbar \partial_u}$ with 
\begin{equation}
\lD\,f(u)=f(u+\hbar),\quad f(u)\,\lD=f(u-\hbar)
\end{equation}
and we use arrows on $\lO$ to denote which direction the shift operators act. 

\medskip

The Q-functions $\sfQ_j$ and $\sfQ^j$ possess an infinite number of poles due to the structure 
\begin{equation}
\sfQ_j(u) = \hat{\sfq}_j(u)\times \Gamma\left[\nu_1(u)\right],\quad \hat{\sfq}_j(u)=z_j^{\frac{u}{\hbar}}\times \sfq_j(u)
\end{equation}
and it will be convenient to redefine the difference operators $\lO$ such that it satisfies 
\begin{equation}\label{baxterpol}
\lO\, \hat{\sfq}_j^{[2]} = 0\,.
\end{equation}
A straightforward calculation shows that by redefining 
\begin{equation}
\lO = 1 - \frac{\sfT(u)}{\nu_1(u)}\lD^{-1} + \frac{\nu_2(u)}{\nu_1(u)}\lD^{-2}
\end{equation}
then \eqref{baxterpol} is indeed satisfied. On the other hand, it is not true that $\sfq^j \overleftarrow{\lO}$ is zero now. Instead we have $\lO^\dagger\,\sfq^j=0$ where $\lO^\dagger$ is a different operator given by 
\begin{equation}
\lO^\dagger=1 - \frac{\sfT^{[2]}(u)}{\nu_2^{[2]}(u)}\lD+ \frac{\nu_1^{[2]}(u)}{\nu_2^{[2]}(u)}\lD^{2}\,.
\end{equation}
By appropriate redefinitions it is possible to obtain a scenario where $\lO^\dagger=\lO$. However, this is only the case for $\gl(2)$ and does not persist at higher rank so we do not use such conventions here. 

\medskip

Since the Q-functions carry all the information about a given state it should be possible to formulate the notion of orthogonality of different Bethe states directly at the level of Q-functions. The possibility of writing down such a functional orthogonality relation is based on the existence of an integration measure $K(u)$ such that the two operators $\lO$ and $\lO^\dagger$ are conjugate under an appropriate scalar product. We consider the space of twisted polynomials and equip it with the bilinear form $\left(-,-\right)$ defined by 
\begin{equation}
\left(f,g\right):=\displaystyle \int_{\lC} {\rm d}u\, K(u)\, f(u)\,g(u) 
\end{equation}
for arbitrary twisted polynomials $f$ and $g$ and the contour $\lC$ is yet to be determined. We are going to impose that
\begin{equation}
\left(f,\lO\, g\right)=\left(\lO^\dagger\,f,g\right)
\end{equation}
and use this to constrain the measure $K$ and the contour $\lC$. We will proceed by direct calculation. First, we have
\begin{equation}
\left(f,\lO\, g\right) = \displaystyle \int_{\lC} {\rm d}u\, K(u)\, f(u)\left(g(u) - \frac{\sfT(u)}{\nu_1(u)}g^{[-2]}(u) + \frac{\nu_2(u)}{\nu_1(u)}g^{[-4]}(u)\right)\,.
\end{equation}
We aim to move the shifts from $g$ to $f$ by shifting the integration contour. For the moment, let us assume that we can freely shift the contour, giving 
\begin{equation}
\displaystyle \int_{\lC} {\rm d}u\,\left(K(u)\, f(u)-K^{[2]}(u)\, f^{[2]}(u)\frac{\sfT^{[2]}(u)}{\nu_1^{[2]}(u)}+K^{[4]}f^{[4]}\frac{\nu_2^{[4]}(u)}{\nu_1^{[4]}(u)}\right)g(u)
\end{equation}
which we demand be equal to 
\begin{equation}
\left(\lO^\dagger\,f,g\right)=\displaystyle \int_{\lC} {\rm d}u\,K(u)\, \left(f(u) - \frac{\sfT^{[2]}(u)}{\nu_2^{[2]}(u)}f^{[2]}(u)+ \frac{\nu_1^{[2]}(u)}{\nu_2^{[2]}(u)}f^{[4]}(u)\right)g(u)\,.
\end{equation}
We immediately see that this will be satisfied if 
\begin{equation}
\frac{K^{[2]}(u)}{K(u)} = \frac{\nu_1^{[2]}(u)}{\nu_2^{[2]}(u)}
\end{equation}
which can then be solved by 
\begin{equation}\label{Kmeasure}
K(u) = \frac{\Gamma[\nu_1^{[2]}(u)]}{\Gamma[\nu_2^{[2]}(u)]}\rho(u)
\end{equation}
where $\rho(u)$ is a $\hbar$-periodic function.

\medskip

It is at this point where the distinction between compact and non-compact models will play a role. For now, we are only going to focus on compact models since all of our discussion in the previous sections has been about them. Later we will consider non-compact models. For compact representation the ratio of $\Gamma$-functions in \eqref{Kmeasure} reduces to a rational function. Hence, if we choose the contour $\lC$ to be a large circle containing all of the (finitely-many) poles of $K(u)$ and zeroes of $\nu_1(u)$ then we will have no problems in shifting the contour. To be more precise let $A$ be the set of poles of $K(u)$ and zeroes of $\nu_1(u)$ and put $r = {\rm max} |A|+2\hbar$. Then if $\lC$ is a circle of radius $r+\varepsilon$ with $\varepsilon>0$ then we can shift the contour $\lC$ and conclude that 
\begin{equation}\label{adjoint}
\left(f,\lO\, g\right)=\left(\lO^\dagger\,f,g\right)\,.
\end{equation}
The adjointness property \eqref{adjoint} guarantees that for any twisted polynomial $f$ we have
\begin{equation}
\left(\hat{\sfq}^j,\lO\,f \right)=0
\end{equation}
and hence, for any two states $A$ and $B$ we have 
\begin{equation}
\boxed{\left(\hat{\sfq}^j_A,(\lO^A-\lO^B)\,\hat{\sfq}^B_i\right)=0}
\end{equation}
where the finite-difference operators $\lO^A$ and $\lO^B$ depend on the states $A$ and $B$ through the transfer matrix eigenvalue $\sfT^{A,B}$. This is the key relation which leads to the functional scalar product. We first note that the only difference between the operators $\lO^A$ and $\lO^B$ is given by the term corresponding to the transfer matrix and so 
\begin{equation}
\lO^A -\lO^B = \frac{1}{\nu_1(u)}\left(\sfT^A(u) - \sfT^B(u)\right)\lD^{-1}
\end{equation}
and hence
\begin{equation}
0 = \left(\hat{\sfq}^j_A,\frac{1}{\nu_1(u)}\left(\sfT^A(u) - \sfT^B(u)\right)\hat{\sfq}^B_i \right)\,.
\end{equation}
Next we notice that the transfer matrix can be expanded as
\begin{equation}
\sfT_{1,1}(u) = \chi_{1,1} u^L + \sum_{k=0}^{L-1} u^k I_k
\end{equation}
where $I_k$ are the eigenvalues of the corresponding integrals of motion. Hence, we have
\begin{equation}\label{1storth}
0=\sum_{k=0}^{L-1}(I^A_k-I^B_k)\left(\hat{\sfq}^j_A,\frac{1}{\nu_1(u)}u^k\hat{\sfq}^B_i \right)\,.
\end{equation}
We now recall that the measure $K$ was not unique -- we are free to rescale it by a periodic function $\rho(u)$. Let us introduce a family of rescaled measure $\mu_\alpha$ with 
\begin{equation}
\mu_\alpha = \frac{\Gamma[\nu_1^{[2]}(u)]}{\Gamma[\nu_2^{[2]}(u)]}\frac{\rho_\alpha(u)}{\nu_1(u)}=\frac{\Gamma[\nu_1(u)]}{\Gamma[\nu_2^{[2]}(u)]}\rho_\alpha(u),\quad \alpha=1,2,\dots,L
\end{equation}
with the periodic functions $\rho_\alpha$ to be determined later and to introduce a new bracket $\langle -\rangle_\alpha$ by 
\begin{equation}\label{mubracket}
\langle f\,g \rangle_\alpha := \displaystyle \int_{\lC}{\rm d}u\,\mu_\alpha(u) f(u)g(u)\,.
\end{equation}
Now, the relation \eqref{1storth} becomes
\begin{equation}
0=\sum_{k=0}^{L-1}(I^A_k-I^B_k)\langle\hat{\sfq}^j_A\,u^k\,\hat{\sfq}^B_i \rangle_\alpha,\quad \alpha=1,\dots,L
\end{equation}
which constitutes $L$ equations for the $L$ unknowns $I^A_k-I^B_k$. Since at least one of the differences $I^A_k-I^B_k$ must be non-zero for two distinct states\footnote{Since the Bethe algebra is continuously connected to the Gelfand-Tsetlin algebra which has non-degenerate spectrum.} $A$ and $B$ the determinant of the linear system must vanish and we find 
\begin{equation}\label{scalarprod}
\delta_{AB} \propto \det_{1\leq \alpha,\beta\leq L}\langle\hat{\sfq}^j_A\,u^\beta\,\hat{\sfq}^B_i \rangle_\alpha\,.
\end{equation}
\paragraph{Choosing the periodic functions}
The exact choice of the periodic functions $\rho_\alpha$, like the contour $\lC$, depends on whether we consider the case of compact or non-compact representations and what exactly we hope to reproduce. For starters we can try to reproduce the scalar product produced by the operatorial SoV construction of the previous sections. s to choose 
\begin{equation}
\rho_\alpha(u) = \prod_{\beta\neq \alpha} 1-\exp\left(\frac{2\pi}{\hbar}(u-\theta_\beta)\right)\,.
\end{equation}
This then guarantees that the integration with $\mu_\alpha$ only picks up poles associated with shifts of $\theta_\alpha$ and hence it is natural to conjecture that for two Bethe algebra eigenstates \eqref{scalarprod} defines their scalar product in separated variables since by expanding the determinant and evaluating the integrals by residues we obtain 
\begin{equation}
\det_{1\leq \alpha,\beta\leq L}\langle\hat{\sfq}^j_A\,u^\beta\,\hat{\sfq}^B_i \rangle_\alpha = \sum_{\svx}\lM_{\svx} \prod_{\alpha=1}^L \hat{\sfq}^j_B(\svx^\alpha)\prod_{\beta=1}^L \hat{\sfq}_i^A(\svx^\beta)
\end{equation}
since the poles of the measure $\mu_\alpha$ precisely match the spectrum of separated variables. Indeed, \eqref{scalarprod} already passes a crucial test of the scalar product -- it vanishes for two different transfer matrix eigenstates -- and the factor $\prod_{\beta=1}^L \hat{\sfq}_i^A(\svx^\beta)$ coincides with the wave function in separated variables, see \eqref{wavefnexamples}.

\medskip

Let us note that the choice of $\rho_\alpha$ is not even unique as we can always multiply any of the $\rho_\alpha$ considered above by a $\hbar$-periodic function with no poles or zeroes. Indeed consider the integral 
\begin{equation}
\langle p\, f \rangle_\alpha = \int_{\lC}{\rm d}u \mu_\alpha(u)p(u)
\end{equation}
where $p(u)$ is $\hbar$-periodic without poles and zeroes. Performing the integral by residues and using the periodicity of $p(u)$ guarantees that $\langle p\, f \rangle= p(\theta_\alpha) \langle f \rangle$. Hence the overall effect of modifying the periodic functions in this way is that $\det_{1\leq \alpha,\beta\leq L}\langle\hat{\sfq}^j_A\,u^\beta\,\hat{\sfq}^B_i \rangle_\alpha$ becomes rescaled by a non-zero number. 

\medskip

Another possibility is to consider the homogeneous limit where all $\theta_\alpha\rightarrow 0$. In this case the simple poles of the measure $\mu_\alpha$ collide producing higher-order poles. A simple calculation yields that the result will take the following form in this case 
\begin{equation}
\det_{1\leq \alpha,\beta\leq L}\langle\hat{\sfq}^j_A\,u^\beta\,\hat{\sfq}^B_i \rangle_\alpha = F \times W(\rho_1,\dots,\rho_L)
\end{equation}
where $F$ is a non-zero term which only depends on the Q-functions and all $\rho$'s enter through the Wronskian 
\begin{equation}
W(\rho_1,\dots,\rho_L) = \displaystyle \det_{1\leq \alpha,\beta \leq L} \rho_\alpha^{(\beta)}(\theta)\,.
\end{equation}
One possibility is that we choose 
\begin{equation}
\rho_\alpha(u) = e^{(\alpha-1)\frac{u}{\hbar}},\quad \alpha=1,2,\dots,L
\end{equation}
which obviously produces a non-vanishing Wronskian.

\paragraph{$\su(3)$}

Having exhausted the $\su(2)$ case we now examine the case of $\su(3)$. Our starting point is again the two Baxter equations 

\begin{equation}
\overrightarrow{\lO}\ \frac{\sfQ^{[2]}_i}{\sfQ_\es}=0,\quad \ \frac{\sfQ^i}{\left(\sfQ^\es\right)^{[2]}}\overleftarrow{\lO}=0\,.
\end{equation}

Like before we introduce two finite-difference operators $\lO$ and $\lO^\dagger$ defined by
\begin{equation}
\lO = 1 - \frac{\sfT_{1,1}(u)}{\nu_1(u)}\lD^{-1}+\frac{\sfT_{2,1}(u)}{\nu_1(u)\nu_1^{[-2]}(u)}\lD^{-2}-\frac{\sfT_{3,1}(u)}{\nu_1(u)\nu_1^{[-2]}(u)\nu_1^{[-4]}(u)}\lD^{-3}
\end{equation}
and 
\begin{equation}\label{daggerbaxter}
\lO^\dagger = 1 - \frac{\sfT_{1,1}^{[2]}(u)}{\nu_3^{[2]}(u)}\lD^{1}+\frac{\sfT_{2,1}^{[4]}(u)}{\nu_3^{[2]}(u)\nu_3^{[4]}(u)}\lD^{2}-\frac{\sfT_{3,1}^{[6]}(u)}{\nu_3^{[2]}(u)\nu_3^{[4]}(u)\nu_3^{[6]}(u)}\lD^{3}
\end{equation}
and satisfy
\begin{equation}
\lO\, \hat{\sfq}_i^{[2]}(u)=0,\quad \lO^\dagger\, \hat{\sfq}^j(u)=0\,.
\end{equation}
By the method described previously it is straightforward to check that we have
\begin{equation}\label{adjointness}
\displaystyle \int_{\lC} {\rm d}u\,\frac{\Gamma[\nu_1^{[2]}(u)]}{\Gamma[\nu_3^{[2]}(u)]}\, g\, \lO\, f = \displaystyle \int_{\lC} {\rm d}u\,\frac{\Gamma[\nu_1^{[2]}(u)]}{\Gamma[\nu_3^{[2]}(u)]}\, f\, \lO^\dagger\, g
\end{equation}
and so for any two states $A$ and $B$ we have
\begin{equation}
0 = \displaystyle \int_{\lC} {\rm d}u\,\frac{\Gamma[\nu_1^{[2]}(u)]}{\Gamma[\nu_3^{[2]}(u)]}\, \hat{\sfq}^j_A\, \left(\lO^A - \lO^B\right)\, \hat{\sfq}^{B\,[2]}_i=\displaystyle \int_{\lC} {\rm d}u\,\frac{\Gamma[\nu_1^{[2]}(u)]}{\Gamma[\nu_3^{[2]}(u)]}\,  \hat{\sfq}^{B\,[2]}_i\, \left(\lO^\dagger_A - \lO^\dagger_B\right)\, \hat{\sfq}^j_A\,.
\end{equation}

\paragraph{Fundamental representation}
We will now proceed with a simple example which is the case of the defining representation on each spin chain site. Hence, we have
\begin{equation}
\nu_1(u) = Q_\theta^{[-2]}(u),\ \nu_2(u) = \nu_3(u) = Q_\theta(u),\quad Q_\theta(u)=\prod_{\alpha=1}^L (u-\theta_\alpha)
\end{equation}
and so 
\begin{equation}
\frac{\Gamma[\nu_1^{[2]}(u)]}{\Gamma[\nu_3^{[2]}(u)]} = \frac{1}{Q_\theta(u)}\,.
\end{equation}
A point to note for this representation which will also be used for $\sua(\gn)$ case is that the higher transfer matrices $\sfT_{a,1}$ contain overall trivial factors of $Q_\theta$ which is not true for general representations. Indeed, we write 
\begin{equation}
\sfT_{2,1} = Q_\theta(u) \sft_{2}(u)
\end{equation}
where $\sft_{2}(u)$ is a polynomial of degree $L$. To make the notation uniform we will also denote $\sft_{1}(u) = \sfT_{1,1}(u)$.

\medskip

From the operator SoV construction we know from \eqref{wavefnexamples} that for this representation the right wave functions $\Psi(\svx)$ are given by the product 
\begin{equation}
\prod_{\alpha=1}^L \hat{\sfq}_1(\svx^\alpha_{11})\hat{\sfq}_1(\svx^\alpha_{21})\,.
\end{equation}
In order to attempt to reproduce this from the Baxter equation we will put all of the shifts acting on the Hodge dual Q-functions $\sfq^i$ and so use
\begin{equation}
0 = \displaystyle \int_{\lC} {\rm d}u\,\frac{\rho_\alpha(u)}{Q_\theta(u)}\, \hat{\sfq}_1^{A\ [2]} \left(\lO^\dagger_A - \lO^\dagger_B\right)\hat{\sfq}^{j}_B\,.
\end{equation}
We now expand the difference operators and perform a shift $u\mapsto u-\hbar$. The result is 
\begin{equation}
0 = \displaystyle \int_{\lC} {\rm d}u\,\,\mu_\alpha(u) \hat{\sfq}_1^A \left(\left(\sft_1^B - \sft_1^A\right)D-\left(\sft_2^{B\, [2]} - \sft_2^{A\, [2]}\right)D^2\right)\hat{\sfq}^{j\ [-2]}_B\,.
\end{equation}
where the measure $\mu_\alpha(u)$ is given as before by 
\begin{equation}
\mu_\alpha(u)=\frac{\rho_\alpha(u)}{Q_\theta^{[-2]}(u)Q_\theta(u)}\,.
\end{equation}
Next we expand the transfer matrices 
\begin{equation}
\sft_a^A(u) = \sum_{\beta=0}^L u^\beta\, I^A_{a,\beta}
\end{equation}
and so finally obtain
\begin{equation}\label{rowsum}
0=\displaystyle \sum_{\beta=0}^{L-1} \langle \hat{\sfq}^A_1 \hat{\sfq}^j_B u^\beta \rangle_\alpha I^{AB}_{1,\beta} + \displaystyle \sum_{\beta=0}^{L-1} \langle \hat{\sfq}^A_1 \hat{\sfq}^{j\, [2]}_B u^\beta \rangle_\alpha I^{AB}_{2,\beta}
\end{equation}
where we have denoted $I^{AB}_{a,\beta}:= (-1)^{a-1}\left(I^{A}_{a,\beta}-I^{B}_{a,\beta}\right)$. 

\medskip

The requirement that this linear system has a non-trivial solution then imposes that vanishing of the following $2L\times 2L$ determinant 
\begin{equation}
\delta^{AB}\propto\det_{(\alpha,i),(\beta,j)}\langle \hat{\sfq}^A_1\hat{\sfq}_B^{i+1\, [2(j-1)]} u^{\beta-1} \rangle_\alpha\,.
\end{equation}
The notation is as follows: the indices $\alpha,\beta$ range over $1,2,\dots,L$ and $i,j$ range over $1,2$. A row in the matrix is labelled by a pair $(\alpha,i)$ and the entries in each row correspond to the coefficients in the expression \eqref{rowsum} which are labelled by a pair $(\beta,j)$. In the simplest case of length $L=1$ this corresponds to 
\begin{equation}
\delta^{AB} \propto \left| \begin{array}{cc}
\langle \hat{\sfq}^A_1\hat{\sfq}_B^2\rangle_1 & \langle \hat{\sfq}^A_1\hat{\sfq}^{2\ [2]}_B\rangle_1\\
\langle \hat{\sfq}^A_1\hat{\sfq}_B^3\rangle_1 & \langle \hat{\sfq}^A_1\hat{\sfq}^{3\ [2]}_B\rangle_1
\end{array} \right|
\end{equation}
while in the case $L=2$ we have
\begin{equation}
\delta^{AB} \propto \left| \begin{array}{cccc}
\langle \hat{\sfq}^A_1\hat{\sfq}_B^2\rangle_1 & \langle \hat{\sfq}^A_1\hat{\sfq}_B^2\, u\rangle_1 & \langle \hat{\sfq}^A_1\hat{\sfq}^{2\ [2]}_B\rangle_1 & \langle\hat{\sfq}^A_1\hat{\sfq}^{2\ [2]}_B\, u\rangle_1\\
\langle \hat{\sfq}^A_1\hat{\sfq}_B^3\rangle_1 & \langle \hat{\sfq}^A_1\hat{\sfq}_B^3\, u\rangle_1 & \langle \hat{\sfq}^A_1\hat{\sfq}^{3\ [2]}_B\rangle_1 & \langle\hat{\sfq}^A_1\hat{\sfq}^{3\ [2]}_B\, u\rangle_1 \\
\langle \hat{\sfq}^A_1\hat{\sfq}_B^2\rangle_2 & \langle \hat{\sfq}^A_1\hat{\sfq}_B^2\, u\rangle_2 & \langle \hat{\sfq}^A_1\hat{\sfq}^{2\ [2]}_B\rangle_2 & \langle\hat{\sfq}^A_1\hat{\sfq}^{2\ [2]}_B\, u\rangle_2 \\
\langle \hat{\sfq}^A_1\hat{\sfq}_B^3\rangle_2 & \langle \hat{\sfq}^A_1\hat{\sfq}_B^3\, u\rangle_2 & \langle \hat{\sfq}^A_1\hat{\sfq}^{3\ [2]}_B\rangle_2 & \langle\hat{\sfq}^A_1\hat{\sfq}^{3\ [2]}_B\, u\rangle_2
\end{array} \right|\,.
\end{equation}

\paragraph{Matching with wave functions}
We now expand the determinant for the case $L=1$ and obtain 
\begin{equation}
\delta^{AB} \propto \displaystyle \int {\rm d}u_1{\rm d}u_2 \mu_1(u_1)\mu_1(u_2)\times \hat{\sfq}^A_1(u_1) \hat{\sfq}^A_1(u_2)\times \det_{1\leq i,j\leq 2}\hat{\sfq}^{i+1\ [2(j-1)]}_B(u_j)
\end{equation}
which indeed matches the type of expression we expect from the operator SoV construction \eqref{wavefnexamples} -- when the integral is calculated by residues we will obtain a sum of the form 
\begin{equation}\label{wavesum}
\sum_{\svx} F(\svx) \times \prod_{\alpha=1}^L \hat{\sfq}_1(\svx_{21}^\alpha)\hat{\sfq}_1(\svx_{11}^\alpha)
\end{equation}
where the sum is over all configurations $\svx_{j1}^\alpha=\theta_\alpha,\theta_\alpha+\hbar$ and $F(\svx)$ denotes all other terms. The product of Q-functions in \eqref{wavesum} precisely matches the wave functions built in the previous Part, see \eqref{wavefnexamples}.

\medskip

Interestingly the dual wave functions corresponding to the term 
\begin{equation}
\det_{1\leq i,j\leq 2}\hat{\sfq}^{i+1\ [2(j-1)]}_B(u_j)
\end{equation}
look strikingly similar to the wave functions when the physical space is the \textit{anti-fundamental} representation \eqref{wavefnexamples}.  Naturally, we can then expect that the dual wave functions will be simple products of Hodge dual Q-functions and hence the left eigenstates can be built using $\bB$ or perhaps some yet-to-be-determined operator. Let us examine what happens now. 

\paragraph{Anti-fundamental representation} 

This case is largely the same as the previous case except since we expect the right wave functions to be given by a $2\times 2$ determinant we put the shifts on $\hat{\sfq}_i$ instead of $\hat{\sfq}^j$. Hence we consider
\begin{equation}
0 = \displaystyle \int_{\lC} {\rm d}u\,\frac{\rho_\alpha(u)}{Q_\theta(u)}\, \hat{\sfq}^{j}_B \left(\lO^A - \lO^B\right)\hat{\sfq}_i^{A\ [2]}\,.
\end{equation}
We now proceed exactly as before, the only difference being that the non-dynamical factor in $\sfT_{2,1}(u)$ is different. We put 
\begin{equation}
\sfT_{1,1}(u)=\sft_1(u),\quad \sfT_{2,1}(u) = \sft_{2}(u) Q_\theta^{[-4]}(u)
\end{equation}
and are hence led to the system of equations 
\begin{equation}
0=\displaystyle \sum_{\beta=0}^{L-1} \langle \hat{\sfq}^1_B \hat{\sfq}^A_i u^\beta \rangle_\alpha I^{AB}_{1,\beta} + \displaystyle \sum_{\beta=0}^{L-1} \langle \hat{\sfq}^1_B \hat{\sfq}_i^{A\, [-2]} u^\beta \rangle_\alpha I^{AB}_{2,\beta}
\end{equation}
where we have fixed $j=1$. We then obtain 
\begin{equation}
\delta^{AB} \propto \displaystyle \det_{(\alpha,i),(\beta,j)} \langle \hat{\sfq}^1_B u^{\beta-1} \hat{\sfq}_i^{A\, [2(1-j)]} \rangle_\alpha 
\end{equation}
which in the case $L=1$ produces 
\begin{equation}\label{antifundsp}
\delta^{AB} \propto \left| \begin{array}{cc}
\langle \hat{\sfq}^1_B  \hat{\sfq}_1^A \rangle_1 & \langle \hat{\sfq}^1_B  \hat{\sfq}_1^{A\,[-2]} \rangle_1\\
\langle \hat{\sfq}^1_B  \hat{\sfq}_2^A \rangle_1 & \langle \hat{\sfq}^1_B  \hat{\sfq}_2^{A\,[-2]} \rangle_1
\end{array} \right|\,.
\end{equation}
We will examine the implications of this result in the next section.

\paragraph{$\sua(\gn)$}
Having extensively treated the $\sua(2)$ and $\sua(3)$ cases we will now consider the general $\sua(\gn)$ case. For simplicity we will consider the same representation each site and specifically consider representations corresponding to symmetric powers of the defining representations with highest weight 
\begin{equation}
\nu^\alpha= [\lambda,0,\dots,0]\,.
\end{equation}
Hence the weight functions are given by 
\begin{equation}
\nu_1(u) = Q_\theta^{[-2\lambda]}(u),\quad \nu_k(u)=Q_\theta(u),\ k\geq 2\,.
\end{equation}
As in the $\sua(3)$ case the anti-symmetric transfer matrices $\sfT_{a,1}(u)$ contain trivial non-dynamical overall factors. We have
\begin{equation}
\sfT_{a,1}(u) = \sft_a(u) \prod_{k=1}^{a-1} Q_\theta^{[-2(k-1)]}(u)\,.
\end{equation}
We expand each of these transfer matrices as 
\begin{equation}
\sft_a(u) =\sum_{\beta=0}^L u^\beta\, I_{a,\beta}\,.
\end{equation}
From our experience with the operator SoV construction we know that the transfer matrix wave functions $\Psi(\svx)$ in this representation are given simply by 
\begin{equation}
\Psi(\svx) = \prod_{\alpha=1}^L \prod_{k=1}^{\gn-1} \hat{\sfq}_1(\svx^\alpha_{k1})
\end{equation}
and so in order to reproduce this result we put all the shifts on the Hodge dual Q-functions and hence consider
\begin{equation}\label{diffeqn}
0 = \displaystyle \int_{\lC} {\rm d}u\,K(u)\, \hat{\sfq}_1^{A\ [2]} \left(\lO_A^\dagger - \lO^\dagger_B\right)\hat{\sfq}^{j}_B\,.
\end{equation}
where the measure factor $K(u)$ is given by 
\begin{equation}
K(u) = \frac{\Gamma[\nu_1^{[2]}(u)]}{\Gamma[\nu_\gn^{[2]}(u)]}\rho(u) = \frac{\rho(u)}{\displaystyle\prod_{k=1}^{\lambda}Q_\theta^{[-2(k-1)]}}\,.
\end{equation}
The finite-difference operator $\lO^\dagger$ is given by 
\begin{equation}
\lO^\dagger=\displaystyle \sum_{a=0}^{\gn}(-1)^a \frac{\sft_{a,1}}{Q_\theta^{[2]}(u)}
\end{equation}
and as a result we find that \eqref{diffeqn} becomes, after again performing a shift $u\rightarrow u-\hbar$,
\begin{equation}\label{diffeqn2}
0 = \displaystyle \int_{\lC} {\rm d}u\, \mu_\alpha(u) \sum_{a=1}^{\gn-1}(-1)^a \hat{\sfq}_1^A \hat{\sfq}^{j\, [2(a-1)]} \sft^{AB\, [2(a-1)]}_a(u) 
\end{equation}
where we have now denoted the integration measure 
\begin{equation}
\mu_\alpha(u) = \frac{\rho_\alpha(u)}{\displaystyle\prod_{k=0}^\lambda Q_\theta^{[-2k]}}
\end{equation}
and $\sft^{AB}_a = \sft^A_a-\sft^B_a$. If we now expand the difference of transfer matrices into integrals of motion 
\begin{equation}
\sft_a^{[2(a-1)]}=(-1)^a\sum_{\beta=0}^L u^\beta I_{a,\beta}
\end{equation}
then \eqref{diffeqn2} constitutes a linear system on the differences $I_{a,\beta}^A - I_{a,\beta}^B$ and the requirement that this linear system has a non-trivial solution then imposes that vanishing of the following $(\gn-1)L\times (\gn-1)L$ determinant 
\begin{equation}
\delta^{AB}\propto\det_{(\alpha,i),(\beta,j)}\langle \hat{\sfq}^A_1\hat{\sfq}_B^{i+1\, [2(j-1)]} u^{\beta-1} \rangle_\alpha\,.
\end{equation}
where now the indices $i$ and $j$ range over $1,2,\dots,\gn-1$. 

\medskip

This completes our study of the functional scalar product for compact $\su(\gn)$ spin chains. Next we will construct an operator realisation of the dual wave functions using a new operator $\bC$ which plays a similar role in the construction of left wave functions as $\bB$ did for right wave functions. We will return to the functional scalar product later when we consider non-compact spin chains. Effectively the only difference is that in the compact case the ratio of Gamma functions in $K(u)$ cancelled to produce a function with finitely many poles while in the non-compact case there is no cancellation and the function $K(u)$ has an infinite number of poles and zeros which requires a careful analysis.

\section{Dual separated variables}

We saw in the previous section that when the physical space is in anti-fundamental representation the dual wave functions are given by a simple product of Q-functions. When the physical space is in the fundamental representation the wave functions were also given by a very similar product and this coincided with the fact that we could create right Bethe algebra eigenstates using $\bB$. We are then led to conjecture that when the physical space is in the anti-fundamental representation we can build \textit{left} Bethe algebra eigenstates using a new operator, naturally denoted $\bC$. 

\subsection{Determining $\bC$ from the functional scalar product}
We can already say quite a lot about $\bC$ from the functional scalar product. Expanding the determinant \eqref{antifundsp} we find 
\begin{equation}
\begin{split}
 \delta^{AB} \propto &  \displaystyle \int{\rm d}u_1 {\rm d}u_2 \mu_1(u_1)\mu_2(u_2) \\
 &\times \hat{\sfq}^1_B(u_1)\hat{\sfq}^1_B(u_2) \times \left(\hat{\sfq}^A_1(u_1)\hat{\sfq}^{A\, [-2]}_2(u_2)-\hat{\sfq}^A_2(u_1)\hat{\sfq}^{A\, [-2]}_1(u_2) \right)\,.
\end{split}
\end{equation}
Now we evaluate the integral as a sum over residues, picking up poles at $\theta$, $\theta+\hbar$, providing three independent terms
\begin{equation}
\begin{split}
& \hat{\sfq}^1_B(\theta)\hat{\sfq}^1_B(\theta) \times \left(\hat{\sfq}^A_1(\theta)\hat{\sfq}^{A\, [-2]}_2(\theta)-\hat{\sfq}^A_2(\theta)\hat{\sfq}^{A\, [-2]}_1(\theta) \right) \\
& \hat{\sfq}^1_B(\theta+\hbar)\hat{\sfq}^1_B(\theta) \times \left(\hat{\sfq}^{A\,[2]}_1(\theta)\hat{\sfq}^{A\, [-2]}_2(\theta)-\hat{\sfq}^{A\,[2]}_2(\theta)\hat{\sfq}^{A\, [-2]}_1(\theta) \right) \\
& \hat{\sfq}^1_B(\theta+\hbar)\hat{\sfq}^1_B(\theta+\hbar) \times \left(\hat{\sfq}^{A\,[2]}_1(\theta)\hat{\sfq}^{A}_2(\theta)-\hat{\sfq}^{A\,[2]}_2(\theta)\hat{\sfq}^{A}_1(\theta) \right)\,.
\end{split}
\end{equation}
We now compare with the known wave function of $\ket{\Psi}$ which is 
\begin{equation}
\det_{1\leq i,j\leq 2} \hat{\sfq}_i(\svx^\alpha_{2j}),\quad \svx^\alpha_{2j}=\theta_\alpha+\hbar(\mu^\alpha_{2j}-j+1)\,.
\end{equation}
The Gelfand-Tsetlin basis vectors are associated to the spectra of $\svx$'s as follows:
\begin{equation}
\begin{array}{cccccccc}
1 & \ & 1 & \ & 0 \\
\ & 1 & \ & 1 \\
\ & \ & 1 \\ 
\end{array} = \left( \begin{array}{c}
1 \\
0 \\
0
\end{array}
\right),\quad
\begin{array}{cccccccc}
1 & \ & 1 & \ & 0 \\
\ & 1 & \ & 0 \\
\ & \ & 1 \\ 
\end{array} = \left( \begin{array}{c}
0 \\
1 \\
0
\end{array}
\right),
\begin{array}{cccccccc}
1 & \ & 1 & \ & 0 \\
\ & 1 & \ & 0 \\
\ & \ & 0 \\ 
\end{array} = \left( \begin{array}{c}
0 \\
0 \\
1
\end{array}
\right)\,.
\end{equation}
We would like to interpret the functions $\hat{\sfq}^1_B(u_1)\hat{\sfq}^1_B(u_2)$ as the wave functions of states $\bra{\Psi}$ generated by a $\bC$ operator such that
\begin{equation}
\bra{\Psi} = \bra{\Omega}\bC(v)
\end{equation}
where $v$ is a root of $\hat{\sfq}^1$. In order to reproduce the sum over wave functions produced by the integral formula then in the Gelfand-Tsetlin basis diagonalising $\bB$ the operator $\bC$ must be given by 
\begin{equation}
\bC(u) \ \sim \ {\rm diag}\left((u-\theta-\hbar)^2,(u-\theta)(u-\theta-\hbar),(u-\theta)^2 \right)\,.
\end{equation}
We now ask ourselves -- is there a product of Gelfand-Tsetlin generators with precisely this spectrum? The answer is yes and turns out to be given by $T_{11}(u)T\left[^{12}_{12} \right]$ which coincides with $\bB^{\rm GT}$ up to a different shift in the second minor. Hence, recalling that $\bB$ is given by 
\begin{equation}
\bB(u)=T_{11}(u)^{[2]}T\left[^{12}_{12} \right] + T_{21}(u)T^{[2]}\left[^{12}_{13} \right]
\end{equation}
the most natural guess for the $\bC$ operator for $\gl(3)$ is 
\begin{equation}
\bC(u) = T_{11}(u)T\left[^{12}_{12} \right] + T_{21}(u)T\left[^{12}_{13} \right]\,.
\end{equation}
This operator should satisfy several properties. To produce a set of separated variables in a way analogous to $\bB$ it should be diagonalised by transfer matrices on some appropriate reference state. As well as this, for spin chains carrying the anti-fundamental representation of $\gl(3)$ it should diagonalise the Bethe algebra as 
\begin{equation}\label{Cstates}
\bra{\Psi}=\bra{\Omega}\prod_{j=1}^M \bC(u_j)
\end{equation}
where $u_j$ in this case correspond to the roots of $\sfq_{12}$ instead of $\sfq_{1}$. These facts can indeed be verified for chains of small length, with the eigenvectors of $\bC$ being given by 
\begin{equation}
\prod_{\alpha=1}^L \T_{1,1}(\theta_\alpha+\hbar)^{n_\alpha}\ket{\bar{0}},\quad n_\alpha\in \{0,1,2\}
\end{equation}
in the anti-fundamental representation where $\ket{\bar{0}}$ is the highest-weight Gelfand-Tsetlin basis state corresponding to the pattern where all nodes take their maximal value in contrast to the situation with $\bB$ where its eigenvectors were generated from $\bra{0}$, the state where all nodes on GT patterns took their \textit{minimal} value. The wave functions can then be worked out in terms of Q-functions and allow us to demonstrate that $\bC$ indeed creates Bethe states in the proposed way \eqref{Cstates}. 

\medskip

In analogy with $\bB$ we now propose the following higher-rank generalisation of $\bC$:
\begin{equation}\label{Cincomp}
\boxed{
\bC(u)=\sum_{J_1,\dots,J_{\gn-1}}T\left[^{J_{\gn-1}}_{1\ J_{n-2}+1}\right]\dots T\left[^{J_2}_{1\ J_1+1}\right]T\left[^{J_1}_1\right]}
\end{equation}
where the order of minors has been reversed compared to $\bB$ \eqref{Bincomp} for convenience, although similar to $\bB$ this reversal does not effect any properties we will use. 

\subsection{$\bC$ operator and $*$-map}

The Yangian $\lY_{\gn}$ admits a very useful anti-automorphism \cite{molev2007yangians} which acts as 
\begin{equation}
\bT_{ij}(u)\rightarrow \bT_{ij}(-u)
\end{equation}
and on products as 
\begin{equation}
\bT_{ij}(u)\bT_{kl}(v)\rightarrow \bT_{kl}(-v)\bT_{ij}(-u)
\end{equation}
which is trivially equivalent to
\begin{equation}
\bT_a(u)\bT_b(v) \mapsto \bT_b(-v) \bT_a(-u)\,.
\end{equation}
Applying this transformation to the RTT relation, we obtain 
\begin{equation}
R_{ab}(u-v)\bT_b(-v)\bT_a(-u)=\bT_a(-u)\bT_b(-v)R_{ab}(u-v)
\end{equation}
and since $R_{ab}(u-v)^{-1} \propto R_{ab}(-u+v)$ we find that the RTT relation is still satisfied and hence this maps indeed constitutes an anti-automorphism. Note that this map is consistent with twisting -- $\bT_{ij}(u) \rightarrow \bT_{ij}(-u)$ if and only if $T_{ij}(u)\rightarrow T_{ij}(-u)$. 

\medskip

We now introduce a convenient notation. For elements $F(u),G(u)\in \lY_\gn[[u]]$ we define $F^*(u)$ by
\begin{equation}
F^*(-u) = \left(F(u)\right)^*,\quad \left(F(u)G(u)\right)^* = G^*(-u)F^*(-u),\quad \bT_{ij}^*(-u) = \bT_{ij}(-u)\,.
\end{equation}
The image of any element of $\lY_{\gn}[[u^{-1}]]$ can then be reconstructed from this basic definition. In lieu of this notation we refer to the anti-automorphism $\bT(u)\mapsto \bT(-u)$ the $*$-map. Clearly it corresponds to applying $\bT(u)\rightarrow \bT(-u)$ followed by a simple relabelling of $u\rightarrow -u$. 

\medskip

One of the most important properties of the $*$-map is its action on quantum minors. A straightforward calculation immediately yields
\begin{equation}\label{minorstar}
\bT\left[^{i_1\dots i_a}_{j_1 \dots j_a}\right](u+\hbar\,k) \rightarrow \bT\left[^{i_1\dots i_a}_{j_1 \dots j_a}\right](u+\hbar(a-1-k))\,.
\end{equation}
Armed with this transformation law we can now compute the action of $*$ on $\bB$. We consider $\gl(3)$, in the companion twist frame, and have
\begin{equation}
\bB(u)=T_{11}T^{[2]}\left[^{12}_{12}\right]+T_{21}T^{[2]}\left[^{12}_{13}\right]\,.
\end{equation}
By applying $*$ and using \eqref{minorstar} we immediately obtain 
\begin{equation}
\bB(u) \rightarrow T\left[^{12}_{12}\right]T_{11}+T\left[^{12}_{13}\right]T_{21}
\end{equation}
which is none other than $\bC(u)$! For higher rank we then make the \textit{definition} 
\begin{equation}
\boxed{\bC(u):=\bB^*(u)}
\end{equation}
which reproduces our conjecture \eqref{Cincomp}.
This simple relation allows us to apply essentially all of the technology developed for studying $\bB$ to $\bC$, in particular we can easily construct its eigenvectors. 

\medskip

We will apply the $*$-map to the commutation relation between $\bB$ and transfer matrices $\T_{\lambda}$. We have
\begin{equation}
\T_{\lambda}(v)\bB(u) = f_\lambda(u,v) \bB(u)\T_{\lambda}(v) + \sum_{j=1}^{\gn} T_{j1}(v)\times \dots
\end{equation}
which leads immediately to 
\begin{equation}\label{CTcom}
\boxed{\bC(u)\T_{\lambda}^*(v) = f_\lambda(-u,-v) \T^*_{\lambda}(v)\bC(u) + \sum_{j=1}^{\gn} \dots \times T_{j1}(v)}\,.
\end{equation}
This relation will serve to diagonalise $\bC$ in the same way we diagonalised $\bB$, except now we must act on $\ket{\bar{0}}$, the highest-weight Gelfand-Tsetlin state satisfying $T_{j1}(u)\ket{\bar{0}}=\delta_{j1}\nu_1(u)\ket{\bar{0}}$. The only other remaining thing to look at are the objects $\T_\lambda^*(v)$ obtained by applying the $*$-operation to transfer matrices. In order to make any kind of conclusions about eigenvectors of $\bC$ factorising Bethe algebra wave functions we would like for these to also be transfer matrices. 

\subsection{$*$-map and Bethe algebra}\label{sec:skewdiagrams}

By virtue of the relation \eqref{minorstar} transfer matrices $\T_{a,1}$ in anti-symmetric representations are only modified by simple shifts under application of $*$ and we have
\begin{equation}
\T_{a,1}^*(u) = \T_{a,1}(u+\hbar(a-1))\,.
\end{equation}
Since these transfer matrices generate the Bethe algebra via the CBR formula it immediately follows that the $*$-map preserves the Bethe algebra. In general however $\T_\lambda(u)$ does not get mapped to itself with simple shifts. Let's take a look at the transfer matrix in the rep $\lambda=[2,1,0]$. By using the CBR formula this can be expressed in terms of anti-symmetric transfer matrices as 
\begin{equation}
\T_{[2,1,0]} = \T_{2,1}\T_{1,1}^{[2]} - \T_{3,1}^{[2]}\,.
\end{equation}
By applying $*$ we find
\begin{equation}
\T_{[2,1,0]}^* = \T_{2,1}^{[2]}\T_{1,1}^{[-2]} - \T_{3,1}^{[2]}
\end{equation}
which does not coincide with $\T_{[2,1,0]}$ for any choice of shift, and indeed does not coincide with $\T_{\lambda}$ for \textit{any} choice of shifts which is easily seen by writing down the CBR formula for all possible Young diagrams with $3$ boxes, of which there are precisely $3$. 

\medskip

Apparently $\T_{[2,1,0]}^*$ does not coincide with a transfer matrix corresponding to an irreducible representation of $\gl(3)$ in the auxiliary space and so the other likely candidate is that it corresponds to some reducible representation. Of course the other alternative is that it simply does not have a simple interpretation but for the moment lets remain optimistic. Irreducible finite-dimensional representations of $\gl(\gn)$ are classified by their characters and so a natural starting point is to examine $\T_{[2,1,0]}^*$ in the limit $u\rightarrow \infty$. In this limit shifts are irrelevant and so both $\T_{[2,1,0}$ and $\T_{[2,1,0]}^*$ coincide in this limit and hence the two representations are isomorphic representation of $\gl(3)$! Hence, the difference between $\T_{[2,1,0]}$ and $\T_{[2,1,0]}^*$ must be a purely quantum effect obtained from taking $\hbar$ contributions into account. 

\medskip

\begin{figure}[htb]
\centering
  \includegraphics[width=10mm,scale=10]{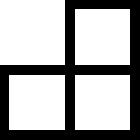}
  \caption{.}
  \label{simpleskew2}
\end{figure}

Further intuition can be obtained from expressing the relevant transfer matrices in terms of quantum semi-standard Young tableaux \eqref{cbrsoln}. We recall that $\T_{1,1}$, $\T_{2,1}$ and $\T_{3,1}$ can be expressed as 
\begin{equation}
\begin{split}
& \T_{1,1} = \Lambda_1+\Lambda_2+\Lambda_3 \\
& \T_{2,1} = \Lambda_3 \Lambda_2^{[-2]}+\Lambda_3 \Lambda_1^{[-2]}+\Lambda_2 \Lambda_1^{[-2]} \\
& \T_{3,1} = \Lambda_3 \Lambda_2^{[-2]}\Lambda_1^{[-4]}\,. \\
\end{split}
\end{equation}
We can then express $\T_{[2,1,0]}^*$ in terms of $\Lambda$'s. If we demand that it corresponds to a table of boxes with the conditions that numbers in a row weakly increase and numbers in a column strictly decrease then the only possible shape we can draw with these properties and producing the correct sum over $\Lambda$'s is given by Figure \ref{simpleskew2}, a \textit{skew} Young diagram. 

\paragraph{Skew diagram representations}

\begin{figure}[htb]
\centering
  \includegraphics[width=100mm,scale=25]{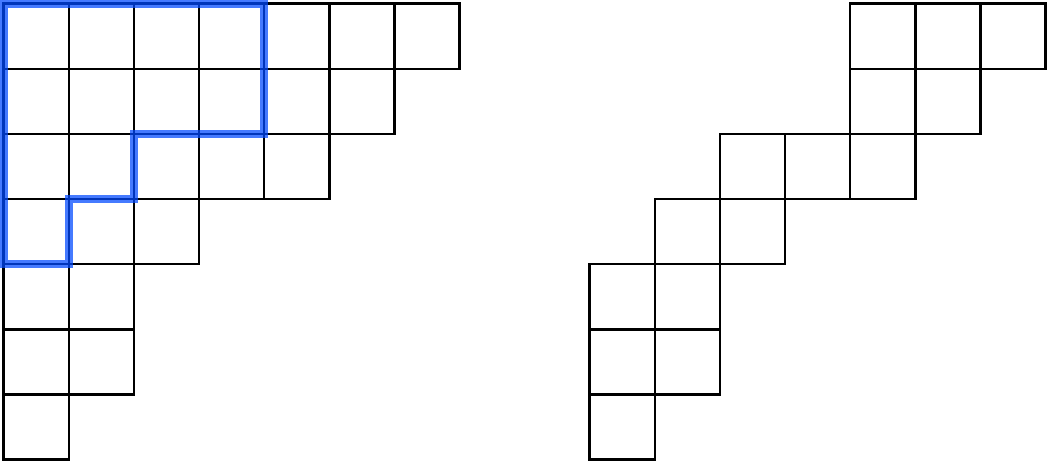}
  \caption{The Young diagram $\mu$ (bold boundary) is contained in the Young diagram $\lambda$ (left). The resulting skew diagram $\mu/\lambda$ is on the right.}
  \label{skewdiagram}
\end{figure}

A skew diagram $\lambda/\mu$ is defined by a pair of Young diagrams $\lambda$ and $\mu$ such that $\mu$ is contained in $\lambda$ when the top left corners of both are aligned. Graphically, $\lambda/\mu$ is obtained by removing the boxes of $\mu$ from $\lambda$. As an example, let $\lambda=[7,6,5,3,2,2,1]$ and $\mu=[4,4,2,1]$. The skew diagram $\mu/ \lambda$ is displayed in Figure \ref{skewdiagram}. 

At the level of representations skew diagrams appear as follows. We very closely follow the exposition in \cite{nazarov2004representations}. Choose an $m\in\ZZ_{\geq 0}$ and consider $\gl(\gn+m)$ with basis $\sfE_{ij}$, $i,j\in\{1,\dots,\gn+m\}$. $\gl(\gn+m)$ has natural subalgebras identified with $\gl(\gn)$ and $\gl(m)$ spanned by $\sfE_{ij}$ with $i,j\in\{1,\dots,\gn\}$ and $i,j\in\{\gn+1,\dots,\gn+m\}$ respectively. 

\medskip

Let $\lambda$ and $\mu$ be $\gl(\gn+m)$ and $\gl(m)$ Young diagrams respectively which give rise to irreps $\lV^\lambda$ and $\lV^\mu$. Consider 
\begin{equation}\label{Homspace}
{\rm Hom}_{\gl(m)}\left(\lV^\mu,\lV^\lambda \right)
\end{equation}
which is the set of all linear maps from $\lV^\mu$ to $\lV^\lambda$ which commute with the action of $\gl(m)$. This space carries a natural action of $\gl(\gn)$ defined by
\begin{equation}
(x.\phi)(v):=x.\phi(v),\quad x\in\gl(\gn),\, v\in\lV^{\mu},\, \phi\in {\rm Hom}_{\gl(m)}\left(\lV^\mu,\lV^\lambda \right)\,.
\end{equation}
It can be shown that the space is non-zero if and only if $\lambda_k\geq \mu_k$ and $\lambda_k^\prime-\mu_k^\prime\leq \gn$ i.e. $\lambda/\mu$ defines a skew diagram and furthermore the representation is independent on the choice of $m$ up to isomorphism. Hence this representation is completely determined by the skew diagram $\lambda/\mu$ and can furthermore be extended to a representation of the Yangian $\lY_{\gn}$. Representations of $\lY_{\gn}$ obtained in this way are referred to as \textit{elementary} representations and are irreducible Yangian representations. Nazarov \cite{nazarov2004representations} provides an explicit construction of them in terms of a fusion procedure similar to what we presented in \ref{Fusionsec}. The transfer matrices $\T_{\lambda/\mu}$ have been extensively studied, in particular their expressions in terms of quantum eigenvalues in the analytical Bethe ansatz framework, see \cite{Tsuboi:1997iq,Tsuboi:1998ne,Tsuboi:1998sc}.

\paragraph{$*$-map and skew diagrams}

Now that we understand that the $*$-operation relates transfer matrices $\T_{\lambda}$ to transfer matrices corresponding to some skew diagram it would be nice to be able to write down an action of $*$ directly on the level of Young diagrams resulting in a relation
\begin{equation}
\T_{\lambda/\mu}^* = \T_{(\lambda/\mu)^*}
\end{equation}
where $(\lambda/\mu)^*$ is some skew diagram to be determined. The key tool we will use is the CBR formula which is also valid \cite{Tsuboi:1997iq,lu2020jacobi} for skew-diagram transfer matrices 
\begin{equation}
\T_{\lambda/\mu}(u)=\det_{1\leq i,j\leq \lambda_1} \T_{\lambda_j^\prime+i-j-\mu_i^\prime}(u+\hbar(i-1-\mu_i^\prime))\,.
\end{equation}
In order to determine $\left(\lambda/\mu\right)^*$ it is convenient to introduce a certain redundant parameterisation of the skew Young diagram $\lambda/\mu$. Indeed, any Young diagram can be viewed as a square with boxes removed from the lower right corner and a skew Young diagram can be viewed as a square with boxes removed from the upper left and lower right corners. Let $\lS$ be a square of size $r\times r$ containing the Young diagram $\lambda$ and embed $\lambda$ in the top left corner of $\lS$. We define another Young diagram $\nu$ by the property $\nu^\prime_i=r - \lambda^\prime_{r+1-i}$. Note that $\nu$ indeed defines a Young diagram since $\lambda_1^\prime \leq r$. 

\medskip

\begin{figure}[htb]
\centering
  \includegraphics[width=80mm,scale=25]{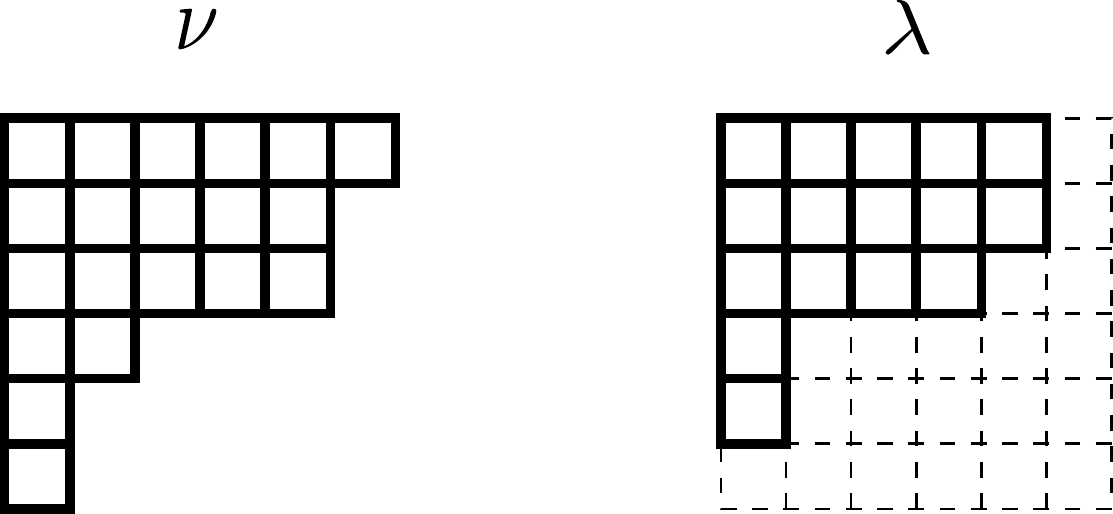}
  \caption{Left: Young diagram $\nu$. Right: Young diagram $\lambda$ obtained by flipping $\nu$ upside down and backwards and removing it from the square $\lS$ of size $6\times 6$.}
  \label{nulambda}
\end{figure}

We use the notation $\nu\backslash\lS$ to denote that $\nu$, after being flipped upside down and backwards, is removed from the lower right corner of $\lS$. Hence, $\lambda = \nu\backslash\lS$ and hence $\lambda/\mu = \nu\backslash\lS/\mu$, see Figure \ref{nulambda}. We will now demonstrate that 
\begin{equation}
\left(\lambda/\mu \right)^* = \left(\nu\backslash\lS/\mu \right)^* = \mu\backslash\lS/\nu\,.
\end{equation}

\paragraph{Proof} The proof is a straightforward consequence of the CBR formula. We have 
\begin{equation}
\T_{\lambda/\mu}(u) = \det_{1\leq i,j\leq \lambda_1} \T_{\lambda_j^\prime+i-j-\mu_i^\prime}(u+\hbar(i-1-\mu_i^\prime))
\end{equation}
which can be rewritten as 
\begin{equation}
\T_{\lambda/\mu}(u) = \det_{1\leq i,j\leq r} \T_{\lambda_j^\prime+i-j-\mu_i^\prime}(u+\hbar(i-1-\mu_i^\prime))
\end{equation}
for any $r\geq 0$ by using the boundary conditions \eqref{bdyconds} and hence, by writing $\lambda=\nu\backslash\lS$, 
\begin{equation}\label{nuSmutransfer}
\T_{\nu\backslash \lS/\mu}(u)= \det_{1\leq i,j\leq r} \T_{r-\nu_{r+1-j}^\prime+i-j-\mu_i^\prime}(u+\hbar(i-1-\mu_i^\prime))\,.
\end{equation}
We can now conjugate the $r\times r$ matrix of transfer matrices we take the determinant of with the matrix $\sigma$ with $\sigma_{ij} = \delta_{1,i+j-r}$ which transforms a matrix with entries $A_{ij}$ as
\begin{equation}
A_{ij} \rightarrow A_{r+1-i,r+1-j}
\end{equation}
and of course does not change the determinant value. Applying this transformation to \eqref{nuSmutransfer}  and performing the transpose $A_{ij} \rightarrow A_{ji}$ we obtain 
\begin{equation}
\T_{\nu\backslash\lS/\mu}(u) =\det_{1\leq i,j\leq r} \T_{r-\nu_{i}^\prime+i-j-\mu_{r+1-j}^\prime}(u+\hbar(r-j-\mu_{r+1-j}^\prime))\,.
\end{equation}
Finally applying the $*$ map we obtain 
\begin{equation}
\T_{\nu\backslash\lS/\mu}^*(u) = \det_{1\leq i,j\leq r} \T_{r-\nu_{i}^\prime+i-j-\mu_{r+j-i}^\prime}(u+\hbar(i-1-\nu_i^\prime)) = \T_{\mu\backslash\lS/\nu}(u)
\end{equation}
and hence identify 
\begin{equation}
(\nu\backslash\lS/\mu)^* = (\mu\backslash\lS/\nu)\,.
\end{equation}
Importantly, it can be checked as a consequence of the boundary conditions \eqref{bdyconds} that the skew diagram $(\lambda/\mu)^*$ is actually independent of the size $r$ of the square $\lS$ as long as $\lS$ is large enough to contain $\lambda$ and hence this construction is well-defined. Graphically it is clear that applying $*$ corresponds to flipping $\lambda/\mu$ upside down and backwards and aligning with the bottom right corner of $\lS$.

\begin{figure}[htb]
\centering
  \includegraphics[width=100mm,scale=25]{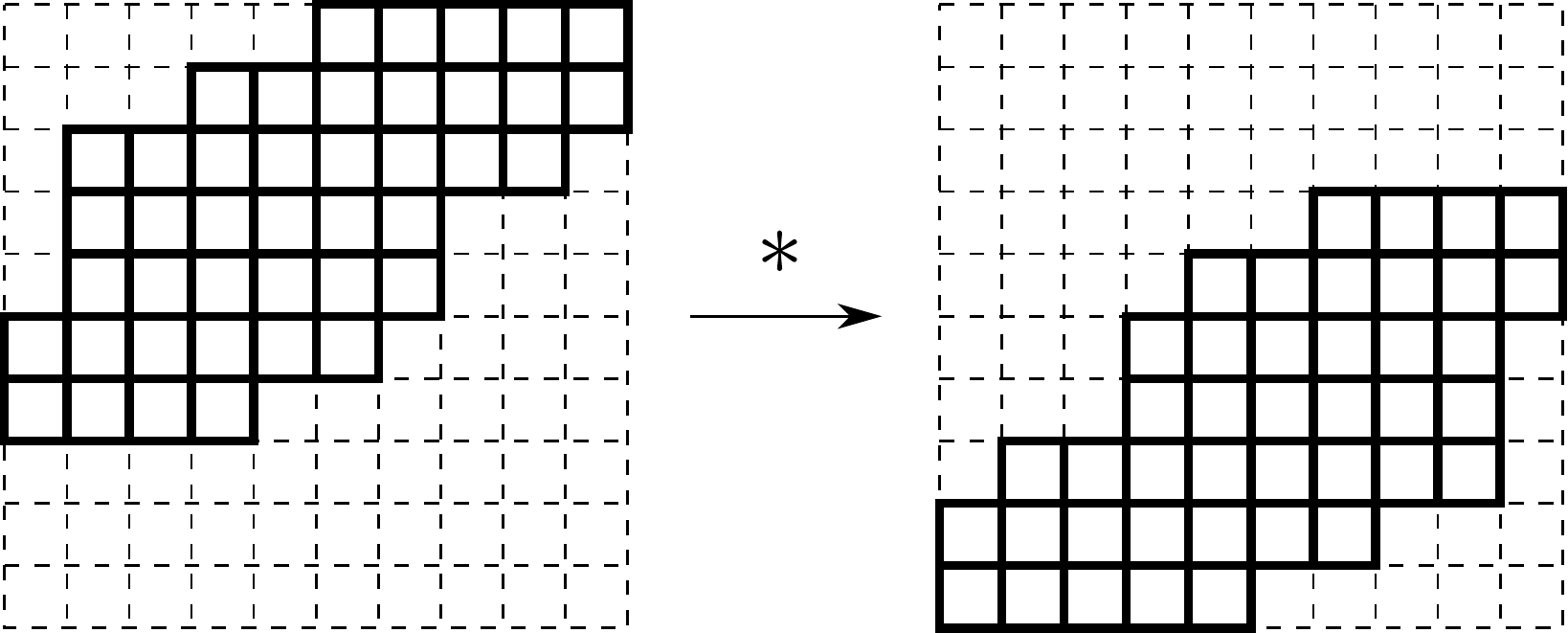}
  \caption{Action of the $*$-map on a skew diagram $\lambda/\mu$ (bold) on the left embedded into the square $\lS$ of size $10\times 10$ (dotted lines). The $*$ operation flips $\lambda/\mu$ upside down and backwards and aligns it with the bottom right corner of $\lS$.}
  \label{parityskew}
\end{figure}

\medskip

\begin{figure}[htb]
\centering
  \includegraphics[width=100mm,scale=25]{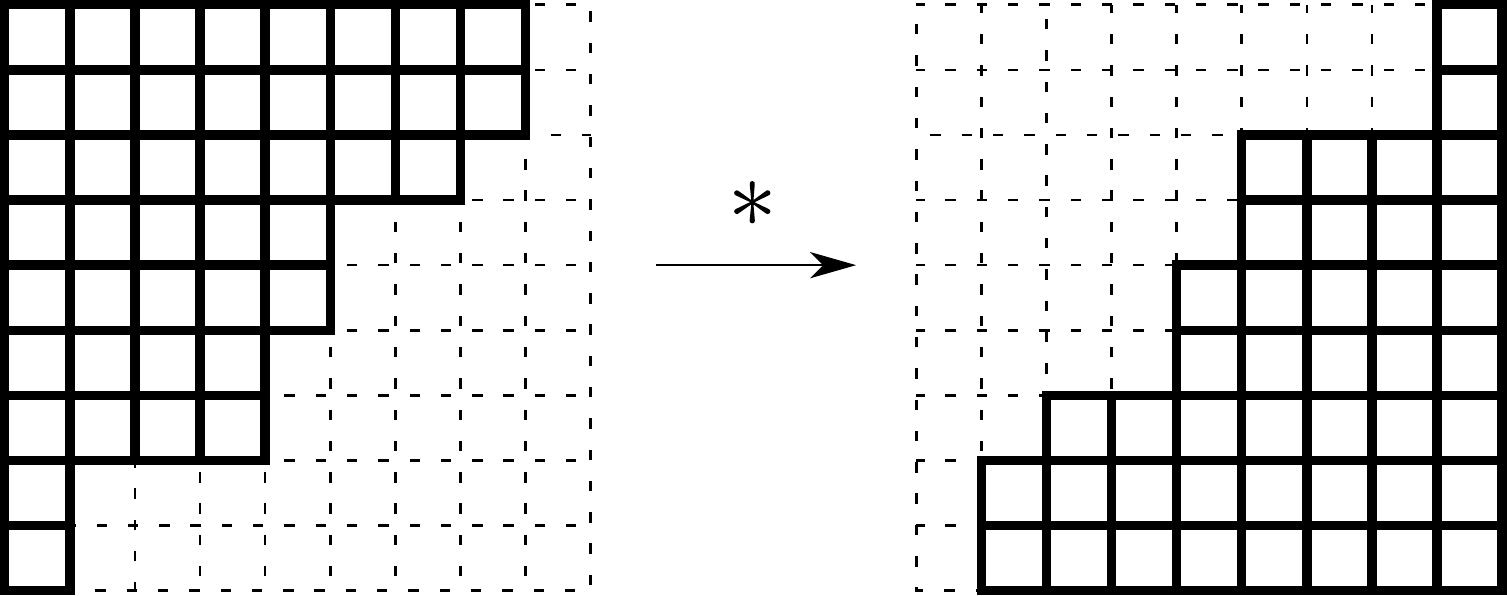}
  \caption{Action of the $*$-map on a non-skew Young diagram.}
  \label{ParityYD}
\end{figure}

A special case of this formula concerns the case when $\mu=\es$. In this case the resulting skew diagram $\lambda^*$ is determined as in Figure \ref{ParityYD}.
Let us point out that this special case was previously obtained in the $\lY_\gn$ case in the paper \cite{Gromov:2020fwh} of the author albeit in the framework of the analytical Bethe ansatz. Here we have presented a derivation for any skew diagram which only requires the CBR formula and hence is true as a statement in $\lY_{\gn}$ and not of any particular representation. 

\medskip

\begin{figure}[htb]
\centering
  \includegraphics[width=100mm,scale=25]{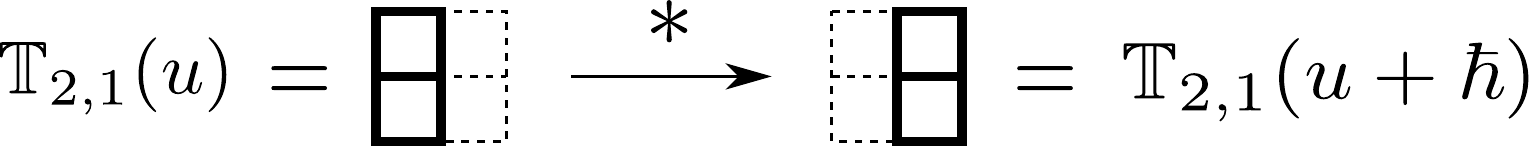}
  \caption{Flipping results in the Young diagram of $\T_{2,1}$ being moved to the right which results in an overall shift of $\hbar$.}
  \label{inversionshifts}
\end{figure}

We should also point out that the flipping procedure also implicitly includes shifts of the spectral parameter. This is done in the following way. To each box $(a,s)$ of the square $\lS$ we associate a value $c_{a,s}=s-a$ (which is also how shifts are associated in the fusion procedure). When a Young diagram is embedded in $\lS$ each of its boxes naturally attains a value $c_{a,s}$ for each box the Young diagram and $\lS$ share in common. Hence, moving the Young diagram around in $\lS$ corresponds to an overall shift in the transfer matrix, as the transfer matrix is defined with left corners aligned. For example, we consider the case of $\T_{2,1}$ and following the flipping procedure the result is $\T_{2,1}(u+\hbar)$, see Figure \ref{inversionshifts}. 

We have presented an in-depth analysis of the transfer matrices $\T_\lambda^*$ which appear in the commutation relation with $\bC$. From a physical point of view it may help to think of them like this. In the classical limit $\hbar\rightarrow 0$ $\bB$ and $\bC$ define the same set of classical set of variables and $\T_\lambda$ and $\T_\lambda^*(u)$ define the same set of integrals of motion. Hence, in the same way that $\bB$ and $\bC$ can both be viewed as alternate quantisations of the classical separated variables $\T_\lambda$ and $\T_\lambda^*$ can be viewed as alternate quantisations of the classical integrals of motion, and this choice of quantisation does not effect their mutual commutativity. 

\paragraph{Physical interpretation of the $*$-map}

Having understood how transfer matrices transform under the $*$-map we can provide a physical interpretation of this map. Since the transfer matrices $\T_\lambda(u)$ are simply mapped to other transfer matrices and the separated variables generated by $\bB$ are mapped to another set of separated variables, generated by $\bC$, the $*$-map has the interpretation of mapping between two distinct, but equivalent, quantisations of the classical spin chain. Indeed, in the classical limit shifts of $\hbar$ are ignored and so $\bB$ and $\bC$ coincide in this limit and in the same way $\T_\lambda(u)$ and $\T_\lambda^*(u)$ also coincide in this limit. 

\paragraph{Baxter equation and Q-operators}
We now have a good understanding of the transfer matrices $\T_\lambda^*$ but in order to write wave-functions in the SoV basis we need to know how to express them in terms of Baxter Q-operators. We know examine the effect of the $*$-operation on the operatorial Baxter equation 
\begin{equation}\label{1Bax1}
\left(\sum_{a=0}^\gn (-1)^a \T_{a,1}(u)\overrightarrow{\lD}^a \right)F(u)=\sum_{a=0}^\gn (-1)^a \T_{a,1}(u) F(u-\hbar\,a)=0
\end{equation}
where the generic solution $F(u)$ is a linear combination of Q-operators of the form 
\begin{equation}
F(u)=\sum_{j=1}^\gn c_j\,\frac{\QQ_j^{[2]}}{\QQ_{\es}}
\end{equation}
where $c_j$ are some constants. Recall also the dual Baxter equation
\begin{equation}
0 = G(u)\left(\sum_{a=0}^\gn (-1)^a\overleftarrow{\lD}^a\T_{a,1}(u+a\hbar)\right)=\sum_{a=0}^\gn (-1)^aG(u+a\hbar)\T_{a,1}(u+a\hbar)
\end{equation}
where the solution $G(u)$ is given by certain Hodge-dual Q-operators
\begin{equation}
G(u) =\sum_{j=1}^{\gn} c_j\,\frac{\QQ^j}{\left(\QQ^\es\right)^{[2]}}\,.
\end{equation}
We now apply the $*$-operation to the Baxter equation \eqref{1Bax1}, obtaining 
\begin{equation}
\sum_{a=0}^\gn (-1)^a \T_{a,1}(u+\hbar(a-1)) F^*(u+a\, \hbar)=0\,.
\end{equation}
Next, perform a shift of the spectral parameter $u\rightarrow u+\hbar$ to obtain 
\begin{equation}
\sum_{a=0}^\gn (-1)^a \T_{a,1}(u+a\,\hbar) F^*(u+(a+1)\,\hbar)=0\,.
\end{equation}
Hence, we clearly see that $F^*(u+\hbar)$ is a solution of the dual Baxter equation. By using the basis of solutions in terms of Q-operators we then can choose, without loss of generality (i.e. up to symmetries of the Q-system which preserve the Baxter equation)
\begin{equation}
\frac{\QQ_i^{[2]}}{\QQ_\es}\mapsto \frac{\left(\QQ^i\right)^{[-2]}}{\QQ^\es}\,.
\end{equation}
By carefully comparing signs in the QQ-relations this suggests that
\begin{equation}
\boxed{\QQ_A(u) \mapsto \QQ^{\overleftarrow{A}}(u)}
\end{equation}
where $\overleftarrow{A}$ denotes the reversal of the indices constituting the set $A$. For example, $\QQ_{123}(u) \mapsto \QQ^{321}(u)$. 

\medskip

Let us note that $\QQ_i \mapsto \QQ^i$ is only one possible transformation consistent with the Baxter equation. In general we can have an extra $H$-transformation
\begin{equation}
\QQ_i \mapsto \displaystyle \sum_{j=1}^\gn H_{ij}\, \QQ^j,\quad H\in \GL(\gn)\,.
\end{equation}
This extra $H$-symmetry drops out of transfer matrices which are our primary objects of interest so we will ignore it and simply put $H_{ij}=\delta_{ij}$. 

\paragraph{Transfer matrices}
Now that we understand how Q-operators transform we can write wronskian expressions for transfer matrices $\T_\lambda^*$ in terms of $\QQ^i$. Starting from 
\begin{equation}
\T_\lambda(u)=\frac{\QQ_\es^{[-2\gn]}}{\QQ_\fs}\displaystyle\det_{1\leq i,j\leq \gn}\left(\frac{\QQ_i}{Q_\es^{[-2]}} \right)^{[2\hat{\lambda}_j]}
\end{equation}
it immediately follows that
\begin{equation}
\T_\lambda^*(u)=(-1)^{\frac{\gn}{2}(\gn-1)}\frac{\left(\QQ^\es\right)^{[2\gn]}}{\QQ^\fs}\displaystyle\det_{1\leq i,j\leq \gn}\left(\frac{\QQ^i}{\left(Q^\es\right)^{[2]}} \right)^{[-2\hat{\lambda}_j]}
\end{equation}
where the overall sign has arisen from using $\QQ^{\gn\dots 21}=(-1)^{\frac{\gn}{2}(\gn-1)} \QQ^{12\dots\gn} = (-1)^{\frac{\gn}{2}(\gn-1)} \QQ^{\fs}$.

\paragraph{Quantum eigenvalues}
Next we examine how quantum eigenvalues transform 
\begin{equation}
\Lambda_k(u)  = \frac{\QQ_{\leftarrow k-1}^{[-2]}}{\QQ_{\leftarrow k-1}}\frac{\QQ_{\leftarrow k}^{[2]}}{\QQ_{\leftarrow k}} \mapsto\frac{\left(\QQ^{\leftarrow k-1}\right)^{[2]}}{\QQ^{\leftarrow k-1}}\frac{\left(\QQ^{\leftarrow k}\right)^{[-2]}}{\QQ^{\leftarrow k}}
\end{equation}
Upon resolving $\Gamma$-functions we see that the \rhs has a factor $\nu_{\gn-k+1}(u)$ and so it is natural to denote this image as 
\begin{equation}
\Lambda_k(u) \mapsto \Lambda^{\gn-k+1}(u):=\frac{\left(\QQ^{\leftarrow k-1}\right)^{[2]}}{\QQ^{\leftarrow k-1}}\frac{\left(\QQ^{\leftarrow k}\right)^{[-2]}}{\QQ^{\leftarrow k}}=\nu_{\gn-k+1}(u) \frac{\left(\sfq^{\leftarrow k-1}\right)^{[2]}}{\sfq^{\leftarrow k-1}}\frac{\left(\sfq^{\leftarrow k}\right)^{[-2]}}{\sfq^{\leftarrow k}}\,.
\end{equation}
We can now easily write down a recipe for expressing transfer matrices $\T_\lambda^*$ in terms of the new quantum eigenvalues $\Lambda^k$. Previously, we had 
\begin{equation}\label{transfertableaux}
\T_\lambda(u) = \displaystyle\sum_{\lT}\prod_{(a,s)\in \lambda} \Lambda_{\#(a,s)}(u+\hbar(s-a))
\end{equation} 
where $\#(a,s)$ denotes the number in box $(a,s)$ of the semi-standard Young tableaux $\lT$ with the rule that the content of the tableaux is such that numbers strictly decrease in columns and weakly decrease in rows. We see that we can immediately write 
\begin{equation}\label{startableaux}
\T^*_\lambda(u) = \displaystyle\sum_{\lT}\prod_{(a,s)\in \lambda} \Lambda^{\#(a,s)}(u+\hbar(a-s))
\end{equation}
but now with the rule that the content of the tableaux is such that numbers strictly increase in columns and weakly increase in rows (which corresponds to how semi-standard Young tableaux are usually defined).

\paragraph{B\"acklund Flow}
B\"acklund transformed transfer matrices $\T^{(k)}_\lambda$ were constructed as in \eqref{transfertableaux} with the requirement that we only filled the tableaux of shape $\lambda$ with numbers from $\{1,2,\dots,k\}$. A natural extension of this is to define a new set of B\"acklund transformed transfer matrices $\T_\lambda^{*\,(k)}$ as in \eqref{startableaux} but now with the property that we only fill tableaux with the numbers $\{\gn-k+1,\dots,\gn\}$. These transfer matrices admit the Wronskian expression
\begin{equation}
\T_\lambda^{*\,(k)} = (-1)^{\frac{k}{2}(k-1)}\frac{\left(\QQ^\es\right)^{[2k]}}{\QQ^{1\dots k}}\det_{1\leq i,j\leq k} \left(\frac{\QQ^i}{\left(\QQ^\es \right)^{[2]}}\right)^{[-2\hat{\lambda}_j]}
\end{equation}
and satisfy the property 
\begin{equation}
\T_\lambda^{*\,(k)}(\theta_\alpha+\hbar\, \nu^\alpha_k) = \T_\lambda^{*\,(k-1)}(\theta_\alpha+\hbar\, \nu^\alpha_k)
\end{equation}
similar to \eqref{bcklndprp}.

\subsection{Diagonalising $\bC$ and dual wave functions}
The procedure for diagonalising $\bC$ is totally analogous to what was done for $\bB$. As such we will not derive all details but simply sketch the main results. 

\medskip

We start by introducing a dual embedding morphism $\bar{\phi}:\lH_k \rightarrow \lH_{k+1}$ with $\bar{\phi}(T_{ij})=T_{1+i,1+j}$ defined by the property
\begin{equation}
\bar{\phi}: \mathcal{J} \ket{\bar{0}_k} \rightarrow \bar{\phi}\left(\mathcal{J} \right)\ket{\bar{0}_k}
\end{equation}
where $\mathcal{J}$ is an element of $\lY_k$ and $\ket{\bar{0}_k}$ is the highest-weight state of the $\gla(k)$ spin chain. Introduce subspaces $\bar{\lV}_{(k)}:=\bar{\phi}(\lH_k)$. In exactly the same way as with $\bB$ it is possible to show that 
\begin{equation}
\bC^{(k+1)}\ \sim \ \bar{\phi}\left(\bC^{(k)} \right)|_{\bar{\lV}_{(k)}}
\end{equation}
and hence $\bC$ can be diagonalised by repeatedly acting with lower-rank transfer matrices embedded into $\lY_{\gn}$ as was done with $\bB$. A generic $\bC$ eigenvector $\ket{\Lambda^{\bC}}$ can be constructed as 
\begin{equation}
\ket{\Lambda^{\bC}} = \displaystyle \prod_{\alpha=1}^L \prod_{k=1}^{\gn-1} \bar{\phi}^{\gn-k-1}\left(\T^*_{\bar{\mu}^\alpha_k}(\theta_\alpha+\hbar\,\nu^\alpha_{\gn-k} \right)\ket{\bar{0}}
\end{equation}
for some Young diagrams $\bar{\mu}_k^\alpha$. 

\medskip

Next, we can use the Backlund flow transfer matrices $\T^{*\, (k)}_\lambda$ to rewrite the constructed eigenvectors as 
\begin{equation}
\ket{\Lambda^{\bC}} = \displaystyle \prod_{\alpha=1}^L \prod_{k=1}^{\gn-1} \T^{*\,(k)}_{\bar{\mu}^\alpha_k}(\theta_\alpha+\hbar\,\nu^\alpha_{\gn-k}) \ket{\bar{0}}
\end{equation}
which immediately implies separation of variables for left Bethe algebra eigenstates $\bra{\Psi}$. By choosing to normalise 
\begin{equation}
\braket{\Psi|\bar{0}}=\prod_{\alpha=1}^L \prod_{k=1}^{\gn-1} \hat{\sfq}^{12\dots k}(\theta_\alpha+\hbar\,\nu^\alpha_{\gn-k})
\end{equation}
and appropriately rescaling $\ket{\Lambda^{\bC}}\rightarrow \ket{\svy}$ we obtain
\begin{equation}\label{Ywavefns}
\boxed{\Psi(\svy)=\braket{\Psi|\svy} = \prod_{\alpha=1}^L \prod_{k=1}^{\gn-1}\det_{1\leq i,j \leq k}\sfq^i(\svy^\alpha_{kj})}
\end{equation}
where $\svy^\alpha_{kj}= \theta_\alpha + \hbar(\nu^\alpha_{\gn-k}-\bar{\mu}^\alpha_{kj}+j-1)$. Note that the labelling for dual GT patterns is different. Traditionally, GT patterns are labelled as 
\begin{equation}
\begin{array}{ccccccccccc}
\nu_{1}^\alpha & \ & \nu_{2}^\alpha & \ & \dots & \ & \nu_{\gn}^\alpha \\
\ & \lambda_{\gn-1,1}^\alpha & \ & \dots & \ & \lambda_{\gn-1,\gn-1}^\alpha \\
\ & \ & \dots & \ & \dots \\
\ & \ & \lambda_{21}^\alpha & \ & \lambda_{22}^\alpha \\
\ & \ & \ & \lambda_{11}^\alpha
\end{array}\,.
\end{equation}
When dealing with $\bra{\svx}$ we introduced a labelling of dual diagonals with $\mu_{kj}= \lambda_{\gn-k+j-1,j}$. For $\ket{\svy}$ we introduce a natural labelling for the \textit{main diagonals} $\mu^\alpha_{kj} = \lambda_{\gn+j-k-1,\gn-k}$ For example, in these coordinates a $\gl(4)$ GT pattern looks like
\begin{equation}\label{mainmupattern}
\begin{array}{ccccccccccc}
\nu_{1}^\alpha & \ & \nu_{2}^\alpha & \ & \nu_{3}^\alpha & \ & \nu_{4}^\alpha \\
\ & \mu_{33}^\alpha & \ & \mu_{22}^\alpha & \ & \mu_{11}^\alpha \\
\ & \ & \mu_{32}^\alpha & \ & \mu_{21}^\alpha \\
\ & \ & \ & \mu_{31}^\alpha
\end{array}\,.
\end{equation}
For $\bra{\svx}$ the parameter $\bar{\mu}^\alpha_{kj}$ labelled how much a given node was excited above its minimum value, whereas here it labels how much a node has been decreased from its maximum value. Furthermore, transfer matrices now act by decreasing the value of nodes along the main diagonal instead of increasing the values along the dual diagonals. For example, if we consider $\gl(3)$, $L=1$ a generic GT pattern is labelled as 
\begin{equation}
\begin{array}{cccccc}
\nu_1 & \ & \nu_2 & \ & \nu_3 \\
\ & \mu_{22} & \ & \mu_{11} \\
\ & \ & \mu_{21} \\
\end{array}
\end{equation}
where $\bar{\mu}_{kj}= \nu_{\gn-k}-\mu_{kj}$. Let's now specialise to the representation $\nu=[3,0,0]$. $\ket{\bar{0}}$ corresponds to the pattern
\begin{equation}
\begin{array}{cccccc}
3 & \ & 0 & \ &  0 \\
\ & 3 & \ & 0 \\
\ & \ & 3 \\
\end{array}\,.
\end{equation}
By acting with the transfer matrix $\T^*_\lambda(\theta_\alpha+\hbar\,\nu_1)=\T^{*\,(2)}_\lambda(\theta_\alpha+\hbar\,\nu_1)$ with $\lambda=[3,1,0]$ we obtain the GT pattern
\begin{equation}
\begin{array}{cccccc}
3 & \ & 0 & \ &  0 \\
\ & 2 & \ & 0 \\
\ & \ & 0 \\
\end{array}\,.
\end{equation}

\paragraph{SoV measure}

We have succeeded in obtaining highly compact wave functions for the transfer matrix eigenstates. However, for these to actual be useful we need to know the measure $\lM$ in the SoV basis defined as the inverse of the matrix of overlaps $\braket{\svx|\svy}$
\begin{equation}
\lM= \left(\braket{\svx|\svy} \right)^{-1}_{\svx,\svy}\,.
\end{equation}
Computing this directly using the definition of the SoV bases in terms of transfer matrices is a formidable task which was done for $\gl(3)$ in the defining representation in \cite{Maillet:2020ykb}. Part of the trouble comes from the fact that the measure is non-diagonal meaning for a given $\bra{\svx}$ there can be more than one $\ket{\svy}$ such that $\braket{\svx|\svy}$ is non-zero. This is in contrast to the $\gl(2)$ case where the measure is indeed diagonal. This computation can however be avoided if we use the functional integral approach which completely determines the measure. For simplicity we will also consider the defining representation, although the argument goes through in general. For the defining representation the space has dimension $\gn^L$ and hence the matrix of overlaps is size $\gn^L\times \gn^L$. On the other hand some overlaps $\braket{\svx|\svy}$ are zero since $\bra{0}$ is an eigenstate of both $\bB$ and $\bC$ and so the following overlaps vanish:
\begin{equation}
\braket{0|\svy}\propto\delta_{0,\svy}
\end{equation}
where the constant of proportionality can be easily worked out. This reduces the number of unknowns in the measure to $\gn^L \times (\gn^L-1)$. As well as this we can consider the following transfer matrix eigenstate overlaps $\braket{\Psi^A|\Psi^B}\propto \delta^{AB}$ with $A\neq B$. This can be expanded into a basis of Q-functions of the form precisely matching the SoV basis wave functions and there are precisely $\gn^L \times (\gn^L-1)$ possible pairs, matching the number of unknowns in the measure. Hence, assuming these equations are indeed all independent (which we have confirmed for low length) the measure can be fully reconstructed from the functional scalar product. 

\medskip

We would like to point out that in the spirit of \cite{Maillet:2018bim} the SoV basis is not unique and we are free to create it with any set of conserved charges we like. In \cite{Maillet:2020ykb} an alternate left and right SoV basis was constructed to fulfil the purpose of obtaining a \textit{diagonal} SoV measure for $\gl(3)$. This was achieved in the following way. Consider the case where the (diagonal) twist matrix $g$ has one zero eigenvalue but still has simple spectrum. Then for generic choice of vectors $\bra{L}$ and $\ket{R}$ the following form SoV bases with diagonal measure
\begin{equation}
\begin{split}
& \bra{L}\prod_{\alpha=1}^L \T_{1,1}(\theta_\alpha)^{\delta_{n_\alpha,2}}\T_{2,1}(\theta_\alpha+2\hbar)^{\delta_{n_\alpha,0}}\\ 
& \prod_{\alpha=1}^L \T_{1,1}(\theta_\alpha)^{\delta_{n_\alpha,2}}\T_{2,1}(\theta_\alpha+\hbar)^{\delta_{n_\alpha,1}}\ket{R},\quad n_\alpha=0,1,2\,.
\end{split}
\end{equation}
The only requirement in demonstrating that the measure is diagonal is the fact that as a result of having a zero twist eigenvalue the quantum determinant vanishes identically leading to the reduced fusion relations 
\begin{equation}\label{redfusion}
\begin{split}
& \T_{1,1}(\theta_\alpha)\T_{2,1}(\theta+2\hbar)=0=\T_{2,1}(\theta_\alpha+\hbar)\T_{2,1}(\theta+2\hbar)\\
& \T_{1,1}(\theta_\alpha)\T_{1,1}(\theta_\alpha+\hbar)=\T_{2,1}(\theta_\alpha+\hbar)\,.
\end{split}
\end{equation}
We then construct the following set of conserved charges for the case where the twist is invertible. Define
\begin{equation}
\tilde{T}_{a,1}(u) = \sum_{j=1}^{3^L} \tilde{t}^{(j)}_{a,1}\frac{\ket{t_j}\bra{t_j}}{\braket{t_j|t_j}}
\end{equation}
where $\tilde{t}_{a,1}^{(j)}$ are the eigenvalues of the transfer matrices with one-vanishing eigenvalue as described above and $\ket{t_j}$ are eigenvectors of the full transfer matrices with invertible twist. These transfer matrices commute with the usual transfer matrices and furthermore satisfy the reduced fusion relations \eqref{redfusion} meaning that one can use them to construct a diagonal measure. It would be very interesting if this measure could be extracted from some functional scalar product approach as we have done with our off-diagonal measure and this certainly deserves further investigation. 

\paragraph{Symmetric powers}
Before finishing this section let's consider representations of the form $[S,0,\dots,0]$. The resulting formulas will be useful in the next Part of this work when we consider a non-compact generalisation of such representations. For these representations only the first main diagonal of a given GT pattern is dynamical and as such we only need to act with a single transfer matrix. The eigenstates of $\bC$ are then constructed as 
\begin{equation}
\ket{\Lambda^{\rm \bC}} = \prod_{\alpha=1}^L \T_{\bar{\mu}^\alpha_1}(\theta_\alpha+\hbar\,S)\ket{\bar{0}}\,.
\end{equation}
As in the case with $\bra{\svx}$ wave functions we have the freedom to perform a permutation on the index $i$ in \eqref{Ywavefns}. It is convenient to perform $i\mapsto i+1\, {\rm mod}\,\gn$ which selects the Q-functions most useful in the non-compact case. Hence, by choosing to normalise $\bra{\Psi}$ with 
\begin{equation}
\braket{\Psi|\bar{0}} = \prod_{\alpha=1}^L \sfq^{23\dots \gn-1}(\theta_\alpha+\hbar\,S)
\end{equation}
and normalising $\ket{\svy}$ appropriately we obtain only a single determinant for each $\alpha$
\begin{equation}\label{Cwavefn}
\braket{\Psi|\svy} = \prod_{\alpha=1}^L \det_{1\leq i,j \leq \gn-1}\sfq^{i+1}(\svy^\alpha_{\gn-1,j})\,.
\end{equation}
\part{Non-compact spin chains}
\section{Representation theory}
The non-compact highest-weight case is not a trivial modification of the compact case. As we will see, new tools are required to be developed. Indeed, as we are now dealing with infinite-dimensional representations the counting of states is more subtle. As well as this, one of the main tools we used in the compact case, namely the GT basis, has not yet been developed for non-compact representations. Finally, when constructing the SoV bases for the compact case we constructed the right states from the highest weight and left states from lowest, and so we need to modify this procedure to create both states from highest since we do not have a lowest-weight state for non-compact representations.
\subsection{Representations}
It is well known that a generic highest-weight representation of $\gl(\gn)$ can be constructed in terms of first order differential operators acting on polynomials of $\frac{\gn}{2}(\gn-1)$ variables, see \cite{Derkachov:2006fw} for a general construction, also \cite{gel1950unitary,nla.cat-vn1878494}. We present a brief review of generic highest-weight representations with examples on $\gl(2)$ and $\gl(3)$. Note that highest-weight reps can also be constructed in a simple manner using oscillator algebras which are also a useful tool for classifying unitary representations \cite{Gunaydin:2017lhg}. 

\paragraph{$\gl(2)$}
A representation of $\gl(2)$ on polynomials $\CC[x]$ of one variable $x$ is given as follows, where we recall $\CC[x]$ is the algebra of polynomials in $x$ with complex coefficients. It is important to emphasise that while a given polynomial has finite degree the algebra $\CC[x]$ is infinite-dimensional and is spanned by $1,x,x^2,\dots$. The Cartan subalgebra is represented by
\begin{equation}\label{gl2diffops1}
\begin{split}
\sfE_{11}& =\lambda_1 - x\partial_x\\
\sfE_{22}& =\lambda_2 + x\partial_x\,.
\end{split}
\end{equation}
The raising operators are given by
\begin{equation}\label{gl2diffops2}
\begin{split}
\sfE_{12}& =\partial_x
\end{split}
\end{equation}
and act by lowering the polynomial degree in $x$ and annihilate the highest-weight state which is simply given by the constant polynomial $1$. Finally, the lowering operators are given by
\begin{equation}\label{gl2diffops3}
\begin{split}
\sfE_{21}& =(\lambda_1-\lambda_2)x -x^2 \partial_x
\end{split}
\end{equation}
and raise the degree in $x$. The highest-weight of the representation is $\lambda=[\lambda_1,\lambda_2]$.

\paragraph{$\gl(3)$}
A generic highest-weight representation of $\gl(3)$ can be constructed on the space of polynomials $\CC[x,y,z]$ in three variables $x,y,z$. The representation is completely fixed by the highest-weights $\lambda_1,\lambda_2,\lambda_3$ - the generators are then represented as differential operators in the variables $x,y,z$.

\medskip

The Cartan subalgebra is given by
\begin{equation} \label{gl3diff}
\begin{split}
\sfE_{11}& =\lambda_1 - x\partial_x -y \partial_y \\
\sfE_{22}& =\lambda_2 + x\partial_x -z \partial_z\\
\sfE_{33}& =\lambda_3 + y\partial_y +z \partial_z\,,
\end{split}
\end{equation}
the raising operators are
\begin{equation}
\begin{split}
\sfE_{12}& =\partial_x \\
\sfE_{13}& =\partial_y \\
\sfE_{23}& =x\partial_y - \partial_z
\end{split}
\end{equation}
and the lowering operators are
\begin{equation}
\begin{split}
\sfE_{21}& =(\lambda_1-\lambda_2)x -x^2 \partial_x -xy\partial_y +(y+xz)\partial_z\\
\sfE_{31}& =y(\lambda_1-\lambda_3)+xz(\lambda_2-\lambda_3)-yx\partial_x-y^2\partial_y-z(y+xz)\partial_z\\
\sfE_{32}& =-z(\lambda_2-\lambda_3)+y\partial_x+z^2 \partial_z\,.
\end{split}
\end{equation}
The highest-weight is $\lambda=[\lambda_1,\lambda_2,\lambda_3]$. Unlike in the case of compact representations there is no restriction on the highest-weights and they are free to take any value. For generic values the representation is infinite-dimensional and irreducible. When we consider $\lambda_j-\lambda_{j+1}$ a non-negative integer the representation is still infinite-dimensional (since the polynomial ring is infinite dimensional) but is now reducible -- the representation now contains an invariant subspace (which contains the highest-weight state) and defines the finite-dim irrep with highest weight $\lambda$. 

\subsection{Non-compact Gelfand-Tsetlin patterns }

For compact representations we made extensive use of the fact that the spectrum of separated variables coincided with the spectrum of the Gelfand-Tsetlin algebra. If we wish to develop SoV for non-compact reps an obvious starting point is to construct a Gelfand-Tsetlin basis and analyse the spectrum of the GT algebra. We will restrict our attention to a length $1$ spin chain in an evaluation representation corresponding to a generic highest-weight representation. 

\paragraph{$\gl(2)$}

In this case the Gelfand-Tsetlin algebra is generated by $\sfE_{11}$ whose eigenfunctions $h_\Lambda$ with eigenvalue $\lambda_{11}$ are trivially worked out to be 
\begin{equation}
h_\Lambda(x) = x^{\lambda_1 - \lambda_{11}}
\end{equation}
where $\Lambda$ denotes a GT pattern, where for the moment there are no branching rules and hence no restriction on $\lambda_{11}$. In order to be in our representation space we need that $h_{\Lambda}(x)$ is a polynomial. Clearly, the only way that this can be a polynomial is that $\lambda_1-\lambda_{11}$ be a non-negative integer and hence $\lambda_{11}$ should have the form 
\begin{equation}
\lambda_{11}=\lambda-n,\quad n\in\{0,1,2,\dots\}
\end{equation}
where $n$ is free to take arbitrarily large values. 

\paragraph{$\gl(3)$}
In this case the GT generators are $\sfE_{11}$ again along with $\sfE_{22}$ and $\sfE_{11}\sfE_{22}-\sfE_{12}\sfE_{21}$. We let $h_\Lambda(x,y,z)$ denote their joint eigenfunctions and have
\begin{equation}
\begin{split}
& \sfE_{11}h_{\Lambda} = \lambda_{11} h_{\Lambda} \\
& \left(\sfE_{11}+\sfE_{22}\right)h_{\Lambda} = (\lambda_{21}+\lambda_{22}) h_{\Lambda} \\
& \left(\sfE_{11}\sfE_{22}-\sfE_{12}\sfE_{21}+\sfE_{11}\right)h_{\Lambda} = \lambda_{22}(\lambda_{21}+1) h_{\Lambda}\,.
\end{split}
\end{equation}
We can now solve the resulting differential equations. A joint eigenfunction of the Gelfand-Tsetlin algebra is then given by 
\begin{equation}
h_\Lambda(x,y,z)=x^{\lambda_{21}-\lambda_{11}}y^{\lambda_1-\lambda_{21}}z^{\lambda_2-\lambda_{22}}\ {}_2F_1\left(\lambda_{11}-\lambda_{21},\lambda_{22}-\lambda_2,\lambda_{22}-\lambda_{21},-\frac{y}{xz}\right)\,.
\end{equation}
The eigenfunction is the unique eigenfunction of the GT subalgebra corresponding to the GT pattern $\Lambda$ for a highest-weight representation with highest weight $[\lambda_1,\lambda_2,\lambda_3]$ and highest-weight state corresponding to the polynomial $1$. 

\medskip

A rather nice feature of the hypergeometric representation of the eigenfunction is that it makes the branching rules for any highest-weight representation easy to derive -- we simply need that the hypergeometric function to be a polynomial for any physical (in the representation space) state and we will now work them out. 

\paragraph{Hypergeometric function ${}_2F_1$}
We will briefly recall some features of the hypergeometric function ${}_2F_1(a,b;c;z)$, see \cite{andrews1999special}. ${}_2F_1$ is defined, for $|z|<1$, by
\begin{equation}\label{Hypergeometricfn}
{}_2F_1(a,b;c;z):=\displaystyle\sum_{n=0}^\infty \frac{(a)_n(b)_n}{(c)_n}\frac{z^n}{n!}
\end{equation}
where $(q)_n$ is the Pochhammer symbol defined by 
\begin{equation}
(q)_n = \prod_{k=0}^{n-1}(q-k)\,.
\end{equation}
The most crucial property of ${}_2F_1$ for us is that if either $a$ or $b$ is a non-positive integer $-m$ then the infinite series expansion in \eqref{Hypergeometricfn} terminates, resulting in a polynomial 
\begin{equation}
\begin{split}
{}_2F_1(-m,b;c;z)& =\displaystyle\sum_{n=0}^m(-1)^n \left(\begin{array}{c}
m \\
n
\end{array} \right) \frac{(b)_n}{(c)_n}z^n \\
{}_2F_1(a,-m;c;z)& =\displaystyle\sum_{n=0}^m(-1)^n \left(\begin{array}{c}
m \\
n
\end{array} \right) \frac{(a)_n}{(c)_n}z^n\,.
\end{split}
\end{equation} 
It immediately follows that if \textit{both} $a$ and $b$ are non-positive integers, $a=-m_1, b=-m_2$ then 
\begin{equation}
{}_2F_1(-m_1,-m_2;c;z)= \sum_{n=0}^m \#_n z^n,\quad m:=\min\left(m_1,m_2\right)
\end{equation}
Since the prefactor of ${}_2F_1$ must be polynomial we immediately get the constraints 
\begin{equation}\label{noncompact}
\begin{split}
\lambda_1-\lambda_{21}& \in \ZZ_{\geq 0}\\
\lambda_{21}-\lambda_{11}& \in \ZZ_{\geq 0}\\
\lambda_2-\lambda_{22}& \in \ZZ_{\geq 0}\\
\end{split}
\end{equation}
and it is easy to check that these are necessary and sufficient conditions. These branching rules should be compared with those of the compact case \eqref{compactbranch} which are naturally more restrictive. Note that similar formulae appeared in \cite{Dobrev:1996rv} where Gelfand-Tsetlin eigenvectors for the quantum algebra $U_q(\sla(3))$ were constructed in terms of a $q$-deformed analogue of ${}_2F_1$. 

\paragraph{Counting of states}
We have successfully manage to construct a family of eigenvectors of the Gelfand-Tsetlin algebra. We now need to ask if we actually form a basis in the infinite-dimensional space. Unlike the finite-dimensional case the counting of states is more complicated in the non-compact case. Certainly all of the states we have constructed are non-zero and are linearly independent since they correspond to different eigenvalues of the GT algebra. It is not clear however if every vector in our space can be written as a finite linear combination of GT eigenvectors. Note that we use the word ``basis" in the algebraic sense where only finite linear combinations are allowed. Later we will also need a basis in the analysis sense, meaning every vector can be written as a convergent infinite series and it is important to distinguish between the two notions.

\medskip

The GT eigenvectors we have constructed do indeed form a (algebraic) basis of the representation space. We can decompose the representation space into weight subspaces of the Cartan generators $\sfE_{11}$ and $\sfE_{22}$ 
\begin{equation}
\lV^\lambda =\bigoplus_{\mu}\lV^\lambda_{\mu_1,\mu_2}
\end{equation}
where $\lV^\lambda_{\mu_1,\mu_2}$ is the joint eigenspace of $\sfE_{11}$ and $\sfE_{22}$ corresponding to the eigenvalue $\mu_1$ and $\mu_2$ respectively. For a monomial $x^{n_1}y^{n_2}z^{n_3}$ the $\sfE_{11}$ eigenvalue is $\lambda_1-n_1-n_2$ and the $\sfE_{22}$ eigenvalue is $\lambda_2+n_1-n_3$ and hence the corresponding eigenspace is clearly finite dimensional since $n_1,n_2,n_3\geq 0$. Hence, in order to prove that we have enough GT eigenstates to form a basis we simply need to check that we have enough to span each of the subspaces $\lV^\lambda_{\mu_1,\mu_2}$ since every element of $\CC[x,y,z]$ has a non-zero projection onto only finitely many $\lV^\lambda_{\mu_1,\mu_2}$ and a straightforward counting exercise verifies that this is the case. 

\paragraph{Completion of $\gl(\gn)$ representations to $\GL(\gn)$}

So far in this work when dealing with representations spaces we have interchangeably considered them as representations of both a Lie algebra and a Lie group -- Lie algebra representations were used to construct representations of the Yangian and Lie group representations were needed to rotate the transfer matrix between diagonal and companion twist frames. However, now that we are considering infinite-dimensional representations we must be more careful.

\medskip

When we consider representations of $\GL(\gn)$ we encounter objects such as (consider $\GL(2)$ for example)
\begin{equation}\label{grpelement}
\exp\left(t\, \sfE_{21} \right),\quad t\in \CC\,.
\end{equation}
Suppose we consider a representation with $\lambda_1-\lambda_2=-2\bs$ and consider the action of \eqref{grpelement} on the highest-weight state $1$ in the space of polynomials. We find 
\begin{equation}
\exp\left(t\, \sfE_{21} \right).1 = (1+t\,x)^{-2\bs}\,.
\end{equation} 
When $-2\bs$ is a positive integer there are no problems - the resulting state is a polynomial. However, for generic values of $\bs$ the \rhs should be expanded into an infinite series. The representation space should then be understood as a completed space of polynomials, where such analytic functions which are regular at the origin are included. 

\section{Non-compact functional scalar product}
Having discussed the fundamentals of non-compact representations we turn to the generalisation of the functional scalar product developed in section \ref{functionalsp}. Our first task will be to understand the analytic structure of Q-functions which are no longer gauge-equivalent to polynomials. 

\subsection{Polynomial Q-functions}

In general for non-compact representations Q-functions are not polynomial but it is still possible to find at least a few, see \cite{Frassek:2017bfz} for their explicit construction from Baxter Q-operators. If one performs the nested Bethe ansatz around the highest-weight state then all Q-functions describing the excitations around this state will be polynomial by construction. The remaining Q-functions will not be polynomial however. Consider the example of $\gl(2)$, length $L=1$ and highest-weights $\lambda_1,\lambda_2$. Then by analysing the QQ-relations at large $u$ we find that if for an integer $M$ we have $\sfq_1(u)\sim u^M$ then 
\begin{equation}
\sfq_2(u)\sim u^{\lambda_1-\lambda_2-M}
\end{equation}
and hence can only be polynomial in the case where $\lambda_1-\lambda_2$ is a positive integer, i.e. the representation is finite-dimensional. 

\medskip

At higher-rank the situation is similar. We restrict ourselves to representations of the form $[S,0,\dots,0]$ where $S$ is free to take any value. By performing the Bethe ansatz we obtain (twisted) polynomial Q-functions $\hat{\sfq}_1,\hat{\sfq}_{12},\dots, \hat{\sfq}_{12\dots\gn-1}$. It can be checked as a consequence of the QQ-relations that the Hodge dual Q-functions $\hat{\sfq}^2,\hat{\sfq}^3,\dots, \hat{\sfq}^\gn$ are also polynomial. Consider for example $\gl(3)$. By construction $\hat{\sfq}_1$ and $\hat{\sfq}_{12} = \hat{\sfq}^3$ are polynomial and by the QQ-relations we have
\begin{equation}
\hat{\sfq}_1 \propto \hat{\sfq}^2 \hat{\sfq}^{3\, [2]}- \hat{\sfq}^3 \hat{\sfq}^{2\, [2]}
\end{equation}
which can be solved for polynomial $\hat{\sfq}^3$. The proportionality factor is a simple function of the twist eigenvalues. 

\medskip

Note that the set of polynomial Q-functions matches the structure of the SoV wave functions very closely -- we need precisely one Q-function to construct the right wave functions $\Psi(\svx)$ and $\gn-1$ dual Q-functions to construct the left wave functions $\Psi(\svy)$. A natural choice is then to choose the Q-functions entering the wave functions so that both wave-functions are (twisted) polynomial functions of the SoV coordinates $\svx$ and $\svy$. 

\subsection{Analytic requirements}
In order to simplify our analysis we will make certain assumptions. First, we choose all spin chain sites to be in the same representation $\nu_1^\alpha=-2\bs$, $\nu_2^\alpha=\dots=\nu^\alpha_\gn=0$ and perform a shift $\theta_\alpha\rightarrow \theta_\alpha+\sfi\,\bs$ and we assume $\bs>1$\footnote{In the publication \cite{Gromov:2020fwh} we used the condition $\bs>0$. This difference arises because in that paper symmetric conventions are used for shifts in the difference operators, whereas here we put all shifts in a given direction.}, $\theta_\alpha\in\RR$. Hence the highest-weight polynomials are given by 
\begin{equation}
\nu_1(u) = Q_\theta^{[2\bs]},\quad \nu_j(u)=Q_\theta^{[-2\bs]}, j\geq 2
\end{equation}
where we have introduced $Q_\theta(u) = \prod_{\alpha=1}^L (u-\theta_\alpha)$. It is most convenient to put $\hbar=\mathsf{i}=\sqrt{-1}$ \footnote{The paper \cite{Gromov:2020fwh} uses the convention $\hbar =-\sfi$.}. 

\medskip

We will start our considerations with the $\sla(2)$ case. Recall that the measure takes the general form 
\begin{equation}
\mu(u) = \frac{\Gamma[\nu_1(u)]}{\Gamma[\nu_2^{[2]}(u)]}\rho(u)
\end{equation}
which now, if $\rho$ were analytic, has an infinite number of poles located at 
\begin{equation}\label{measpoles}
\theta_\alpha -\sfi(\bs+m),\quad m\geq 0,\quad \alpha=1,\dots,L
\end{equation}
and an infinite number of zeroes at 
\begin{equation}\label{measzeroes}
\theta_\alpha+\sfi(\bs-n),\quad n\geq 1,\quad \alpha=1,\dots,L\,.
\end{equation}
In order to deal with this we will choose the contour of integration to be over the whole real line and perform the integral by residues by closing the contour in the lower-half plane. The periodic functions $\rho_\alpha$ must be chosen so that 1) the integral converges, 2) there are no unwanted contributions arising when we shift the contour.

\paragraph{Contour shifts}

Let's consider what happens when we try to shift the contour. We demand that 
\begin{equation}\label{conshift}
\displaystyle \int {\rm d}u\, K(u) f \lO g = \displaystyle \int {\rm d}u\, K(u) g \lO^\dagger f
\end{equation}
where we remind the reader that the finite difference operators $\lO$ and $\lO^\dagger$ are given by 
\begin{equation}
\begin{split}
& \lO = 1 -\frac{\sfT}{Q_\theta^{[2\bs]}}\lD^{-1} + \frac{Q_\theta^{[-2\bs]}}{Q_\theta^{[2\bs]}}\lD^{-2} \\
& \lO^\dagger = 1 -\frac{\sfT^{[2]}}{Q_\theta^{[-2(\bs-1)]}}\lD + \frac{Q_\theta^{[2(\bs+1)]}}{Q_\theta^{[-2(\bs-1)]}}\lD^{2}
\end{split}
\end{equation}
where $\sfT$ denotes the eigenvalue of the transfer matrix $\T_{1,1}$ and as before we put $\mu=K(u)\rho(u)/Q_\theta^{[2\bs]}$.

\medskip

The \lhs of \eqref{conshift} contains a term $\mu\, f\,\sfT\, g^{[-2]}$ in the integrand. In order to be able to shift the contour and produce the term $\mu\,g\,\sfT^{[2]}f^{[2]}$ on the \rhs we need that this term does not contain poles in the strip $0\leq {\rm Im}\,u\leq 1$. As a result of \eqref{measzeroes} for any $\bs>1$ we will always have a zero in this region for some $n\geq 1$ and hence we can allow for $\rho$ to have poles at $\theta_\alpha+\sfi(\bs+m)$, $m\in\ZZ$ and the single pole in the region will be cancelled by a zero allowing us to shift the contour. 

\medskip

The next term in the \lhs of \eqref{conshift} is $\mu\, f\, g^{[-4]} Q_\theta^{[-2\bs]}$ and we require that it does not have poles in the strip $0\leq {\rm Im}\,u \leq 2$. From the previous analysis we know things will work out in $0\leq {\rm Im}\,u \leq 1$ and so we need to consider $1\leq {\rm Im}\,u \leq 2$. Now we might have a problem since one of the previously allowed poles $\theta_\alpha+\sfi\,\bs$ could pose a problem, but thankfully this is cancelled by the factor $Q_\theta^{[-2\bs]}$. Hence, we are free to shift the contour.

\paragraph{Convergence}
We require that the integral
\begin{equation}
\displaystyle \int^{\infty}_{-\infty} {\rm d}u\,\mu_\alpha(u) f(u)g(u)\, u^k,\quad k=0,1,\dots,L-1
\end{equation}
converges where the test functions $f$ and $g$ are twisted polynomials behaving as 
\begin{equation}
f(u)\sim z_1^{-\sfi\,u} u^{t_f},\quad g(u)\sim z_1^{-\sfi\,u} u^{t_g}
\end{equation}
for some integers $t_f$ and $t_g$. Let $\phi_1 = {\rm arg}\,z_1$. We will assume for the moment that $\phi_1\in (0,\pi]$ (we will comment on $\phi_1\in(-\pi,0)$ later). The measure $\mu_\alpha(u)$ behaves as
\begin{equation}
\mu_\alpha(u)\sim u^{2\bs} \rho_\alpha^{\infty}(u)
\end{equation}
where $\rho^\infty_\alpha$ denotes the $u\rightarrow +\infty$ asymptotic of $\rho_\alpha(u)$, $\rho_\alpha(u)\sim \rho^\infty_\alpha(u)$.
Hence, we see 
\begin{equation}
\mu_\alpha(u)f(u)g(u)u^k \sim z_1^{-2\sfi\,u}u^{t} \rho_\alpha^{\infty}(u)
\end{equation}
where we have dropped the irrelevant constant factor and denoted by $t$ some irrelevant number whose value will not spoil convergence. Convergence of the integral then requires that $\rho_\alpha$ decays exponentially and faster than $z_1^{-2\sfi\,u}$. Similarly it should decay faster than $z_1^{2\sfi\,u}$ at $u\rightarrow -\infty$. As a result of these conditions we see that $z_1$ cannot be real and an imaginary part is required to ensure the required behaviour at infinity. 

\medskip

Combining the requirements of convergence and shifting the contour we see that an $\sfi$-periodic function with the required properties is 
\begin{equation}
\sum_{\alpha=1}^L \frac{C_\alpha}{1-e^{2\pi(u-\theta_\alpha-\sfi\, \bs)}}
\end{equation}
where $C_\alpha$ is a constant. We will then choose the $L$ independent functions 
\begin{equation}\label{noncommeasure}
\mu_\alpha(u) = \frac{\Gamma[Q_\theta^{2\bs]}]}{\Gamma[Q_\theta^{[-2(\bs-1)]}]}\frac{\rho_\alpha}{1-e^{2\pi(u-\theta_\alpha-\sfi\, \bs)}}
\end{equation}
as our measures as these are sufficient to reproduce the scalar product from the operatorial construction of the SoV wave functions which will be done in the next section. Note that if we were to choose $\phi_1\in(-\pi,0)$ we would simply need to change the sign in the exponent in the denominator of \eqref{noncommeasure} to ensure convergence. The poles of $\mu_\alpha$ in the lower half plane are at $\theta_\alpha-\sfi(\bs+n)$, $n\geq 0$ which as we will see precisely matches the spectrum of separated variables. In fact that it already be inferred by simply analytically continuing the results of the compact case. 

\medskip

By repeating the steps from the non-compact case we then obtain that for two different states $A$ and $B$ we have
\begin{equation} 
\delta^{AB} \propto \braket{\Psi^A|\Psi^B} = \frac{1}{\lN} \det_{1\leq \alpha,\beta\leq L} \langle \hat{\sfq}_1 u^{\beta-1} \hat{\sfq}_1\rangle_\alpha
\end{equation}
where $\lN$ is chosen so that the sum over residues is of the form 
\begin{equation}
\braket{\Psi^A|\Psi^B} = \prod_{\alpha=1}^L \hat{\sfq}_1(\theta_\alpha-\sfi\bs)^2 +\dots
\end{equation}
similar to what was done in the compact case. 

\paragraph{$\sla(3)$}

We now examine the $\sla(3)$ case, which is largely similar to the $\sua(3)$ case. Let us fix the measure $\mu_\alpha$ as in the $\sla(2)$ case and put
\begin{equation}
\mu_\alpha(u) = \frac{\Gamma[Q_\theta^{[2\bs]}]}{Q_\theta^{[-2(\bs-1)]}}\frac{1}{1-e^{2\pi(u-\theta_\alpha-\sfi\, \bs)}}\,.
\end{equation}
The adjointness property of the operator $\lO$ we seek then amounts to showing that we can write 
\begin{equation}
\begin{split}
& \displaystyle \int^\infty_{-\infty}{\rm d}u\,\mu_\alpha\, f\left(Q_\theta^{[2\bs]}-\sft_1 \lD^{-1}+
\frac{Q_\theta^{[-2\bs]}}{Q_\theta^{[2(\bs-1)]}}\sft_2\lD^{-2} -\frac{Q_\theta^{[-2\bs]}Q_\theta^{-2(\bs+1)}}{Q_\theta^{[2(\bs-1)]}}\lD^{-3}\right) g \\
& = .\displaystyle \int^\infty_{-\infty}{\rm d}u\,\mu_\alpha\,\frac{Q_\theta^{[2\bs]}}{Q_\theta^{-2(\bs-1)}} g\left(Q_\theta^{[-2(\bs-1)]}-\sft_1^{[2]}\lD+\sft_2^{[4]}\lD^2-Q_\theta^{[2(\bs+1)]}\lD^3\right)f
\end{split}
\end{equation}
where we have written $\sfT_{2,1}=Q_\theta^{[-2\bs]}\sft_2$ and $\sft_2$ is a polynomial of degree $L$. By our analysis in the $\sla(2)$ case we know there will be no problems with shifting the contour in the region $0\leq {\rm Im}\,u\leq 2$ so we just need to check the region $2\leq {\rm Im}\,u\leq 3$. As before we indeed have a pole in this region coming from $\rho_\alpha$ but it is cancelled by the factor $Q_\theta^{[2(\bs-1)]}$ and hence the adjointness property is guaranteed. 

\medskip

We turn our attention to integral convergence where now we require that 
\begin{equation}\label{sl3int}
\displaystyle \int^\infty_{-\infty}{\rm d}u\, \mu_\alpha(u)\,\hat{\sfq}_1 u^k \hat{\sfq}^{i},\ i=2,3
\end{equation}
converges where the Q-function asymptotics are now given by 
\begin{equation}\label{sl3twists}
\hat{\sfq}_1 \sim z_1^{-\sfi\, u} u^{t_1},\quad \hat{\sfq}^2 \sim (z_1 z_3)^{-\sfi\, u} u^{t_2},\quad\hat{\sfq}^3 \sim (z_2 z_3)^{-\sfi\, u} u^{t_3}
\end{equation}
for some integers $t_1,t_2,t_3$. Similarly to the $\sla(2)$ case we will assume for definiteness that
\begin{equation}
0 < {\rm arg}\,z_2 - {\rm arg}\, z_1<\pi,\quad 0 < {\rm arg}\,z_3 - {\rm arg}\, z_1<\pi\,.
\end{equation}
These conditions ensure that the integral in \eqref{sl3int} will be convergent for both choices of
$i=2,3$. Also, like for $\sla(2)$, if e.g. the first inequality in \eqref{sl3twists} is violated, we
should redefine $\mu_\alpha$ by flipping the sign in the exponent in the denominator of \eqref{noncommeasure}.

\medskip

Finally, the scalar product is then given by precisely the same formula as in the $\sua(3)$ case
\begin{equation}\label{sl3SP}
\delta^{AB}\propto\det_{(\alpha,i),(\beta,j)}\langle \hat{\sfq}^A_1\hat{\sfq}_B^{i+1\, [2(j-1)]} u^{\beta-1} \rangle_\alpha
\end{equation}
the only difference being the integration contour and measure. The generalisation to $\sla(\gn)$ is immediate. 

\paragraph{Analytic continuation in the spin $\bs$}

So far we restricted ourselves to the case $\bs>1$. When we consider the case $\bs\leq 1$ we need to be careful that we do not have unwanted contributions of poles when trying to shift the integration contour. The simplest way around this as follows. One can start by rewriting the integrals as a sum over poles and with appropriate overall normalisation (as we will later see) the coefficients of the Q-functions in this infinite series are polynomial functions of the spin $\bs$ and can be trivially analytically continued to $\bs\leq 1$ without spoiling any properties such as convergence or orthogonality of different Bethe states.

\paragraph{Restriction to finite-dimensional case}

We now consider the finite-dimensional case $\bs= -\frac{m}{2},\,m\in \ZZ_{\geq 0}$. It is not totally obvious that our integral expression reduces to what we had in the compact case. However, as we explained above we can rewrite the integrations as an infinite sum over poles. Upon restriction to the finite-dimensional case this infinite series will truncate leaving only finitely-many terms, and this finite series can then be re-expressed using contour integrals around finitely many poles as was done in the compact case. 

\section{Operatorial SoV}\label{operatornoncom}

We now turn to the question of constructing the SoV bases for infinite-dimensional representations. From the get-go we immediately run into two issues. 

\medskip

The first regards the explicit construction of states. Previously $\bC$ was diagonalised starting from $\ket{\bar{0}}$ while $\bB$ was diagonalised starting from $\bra{0}$. For an infinite-dimensional highest-weight representation we no longer have access to both of these vectors -- we only have $\ket{\bar{0}}$. On the other hand, $\bra{\bar{0}}$ is also an eigenvector of $\bB$ in the compact case. Hence, we need to modify our diagonalisation procedure to enable us to diagonalise $\bB$ starting from $\bra{\bar{0}}$. 

\medskip

The second issue involves the explicit counting of states. In the compact setting we made extensive use of the fact that in the auxiliary singular twist limit the SoV basis reduces to the Gelfand-Tsetlin basis. This was crucial for representations where the spectra of $\bB$ and $\bC$ are degenerate. Unfortunately, to our knowledge Gelfand-Tsetlin bases have not yet been constructed for generic highest-weight representations of the Yangian $\lY_{\gn}$ and so in order to be able to make accurate statements regarding whether or not a family of eigenvectors of $\bB$ and $\bC$ constitute a basis we must restrict ourselves to representations where they have non-degenerate spectra. After that, we need to perform a counting of states in some controlled manner. The family of representations we will consider are those with $\nu_1=-\bs$ and $\nu_2=\dots=\nu_{\gn}=\bs$ and hence the weight functions are given by 
\begin{equation}
\nu_1(u) = Q_\theta^{[2\bs]}(u),\quad \nu_k(u) = Q^{[-2\bs]}_\theta(u),\ k=2,\dots,\gn
\end{equation}
i.e the same class of representations discussed in the previous section for the functional orthogonality approach. This class of representations can be thought of as a non-compact analogue of the symmetric powers of the defining representation where the length of the single row of the Young diagram is free to take any desired value, including negative values. For this class of representations, for $\sla(3)$, the representation space, initially defined on $\CC[x,y,z]$, reduces to $\CC[x,y]$ since it is not possible to create $z$ excitations. Hence for practical purposes one can use the differential operator realisation \eqref{gl3diff} and send all $z$-terms to $0$. Similarly for $\sla(\gn)$ we reduce from polynomials in $\frac{\gn}{2}(\gn-1)$ variables to $\gn-1$ variables. As a result of this reduction $\bB$ and $\bC$ contain a number of overall trivial factors. It is convenient to introduce operators $\bb(u)$ and $\bc(u)$ which have all trivial factors stripped out 
\begin{equation}\label{smallbc}
\bB(u) = \bb(u)\, \prod_{a=2}^{\gn-1}\prod_{k=2}^a \nu_{\gn}^{[2(k-1)]},\quad \bC(u) = \bc(u) \, \prod_{a=2}^{\gn-1}\prod_{k=2}^a \nu_\gn^{[2(k-a)]}\,.
\end{equation}

\medskip

Before tackling the discussed issues let us construct the eigenvectors of $\bC$ in the non-compact case.

\subsection{Diagonalising $\bC$}

The procedure for diagonalising $\bC$ is essentially identical to the compact case when dealing with symmetric powers of the defining representation and the corresponding wave functions written down in \eqref{Cwavefn} are given by 
\begin{equation}
\braket{\Psi|\svy} = \prod_{\alpha=1}^L \det_{1\leq i,j \leq \gn-1}\sfq^{i+1}(\svy^\alpha_{\gn-1,j})\,.
\end{equation}
The eigenstates $\ket{\svy}$ are constructed up to normalisation as 
\begin{equation}
\ket{\svy} \sim \prod_{\alpha=1}^L \T_{\bar{\mu}^\alpha}^*(\theta_\alpha-\hbar \nu_1^\alpha)\ket{\bar{0}}\,.
\end{equation}
This construction continues to work perfectly in the non-compact case with $\nu_1^\alpha=-\bs$ since the commutation relation \eqref{CTcom} is unchanged the fact that $T_{j1}(\theta_\alpha-\sfi\,\bs)\ket{\bar{0}}=0$. The only difference between the compact and non-compact case is the restriction on the Young diagrams $\bar{\mu}^\alpha$ in order to ensure that the eigenvector is actually non-zero. The analysis presented in the compact case can be easily extended to the case when $\bs$ is generic -- for \text{any} Young diagram $\bar{\mu}^\alpha$ the resulting eigenvectors are non-zero, unlike in the compact case where these eigenvectors would only be non-zero for $\bar{\mu}^\alpha$ contained in the rectangle of size $\nu_1 \times \gn-1$. 

\medskip

In the next section we will demonstrate that these states indeed form a basis.

\subsection{Counting with the SoV charge operator}
For notational simplicity we denote by $d$, for fixed $\gn$ and $L$, the number
\begin{equation}
d= L(\gn-1)\,.
\end{equation}
Classically this is degrees of freedom of the spin chain, that is half of the dimension of the phase space of the corresponding classical $\sla(\gn)$ spin chain, for the reduced representation space on $\gn-1$ variables instead of $\frac{\gn}{2}(\gn-1)$. 

\medskip

We already saw that $\bB$ and $\bC$ do not commute and hence in general do not share eigenvectors except for a few special states. On the other hand, since $\bB$ and $\bC$ only differ by shifts of the spectral parameter $u$ they become related at large $u$ where shifts become inconsequential. Expanding $\bb$ at large $u$ we have 
\begin{equation}
\bb(u) = u^d-u^{d-1}\left((\gn-1)\sum_{\alpha=1}^L \theta_\alpha-2\hbar\, d\, \bs-\hbar\, \bN \right)+\mathcal{O}\left(u^{d-2} \right)\,.
\end{equation}
The operator $\bN$ is known as the SoV charge operator and it can also be obtained in the expansion of $\bc(u)$ in the same way. It satisfies the following very useful property
\begin{equation}
[\bb(u),\bN] = [\bc(u),\bN]=0
\end{equation}
and also commutes with all Gelfand-Tsetlin generators
\begin{equation}
[\bN,\GT_a(u)]=0,\ a=1,2,\dots,\gn-1
\end{equation}
and hence $\bN$ acts diagonally on any eigenvector of $\bB$, $\bC$ or the Gelfand-Tsetlin algebra. 

\medskip

We saw in the previous section that a generic eigenvector of $\bC$ can be constructed as 
\begin{equation}
\ket{\Lambda^{\bC}} = \prod_{\alpha=1}^L \T_{\bar{\mu}^\alpha}^*(\theta_\alpha+\hbar \lambda_1^\alpha)\ket{\bar{0}},\quad \lambda_1^\alpha = -\bs\,.
\end{equation}
By using the expression of the SoV charge operator extracted from $\bc(u)$, like we did above with $\bb(u)$, we see that for such a state we have
\begin{equation}
\bN \ket{\Lambda^{\bC}}=\left(\displaystyle\sum_{j=1}^{\gn-1}\sum_{\alpha=1}^L \bar{\mu}^\alpha_j\right)\ket{\Lambda^{\bC}}\
\end{equation}
and so the SoV charge operator counts the number of ``excitations" above the SoV vacuum state $\ket{\bar{0}}$. From this we see the following crucial property -- the eigenspaces of $\bN$ are finite-dimensional. Indeed, since each $\bar{\mu}^\alpha_j$ is a positive integer and the eigenvalue of $\bN$ is a sum of such terms there are only finitely many choices for $\bar{\mu}^\alpha_j$ to produce the same eigenvalue. Hence, in each subspace of fixed SoV charge we can express each vector as a finite linear combination of basis monomials which can then be expressed as a finite linear combination of $\bra{\Lambda^{\bC}}$ and hence the eigenvectors of $\bC$ (in the MCT frame) form a basis of the representation space of polynomials. 

\medskip

On the other hand, if we rotate back to the frame with diagonal twist then an eigenvector of $\bC$ (a polynomial) will be mapped to a convergent infinite series. In this case need to ask if we can write any vector $\ket{v}$ in this completed space as 
\begin{equation}
\ket{v}=\sum_{\Lambda}c_\Lambda \ket{\Lambda^{\bC}}
\end{equation}
where the sum is over all the eigenvectors $\ket{\Lambda^{\bC}}$ of $\bC$ and $c_\Lambda$ are some finite coefficients. This representation of $\ket{v}$ does indeed exist. By definition we have
\begin{equation}
\ket{v} = \lim_{k\rightarrow \infty} \sum_{r=0}^k c_{(r)} e_{(r)}
\end{equation}
where we sum over basis monomials $e_{(r)}$ of the polynomial space up to degree $r$. By choosing $k$ to be sufficiently large we can include all basis monomials up to a given SoV charge and no others, which can then be expressed in terms of the eigenvectors $\ket{\Lambda^{\bC}}$ of $\bC$. 

\medskip

To demonstrate the procedure, let us consider $\sla(3)$ length $L=1$. The basis monomials are $x$ and $y$ with $x$ contributing $1$ unit to SoV charge and $y$ 2 units. That is 
\begin{equation}
\bN\,x^{n_1} y^{n_2} = (n_1+2 n_2)x^{n_1}y^{n_2}\,.
\end{equation}
We can write any vector in the completed space as 
\begin{equation}
\ket{v} = \lim_{k\rightarrow \infty} \displaystyle \sum_{r_1=0}^{2k}\sum_{r_2=0}^k c_{r_1,r_2} x^{r_1} y^{r_2}
\end{equation}
and the finite sum $\displaystyle \sum_{r_1=0}^{2k}\sum_{r_2=0}^k c_{r_1,r_2} x^{r_1} y^{r_2}$ contains all terms up to an including SoV charge $2k$. We then rewrite this finite sum in terms of $\bC$ eigenvectors, completing the construction. This guarantees that any element of the completed space of polynomials can be expressed as an infinite linear combination of the SoV basis elements. That is, we can write, for any  vector $\ket{v}$,
\begin{equation}
\ket{v}=\sum_{n=0}^\infty \ket{\svy^n}
\end{equation}
where $\ket{\svy^n}$ is a \textit{finite} linear combination of right SoV basis states $\ket{\svy}$ with the property 
\begin{equation}
\bN\ket{\svy}=n\ket{\svy}\,.
\end{equation}

\subsection{Diagonalising $\bB$ and antipode}

We now explain how to diagonalise $\bB$ starting from $\bra{\bar{0}}$. We recall that in the compact version of the representations we are considering the eigenvectors of $\bB$ are constructed, for length $L=1$, as 
\begin{equation}
\bra{\Lambda^{\bB}} =\bra{0} \prod_{j=1}^{\gn-1}\T_{s_j,1}(\theta) \propto \bra{0} \prod_{j=1}^{\gn-1}\frac{\hat{\sfq}_1^{[2s_j]}(\theta)}{\hat{\sfq}_1(\theta)},\quad s_j\in\{0,1,\dots,S\}\,.
\end{equation}
The highest-weight state $\bra{\bar{0}}$ then corresponds to all $s_j=S$ and so 
\begin{equation}
\bra{\bar{0}}\propto \bra{0} \left(\T_{S,1}(\theta)\right)^{\gn-1}\propto \bra{\bar{0}}\left( \frac{\hat{\sfq}_1^{[2S]}(\theta)}{\hat{\sfq}_1(\theta)}\right)^{\gn-1}\,.
\end{equation}
We can now move back down the chain of eigenvectors by repeatedly acting with 
\begin{equation}
\frac{\hat{\sfq}_1^{[2(S-s_j)]}(\theta)}{\hat{\sfq}_1^{[2S]}(\theta)},\quad s_j\in\{0,1,\dots,S\}\,.
\end{equation}
Seemingly we can diagonalise $\bB$ starting from $\bra{\bar{0}}$ by acting with simple ratios of Q-operators. On the other hand to be able to make formal statements regarding the construction it is preferable to be able to diagonalise $\bB$ using transfer matrices instead of Q-operators since the former are usually easier to work with. We hence ask ourselves: does this ratio of Q-operators coincide with a transfer matrix evaluated at some particular point? The answer is yes and it coincides with the transfer matrix $\T_{\gn-1,s_j}(\theta+\hbar\,(S+\gn-s_j))$. This can be easily seen by using quantum eigenvalues. Consider $s_j=1$ and the expansion 
\begin{equation}
\T_{\gn-1,1}(u) = \sum_{1\leq i_1<\dots i_{\gn-1}\leq \gn} \Lambda_{i_{\gn-1}}\Lambda_{i_{\gn-2}}^{[-2]}\dots \Lambda_{i_{1}}^{[-2(\gn-1)]}.
\end{equation}
If $\Lambda_1$ appears in a term it can only be in the right-most position which carries a shift $\Lambda_1^{[-2(\gn-1)]}$ and so vanishes at $\theta+\hbar(S+\gn-1)$ due to the fact that 
\begin{equation}
\Lambda_1(u) = Q_\theta^{[-2S]} \frac{\sfq_1^{[2]}}{\sfq_1}\,.
\end{equation}
Hence 
\begin{equation}
\T_{\gn-1,1}(\theta+\hbar(S+\gn-1)) = \Lambda_{\gn}\Lambda^{[-2]}_{\gn-1}\dots \Lambda_{2}^{[-2(\gn-1)]}
\end{equation}
where all terms on the \rhs are evaluated $\theta+\hbar(S+\gn-1)$. By using the known expressions for quantum eigenvalues in terms of Q-functions \eqref{QEVq} we then obtain 
\begin{equation}
\T_{\gn-1,1}(\theta+\hbar(S+\gn-1)) \propto \frac{\hat{\sfq}_1^{[2(S-1)]}(\theta)}{\hat{\sfq}_1^{[2S]}(\theta)}\,.
\end{equation}

\medskip

We will now formalise this argument. The key tool we will use is the Yangian antipode map $S$ which is known to map the transfer matrix $\T_{1,1}$ to $\T_{\gn-1,1}$ \cite{molev2007yangians} and so seems like a natural starting point. The antipode $S$ is defined by 
\begin{equation}
S:\, T(u) \rightarrow T^{-1}(u)\,.
\end{equation}
Note that this is clearly compatible with twisting -- if $\bT(u)=T(u)G$ then we can extend $S$ to the twisted case by defining 
\begin{equation}
S:\,T(u)G \rightarrow G^{-1}T^{-1}(u)=\bT^{-1}(u)
\end{equation} 
and hence $S$ acts the same on both twisted and untwisted monodromy matrix elements. For notational simplicity we will denote $S(u)$ and $\bS(u)$ by 
\begin{equation}
S(u) = T^{-1}(u),\quad \bS(u) = \bT^{-1}(u)\,.
\end{equation}
We will need to perform fusion with the inverse monodromy matrix. $\bT$ satisfies the RTT relation 
\begin{equation}
R_{ab}(u,v)\bT_a(u)\bT_b(v) = \bT_b(v)\bT_a(u)R_{ab}(u,v)
\end{equation}
and hence $\bS$ satisfies 
\begin{equation}
\bar{R}_{ab}(u,v)\bS_a(u)\bS_b(v)=\bS_b(v)\bS_a(u)\bar{R}_{ab}(u,v),\quad \bar{R}(u,v) = u-v +\hbar\,P\,.
\end{equation}
As a result, fusion for $\bS$ is performed in precisely the same way as for $\bT$ but now we use the opposite sign of $\hbar$. Let us denote by $\Sa_\lambda(u)$ the transfer matrix constructed from $\bS$ in the irrep $\lambda$. In analogy with fusion for $\bT$ is satisfies the a slightly modified CBR formula compared to \eqref{CBRfla} and reads
\begin{equation}\label{Scbr}
\Sa_\lambda(u)=\displaystyle\det_{1\leq i,j\leq \lambda_1}\Sa_{\lambda^{\prime}_j+i-j,1}(u-\hbar(i-1))\,.
\end{equation}
Let's examine the structure of these transfer matrices in a bit more detail. The transfer matrices $\Sa_{a,1}$ are linear combinations of quantum minors of size $a$ built from $\bS$ 
\begin{equation}
\Sa_{a,1}(u) = \sum_{i,j} \bS\left[^{i_1\dots i_a}_{j_1\dots j_a} \right]
\end{equation}
where we sum over all indices with $1\leq i_1 < \dots < i_a \leq \gn$ and similarly for $j$ and the quantum minor built from $\bS$ is defined as 
\begin{equation}
\bS\left[^{i_1\dots i_a}_{j_1\dots j_a} \right] = \sum_{\sigma\in\mathfrak{S}_a} \bS_{i_{\sigma(1)}j_1}\bS_{i_{\sigma(2)}j_2}^{[2]} \dots \bS_{i_{\sigma(a)}j_a}^{[2(a-1)]}\,.
\end{equation}
Notice the sign change compared to \eqref{quantumminor}. 

\medskip

We are now in a position to relate the transfer matrices $\Sa_{\lambda}$ to the transfer matrices $\T_\lambda$. The main tool needed for this is the known formula \cite{molev2007yangians} for the action of the antipode on quantum minors. Let $p=\{i_1,\dots,i_{\gn}\}$ and $q=\{j_1,\dots,j_{\gn}\}$ be two permutations of $\{1,\dots,\gn\}$. Then we have
\begin{equation}\label{antipodeminor}
\qdet\,\bT(u)\ S\left(\bT\left[^{j_{m+1}\dots j_\gn}_{i_{m+1}\dots i_{\gn}} \right](u-\hbar\,m) \right) = {\rm sgn}p\, {\rm sgn}\,q\,\bT\left[^{i_1\dots i_m}_{j_1 \dots j_m} \right]\,.
\end{equation}
It is trivial to work out the action of $S$ on the minor constructed from $\bT$ to produce 
\begin{equation}
S\left(\bT\left[^{i_1\dots i_a}_{j_1\dots j_a}\right] \right)(u) = \bS\left[^{i_1\dots i_a}_{j_1\dots j_a}\right](u-\hbar(a-1))
\end{equation}
and so by summing over indices in \eqref{antipodeminor} we obtain 
\begin{equation}\label{antitransfer}
\Sa_{a,1}(u) = \frac{\T_{\gn-a,1}^{[2(\gn-1)]}}{\T_{\gn,1}^{[2(\gn-1)]}}\,.
\end{equation}
Hence, the Bethe algebra is closed under the action of the antipode and as a result all transfer matrices $\Sa_\lambda$ built from the inverse monodromy matrix can all be expressed in terms of the transfer matrices constructed with $\bT(u)$. 

\medskip

We now consider the commutation relation \eqref{BTcom}
\begin{equation}
\T_\lambda(v)\bB(u) = f_\lambda(u,v)\T_\lambda(v)\bB(u) +\lR(u,v)\,.
\end{equation}
We are going to apply some transformations to it and end up with a commutation relation intertwining $\Sa_\lambda$ and $\bB$ which will allow us to generate $\bB$ eigenstates from $\bra{\bar{0}}$. To do this we need a map $\omega = S\circ *$ which, as a composition of Yangian anti-automorphisms, is an automorphism and satisfies \cite{molev2007yangians}
\begin{equation}\label{omegaant}
\omega\left(T\left[^{i_1\dots i_a}_{j_1\dots j_j}\right](u)\right) = S\left[^{i_1\dots i_a}_{j_1\dots j_j}\right](u)\,.
\end{equation}
We also need the matrix $\sigma$ with $\sigma_{ij}=\delta_{\gn+1-i,j}$ and maps $T_{ij}\rightarrow T_{\gn+1-i,\gn+1-j}$. Note that these monodromy matrix elements are bare, i.e. untwisted. Note that there is no argument in the \rhs of \eqref{omegaant} due to how the $*$-map was defined with an additional relabelling of $u\rightarrow -u$ compared with \cite{molev2007yangians}. 

\medskip

We now show that the composition $\omega \circ \sigma$ transforms our commutation relation in the desired way. We will demonstrate it first for $\gl(3)$. Let's apply the sequence of maps to $\bB$, where we have set all $w$'s to $1$ for convenience. We have
\begin{equation}
\bB(u)=T\left[^{1}_{1} \right]T^{[2]}\left[^{12}_{12} \right]+T\left[^{2}_{1} \right]T^{[2]}\left[^{12}_{13} \right]
\end{equation}
which becomes, after applying $\sigma$, 
\begin{equation}
\bB(u)\rightarrow T\left[^{3}_{3} \right]T^{[2]}\left[^{23}_{23} \right]+T\left[^{2}_{3} \right]T^{[2]}\left[^{23}_{13} \right]
\end{equation}
where we have also used the symmetry properties of quantum minors. We now apply $*$, the first map in $\omega$ to obtain 
\begin{equation}
\bB(u)\rightarrow T\left[^{23}_{23} \right]T\left[^{3}_{3} \right]+T\left[^{23}_{13} \right]T\left[^{2}_{3} \right]\,.
\end{equation}
Finally, we apply $S$ and use \eqref{antipodeminor} to obtain 
\begin{equation}
\bB(u)\rightarrow T^{[4]}\left[^{12}_{12} \right]T^{[2]}\left[^{1}_{1} \right]+T^{[4]}\left[^{2}_{1} \right]T^{[2]}\left[^{12}_{13} \right]=\bB^{[2]}
\end{equation}
where we have used the fact that minors and single monodromy matrix elements in $\bB$ commute with each other and have ignored overall factors of the quantum determinant which will drop out of the resulting commutations relation. The calculation can easily be repeated for $\gl(\gn)$ -- the end conclusion is the same, $\bB(u) \rightarrow \bB^{[2]}(u)$ except this time we cannot freely reverse the order of minors. On the other hand we argued earlier that this reversal does not effect the commutation relation of interest, and so we obtain the result. 

\medskip

Next we consider the remainder term $\lR(u,v)=\sum_{j=1}^\gn T\left[^{j}_{1}\right](v)\times \dots$. We clearly end up with 
\begin{equation}\label{antiremainder}
\lR(u,v) \rightarrow \sum_{j=1}^\gn S\left[^{j}_\gn\right](v)\times \dots
\end{equation}
where we also performed $j\rightarrow \gn+1-j$. For the moment it is convenient to write the result like this instead of re-expressing in terms of $T_{ij}$. 

\medskip

Finally, we consider the action of our sequence of maps on the transfer matrix $\T_\lambda(u)$. For illustrative purposes let us consider the case where $\lambda$ is the defining representation. We have
\begin{equation}
\T_{1,1}(u) = {\rm tr}\left(T(u)\,G \right)\rightarrow {\rm tr}\left(S\,\sigma^{-1}G\sigma \right)\,.
\end{equation}
Unfortunately the object on the \rhs is not $\Sa_{1,1}(u)$ since $\sigma^{-1}G \sigma \neq G^{-1}$. However, the difference is quite manageable. Consider $\gl(3)$. We have
\begin{equation}
\Sa_{1,1}=S_{21}+S_{32}+\frac{1}{\chi_3}S_{13}-\frac{\chi_1}{\chi_3}S_{23}+\frac{\chi_2}{\chi_3}S_{33}
\end{equation}
whereas
\begin{equation}
{\rm tr}\left(S\,\sigma^{-1}G\sigma \right) = S_{21}+S_{32}+\chi_3 S_{13}-\chi_2 S_{23}+\chi_1 S_{33}\,.
\end{equation}
We see that the only difference between these two objects is arising from terms $S_{j \gn}$, but these are precisely the type of terms in the remainder \eqref{antiremainder}! We can then repeat the argumentation we used earlier in deriving the commutation relation between $\bB$ and $\T_\lambda$ to move all of these terms into the remainder and hence for the purposes of the commutation relation we can simply replace $\sigma^{-1}G\sigma$ with $G^{-1}$ and hence obtain $\Sa_{1,1}$ in the commutation relation. 

\medskip

We can now repeat the argument for other transfer matrices. First we look at $\T_{a,1}$ and again obtain $\Sa_{a,1}$ up to terms which do not effect the commutation relation and then by using the CBR formula the same statement is true for all $\T_{\lambda}$. Under our transformation $f_\lambda(u,v)\rightarrow f_\lambda(-u,-v)$ and hence we finally obtain 
\begin{equation}
\Sa_\lambda(v) \bB^{[2]}(u) = f_\lambda(-u,-v) \bB^{[2]}(u)\Sa_\lambda(v) + \sum_{j=1}^\gn S_{j\gn}(v)\times \dots\,.
\end{equation}
We will now prove that $\Sa_{1,s}\propto \T_{\gn-1,s}$. We use the CBR formulae 
\begin{equation}
\T_{\gn-1,s} = \det_{1\leq i,j\leq s} \T_{\gn-1+i-j,1}(u+\hbar(i-1))
\end{equation}
\begin{equation}
\Sa_{1,s} = \det_{1\leq i,j\leq s} \Sa_{1+i-j,1}(u-\hbar(i-1))
\end{equation}
and use the relation \eqref{antitransfer} to rewrite the second equation as 
\begin{equation}
\Sa_{1,s}=\prod_{i=1}^s\frac{1}{\T_{\gn,1}^{[2(\gn-i)]}} \times \det_{1\leq i,j\leq s}\T_{\gn-1-i+j,1}(u+\hbar(\gn-i))\,.
\end{equation}
We can conjugate the matrix inside the determinant with the matrix which transforms $(i,j)\rightarrow (s+1-i,s+1-j)$ without changing the determinant value, obtaining 
\begin{equation}
\Sa_{1,s}=\prod_{i=1}^s\frac{1}{\T_{\gn,1}^{[2(\gn-i)]}} \times \det_{1\leq i,j\leq s}\T_{\gn-1+i-j,1}(u+\hbar(i-1+\gn-s))
\end{equation}
and hence conclude
\begin{equation}
\Sa_{1,s}=\prod_{i=1}^s\frac{1}{\T_{\gn,1}^{[2(\gn-i)]}} \times \T_{\gn-1,s}(u+\hbar(\gn-s))\,.
\end{equation}
It is straightforward to verify that the remainder term will vanish for $v=\theta_\alpha+\hbar\,S$ and hence $\bB$ eigenstates are created by action of $\T_{\gn-1,s}(\theta_\alpha+\hbar(S+\gn-s)$, as claimed. Hence, the eigenvectors of $\bB$ are constructed as 
\begin{equation}
\bra{\Lambda^{\bB}} = \bra{\bar{0}}\prod_{\alpha=1}^L \prod_{j=1}^{\gn-1} \T_{\gn-1,1}(\theta_\alpha+ \hbar(S+\gn-s_j))\,.
\end{equation}
Hence, by normalising $\ket{\Psi}$ so that
\begin{equation}
\braket{\bar{0}|\Psi} = \prod_{\alpha=1}^L \prod_{j=1}^{\gn-1} \hat{\sfq}_1^{[2S]}(\theta_\alpha)
\end{equation}
and renormalising $\bra{\Lambda^{\bB}}\rightarrow \bra{\svx}$ we obtain 
\begin{equation}
\braket{\svx|\Psi} = \prod_{\alpha=1}^L \prod_{j=1}^{\gn-1} \hat{\sfq}_1^{[2(S-s_j)]}(\theta_\alpha)
\end{equation}
which holds regardless of what value $S$ takes. Finally we perform the usual $S\rightarrow -2\bs$ and $\theta_\alpha\rightarrow \theta_\alpha+\sfi\bs$ to obtain 
\begin{equation}\label{vacvac}
\braket{\svx|\Psi} = \prod_{\alpha=1}^L \prod_{j=1}^{\gn-1} \hat{\sfq}_1^{[-2(\bs+s_j)]}(\theta_\alpha)\,.
\end{equation}
\subsection{Explicit examples in $\sla(2)$ and $\sla(3)$ spin chains}

In this section we will demonstrate the developed techniques on some explicit examples of low rank and low length $L$. We will start with the $\sla(2)$ case and $L=1$. For simplicity when dealing with Hodge dual Q-functions we impose that $\det\,G=1$ and hence for $\sla(2)$ have $z_1z_2 = 1$ and for $\sla(3)$ have $z_1 z_2 z_3=1$. This results in Q-functions having the structure 
\begin{equation}
\hat{\sfq}_j = z_j^{-\sfi\,u}\times \dots,\quad \hat{\sfq}^j = z_j^{\sfi\,u}\times \dots\,.
\end{equation}

\paragraph{$\sla(2)$ spin chain}

We start by determining the vacuum state $\ket{\Omega}$ in the companion twist frame. There are two ways to achieve this. The first is to construct a rotation matrix which brings the diagonal twist matrix $g$ to the companion matrix $G$, write it as a product of exponentials of Lie algebra generators and then evaluate its action on the highest-weight state for diagonal twist. An alternate approach is to write the transfer matrix in the companion twist frame explicitly and solve the resulting differential equations using its known vacuum eigenvalue. The result is of course the same in both cases, but we follow the second approach for simplicity and solve
\begin{equation}
\T_{1,1}(u)\ket{\Omega} = \sfT_{1,1}(u)\ket{\Omega},\quad \sfT_{1,1}(u) = z_1 Q_\theta^{[2\bs]} + z_2 Q_\theta^{[-2\bs]}.
\end{equation}
The result is 
\begin{equation}
\ket{\Omega} = z_1^{-\sfi\,\theta-\bs} \left(1+ \frac{x}{z_1} \right)^{-2\bs}
\end{equation}
and is clearly not a polynomial for generic $\bs$ as expected. The normalisation factor has been chosen to ensure that $\braket{\bar{0}|\Omega}$ produces the expected value \eqref{vacvac}.

\medskip

In order to compute the dual vacuum state $\bra{\Omega}$ we need to equip our representation space with a scalar product and notion of transpose. Transposition ${\rm T}$ is naturally introduced by the property 
\begin{equation}
\braket{\Psi_1|\lO\Psi_2}=\braket{\lO^{\rm T}\Psi_1|\Psi_2}\,.
\end{equation}
To proceed, we introduce an orthonormal basis of states $\ket{e_n}$ on the representation which by definition satisfy 
\begin{equation}
\braket{e_n|e_m}=\delta_{nm},\quad n,m\geq 0\,.
\end{equation}
A natural guess would be to simply use $e_n= x^n$ but this is not compatible with the explicit realisation of our Lie algebra generators. It is common to choose an explicit realisation of generators with the property 
\begin{equation}\label{transposecond}
\sfE_{ij}^{\rm T}=\sfE_{ji}
\end{equation}
but it is more convenient in order to avoid messy signs to define the transpose with 
\begin{equation}
\sfE_{jj}^{\rm T} = \sfE_{jj},\quad \sfE_{12}^{\rm T} = - \sfE_{21}\,.
\end{equation}
Together with the requirement $e_0=1$ this fixes the basis to be given by
\begin{equation}
e_n=x^n\sqrt{\frac{\Gamma (n+2 s)}{\Gamma (n+1) \Gamma (2
   \bs)}}\;.
\end{equation}
At this point we should mention that our scalar product does not involve any complex conjugation and the scalar product is linear in both arguments. To promote it to Hermitian conjugation we need to impose that $\bs$ is real. Furthermore, in order for Hermitian conjugation to lift to the Yangian generators in a natural way one should make certain choices on the reality of the parameters of the model such as inhomogeneities $\theta_\alpha$ and twists $z_i$. Instead we will view our scalar product as simply defining the action of a dual vector on a vector. 

\medskip

Equipped with our scalar product we can now calculate $\bra{\Omega}$ using the differential equation approach as before and we find 
\begin{equation}
\bra{\Omega} =z_1^{-\sfi \theta-\bs} \left(1+\frac{x}{z_1} \right)^{-2\bs}\,.
\end{equation}
We can then calculate the overlap $\braket{\Omega|\Omega}$ which should be expanded into an infinite series. To check convergence one can use the ratio test and we find that convergence is guaranteed provided we take $|z_1|>1$ and so $z_1$ cannot be a pure phase. Computing the overlap we then find 
\begin{equation}\label{directoverlap}
\braket{\Omega|\Omega} = z_1^{-2(\sfi \theta+\bs)} \left(1-\frac{1}{z_1^2} \right)^{-2\bs}\,.
\end{equation}

\medskip

We now compare our results with the functional integral approach. We have that
\begin{equation}
\braket{\Psi^A|\Psi^B}:=\frac{1}{\lN}\langle \hat{\sfq}_1^A(u)\,\hat{\sfq}_1^A(u) \rangle
\end{equation}
where $\lN$ is a normalisation fixed as follows. The integral can be expanded into a sum over residues 
\begin{equation}\label{sumSoV}
\braket{\Psi^A|\Psi^B}:=\frac{2\pi\sfi}{\lN} \sum_{n=0}^\infty \lM^\prime_{n}\hat{\sfq}_1^A(\theta-\sfi(\bs+n))\,\hat{\sfq}_1^B(\theta-\sfi(\bs+n))
\end{equation}
where $\lM^\prime_n$ correspond to the residues of the measure at $\theta-\sfi(\bs+n)$. The normalisation $\lN$ is then given by $\lN=2\pi\sfi\lM^\prime_0$ in order to match the normalisation of the SoV measure which starts with the leading term $\braket{\bar{0}|\bar{0}}=1$ and by putting $\lM_n = 2\pi\sfi\lM^\prime_n/\lN$ we have
\begin{equation}\label{intmeas}
\lM_n= \frac{1}{n!} \frac{\Gamma[2\bs+n]}{\Gamma[2\bs]}\,.
\end{equation}
We now compare the results from the integral with our explicit computation of $\braket{\Omega|\Omega}$. The Q-function for this state is given by 
\begin{equation}
\hat{\sfq}_1 = z_1^{-\sfi\,u}
\end{equation}
and if we plug it into \eqref{sumSoV} we find 
\begin{equation}
\braket{\Omega|\Omega} = z_1^{-2\bs}\left(1-\frac{1}{z_1^2} \right)^{-2\bs}
\end{equation}
which perfectly reproduces the result of our direct computation and like in the case of the direct computation \eqref{directoverlap} we find that for the infinite series \eqref{sumSoV} to converge we need $|z_1|>1$. 

\medskip

At this point it is useful to note that there is another, rather non-trivial, quantity we can compute using our techniques. We made extensive use of the fact that in the companion twist frame the SoV bases were independent of the twist eigenvalues and hence they serve to separate the wave functions of a transfer matrix with a twist of companion form but with different eigenvalues. The SoV measure is obviously unaffected by this and so we expect that our integral can also be used to compute overlaps between states of different twists. 

\medskip

Let $\bra{\tilde{\Psi}}$ and $\ket{\Psi}$ be eigenstates of transfer matrices $\tilde{\T}$ and $\T$ built with companion twist matrices $\tilde{G}$ and $G$ respectively different only in their eigenvalues. Then we claim that 
\begin{equation}
\braket{\tilde{\Psi}|\Psi} = \frac{1}{\lN}\langle \tilde{\hat{\sfq}}_1(u)\,\hat{\sfq}_1(u) \rangle
\end{equation}
where $\tilde{\hat{\sfq}}_1$ is the Q-function corresponding to the state $\bra{\tilde{\Psi}}$. As before we will compare the integral result with direct computation for the overlap $\braket{\tilde{\Omega}|\Omega}$ where 
\begin{equation}
\bra{\tilde{\Omega}} = \bra{\Omega} =\tilde{z}_1^{-\sfi \theta-\bs} \left(1+\frac{x}{\tilde{z}_1} \right)^{-2\bs}\,.
\end{equation}
A direct computation of the overlap yields
\begin{equation}
\braket{\tilde{\Omega}|\Omega} = z_1^{-\sfi\theta-\bs}\tilde{z}_1^{-\sfi\theta-\bs}\left(1-\frac{1}{z_1\tilde{z}_1} \right)^{-2\bs}\,.
\end{equation}
Comparing with the integral we find precisely the same result!

\medskip

So far we have performed a few non-trivial checks that the integral approach indeed reproduces the correct expression for the overlaps. In this simple set-up we can show however that the integral produces the correct SoV measure precisely without needing to consider special states which amounts to computing the overlaps $\braket{\svx|\svx}$. We start by computing $\bra{\svx}$. Since these are eigenvectors of the GT algebra they must be simple monomials $x^n$, $n\in \ZZ_{\geq 0}$. Imposing the normalisation 
\begin{equation}
\braket{\svx|\Omega} = z_1^{-\sfi \theta-(\bs+n)}
\end{equation} 
immediately yields $\bra{\svx} = (-1)^n x^n=\ket{\svx}$. We can now easily work out the overlap $\braket{\svx|\svx}$ which is given by 
\begin{equation}
\braket{\svx|\svx}^{-1}=\frac{1}{n!}\frac{\Gamma[2\bs+n]}{\Gamma[2\bs]}
\end{equation}
which precisely matches the measure \eqref{intmeas} $\lM_n$ obtained from the functional integral approach. and demonstrating the equivalence between the two approaches.

\medskip

\paragraph{$\sla(3)$}

We now repeat the previous computations for the $\sla(3)$ $L=1$ case which is a bit more involved but nevertheless is straightforward.

\medskip

Our starting point is to introduce a scalar product $\langle -|-\rangle$ such that the $\gl(3)$ Lie algebra generators satisfy $\sfE_{ij}^{\rm T} \propto \sfE_{ji}$ where $\propto$ indices a possible sign. We impose that the sign is $+$ for the Cartan generators while $\sfE_{12}^{\rm T}=-\sfE_{21}$, $\sfE_{13}^{\rm T}=-\sfE_{31}$ and $\sfE_{23}^{\rm T}=\sfE_{32}$. The representation space is on $\CC[x,y]$ and we introduce a basis $e_{n,k} = c_{n,k} x^n y^k$ with the property $\langle e_{n,k}|e_{n^\prime,k^\prime}\rangle = \delta_{nn^\prime}\delta_{kk^\prime}$. Our requirements then fix 
\begin{equation}
e_{n,k}=x^ny^k \sqrt{\frac{\Gamma[2\bs+n+k]}{\Gamma[n+1]\Gamma[k+1]\Gamma[2\bs]}}
\end{equation}
once we impose $_{0,0}=1$. 

\medskip

As in the $\sla(2)$ case we can compute the transfer matrix vacuum eigenvectors $\ket{\Omega}$ and $\bra{\Omega}$ by explicitly writing the transfer matrix as a differential operator and using its known eigenvalues. We find 
\begin{equation}
\ket{\Omega} = \left(1+\frac{x}{z_1}+\frac{y}{z_1^2} \right)^{-2\bs}\times z_1^{-2(\sfi\,\theta+\bs)}
\end{equation}
\begin{equation}
\bra{\Omega} =  \left(1+x(z_2+z_3)-\frac{y}{z_1} \right)^{-2\bs} \times z_1^{-\sfi\,\theta-\bs+1}(z_2-z_3)\,.
\end{equation}
As before the normalisation factors are chosen to ensure we obtain the correct results for $\braket{\bar{0}|\Omega}$ and $\braket{\Omega|\bar{0}}$. 

\medskip

Like in the $\sla(2)$ case we can compute the overlap $\braket{\tilde{\Omega}|\Omega}$ where $\bra{\tilde{\Omega}}$ is the vacuum state for a transfer matrix with twist eigenvalues $\tilde{z}_j$, $j=1,2,3$. We find 
\begin{equation}\label{directoverlap2}
\braket{\tilde{\Omega}|\Omega} = \tilde{z}_1^{-\sfi\,\theta-\bs+1}(\tilde{z}_2-\tilde{z}_3)z_1^{-2(\sfi\,\theta+\bs)}\left(1-\frac{\tilde{z}_2}{z_1} \right)^{-2\bs} \left(1-\frac{\tilde{z}_3}{z_1} \right)^{-2\bs}\,.
\end{equation}

We will now attempt to reconstruct this result using the functional integral approach. We need to compute 
\begin{equation}
\frac{1}{\lN} \left| \begin{array}{cc}
\langle \hat{\sfq}^A_1\hat{\sfq}_B^2\rangle_1 & \langle \hat{\sfq}^A_1\hat{\sfq}^{2\ [2]}_B\rangle_1\\
\langle \hat{\sfq}^A_1\hat{\sfq}_B^3\rangle_1 & \langle \hat{\sfq}^A_1\hat{\sfq}^{3\ [2]}_B\rangle_1
\end{array} \right|
\end{equation}
where $\hat{\sfq}_1^A = z_1^{-\sfi\,u}$ and $\hat{\sfq}_B^2= \tilde{z}_2^{\sfi\,u}$, $\hat{\sfq}_B^3= \tilde{z}_3^{\sfi\,u}$. Expanding the determinant we obtain 
\begin{equation}
\frac{1}{\lN}\displaystyle \int {\rm d}u {\rm d}v\,\mu_1(u)\mu_1(v)\,f(u,v)
\end{equation}
where $f(u,v)$ is the term containing Q-functions. Like before the normalisation is fixed so that in the expansion over residues the coefficient of the $f(\theta-\sfi\bs,\theta-\sfi\bs)$ term is $1$ and so for the case at hand 
\begin{equation}
\lN=\frac{4 \pi^2 e^{8 \sfi \pi  \bs}}{\left(-1+e^{-4\sfi \pi  \bs}\right)^2 \Gamma (1-2 \bs)^2}\,.
\end{equation}
It is straightforward to perform the sum over residues yielding 
\begin{equation}
\frac{1}{\lN}\displaystyle \int {\rm d}u {\rm d}v\,\mu_1(u)\mu_1(v)\,f(u,v) = \sum_{n,m=0}^\infty \frac{\Gamma[2\bs+n]\Gamma[2\bs+m]}{\Gamma[2\bs]^2}f(\theta-\sfi(\bs+m),\theta-\sfi(\bs+n))\,.
\end{equation}
By plugging in the simple expression for $f(u,v)$ in terms of twist eigenvalues and performing the sum we find that we perfectly reproduce \eqref{directoverlap2}!

\medskip

\begin{figure}
    \centering
    \includegraphics[scale=0.5]{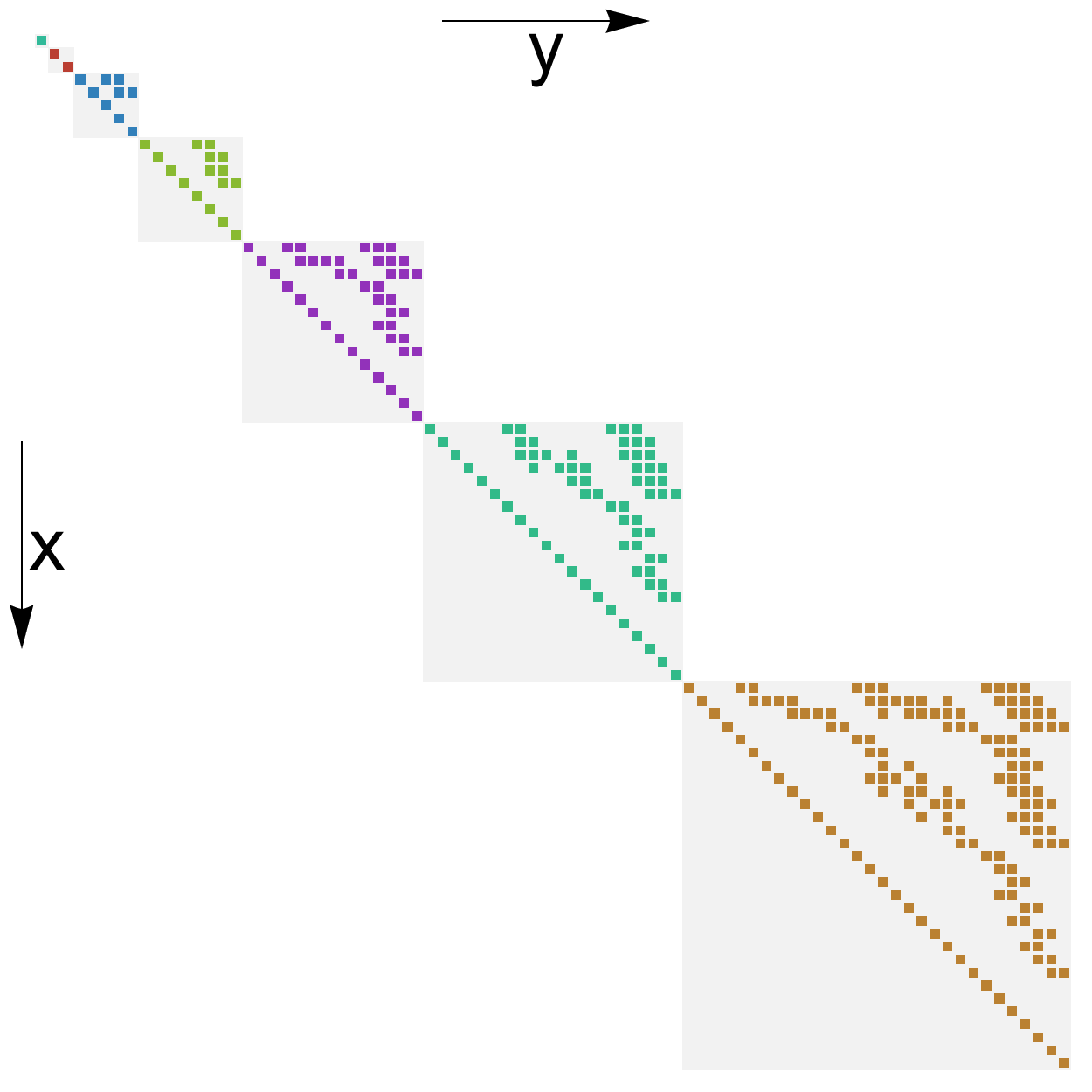}\quad\quad
        \includegraphics[scale=0.5]{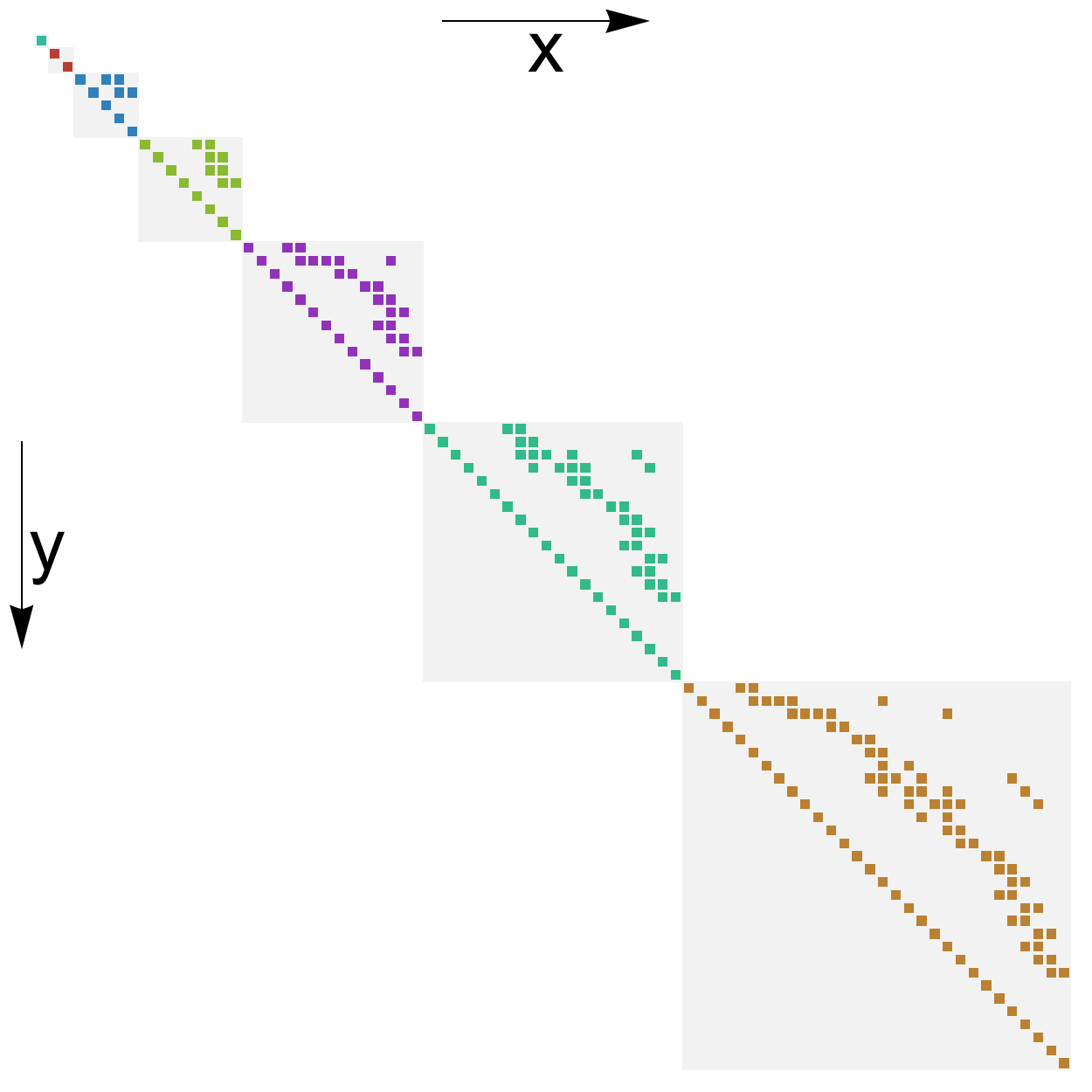}
    \caption{Non-zero elements of the matrix $\langle \svx|\svy \rangle$ (Left) and its inverse (Right) up to SoV charge $6$. Blocks indicate fixed SoV charges.}
    \label{fig:measurepic}
\end{figure}

The fact that the such overlaps can be expressed as a simple determinant in Q-functions is extremely powerful. As we already said, the SoV measure obtained from the matrix of overlaps $\braket{\svx|\svy}$ is non-diagonal in general means not having to compute it directly is a huge advantage. In Figure \ref{fig:measurepic} we present an example of the matrix of overlaps and associated SoV measure for $\sla(3)$ length $L=2$ to demonstrate its structure, keeping in mind that it is block diagonal since both $\bB$ and $\bC$ commute with the SoV charge operator.

\section{Computing observables -- determinant representations for overlaps and expectation values}\label{detrep}

In this section we will extend the previous results by deriving SoV-based determinant representations for overlaps and expectation values of various operators.

\subsection{Defining det-product and its relation to SoV}

Here we discuss the main tools for computing some physical observables with the help of the SoV approach we developed in the previous sections. For simplicity we will mostly demonstrate the method on the $\sla(3)$ example but in all cases the generalisation to $\sla(\gn)$ is clear.

\medskip

One of the key quantities we will compute in this section is the overlap between transfer matrix eigenstates corresponding to spin chains with different twist eigenvalues. Indeed, in the MCT frame the SoV basis states are twist independent and hence factorise the wave functions of transfer matrix eigenstates built with \text{any} twist. This implies that the integral representation we derived in the previous section for states of the same spin chain can be, very non-trivially, used to compute overlaps between the eigenstates of different transfer matrices. Such overlaps were recently considered in the context of AdS/CFT correspondence~\cite{Cavaglia:2020hdb} and can be interpreted as $3$-point correlation functions involving so-called color twist operators and have also been studied in the Bethe ansatz framework \cite{belliard2021overlap}.

\medskip

For what follows it will be convenient to introduce notation for what refer to as the {\it det-product},
\begin{equation}\label{detproduct}
\detl G|F\detr = \frac{1}{\lN_0} \det_{(\alpha,i),(\beta,j)} \langle F_\alpha(u) u^{\beta-1} \lD^{2(j-1)} G_\beta^{i+1} \rangle_\alpha
\end{equation}
where the notation $\langle f\rangle_\alpha$ initially referred to an integration
\begin{equation}
\langle f\rangle_\alpha = \displaystyle \int^\infty_{-\infty} {\rm d}u\,\mu_\alpha(u)\,f(u)
\end{equation}
we now understand more generally as a sum over the poles of the factors $\mu_\alpha$ appearing in the integration measure. The normalisation $\lN$ is fixed in the way described in the previous Section. 

\medskip

For the case when $G$ and $F$ in \eqref{detproduct} are Q-functions describing two spin chain states, the det-product gives the overlap of these states we presented above in \eqref{sl3SP}. 
By performing the integrations and sum over residues the expression \eqref{detproduct} can be expressed
\begin{equation}\label{detsov}
\detl G_{\alpha}^a|F_\alpha\detr=\sum_{\svx,\svy}M_{\svy,\svx}\prod_{\alpha,a}F_\alpha(\svx^\alpha_a)
\prod_{\alpha}\left(G_{\alpha}^2(\svy^{\alpha}_1)G_{\alpha}^3(\svy^{\alpha}_2)-G_{\alpha}^3(\svy^{\alpha}_1)G_{\alpha}^2(\svy^{\alpha}_2)\right)
\end{equation}
where $\lM_{\svx,\svy}$ denotes the measure. We demonstrated the validity of this formula in some simple $L=1$ examples in the previous section but it holds generally. We will not present the somewhat involved combinatorics here put refer the reader to the original publication \cite{Gromov:2020fwh} of the author. 

\medskip

We will see that a number of correlators can be expressed in terms of the det-product, bypassing the need to explicitly compute the measure components $\lM_{\svx,\svy}$. In order for two states $\ket{\Theta}$ and $\bra{\Phi}$ to have a scalar product which can be written in the det-product form, we have to require what we call {\it separability} property from these states, which can be expressed as
\begin{equation}\label{PsiPhi}
\begin{split}
& \braket{\svx|\Theta} = \prod_{\alpha=1}^L F_\alpha(\svx^\alpha_1)F_\alpha(\svx^\alpha_2)\\
& \braket{\Phi|\svy} = \prod_{\alpha=1}^L \left[G_\alpha^2(\svy^\alpha_1)G_\alpha^3(\svy^\alpha_3)-G_\alpha^3(\svy^\alpha_1)G_\alpha^3(\svy^\alpha_3) \right]\,.
\end{split}
\end{equation}
If that is the case, then as a consequence of the completeness of both SoV bases $\{\svx\}$ and $\{\svy\}$ and due to the relation \eqref{detsov} we immediately get
\begin{equation}
\braket{\Phi|\Theta} = \detl G|F\detr\,.
\end{equation}
In what follows we explore a few examples when \eqref{PsiPhi} does hold. One immediate example is when both states are  transfer matrix eigenstates. In this case of course we simply have $F_{\alpha}=\hat{\sfq}_1(\svx^\alpha)$ and $G_\alpha^{i+1}=\hat{\sfq}^{i+1}(\svy^\alpha_i)$, so that
\begin{equation}\label{ABover}
\braket{\Psi^A|\Psi^B} = \detl \hat{\sfq}_A^{i+1}| \sfq^B_1\detr \propto \delta^{AB}
\end{equation}
In the above expression the left and right wave functions are normalised according to our conventions from section~\ref{operatornoncom}. 

\subsection{Overlaps between wave functions with different twists}
Another quite obvious example where the separability property~\eqref{PsiPhi} is satisfied for both states but gives much less trivial overlap than~\eqref{ABover} is the case when both states are eigenstates of transfer matrices with {\it different} sets of twists eigenvalues $z_a$ and $z_i$. As we emphasised before, the SoV states do not depend on $z$'s and thus should separate wave functions corresponding to spin chains with arbitrary twist eigenvalues $z_j$ (provided the twist matrix is of MCT form) and indeed we already saw an example of this in the previous section based on explicit $L=1$ examples.

\medskip

We thus conclude that the overlap between the states of the spin chains with different twist eigenvalues can we written in the form
\begin{equation}
\braket{\tilde{\Psi}|\Psi} = \detl \tilde{G}|F\detr
\end{equation}
where $\tilde{G}$ and $F$ are appropriate Q-functions. In the above expression we still assume that the states are normalised in agreement with our conventions. However, we can also form a normalisation independent combination, for example
\begin{equation}
\frac{\braket{\tilde{\Psi}|\Psi}\braket{\Psi|\tilde{\Psi}}}{\braket{\tilde{\Psi}|\tilde{\Psi}}\braket{\Psi|\Psi}}
\end{equation}
which we will make use of in a moment.

\paragraph{Probing the transition matrix.}
The overlap between two eigenstates of the transfer matrix in different frames is $\sla(\gn)$ invariant.
This means that one can diagonalise either one of the two twist matrices appearing in the transfer matrices. The matrix which relates the two frames that diagonalises one of these two twist matrices has the following general form, valid for $\sla(\gn)$:
\begin{equation}
S_{ab} = \prod_{i\neq a} \frac{z_a-\tilde{z}_a}{z_i-z_a},\quad S_{ab}^{-1} = \prod_{i\neq a} \frac{\tilde{z}_i-z_b}{\tilde{z}_i-\tilde{z}_a}\,.
\end{equation}
Let us show that the above transformation is hard-wired into the SoV construction and into the det-product in particular. Consider the normalisation independent combination
of the scalar products of two twisted vacua,
\begin{equation}\label{vactvac2}
\frac{\braket{\tilde{\Omega}|\Omega}\braket{\Omega|\tilde{\Omega}}}{\braket{\tilde{\Omega}|\tilde{\Omega}}\braket{\Omega|\Omega}}=\frac{
{(z_1-\tilde{z}_2)^{-2\bs} (z_1 -\tilde{z}_3)^{-2\bs}}
{(\tilde{z}_1-z_2)^{-2\bs} (\tilde{z}_1 -z_3)^{-2\bs}}
}
{
{(z_1-z_2)^{-2\bs} (z_1 -z_3)^{-2\bs}}
{(\tilde{z}_1-\tilde{z}_2)^{-2\bs} (\tilde{z}_1 -\tilde{z}_3)^{-2\bs}}
}\,.
\end{equation}
Let's now focus on the defining representation, i.e. $\bs=-1/2$. Let's assume that $|\Omega\rangle$
is in the diagonalised frame. We know that for the diagonal twist the ground is simply the highest weight state $|\Omega\rangle = \vec \sfe_1$, whereas the other state reads $|\tilde\Omega\rangle = S^{-1}|\Omega\rangle = S_{11}^{-1} \vec \sfe_1+ S_{21}^{-1}\vec \sfe_2+ S_{31}^{-1}\vec \sfe_3$. Similarly for the left
states $\langle\Omega| = \vec \sfe_1$
and $\langle\tilde\Omega| = \langle\Omega|S = 
S_{11} \vec \sfe_1+ S_{21}\vec \sfe_2+ S_{31}\vec \sfe_3$, from where we would expect that for $\bs=-1/2$ we should get
\begin{equation}
\frac{\braket{\tilde{\Psi}|\Psi}\braket{\Psi|\tilde{\Psi}}}{\braket{\tilde{\Psi}|\tilde{\Psi}}\braket{\Psi|\Psi}}=S_{11}S_{11}^{-1}
\end{equation}
which is indeed the case as we see from \eqref{vactvac2}. Note that one can further interchange the order of the eigenvalues, changing the vacua accordingly, to deduce  any combination of the form $S_{ab}S_{ba}^{-1},\;a,b=1,2,3$.
One can invert the logic and verify that the knowledge of all $S_{ab}S_{ba}^{-1},\;a,b=1,2,\dots,\gn$
allows one to reconstruct $S_{ab}$ modulo the transformation $S\to D_1.S.D_2$, where $D_1,D_2$ are two independent diagonal matrices. The diagonal matrices will commute with the twist matrices and they reflect the freedom in the definition of $S$ in the first place.

\subsection{On-shell off-shell overlap}
In this section we explore the effect of the action by  ${\bf B}(u)$ or ${\bf C}(u)$ operators on factorisable states. Assuming the state $\ket{\Theta}$ is separated by the SoV basis like in \eqref{PsiPhi}, we have
\begin{equation}\label{PsiPhiB}
\bra{\svx}{\bf b}(w)\ket{\Theta} = \braket{\svx|\Theta}\prod_{\alpha=1}^L (w-\svx^\alpha_1)(w-\svx^\alpha_2)
\end{equation}
where ${\bf b}(w)$ is the non-trivial part of the ${\bf B}(w)$ operator defined in \eqref{smallbc} and we have introduced the shorthand notation for its roots $\svx^\alpha_{j1}\rightarrow \svx^\alpha_j$. We see that the action by ${\bf b}(w)$ simply translates into the replacement $F_\alpha(u)\to (w-u)F_\alpha(u)$. It is clear that there is a potential to generalise this further. We can define a ``local" ${\bf b}_\alpha$ operator so that, for $\gl(\gn)$,
\begin{equation}
{\bf b}_\alpha = \prod_{j=1}^{\gn-1} (u-\svX^\alpha_{j})
\end{equation}
Repeating the same calculation as in \eqref{PsiPhiB} we see that ${\bf b}_\beta(w)$ acts on $F_\alpha$ as\footnote{One should be careful with the $\circ$ notation, as there is no linearity in the first argument, e.g. the sum of two operators does not necessarily produce a factorisable state and thus does not have any well defined action on individual $F_\alpha$s. However, $\circ$ is an associative operation and does support an action by several operators.}
\begin{equation}
{\bf b}_\beta(w)\circ F_\alpha(u) = (w-u)^{\delta_{\beta\alpha}} F_\alpha(u)\,.
\end{equation}
To summarise, this means that multiple action of any ${\bf b}_\beta(w)$ operators does not spoil the separability property of the wave function. This means that we can compute a set of rather non-trivial form factors in a determinant form,
\begin{equation}\label{FBBT}
\frac{\bra{\Phi}{\bf b}_{\beta_1}(v_1)\dots {\bf b}_{\beta_K}(v_K)\ket{\Theta}}{\braket{\Phi|\Theta}} = \frac{\detl G_\alpha^a|\displaystyle \prod_{i=1}^K (v_i-u)^{\delta_{\beta_i \alpha}} F_\alpha\detr}{\detl G_{\alpha,a}|F_\alpha\detr}\,.
\end{equation}
A particularly important case involves the following state
\begin{equation}\label{offdef}
\ket{\Psi}_{\rm off-shell}:={\bf b}(v_1)\dots {\bf b}(v_K)\ket{\Omega}
\end{equation}
which in analogy with $\sla(2)$ one could call an {\it off-shell} Bethe state.
To distinguish it from some other {\it off-shell} Bethe states existing in the literature, one could call it {\it algebraic off-shell} Bethe states as opposed to the hybrid coordinate-algebraic way of building eigenstates of transfer matrix in the nested Bethe ansatz approach. It follows immediately from \eqref{FBBT}
that the overlap between \eqref{offdef} and any separable state, and in particular with an eigenstate $\bra{\Phi}$ of the transfer matrix, is of a determinant form
\begin{equation}
\braket{\Phi|\Psi}_{\rm off-shell} = \detl\hat{\sfq}_{1,a+1}| z_1^{-\sfi\,u}\displaystyle \prod_{j=1}^K (u-v_j)\detr\,.
\end{equation}
Note that for this to be true it is not required that $\{v_j \}$ are Bethe roots solving Bethe ansatz equations. As we described before when the parameters $\{v_k \}$ do satisfy the Bethe ansatz equations
the state $|\Psi\rangle_{\rm off\;shell}$ does actually become an eigenstate of the transfer matrix.

\medskip

In analogy with ${\bf b}_\alpha(u)$ we can also define ${\bf c}_{\alpha}(u)$, containing only those roots of ${\bf c}(u)$ that are associated with $\theta_\alpha$. For the insertion of this operator we can use the relation
\begin{equation}
\begin{split}
\bra{\Phi}{\bf c}_\beta(w)\ket{\svy}&  = \braket{\Phi|\svy}(w-\svy_1^\beta)(w-\svy^\beta_2)  \\
& = (w-\svy_1^\beta)(w-\svy^\beta_2) \displaystyle \prod_{\alpha=1}^L (G^2_\alpha(\svy^\alpha_1)G_\alpha^3(\svy^\alpha_2)-G^3_\alpha(\svy^\alpha_1)G_\alpha^2(\svy^\alpha_2))
\end{split}
\end{equation}
implying that $G_\beta^a(u)\to (w-u)G_\beta^a(u)$, leaving other $G_\alpha^a(u)$ with $\alpha\neq\beta$ unchanged. Therefore we can generalise the result~\eqref{FBBT}
as follows:
\begin{equation}
\begin{split}
& \frac{\bra{\Phi}{\bf c}_{\gamma_1}(v_1)\dots {\bf c}_{\gamma_K}(v_K){\bf b}_{\beta_1}(w_1)\dots {\bf b}_{\beta_J}(w_J)\ket{\Theta}}{\braket{\Phi|\Theta}}\\
&  = \frac{\detl G_\alpha^a\displaystyle\prod_{j=1}^K(v_j-u)^{\delta_{\gamma_j\alpha}}|\displaystyle \prod_{i=1}^J (w_i-u)^{\delta_{\beta_i \alpha}} F_\alpha\detr}{\detl G|F\detr}\,.
\end{split}
\end{equation}

\subsection{Form factors of derivatives of the transfer matrices}

In this section we show how our integral SoV approach leads to determinant representations for a large class of diagonal form factors, extending the results of \cite{Cavaglia:2019pow} from $\bs=1/2$ to generic $\bs$. We demonstrate the $\sla(3)$ case, but generalization to $\sla(\gn)$ is immediate as we will see. We also show how to compute matrix elements of some $\textit{local}$ operators from this data.

\medskip

We consider a basis of integrals of motion $\mathbb{I}_{a,\beta}$ defined by
\begin{equation}
\T_{a,1}(u) = Q_\theta(u)^{\delta_{a2}}\sum_{\beta=0}^L u^\beta \mathbb{I}_{a,\beta},\quad a=1,2\,.
\end{equation}
The form factors we consider are the diagonal matrix elements of the derivatives these integrals of motion  defined as
\begin{equation}\label{ffac}
\frac{\bra{\Psi} \frac{\partial \mathbb{I}_{a,\beta}}{\partial p}\ket{\Psi}}{\braket{\Psi|\Psi}} = \frac{I_{b,\beta}}{\partial p}
\end{equation}
where $p$ is a parameter of the model (either an inhomogeneity $\theta_\alpha$ or a twist $z_j$). While the spectrum of the model is under good control and one could in principle compute the derivative in the \rhs of \eqref{ffac} directly (as a ratio of finite differences), here we rather wish to express it in terms of Q-functions evaluated at one fixed value of $p$, and it is nontrivial that such an expression exists at all. We will see that the result has a rather natural form of a ratio of two determinants, with 
the denominator corresponding to the norm \eqref{sl3SP} and the numerator given by the same expression with an extra insertion that we interpret as describing the operator $\partial_p\mathbb{I}_{a,\beta}$ we consider. In the AdS/CFT context correlators of this kind are also important as they correspond to 3-point functions with marginal operators \cite{Costa:2010rz}. 

\medskip

If we consider a small variation of our parameter $p\to p+\delta p$, the Q-functions $\hat{\sfq}^{a+1}$ as well as the difference operator $\mathcal{O}^\dagger$ in the Baxter equation \eqref{daggerbaxter} will change, but the equation will remain satisfied, so that $(\lO^\dagger+\delta\lO^\dagger)(\hat{\sfq}^{a+1}+\delta \hat{\sfq}^{a+1})=0$. Recall the bracket $(f,g)_\alpha$ defined by 
\begin{equation}
(f,g)_\alpha :=\displaystyle \int {\rm d}u\, K_\alpha(u)\,f(u)\,g(u)
\end{equation}
where $K_\alpha(u)=\nu_1(u)\mu_\alpha(u)$ which satisfies $(f,\lO^\dagger g)=(\lO\,f,g)$. Using that the original Q-function satisfies $\lO^\dagger \hat{\sfq}^{a+1}=0$, and dropping the terms quadratic in variations, we have
\begin{equation}
0=(\hat{\sfq}_1^{[2]}(\lO^\dagger+\delta\lO^\dagger)(\hat{\sfq}^{a+1}+\delta \hat{\sfq}^{a+1}))_\alpha=( \hat{\sfq}^{[2]}_1\lO^\dagger\delta \hat{\sfq}^{a+1})_\alpha+(\hat{\sfq}^{[2]}_1\delta\lO^\dagger \hat{\sfq}^{a+1})_\alpha \,.
\end{equation}
Now using the adjoint property $(f,\lO^\dagger g)=(\lO\,f,g)$ we see that the first term vanishes so that we get
\begin{equation}
( \hat{\sfq}_1^{[2]},\partial_p\lO^\dagger \hat{\sfq}^{a+1})_\alpha=0 \,.
\end{equation}
It is convenient to introduce a rescaled operator $\bar{\lO}$ with $\bar{\lO}=Q_\theta^{[-2(\bs-1)]} \lO^\dagger$. The benefit of this is that $\bar{\lO}$ is easier to work with, being polynomial in $\theta$s. We still have the property $\bar{\lO}\hat{\sfq}^a=0$ and hence
\begin{equation}\label{Ovar}
\left( \hat{\sfq}_1^{[2]},\frac{1}{Q_\theta^{[-2(\bs-1)]}}\partial_p\bar{\lO} \hat{\sfq}^{a+1}\right)_\alpha=0 \,.
\end{equation}
Explicitly, the variation $\partial_p\bar{\lO}$ of $\bar{\lO}$ reads
\begin{equation}\label{variat}
\partial_p\bar{\lO}=\displaystyle \sum_{(\beta,b)}\partial_p I_{b,\beta-1}(u+\hbar)^{\beta-1}\lD^b - F_p^{[2]}
\end{equation}
with
\begin{equation}
F_p^{[2]}=-\left(\partial_p Q_\theta^{[-2(\bs-1)]}- \partial_p Q_\theta^{[2(\bs+1)]}\lD^3 \right)+\sum_{b}\partial_p I_{b,L}(u+\hbar)^L \lD^b\,.
\end{equation}
We have singled out the integrals of motion $I_{b,L}$ since they are simply proportional to the identity operator and so do not carry non-trivial dynamical information which is only contained in the functions $I_{b,\beta-1}$, $\beta=1,\dots,L$.
Plugging \eqref{variat} into \eqref{Ovar} we get a linear system for the variations $\partial_p I_{b,\beta-1}$ of the form (after performing an overall shift $u\rightarrow u-\hbar$)
\begin{equation}\label{ABN1}
\displaystyle \sum_{(\beta,b)}m_{(\alpha,a),(\beta,b)}\partial_p I_{b,\beta-1}-f_{(\alpha,a)}=0,\quad f_{(\alpha,a)}=\langle \hat{\sfq}_1\,F_p\circ \hat{\sfq}^{a+1}\rangle_\alpha
\end{equation}
where
\begin{equation}
m_{(\alpha,a),(\beta,b)} = \displaystyle\det_{(\alpha,a),(\beta,b)}\langle \hat{\sfq}_1 u^{\beta-1} \lD^{b-1}\circ \hat{\sfq}^{a+1} \rangle_\alpha
\end{equation}
is the same matrix appearing in the $\sla(3)$ scalar product \eqref{sl3SP} with the two states taken to be the same. Note that in \eqref{ABN1} we have switched from the brackets $(f,g)_\alpha$ to the bracket $\langle f\,g\rangle_\alpha)$ defined in \eqref{mubracket} as a consequence of that fact that after a shift of $u\rightarrow u-\hbar$ we have
\begin{equation}
\left(f,\frac{1}{Q_\theta^{[-2(\bs-1)}}g\right)\rightarrow \langle f^{[-2]} g^{[-2]}\rangle_\alpha\,.
\end{equation}
We can write the solution of \eqref{ABN1} using Cramer's formula as
\begin{equation}\label{Ivar}
\partial_p I_{b^\prime,\beta^\prime-1} = \frac{\det_{(\alpha,a)(\beta,b)} \tilde{m}_{(\alpha,a)(\beta,b)}}{\det_{(\alpha,a)(\beta,b)} m_{(\alpha,a)(\beta,b)}}
\end{equation}
where $\tilde{m}_{(\alpha,a)(\beta,b)}$ is the matrix $m_{(\alpha,a)(\beta,b)}$ with the column $(\beta',b')$ replaced with $f_{(\alpha,a)}$ defined in \eqref{ABN1}. This gives a determinant representation for the variation of integrals of motion and the form factor \eqref{ffac}. The generalisation to $\sla(\gn)$ is immediate.

\paragraph{Local spin expectation value}\label{sec:detloc}

One of the key quantities of interest in spin chains are correlators of ``local" operators, i.e. those that act on a particular spin chain site in contrast to ``global" operators such as the transfer matrix. While certain maps from local to global operators are well known (see e.g. \cite{Maillet:1999re} and the review \cite{Slavnov:2019hdn}), here we will demonstrate that our approach offers yet another way to access local quantities.  Namely, there is a remarkable relation between a subset of local operators and derivatives of the integrals of motion ${\partial\mathbb{I}_{a,\beta}}/{\partial\theta_\alpha}$, whose expectation values we computed in the previous section. 

\medskip

The main idea is that when taking the derivative in $\theta_\alpha$ we can single out the $\alpha$-th spin chain site. To make it precise, let us write explicitly the large $u$ expansion of the transfer matrix with fundamental representation in the auxiliary space defined using the form of the Lax matrix from \eqref{generalisedperm},
\begin{equation}\label{Tuexp}
\T_{1,1}(u)=\chi_1 u^L - u^{L-1} \left(\sum_{\alpha=1}^L \chi_1 \theta_\alpha+\hbar\, {\rm tr}\left(\sfE^{(\alpha)t}G \right) \right)+\lO(u^{L-2})\,.
\end{equation}
The trace here is taken over the auxiliary space, and $\sfE^{(\alpha)}$ is an $\gn\times \gn$ matrix whose element at position $(i,j)$ is the operator $\sf_{i,j}$ (the $\gl(\gn)$ generator) acting on the $\alpha$-th site of the spin chain. Note that $\sfE$ in this expression  is transposed w.r.t. the indices $i,j$ as we indicated with the superscript $t$.  We see that in \eqref{Tuexp} we have a sum of local operators over all sites of the spin chain. Now we notice that when we differentiate the transfer matrix in $\theta_\alpha$, the Lax operator at position $\alpha$ in its definition will be simply replaced by minus the identity matrix, so as a result we will get the transfer matrix for the spin chain with the $\alpha$-th site removed. This means that the derivative will be given by the same result \eqref{Tuexp} but with sum taken over all sites except one,
\begin{equation}
\frac{\partial\T_{1,1}(u)}{\partial\theta_\alpha}=-\chi_1 u^{L-1} + u^{L-2} \left(\sum_{\beta\neq\alpha}^L \chi_1 \theta_\beta+\hbar\, {\rm tr}\left(\sfE^{(\beta)t}G \right) \right)+\lO(u^{L-3})\,.
\end{equation}
By combining this with \eqref{Tuexp} we can therefore extract the contribution from the site $\alpha$ only,
\begin{equation}
\T_{1,1}(u)+u\frac{\partial\T_{1,1}(u)}{\partial\theta_\alpha}=-u^{L-1}\left(\chi_1 \theta_\alpha+\hbar\,{\rm tr}\left(\sfE^{(\alpha)t}G \right) \right) + \lO(u^{L-2})
\end{equation}
Taking the coefficient of $u^{L-1}$ in this relation, we finally get
\begin{equation}\label{EErel2}
\mathbb{I}_{1,L-1} + \frac{\partial \mathbb{I}_{1,L-2}}{\partial\theta_\alpha} = -\chi_1 \theta_\alpha-\hbar\,{\rm tr}\left(\sfE^{(\alpha)t}G \right)\,.
\end{equation}
We remind the reader that $\mathbb{I}_{1,\alpha}$ are the operator coefficients in the expansion of the transfer matrices
\begin{equation}
\T_{1,1}(u)=\displaystyle\sum_{\alpha=0}^L u^\alpha \mathbb{I}_{1,\alpha}\,.
\end{equation}
We see that \eqref{EErel2} is a relation between a local operator acting on the $\alpha$-th site (in the l.h.s.) and a global operator acting on all sites (in the r.h.s.).  Sandwiching this relation between left and right transfer matrix eigenstates $\ket{\Psi}$ and $\bra{\Psi}$, we find that the expectation value is given by
\begin{equation}\label{loc1}
-\sfi\,\frac{\bra{\Psi}{\rm tr}\left(\sfE^{(\alpha)t}G \right)\ket{\Psi}}{\braket{\Psi|\Psi}}=\frac{\bra{\Psi} \frac{\partial \mathbb{I}_{1,L-2}}{\partial\theta_\alpha}\ket{\Psi}}{\braket{\Psi|\Psi}}+I_{1,L-1}+\chi_1 \theta_\alpha\,.
\end{equation}
Let us note that this expression does not depend on normalisation of the states $\ket{\Psi}$. The only nontrivial correlator in the r.h.s. is the first term, which is given by the determinant \eqref{Ivar} we derived above in the SoV approach. Thus we find a compact result for the expectation value of the local operator ${\rm tr}\left(\sfE^{(\alpha)t}G \right)$. 

\medskip

We can also repeat a similar argument starting from the transfer matrices in $a$-th antisymmetric  representation in the auxiliary space. We start with $\sla(3)$. A straightforward calculation yields that for the class of representations we consider we have
\begin{equation}
\lL^{\wedge^2}(u-\theta)=(u-\theta-\sfi \bs)\left((u-\theta+\sfi(\bs-1))+\sfi\,\sfE\right)
\end{equation}
from which we express the transfer matrix $\T_{2,1}$ as 
\begin{equation}
\T_{2,1}(u) = {\rm tr}\left(\lL^{\wedge^2}(u-\theta_L)\dots \lL^{\wedge^2}(u-\theta_1)G^{\wedge^2} \right)\,.
\end{equation}
By repeating the same procedure as before we can express the local operator ${\rm tr}\left(\sfE ^{(\alpha)}G^{\wedge^2}\right)$ as 
\begin{equation}\label{loc2}
\sfi\frac{\bra{\Psi}{\rm tr}\left(\sfE^{(\alpha)}G^{\wedge^{2}} \right)\ket{\Psi}}{\braket{\Psi|\Psi}}=\frac{\bra{\Psi} \frac{\partial \mathbb{I}_{2,L-2}}{\partial\theta_\alpha}\ket{\Psi}}{\braket{\Psi|\Psi}}+I_{2,L-1}+\chi_2 \theta_\alpha\,.
\end{equation}
The main point is now that when the twist is diagonal \eqref{loc1} and \eqref{loc2} together with the constraint $\sfE^{(\alpha)}_{11}+\sfE_{22}^{(\alpha)}+\sfE_{33}^{(\alpha)}=-2\bs$ provide an inhomogeneous system of three independent equations for the three form factors $\bra{\Psi}\sfE^{(\alpha)}_{jj}\ket{\Psi}/\braket{\Psi|\Psi}$ which has a unique solution. 

\medskip

The procedure for $\gl(\gn)$ is analogous -- in our particular class of representations in the physical space each of the fused Lax operators $\lL^{\wedge^a}(u)$ contain $(a-1)$ trivial zeroes which multiply a part linear in $\sfE_{ij}$s. The derivatives in $\theta_\alpha$ together with the central charge constraint $\sfE_{11}+\dots \sfE_{\gn\gn}=-2\bs$ provide $\gn$ equations for the $\gn$ unknowns $\sfE_{ii}^{(\alpha)}$ allowing us to solve for them in terms of the form factors $\bra{\Psi}\frac{\partial \mathbb{I}_{a,\alpha}}{\partial \theta_\alpha}\ket{\Psi}/\braket{\Psi|\Psi}$, fully expressible in terms of determinants of Q-functions. 

\medskip

A natural question to ask is how tractable are such determinants of (integrals of) Q-functions from a computational perspective. For low length spin chains with finite dimensional representations there is no problem as Q-functions are simply polynomials and can be computed very easily, see \cite{Marboe:2016yyn}. For non-compact representations the situation is similar provided one is satisfied with some small excitations around the highest-weight state as was demonstrated in the previous section for non-compact $\sla(3)$ spin chains. For representations without highest-weight the determinants in question can be efficiently computed numerically for example using the techniques \cite{Gromov:2015wca}.

\medskip

We note that form factors of exactly the type we can compute here are important e.g. in Landau-Lifshitz models \cite{Gerotto:2017sat}, and it would be interesting to further explore their properties.
Let us also point out that the expectation values of operators like $\partial \T(u)/\partial\theta$ are not straightforwardly accessible by traditional methods of the algebraic Bethe ansatz, but appear to be natural objects in the SoV approach. We believe that exploring the interrelations between the SoV and more standard methods should open the way to computing a still larger class of correlators in the future.

\part{Solving the Yang-Baxter equation}\label{YBEboost}

\section{Local charges and Boost automorphism}

In the previous sections we extensively developed the SoV program for integrable spin chains based on the rational $R$-matrix $R(u,v)=u-v-\hbar\,P$. In this Part we will take a different route and examine more general solutions of the Yang-Baxter equation and develop an efficient procedure for constructing them. 

\medskip

Throughout the history of quantum integrable systems numerous different approaches have been developed for finding solutions of the Yang-Baxter equation. In the early days a very fruitful approach has been through requiring the solutions to have certain symmetries \cite{Kulish:1980ii,Kulish:1981gi,Jimbo:1985ua}. For example, if we wish for the Hamiltonian $\mathbb{H}$ to commute with the generators a of some Lie algebra $\ga$ then one should impose that $[R(u,v),\mathfrak{a}\otimes 1 + 1 \otimes \mathfrak{a}]=0$. More generally given some bialgebra $\lA$ we require that $\Delta^{\rm op}(\mathfrak{a})R(u, v) = R(u, v)\Delta(\mathfrak{a})$ where $\Delta$ and $\Delta^{\rm op}$ denote the coproduct and opposite coproduct on $\lA$, respectively, cf the general discussion in section \eqref{quantumalgebras}. In many cases this is enough to completely fix $R$ up to a small number of functions, drastically simplifying the construction, as was demonstrated in the case of AdS/CFT integrable systems \cite{Beisert:2005tm,Borsato:2014hja,Borsato:2014exa,Lloyd:2014bsa,Borsato:2015mma,Hoare:2014kma,Garcia:2020lrg,Garcia:2020vbz}. Of course, this approach first requires one to know what the corresponding symmetry is and there are $R$-matrices which may have no such symmetry at all. 

\medskip

Still within the realm of algebra, a more abstract  approach is that of Baxterisation which
initially appeared in the realm of knot theory \cite{Turaev:1988eb,Jones:1989ed,jones1990baxterization} and consists of constructing solutions of the YBE as representations of certain algebras, for example Hecke algebras and Temperly-Lieb algebras \cite{Jimbo:1985vd,zhang1991representations,Crampe:2020slf}.

\medskip

A more hands-on approach is to simply try and solve the Yang-Baxter equation directly. The upside to this is that in principle one can obtain all solutions in this way, but this is contrasted with the enormous difficulty of solving cubic functional equations. This approach is usually supplemented with differentiating the YBE and reducing the cubic functional equations to a system of coupled partial differential equations. This approach has recently been used to provide a full classification of $R$-matrices of size $4 \times 4$ so-called 8-and-lower-vertex models \cite{Vieira:2017vnw} satisfying the difference property $R(u,v)=R(u-v)$ and to obtain certain $9\times 9$ models \cite{Vieira:2019vog} whose $R$-matrix satisfies the so-called ice rule but this method quickly becomes unwieldy as the size of the $R$-matrix increases.

\medskip

In this Part we will describe a new approach for constructing solutions of the Yang-Baxter equation which uses a suitably defined \textit{integrable Hamiltonian} as a starting point. 

\subsection{Solutions}

\paragraph{Difference vs non-difference form} 
The $R$-matrix $R(u,v)$ depends on two spectral parameters $u$ and $v$ which in general are totally independent. There is a special class of models for which the $R$-matrix is of \textit{difference form} where $R(u,v)=R(u-v)$. There are numerous physical models which fall into this class such as the XYZ spin chain and its derivatives. As well as this $1+1$-dimensional integrable S-matrices possessing this property correspond to models possessing Poincare invariance such as in the $O(N)$ sigma model. Most models however do not have this property and it is not possessed by the integrable S-matrices associated with the AdS/CFT correspondence, although in certain limits the difference form property can be re-established. Nevertheless it is difference form models are useful and are especially easy to classify owing to the fact that the $R$-matrix only depends on a single independent variable and in this case the Yang-Baxter equation can be written as 
\begin{equation}
R_{12}(u-v)R_{13}(u)R_{23}(v) =R_{23}(v)R_{13}(u)R_{12}(u-v)\,.
\end{equation}

\medskip

We will now give some examples of solutions of the YBE and the integrable systems they correspond to. 

\paragraph{Rational XXX spin chain} 

In the previous sections we studied the Yangian $\lY_{\gn}$ in depth. It has the $R$-matrix 
\begin{equation}
R_{12}(u,v)= (u-v)1_{12} -\hbar P_{12}
\end{equation}
where $P$ is the permutation operator on $\CC^{\gn}$. The Hamiltonian density $\lH_{12}$ is simply given by the permutation operator
\begin{equation}
\lH_{12} = P_{12}\,.
\end{equation}
The corresponding integrable system is the Heisenberg XXX spin chain and possesses $\gl(\gn)$ symmetry. For the simplest case $\gn=2$ the Hamiltonian is given by 
\begin{equation}
\lH_{12}=\frac{1}{2}\left(\sigma_x \otimes \sigma_x + \sigma_y \otimes \sigma_y+ \sigma_z \otimes \sigma_z +1\otimes 1\right)
\end{equation}
where $\sigma_{x,y,z}$ are the usual $\sua(2)$ Pauli matrices. At the point $u=v$ the $R$-matrix becomes proportional to the permutation operator and the $R$-matrix is manifestly of difference form.

\paragraph{Trigonometric XXZ spin chain}

A closely related solution corresponds to the quantum algebra $U_q(\widehat{\sla(\gn)})$ with $R$-matrix given by 
\begin{equation}
\begin{split}
R_{12}(u,v) & = \left(\frac{u}{v}q - \frac{v}{u}q^{-1} \right)\sum_{a=1}^\gn \sfe_{aa}\otimes \sfe_{aa}+ \left(\frac{u}{v} - \frac{v}{u}\right)\sum_{a\neq b}^\gn \sfe_{aa}\otimes \sfe_{bb} \\
& +\left(q-q^{-1}\right)\sum_{a\neq b}^\gn \left(\frac{u}{v} \right)^{\text{sign}(a-b)} \sfe_{ab}\otimes \sfe_{ba}\,.
\end{split}
\end{equation}
The corresponding integrable system is the Heisenberg XXZ spin chain. We will write out the Hamiltonian density explicitly in the simplest case $\gn=2$. Before this, it is convenient to make the transformation
\begin{equation}
R_{12}(u,v) \rightarrow A_1(u) A_2(v) R_{12}(u,v)A_1(u)^{-1}A_2(v)^{-1},\quad A(u) = {\rm diag}\left(u^{-1},1\right)\,.
\end{equation}
This transformation does not modify regularity of the $R$-matrix nor the YBE and leads to an equivalent quantum algebra but leads to a more transparent Hamiltonian which is given by
\begin{equation}
\lH_{12}= \sigma_x \otimes \sigma_x + \sigma_y \otimes \sigma_y+\Delta \sigma_z \otimes \sigma_z + \Delta 1\otimes 1
\end{equation}
where $\Delta =\frac{1}{2}(1+q)(1+q^{-1})$ is the so-called anisotropy parameter and physically corresponds to switching on an external magnetic field in the $z$-direction. Clearly in the $\Delta\rightarrow 1$ limit the model reduces to that of the XXX model up to an overall rescaling of the Hamiltonian. 

\medskip

The quantum algebra $U_q(\widehat{\sla(\gn)})$ contains $U_q(\sla(\gn))$ as a subalgebra but it is not a symmetry of the conserved charges which only have $\mathfrak{u}(1)^{\gn}$ (the Cartan subalgebra) as a symmetry. This is a consequence of the fact that the other conserved charges possess a non-trivial coproduct and so immediately can be seen to not commute with the momentum operator, for example. 

\medskip

Finally, we note that at the point $u=v$ the $R$-matrix again becomes proportional to the permutation operator. The $R$-matrix is also of difference form but in disguise -- it becomes manifestly of difference form by making the replacement $\left( u,v\right) \rightarrow \left( e^u,e^v\right)$.

\paragraph{One-dimensional Hubbard model}
The quantum algebra describing the one-dimensional Hubbard model is based on $\su(2|2)_{\rm ce}:=\su(2|2)\ltimes \RR^2$, where ``${\rm ce}$" refers to ``central extension", owing to the fact that that the algebra $\su(2|2)$ is extended by two central charges. The full quantum algebra is a deformation of the super Yangian $\lY(\sua(2|2))$ \cite{Beisert:2014hya}. The model is not of difference form, but like the previous two examples it becomes proportional to the permutation operator at $u=v$. Owing to its bulky nature we will not write out the $R$-matrix explicitly but it is similar in structure to the model presented in \eqref{AdS2Rmat}.

\medskip

\paragraph{Regularity} The three solutions presented above all share a common feature. Namely, at the point $u=v$ the $R$-matrix simply becomes the permutation operator. This has interesting consequences and allows us to choose the conserved charges, which we label as $\JJ_2,\JJ_3,\dots$, to be such that the charge $\JJ_r$ acts on $r$-neighbouring spin chain sites. The construction is as follows. Any Lax operator $\lL$ (or families of them) allows us to define a quantum integrable system by defining the monodromy matrix $T(u)$ as a product of such Lax operators. Since the $R$-matrix can be itself be viewed as a Lax operator we can define an associated transfer matrix $\T(u,\theta)$ by
\begin{equation}
\T(u,\theta) = {\rm tr}_a\left(R_{aL}(u,\theta)\dots R_{a2}(u,\theta)R_{a1}(u,\theta) \right)
\end{equation}
where we have chosen to specify the dependence on $\theta$ -- we still have the property 
\begin{equation}
[\T(u,\theta),\T(v,\theta)]=0\,.
\end{equation}
The spin chain is homogeneous -- we have used the same inhomogeneity parameter $\theta$ at each site. The crucial point which leads to local and homogeneous conserved charges is that at the point $u=\theta$ the transfer matrix becomes the shift operator $U$ along the chain 
\begin{equation}
\T(\theta,\theta) = U
\end{equation}
which we already saw for the special case of the XXX spin chain in Part \ref{Part1}. It can then be checked that the conserved charges $\JJ_2,\JJ_3,\dots$ defined by 
\begin{equation}
{\rm Log}\,\T(u) = 1+(u-\theta)\,\JJ_2(\theta)+\frac{1}{2}(u-\theta)^2\,\JJ_3(\theta)+\dots 
\end{equation}
are such that the charges $\JJ_r$ act on $r$ neighbouring spin chain sites as a sum of local densities, for example 
\begin{equation}
\JJ_2 = \sum_{\alpha=1}^L \mathcal{J}^{(2)}_{\alpha,\alpha+1},\quad \JJ_3 = \sum_{\alpha=1}^L \mathcal{J}^{(3)}_{\alpha,\alpha+1,\alpha+2}
\end{equation}
where the indices are defined modulo $L$. Usually the charge $\JJ_2$ is taken to be the Hamiltonian $\mathbb{H}$ of the model as is the case with the XXX spin chain. As well as locality and homogeneity another highly useful property of the regularity is that the conserved charges all take a universal form. Let us denote the Hamiltonian density as $\lH_{12}$ and so 
\begin{equation}
\mathbb{H}(\theta) = \sum_{\alpha=1}^L \lH_{\alpha,\alpha+1}(\theta)\,.
\end{equation}
Then the higher conserved charge $\JJ_3(\theta)$ takes the form 
\begin{equation}
\mathbb{J}_3 = \sum_{\alpha=1}^L [\lH_{\alpha,\alpha+1},\lH_{\alpha+1,\alpha+2}] + \frac{{\rm d}}{{\rm d}\theta}\mathbb{H}(\theta)\,.
\end{equation}
We will derive this in the next section using the so-called \textit{boost automorphism}. Notice that a consequence of this construction is that the $R$-matrix can always be expressed in terms of the Hamiltonian density as 
\begin{equation}
R_{12}(u,v) = P_{12}\left(1+(u-v)\lH_{12}\left(\frac{u+v}{2}\right) +\lO\left(u-v\right)^2\right)\,.
\end{equation}

\subsection{Sutherland equations and Boost automorphism}

\paragraph{Sutherland equations}

We will now begin setting up a systematic framework for solving the Yang-Baxter equation for R-matrices satisfying the regularity condition. Attempting to solve the YBE directly is a formidable task. It is more efficient to use the \textit{Sutherland equations} which are obtained from the YBE by differentiating and applying the regularity condition. This results in two equations -- the Sutherland equations \cite{sutherland1970two}-- and reads
\begin{equation}\label{eqn:Sutherland}
\begin{split}
& [R_{13}R_{23},\lH_{12}(u)]=\dot{R}_{13}R_{23}-R_{13}\dot{R}_{23}\\
& [R_{13}R_{12},\lH_{23}(v)]=R_{13}R_{12}^\prime-R_{13}^\prime R_{12}\\
\end{split}
\end{equation}
where $R_{ij}:=R_{ij}(u,v)$ and $\dot{R}$ and $R^\prime$ denote the derivatives of R with respect to the first and second variables, respectively. Clearly this is at least some improvement -- we have reduced a functional equation in $3$ variables to two ODEs in two variables. However, for this to be of any use we need to specify initial conditions on R. We have the regularity condition but also need to know the form of $\lH$ as input. This can be done by consistency and was used to great effect in \cite{Vieira:2017vnw} but the process is much simpler if the initial conditions leading to integrable models are known from the start. In other words, how do we know from the start which function $\lH_{12}$ will lead to an $R$-matrix satisfying the Sutherland equations?

\paragraph{Boost automorphism}

Our starting point is the Sutherland equation 
\begin{equation}
\left[R_{13} R_{12}, \lH_{23}(\theta)\right]  =  R_{13}R^\prime_{12} - R^\prime_{13}R_{12} ,
\end{equation}
We now make the replacement $1\mapsto a$, $2\mapsto k$, $3\mapsto k+1$, obtaining 
\begin{equation}\label{localsuther}
\left[R_{a,k+1} R_{ak}, \lH_{k,k+1}(\theta)\right]  =  R_{a,k+1}R^\prime_{ak} - R^\prime_{a,k+1}R_{ak}.
\end{equation}
We now consider an infinite spin chain with monodromy matrix $T_a(u,\theta)$ given by 
\begin{equation}
T_a(u,\theta)=\dots R_{a1}R_{a0}R_{a,-1}\dots.
\end{equation}
Now take \eqref{localsuther} and multiply from the left with the product of $R$-matrices $\dots R_{a,k+2}$ and from the right with $R_{a,k-1}\dots$. We then multiply the resulting equation by $k$ and sum over $k$ from $-\infty$ to $\infty$. The two terms on the right hand side of \eqref{localsuther} telescopically cancel and we are left with 
\begin{equation}
\sum_{k=-\infty}^\infty k\, [T_a(u,\theta),\lH_{k,k+1}(\theta)]=\frac{dT_a(u,\theta)}{d\theta},
\end{equation}
which gives
\begin{equation}
\sum_{k=-\infty}^\infty k\, [\T(u,\theta),\lH_{k,k+1}(\theta)]=\frac{{\rm d}}{{\rm d}\theta}\T(u,\theta)
\end{equation}
after tracing over the auxiliary space. Finally, using the expansion 
\begin{equation}
{\rm Log}\,\T(u,\theta)=\JJ_1(\theta)+(u-\theta)\JJ_2(\theta)+\frac{1}{2}(u-\theta)^2\JJ_3(\theta)+\dots
\end{equation}
we obtain 
\begin{equation}\label{boostedcharges}
\JJ_{r+1}(\theta)=\sum_{k=-\infty}^\infty k\, [\lH_{k,k+1}(\theta),\JJ_r(\theta)]+\frac{{\rm d}}{{\rm d}\theta}\JJ_r(\theta),\quad r=2,3,\dots\,.
\end{equation}
It is common to use the notation $\mathcal{B}[\mathbb{H}]$ to denote the formal sum
\begin{equation}
\mathcal{B}[\mathbb{H}] = \sum_{k=-\infty}^\infty k\, \lH_{k,k+1}
\end{equation}
which is referred to as the \textit{Boosted Hamiltonian}. As we have just seen the boost operator allows us to generate all conserved charges in a recursive fashion, without ever needing to construct the transfer matrix! Indeed, the only input is the Hamiltonian density $\lH_{12}$ which is determined from the $R$-matrix. 

\medskip

We can now attempt to reverse the logic. Let $\lH_{12}(\theta)$ be a linear operator on $\CC^n\otimes \CC^n$ for some $n\geq 2$. \textit{Define} a Hamiltonian $\mathbb{H}$ as 
\begin{equation}
\mathbb{H} = \sum_{\alpha=1}^L \lH_{\alpha,\alpha+1}
\end{equation}
and impose that the tower $\mathbb{H}=\JJ_2,\JJ_3,\JJ_4,\dots$ mutually commute where $\JJ_{r+1}$ is defined recursively from $\JJ_3$ by \eqref{boostedcharges}. This will place very strong constraints on the matrix elements of the density $\lH_{12}$ and we can then ask ourselves: is this the Hamiltonian density of an integrable system obtain from an $R$-matrix? To check this we have to plug the Hamiltonian density into the Sutherland equations and determine R. Note that although the boosted Hamiltonian is defined on a chain of infinite length it is enough to consider finite length chains with periodic boundary conditions as in this case there is a huge cancellation in the commutators. For example, 
\begin{equation}
\sum_{k=-\infty}^\infty k\, [\lH_{k,k+1}(\theta),\JJ_2] = \sum_{k=-\infty}^\infty [\lH_{k,k+1},\lH_{k+1,k+2}]
\end{equation}
and we see that the overall dependence on $k$ has dropped out leading to an expression which can be reduced consistently to finite length. 

\medskip

In order to fully constrain the Hamiltonian density we require that the full set of commutators $[\JJ_r,\JJ_s]=0$ for all $r,s$. For finite length these cannot all be independent since the spin chain Hilbert space is finite-dimensional. On the other hand it seems likely that we will need to consider at least a few such commutators to constrain the Hamiltonian density. The length we should consider depends on which commutators we are looking at. Since $\JJ_2$ is a sum of range $2$ densities and $\JJ_3$ is a sum of range $3$ densities the non-vanishing terms in their commutator $[\JJ_2,\JJ_3]$ is a sum of densities of range $2+3-1=4$. Hence, if we restrict to a spin chain of length $3$ say, then these non-zero commutators will effectively wrap around the spin chain producing cancellations which do not happen in general. Hence we must consider spin chains of at least length $4$ in order to avoid this happening. In case one needs to consider the commutation relations between higher conserved charges, the length of the spin chain needs to be adjusted accordingly - if one wants to consider the commutator $[\JJ_r, \JJ_s]$ then a spin chain of length $L=r+s-1$ should be considered.

We will now explicitly implement the above-described procedure for spin chains with a local Hilbert space of dimension $2$. This is the simplest non-trivial case and includes the XYZ spin chain and its derivatives. 

\subsection{Symmetries and example}

Since our proposed method for solving the YBE is based on the brute-force solving of commutators $[\JJ_r,\JJ_s]=0$ we would like the Hamiltonian density $\lH_{12}$ to have as few free parameters as possible. In our current set-up $\lH_{12}$ is a $4\times 4$ matrix and so has $16$ free parameters. This is rather unwieldy as even just the first commutator $[\JJ_2,\JJ_3]=0$ amounts to solving a system of first-order differential equations which are cubic in the unknown functions -- a monstrous task. We will now explain how various symmetries of the Yang-Baxter equation can be used to reduce this number of free parameters without any loss of generality. 

\medskip

\paragraph{Local basis transformation} 
If $R(u,v)$ is a solution of the Yang-Baxter equation of size $n^2\times n^2$ and $V(u)$ an invertible $n\times n$ matrix then we can generate another solution $R^{(V)}(u,v)$ of the YBE by defining
\begin{align}
R^{(V)}(u,v) = \Big[V(u)\otimes V(v)\Big] R(u,v)  \Big[V(u)\otimes V(v)\Big]^{-1}.
\end{align}
This new solution is trivially compatible with regularity and just corresponds to a change of basis on each site. On the level of the Hamiltonian it gives rise to a new integrable Hamiltonian which takes the form
\begin{align}\label{LBTlaw}
\lH^{(V)} = \big[\!V\otimes V\big] \lH  \big[\!V\otimes V\big]^{-1}\! - \big[\dot{V} V^{-1}\otimes I - I \otimes \dot{V} V^{-1}\big],
\end{align}
where everything is evaluated at $\theta$ and $I$ is the identity matrix. In particular, we see that terms of the form $A\otimes I-I\otimes A$ in the Hamiltonian can be removed by performing the basis transformation \eqref{LBTlaw} with the matrix $V(u)$ satisfying $\dot{V}=A V$ which can be solved by means of a path-ordered exponential. 

\paragraph{Reparameterization}

If $R(u,v)$ is a solution, then $R(g(u),g(v))$ clearly is a solution of the YBE as well. This transformation affects the normalization of the Hamiltonian since by the chain rule the logarithmic derivative of $R$ will give an extra factor $\dot{g}$, so that
\begin{align}
\lH(u) \mapsto \dot{g} \lH(g(u)).
\end{align}
Notice furthermore that this will similarly affect the derivative term in the boost operator. We are also free to reparameterize any other functions and constants in both the $R$-matrix and Hamiltonian.

\paragraph{Normalization}

We can normalize the $R$-matrix in any way we want since multiplying any solution $R$ of the YBE by an arbitrary function $g$ is clearly allowed.  On the level of the Hamiltonian this corresponds to a simple shift of the Hamiltonian
\begin{align}
\lH \mapsto \lH + \dot{g}\, I
\end{align}
where $ I $ is the identity matrix. We have imposed $g(\theta,\theta)=1$ in order to preserve $R(\theta,\theta)=P$. 

\paragraph{Discrete transformations} It is straightforward to see that for any solution $R(u,v)$ of the Yang-Baxter equation, $PR(u,v)P, R^T(u,v)$ and $PR^T(u,v)P$ are solutions as well. We summarise the relation between these $R$-matricies and Hamiltonians below 
\begin{align}
& R &&\leftrightarrow&& \lH \\
& PRP &&\leftrightarrow&& P\lH P \\
& R^T &&\leftrightarrow&& P\lH^TP \\
& PR^TP &&\leftrightarrow&& \lH^T
\end{align}
and emphasise that the Hamiltonian associated to $R(u)^T$ is $P\lH^TP$ and \textit{not} $\lH^T$. 

\medskip

All the above transformations are universal and hold for any integrable model. Moreover, they have a trivial effect on the spectrum, which means that they basically describe the same physical model. Additionally, there are some transformations called twists that we can use for identifications that are model dependent. Twists generically change the spectrum and more generally the physical properties of the integrable model in a non-trivial way. However, on the level of the $R$-matrix a twist is a simple transformation.

\paragraph{Twists}

If $U(u)$ is an invertible $n\times n$ matrix which satisfies $[U(u)\otimes U(v),R_{12}(u,v)]=0$ then it can be shown that 
\begin{equation}\label{twist}
U_2(u)R_{12}(u,v)U_1(v)^{-1}
\end{equation}
is a solution of the YBE provided $R$ is. Note that much more general transformations which preserve the YBE can be obtained by combining \eqref{twist} together with other transformations. For example, if both $U$ and $V$ are constant invertible matrices satisfying $[U\otimes U,R_{12}]=0=[V\otimes V,R_{12}]=0$ then the following is also a solution
\begin{equation}
U_1 V_2 R_{12} U_2^{-1} V_1^{-1}
\end{equation}
which can be obtained by applying \eqref{twist} together with a similarity transformation and applying \eqref{twist} again. We will refer to any transformation obtained by combining \eqref{twist} with the other transformations mentioned above as a \textit{twist}. 

\medskip

Under the transformation \eqref{twist} the Hamiltonian density $\lH_{12}$ transforms as
\begin{equation}\label{twistedH}
\lH_{12}\mapsto U_1 \lH_{12} U_1^{-1}+\dot{U}_1 U_1^{-1}
\end{equation}
and the analogue of the condition $[U(u)\otimes U(v),R_{12}(u,v)]=0$ for the Hamiltonian density can be easily worked out to be 
\begin{equation}\label{twistcond}
[U_1 U_2,\lH_{12}]=\dot{U}_1U_2-U_1\dot{U}_2.
\end{equation}
Alternatively this relation may be derived by plugging the twisted $R$-matrix \eqref{twist} and Hamiltonian \eqref{twistedH} into the Sutherland equations \eqref{eqn:Sutherland} and sending $v\rightarrow u$, which is not surprising given the striking similarity between \eqref{twistcond} and the Sutherland equations.  

\medskip

Finally, there can be other, model dependent, twists such as Drinfeld twists \cite{drinfeld1983constant,Reshetikhin:1990ep} which we will not consider here. 

\paragraph{Worked example}

As a demonstration of our method let us work out an example in full detail. From here on we will use the following notation:
\begin{itemize}
\item $h_i(u)$ are matrix elements of $\lH(u)$
\item $\dot{h}_i(u) = \partial_u h_i(u)$ 
\item $H_i(u) = \int_0^u h_i$ and $H_i(u,v) = \int_v^u h_i = H_i(u)-H_i(v)$
\item $r_i(u,v)$ are matrix elements of $R(u,v)$
\item $\dot{r}_i(u,v) = \partial_u r_i(u,v)$ and $r^\prime_i(u,v) = \partial_v r_i(u,v)$.
\end{itemize}

\paragraph{Hamiltonian} Let us classify all regular solutions of the YBE whose Hamiltonian densities have the following form
\begin{align}
\lH_{12}(\theta) = \begin{pmatrix}
0 & 0 & 0 & 0 \\
0 & h_1(\theta) & h_3(\theta) & 0 \\
0 & h_4(\theta) & h_2(\theta) & 0 \\
0 & 0 & 0 & 0 
\end{pmatrix}.
\end{align}
From the boost operator construction we find that the corresponding charge $\JJ_3$ has density
\begin{align}
\mathcal{J}_{123}(\theta) = \begin{pmatrix}
 0 & 0 & 0 & 0 & 0 & 0 & 0 & 0 \\
 0 & 0 & -h_1 h_3 & 0 & -h_3^2 & 0 & 0 & 0 \\
 0 & h_1 h_4 & \dot{h}_1 & 0 & \dot{h}_3-h_2 h_3 & 0 & 0 & 0 \\
 0 & 0 & 0 & \dot{h}_1 & 0 & \dot{h}_3+h_1 h_3 & h_3^2 & 0 \\
 0 & h_4^2 & \dot{h}_4+h_2 h_4 & 0 & \dot{h}_2 & 0 & 0 & 0 \\
 0 & 0 & 0 & \dot{h}_4-h_1 h_4 & 0 & \dot{h}_2 & h_2 h_3 & 0 \\
 0 & 0 & 0 & -h_4^2 & 0 & -h_2 h_4 & 0 & 0 \\
 0 & 0 & 0 & 0 & 0 & 0 & 0 & 0 
\end{pmatrix}
\end{align}
and is quadratic in the components $h_i(\theta)$ of the Hamiltonian density $\lH$. We have suppressed the $\theta$ dependence.

The next step is to impose $[\JJ_2(\theta),\JJ_3(\theta)]=0$ which gives the equations
\begin{align}
&\dot{h}_3 (h_1+h_2) = (\dot{h}_1+\dot{h}_2) h_3 ,
&&\dot{h}_4 (h_1+h_2) = (\dot{h}_1+\dot{h}_2) h_4 .
\end{align}
These are solved by
\begin{align}
&h_3 = \frac{c_3}{2} (h_1+h_2),
&& h_4 = \frac{c_4}{2} (h_1+h_2),
\end{align}
for some constants $c_{3,4}$. Thus we find that if $\lH_{12}$ is to be obtained from an $R$-matrix it must have the form
\begin{align}\label{HXXZ}
\lH(\theta) = \begin{pmatrix}
0 & 0 & 0 & 0 \\
0 & h_1 & \frac{c_3}{2} (h_1+h_2) & 0 \\
0 & \frac{c_4}{2} (h_1+h_2) & h_2 & 0 \\
0 & 0 & 0 & 0 
\end{pmatrix}.
\end{align}
\paragraph{$R$-matrix}
We make an ansatz for our $R$-matrix of the following form
\begin{align}
R = \begin{pmatrix}
r_1 & 0 & 0 & 0 \\
0 & r_2 & r_3 & 0 \\
0 & r_4 & r_5 & 0 \\
0 & 0 & 0 & r_6
\end{pmatrix}.
\end{align}
We will first solve the Sutherland equations using brute force before using identifications to greatly simplify the process. 
The Sutherland equations \eqref{eqn:Sutherland} give the following independent set of PDEs
\begin{align}
&c_3 r_2 r_6 = c_4 r_1r_5,
&&\frac{\dot{r}_2}{r_2} = \frac{\dot{r}_3}{r_3} +h_1+h_3\frac{r_6}{r_5},
&&\frac{\dot{r}_4}{r_4} = \frac{\dot{r}_3}{r_3} +h_1-h_2,
&&\frac{\dot{r}_1}{r_1} = \frac{\dot{r}_6}{r_6} ,\\
&\frac{\dot{r}_2}{r_2} = \frac{\dot{r}_5}{r_5} ,
&&\frac{\dot{r}_3}{r_3} = \frac{\dot{r}_1}{r_1} +h_2+h_3\frac{r_2}{r_1},
&&\frac{c_3}{2} \Big[\frac{r_4 r_3}{r_1r_5}- \frac{r_6}{r_5}-\frac{r_2}{r_1}\Big]=1.
\end{align}
From this we see that
\begin{align}
&r_6 = A r_1,
&&r_5 = B r_2
&&\Rightarrow &&A c_3 = B c_4.
\end{align}
Since we need to impose regularity $R(u,u) = P$, we find that $A=1$ and $B = c_3/c_4$. Next, we derive that
\begin{align}
r_4 =  r_3 e^{H_1(u,v)-H_2(u,v)}&&\mathrm{with}&& H_i(u,v) = \int^u_v h_i.
\end{align}
We are then left with three unsolved PDEs
\begin{align}
&\frac{\dot{r}_2}{r_2} = \frac{\dot{r}_3}{r_3} +h_1+h_3\frac{r_6}{r_5},
&&\frac{\dot{r}_3}{r_3} = \frac{\dot{r}_1}{r_1} +h_2+h_3\frac{r_2}{r_1},
&&\frac{c_3}{2} \Big[\frac{r_4 r_3}{r_1r_5}- \frac{r_1}{r_5}-\frac{r_2}{r_1}\Big]=1.
\end{align}
In order to solve these we redefine 
\begin{align}
&r_{1} \mapsto r_3 \Big(\tilde{r}_{1} - \frac{\tilde{r}_2}{c_4}\Big),
& &r_{2} \mapsto r_3 \tilde{r}_{2},
&& r_3 \mapsto r_3
\end{align}
so that the last equation becomes
\begin{align}
& c_4^2 e^{H_1-H_2}=c^2_4 \tilde{r}_1^2 +\omega^2\tilde{r}_2^2,
\end{align}
where $H_i=H_i(u)-H_i(v)$ and we have put $\omega^2=c_3 c_4-1$. 
This equation can now be most conveniently solved by substituting cylindrical coordinates, so that we find
\begin{align}\label{eq:sphXXZ}
&\tilde{r}_1 =  e^{\frac{H_1-H_2}{2}} \cos \phi,
&&\tilde{r}_2 =  e^{\frac{H_1-H_2}{2}} \frac{c_4}{\omega} \sin \phi,
\end{align}
for some function $\phi$ to be determined by the remaining two differential equations. Notice that this is an overdetermined system. Plugging \eqref{eq:sphXXZ} then back into the remaining Sutherland equations gives the following 
\begin{align}
\frac{\dot{\phi}}{\omega} = \frac{h_1+h_2}{2},
\end{align}
which is easily solved upon using the boundary condition that $\phi(u,u) = 0$.
Setting $H_\pm(u,v) = \frac{H_1(u,v) \pm H_2(u,v) }{2}$ and combining everything we are left with the following $R$-matrix
\begin{align}\label{eq:RXXZ}
R =e^{H_+}\begin{pmatrix}
 \cos \omega H_+ -  \frac{\sin \omega H_+}{\omega} & 0 & 0 \\
0 & c_4\frac{\sin \omega H_+}{\omega}  &  e^{-H_-} & 0 \\
0 &  e^{H_-} & c_3 \frac{\sin \omega H_+}{\omega} & 0 \\
0 & 0 & 0 & \cos \omega H_+ - \frac{\sin \omega H_+}{\omega}
\end{pmatrix}
\end{align}
after choosing the overall normalisation $r_3$ to correctly reproduce the Hamiltonian. 
Owing to the dependence on both $H_+$ and $H_-$, this $R$-matrix is manifestly of non-difference form. It is straightforward to check that $R$ indeed satisfies the Yang-Baxter equation and that its logarithmic derivative gives the density Hamiltonian \eqref{HXXZ}. 

\paragraph{Using symmetries}
The above method of finding the $R$-matrix can be greatly simplified if we use some identifications that relate various solutions of the Yang-Baxter equation that we discussed in the previous section.

\medskip

We start from \eqref{HXXZ} and use a local basis transformation to set $h_1=h_2$. This is achieved using the matrix $V(\theta)$ with 
\begin{equation}\label{LBTapply}
V(\theta)={\rm exp}\left(\frac{1}{2}H_-(\theta)\sigma_z\right),\quad H_\pm(\theta)=\frac{1}{2}\left(H_1(\theta)\pm H_2(\theta)\right)
\end{equation}
together with the transformation law \eqref{LBTlaw}. Next, we use reparameterization symmetry to set $h_1=h_2=1$. Thus, it follows that all the entries of the Hamiltonian are constant and the resulting Hamiltonian density has the form
\begin{align}\label{Hc3c4}
\lH(\theta) = \begin{pmatrix}
0 & 0 & 0 & 0 \\
0 & 1 & c_3 & 0 \\
0 & c_4 & 1 & 0 \\
0 & 0 & 0 & 0 
\end{pmatrix}.
\end{align}
Moreover, we can use a twist and set $c_3=c_4=c$. Indeed, it is trivial to check that the twist condition \eqref{twistcond} is satisfied for any constant invertible diagonal matrix $U$ and the matrix
\begin{equation}
U={\rm diag}\left(\sqrt{c_4},\sqrt{c_3}\right),
\end{equation}
can be used to bring the Hamiltonian density to the form
\begin{align}\label{diffformham}
\lH(\theta) = \begin{pmatrix}
0 & 0 & 0 & 0 \\
0 & 1 & c & 0 \\
0 & c & 1 & 0 \\
0 & 0 & 0 & 0 
\end{pmatrix},
\end{align}
after applying $\lH_{12}\mapsto U_1 \lH_{12} U_1^{-1}$. 

The Sutherland equations are now also easily solved since all the coefficients of the Hamiltonian are simply constants. As a consequence, the $R$-matrix is of difference form and is given by the usual XXZ solution. Putting $\omega^2=c^2-1$ we find 
\begin{align}\label{eq:XXZcons}
R =   e^u\begin{pmatrix}
 \cos \omega u -\frac{\sin \omega u}{\omega}  & 0 & 0 \\
0 & c \frac{\sin \omega u}{\omega} &  1 & 0 \\
0 & 1  & c \frac{\sin \omega u}{\omega} & 0 \\
0 & 0 & 0 &  \cos \omega u -\frac{\sin \omega u}{\omega}
\end{pmatrix}.
\end{align}
In order to see that this solution is equivalent to the solution \eqref{eq:RXXZ}, let us undo the identifications that we performed to make the Hamiltonian constant. First we undo the twist and apply $R_{12}\mapsto U_2^{-1}R_{12}U_1$ to \eqref{eq:XXZcons} and put $c=\sqrt{c_3}\sqrt{c_4}$ so that we arrive at the $R$-matrix for the Hamiltonian \eqref{Hc3c4}. Next we reparameterize
\begin{equation}
u\mapsto H_+(u)
\end{equation}
and finally we apply the inverse of the local basis transformation \eqref{LBTapply}, immediately obtaining \eqref{eq:RXXZ}.

\paragraph{Difference vs. Non-difference} After using all the identifications, we see that \eqref{eq:RXXZ} is actually just an $R$-matrix of difference form in disguise. The non-difference nature of the rapidity dependence of the $R$-matrix only resides in local basis transformations, a rescaling and a reparameterization. These can obviously be applied to any solution of difference form to generate a non-difference form solution. In the remainder of this work we will also encounter models which are genuinely of non-difference form, but it is easy to see already at the level of the Hamiltonian if this is the case. More precisely, after solving the integrability condition $[\JJ_2(\theta),\JJ_3(\theta)]=0$ our Hamiltonian will depend on a number of free functions. One will usually correspond to a shift, one can be absorbed in a reparameterization of the spectral parameter and then remains a number that can be absorbed by local basis transformations and potentially twists. The exact number of the latter will depend on the set-up. Thus in case of \eqref{HXXZ}, we count 2 free functions $h_1,h_2$ and we could have already at that point concluded that the underlying model was actually of difference form.

\medskip

Notice that in this example imposing $[\JJ_2,\JJ_3]=0$ was enough to completely fix the Hamiltonian to correspond to an $R$-matrix satisfying the YBE. This is related to a conjecture made in \cite{Grabowski:1994rb} where it is suggested that a Hamiltonian (of range 2) is integrable if it commutes with a single higher charge of range $3$. This is also closely related to the fact that a \textit{single} higher conserved charge in $1+1$-dim QFT is enough to guarantee factorisation of the scattering problem \cite{Parke:1980ki}, although two higher charges are needed in theories which are not parity-symmetric. Remarkably, this observation will turn out to be true for \textit{all} cases considered in this work and so far a counter-example where even higher conserved charges are necessary has not yet been found. 

\section{Non-difference form}

\subsection{Non-difference form models}
We will now apply the proposed method to $4\times 4$ R-matrices of non-difference form. In order to simplify the set-up we will consider so-called \textit{8-vertex models} which have the form
\begin{align}
\begin{split}
R  =\begin{pmatrix}
r_1 & 0 & 0 & r_8 \\
0 & r_2 & r_6 & 0 \\
0 & r_5 & r_3 & 0 \\
r_7 & 0 & 0 & r_4 \\
\end{pmatrix}
\end{split},
\label{generalR8vertex}
\end{align}
and, consequently, the corresponding Hamiltonian densities are of the form
\begin{align}
\begin{split}
\mathcal{H}  =\,& 
h_1  \text{ }1 + h_2 (\sigma _z\otimes 1- 1\otimes \sigma _z) + h_3  \sigma _+\otimes \sigma _-  + 
 h_4 \sigma_-\otimes \sigma _+  \\
& +    h_5 ( \sigma _z \otimes 1 +  1 \otimes \sigma _z ) + 
h_6 \sigma _z\otimes \sigma _z  + h_7 \sigma _-\otimes
\sigma _- + h_8 \sigma _+\otimes \sigma _+.
\end{split}
\label{generalHamiltonian}
\end{align}
It is worth stressing that the restriction to 8-vertex models is based on physical grounds. Indeed, an S-matrix of 8-vertex type is the most general type of S-matrix in a theory containing a single boson and fermion which is consistent with conservation of spin statistics. Indeed, if we allowed other non-zero entries in the $R$-matrix it would allow for scattering processes in which for example two bosons could scatter to produce a boson-fermion pair. 

\medskip

We will now briefly outline the possible solutions. After performing symmetry transformations, one finds only four different types of integrable $4\times 4$ Hamiltonians that solve the integrability condition $[\JJ_2(\theta),\JJ_3(\theta)]=0$. Two of them are of 8-vertex type \eqref{8VBrmatrix} while two more are of so-called 6-vertex type where the $(1,4)$ and $(4,1)$ components in \eqref{8VBrmatrix} are set to $0$. The solutions are given as follows
\begin{itemize}
\item 6-vertex A, $h_6 \neq 0$ and $h_7=h_8 = 0$ 
\item 6-vertex B, $h_6 = h_7=h_8 = 0$ 
\item 8-vertex A, $h_6\neq0,h_7\neq0,h_8\neq0$
\item 8-vertex B, $h_6=0$ and $h_7\neq0,h_8\neq0$.
\end{itemize}
Let us discuss the models in more detail.

\paragraph{6-vertex A} 
Setting $h_7=h_8=0$ and assuming $h_6\neq 0$ we find that $[\JJ_2(\theta),\JJ_3(\theta)]=0$ is satisfied if and only if
\begin{align}
&h_3 = c_3 h_6 e^{4 H_5},
&&h_4 = c_4 h_6 e^{-4 H_5},
\end{align}
where $c_{3,4}$ are constants. The Hamiltonian is actually equivalent to that of the XXZ spin chain. Indeed, by applying a local basis transformation, twist, reparameterization and normalization we can bring the Hamiltonian density to the form 
\begin{equation}
\lH=\left(
\begin{array}{cccc}
0 & 0 & 0 & 0 \\
0 & 1 & c & 0 \\
0 & c & 1 & 0 \\
0 & 0 & 0 & 0 
\end{array}
\right),
\end{equation}
which is precisely the Hamiltonian density \eqref{diffformham}, and so its $R$-matrix is given by \eqref{eq:XXZcons}. Notice that this solution also contains the most general diagonal Hamiltonian since only the off-diagonal elements $h_{3,4}$ are restricted by the integrability condition.

\paragraph{6-vertex B}\label{6vertexB} If we take $h_6=h_7=h_8=0$ then it makes the Hamiltonian satisfy $[\JJ_2,\JJ_3]=0$ for \textit{any} choice of $h_1, \ldots, h_5$. So, the Hamiltonian depends on five free functions. Three of these functions can be absorbed in identifications. In particular, a local basis transformation ($h_2$), a normalization ($h_1$) and a reparameterization of the spectral parameter ($h_3$). Moreover, it is convenient to redefine $h_5 \rightarrow \frac{1}{2} h_4 h_5$.

\medskip

We normalize the $R$-matrix such that $r_5=1$ and then it follows from the Sutherland equations \eqref{eqn:Sutherland} that
\begin{align}
&r_7=r_8=0,
&&r_6=1,
&&\dot{r_2} = 
h_4(r_1-h_5 r_2) ,
&&\dot{r}_4 =  -h_4(r_3+h_5 r_4),
&& r_1 r_4 + r_2 r_3 =1,
\end{align}
while $r_4$ satisfies the second order version of the Riccati equation
\begin{align}
\ddot r_4-\frac{\dot{h}_4}{h_4}\dot{r}_4+h_4 r_4\Big[h_3 + \dot{h}_5-h_4h_5^2\Big]=0.
\end{align}
We now introduce a reparameterization of the spectral parameter 
\begin{align}
u_i  \mapsto x_i = \int^{u_i}  \frac{\dot h_5}{h_4 h_5^2-h_3},
\end{align}
which kills the non-derivative term in the Riccati equation and removes the explicit dependence on $h_3$. It is then straightforward to solve our system of differential equations to find
\begin{align}
r_2(x,y) &= H_4(x,y),\\
r_1(x,y) &= 1 +  h_5(x)H_4(x,y), \\
r_3(x,y) &=   h_5(x) h_5(y) H_4(x,y) -h_5(x) + h_5(y) ,\\
r_4(x,y) &= 1 -h_5(y) H_4(x,y),
\end{align}
where again $H_i(x,y) = \int_y^x h_i$.\\
It is instructive to write the $R$-matrix as
\begin{align}
R = H_4(x,y)
\begin{pmatrix}
h_5(x) & 0 & 0 & 0 \\
0 & 1 & 0 & 0 \\
0 & 0 & h_5(x)h_5(y) & 0 \\
0 & 0 & 0 & -h_5(y) 
\end{pmatrix}
+
\begin{pmatrix}
1 & 0 & 0 & 0 \\
0 & 0 & 1 & 0 \\
0 & 1 & h_5(y) - h_5(x) & 0 \\
0 & 0 & 0 & 1 \\
\end{pmatrix}.
\end{align}
We see that $h_5$ gives rise to the non-difference nature of this solution. In particular, when $h_5$ is constant the $R$-matrix reduces to an $R$-matrix of XXZ type. It is easy to show that it satisfies the Yang-Baxter equation and the correct boundary conditions. This model can be mapped by a twist into the solution A of the pure coloured Yang-Baxter equation considered in \cite{6vColored}.

\paragraph{8-vertex A} In the case $h_6\neq0$, the integrability constraint gives that 
\begin{align}
& h_4=h_3 = c_3 h_6,
&&h_5 =0,
&& h_7 = c_7 h_6 e^{4H_2},
&& h_8 = c_8 h_6 e^{-4H_2},
\end{align}
where $c_i$ are constants. The resulting Hamiltonian is that of the XYZ spin chain \cite{Kulish1982,Vieira:2017vnw} under our symmetry identifications.

\paragraph{8-vertex B} In the case when $h_6=0$, we find the following differential equations
\begin{align}
&   \frac{\dot{h}_7}{h_7} = 4 h_2 +  \frac{\dot{h}_3+\dot{h}_4}{h_3+h_4} + 4 \frac{h_3-h_4}{h_3+h_4} h_5 ,\label{8VB1}\\
&   \frac{\dot{h}_8}{h_8} = -4 h_2 +  \frac{\dot{h}_3+\dot{h}_4}{h_3+h_4} +  4\frac{h_3-h_4}{h_3+h_4} h_5,\\
& \frac{\dot{h}_5}{h_5} =- \frac{h_3^2-h_4^2}{4 h_5}+ \frac{\dot{h}_3+\dot{h}_4}{h_3+h_4} +4\frac{h_3-h_4}{h_3+h_4} h_5\label{8VB2}.
\end{align}
We use a local basis transformation to set $h_2 = 0$ and then these equations are solved by
\begin{align}
&h_5= - \frac{1}{4} (h_3+h_4) \tanh (H_3-H_4 + c_5), \label{8vb1}\\
&h_7=c_7 \frac{h_3+h_4}{ \cosh(H_3-H_4+c_5)},\label{8vb2}\\
&h_8=c_8 \frac{h_3+h_4}{ \cosh(H_3-H_4+c_5)}.\label{8vb3}
\end{align}
By using a local basis transformation we can set $c_8=c_7$ and after applying further identifications the remaining functions can be brought to the following form 
\begin{align}
& h_3 = \frac{1}{2}\csc(\eta(v))(2-\dot{\eta}(v)), \\
& h_4 = \frac{1}{2}\csc(\eta(v))(2+\dot{\eta}(v))
\end{align}
where $\eta$ is some free function. This further results in $h_7=h_8=2 c_7:= k$, which all together imply that $r_5=r_6=1$ and $r_7=r_8$ for the $R$-matrix. The remaining functions are easily determined from the Sutherland equations and we find
\begin{align}
r_8(u,v) = k \frac{
\mathrm{sn} (u-v,k^2) \mathrm{cn}(u-v,k^2)}{\mathrm{dn}(u-v,k^2)},
\end{align}
where $\mathrm{sn,cn,dn}$ are the usual Jacobi elliptic functions with modulus $k^2$ and
\begin{align}\label{8VBrmatrix}
r_1 &= 
\frac{1}{\sqrt{\sin \eta(u)}\sqrt{\sin\eta(v)}}  \bigg[\sin\eta_+\frac{\mathrm{cn}}{\mathrm{dn}} 
-\cos\eta_+ \mathrm{sn}\bigg], \\
r_2 &= 
\frac{1}{\sqrt{\sin \eta(u)}\sqrt{\sin\eta(v)}}  \bigg[\cos\eta_-\mathrm{sn} +\sin\eta_-\frac{\mathrm{cn}}{\mathrm{dn}} 
\bigg],\\
r_3 &= 
\frac{1}{\sqrt{\sin \eta(u)}\sqrt{\sin\eta(v)}}  \bigg[\cos\eta_-\mathrm{sn}  - \sin\eta_-\frac{\mathrm{cn}}{\mathrm{dn}} 
\bigg],\\
r_4 &= 
\frac{1}{\sqrt{\sin \eta(u)}\sqrt{\sin\eta(v)}}  \bigg[\sin\eta_+\frac{\mathrm{cn}}{\mathrm{dn}} 
+\cos\eta_+ \mathrm{sn}\bigg],
\end{align}
where $\eta_\pm = \frac{\eta(u) \pm \eta(v)}{2}$ and all the Jacobi elliptic functions depend on the difference $u-v$, \textit{i.e.} $\mathrm{sn} = \mathrm{sn}(u-v,k^2)$.
This solution indeed satisfies the Yang-Baxter equation and has the correct boundary conditions. Moreover, it is easy to see that in the case where $\eta$ is constant, it becomes 
of difference form and reduces to the well-known solution found in \cite{8v,Khachatryan:2012wy,Vieira:2017vnw}. 

\paragraph{8VB$^\prime$} 
As can be seen from \eqref{8VB1}-\eqref{8VB2}, the cases where $h_5=0$ and $h_3=-h_4$ need special attention due to possible singularities. In particular it is easy to see that by setting $h_5=0$ it follows that the Hamiltonian is constant unless $h_3=-h_4$. And, indeed,  in our final expression the limit $h_5 =0$ corresponds to setting $\eta(x) = \pi/2$.

\medskip

However, the case $h_3=-h_4$ warrants special attention. In this case, the entries of the Hamiltonian are
\begin{align}
& h_1=h_2=h_5=h_6=0,
&&h_7 =c_8\; h_8,
&& h_3 =-h_4.
\label{offdiagmodel}
\end{align}
We see that the Hamiltonian for this model only has off-diagonal entries. It can be shown that it is possible to recover this model, starting from the Hamiltonian of 8-vertex B. Since the procedure is highly non-trivial, we explain the steps of this identification. 

\medskip

In order to recover \eqref{offdiagmodel} we followed the following steps:
\begin{itemize}
\item[1.] To the Hamiltonian density $\lH_{8VB}$ with entries \eqref{8vb1}-\eqref{8vb3}, we apply the off-diagonal constant twist
\begin{align}
U=
\begin{pmatrix}
 0 & a \\
 b & 0 
\end{pmatrix}
\end{align}
to obtain $\tilde \lH_{8VB}= U_1 \lH_{8VB}U_1^{-1}$. In order to make $\tilde \lH_{8VB}$ verify the integrability condition $[\JJ_2,\JJ_3]=0$, we fixed one entry of the twist $a\to s_1 \sqrt{c_8} b$, with $s_1=\pm 1, \pm i$ and we had to impose a constraint on the entries of the Hamiltonian
\begin{align}
&h_3+h_4=\alpha _3'\;,
&&h_3-h_4=\frac{\alpha _3' \sinh \alpha _3}{\sqrt{\cosh ^2\alpha _3+1}},
\end{align}
with $\alpha_3$ some $\theta$-dependent function. Notice that this twist is non-standard as is does not satisfy \eqref{twistcond}. 
\item[2.] We apply a diagonal local basis transformation $V(\theta)$. In particular by using \eqref{LBTlaw}, we first fix $\dot{V}V^{-1}$ to eliminate the elements in the (2,2) and (3,3) positions of the Hamiltonian. Then by solving the differential equations, we fixed the matrix $V(\theta)$.
\item[3.] We get an off-diagonal Hamiltonian density and we checked that the sum of the elements at position 2,3 and 3,2 is zero if $s_1$ (defined in step 1) is $\pm i$. Moreover the ratio between elements in 1,4 and 4,1 is constant.
\end{itemize}
In this way we have recovered model \eqref{offdiagmodel} from $\lH_{8VB}$. Since the twist that we used is non-standard, it is unclear how to easily lift it to the level of the $R$-matrix. Nevertheless, it is easy to solve the Sutherland equations for this model directly and we obtain
\begin{align}
R_{{\rm 8VB}^\prime} = \begin{pmatrix}
 \cosh H_3(u,v) & 0 & 0 & \sin H_7(u,v) \\
 0 & -\sinh H_3(u,v)  & \cos  H_7(u,v)  & 0 \\
 0 & \cos H_7(u,v)  & \sinh  H_3(u,v)  & 0 \\
 \sin H_7(u,v) & 0 & 0 & \cosh H_3(u,v) 
\end{pmatrix}.
\end{align}
We see that it is of quasi-difference form, meaning all of the dependence on the spectral parameters is of the form $H_3(u)-H_3(v)$ and $H_7(u)-H_7(v)$.

\subsection{Difference-form models}

We will now consider the case of difference form where the $R$-matrix satisfies $R(u,v)=R(u-v)$. In this setting a number of simplifications take place, most notably the derivative term vanishes in the recursive expression for the conserved charges using the boost and as a result we obtain 
\begin{equation}
\JJ_{r+1} = [\lB[\mathbb{H}],\JJ_{r}]
\end{equation}
and hence the solving the integrability condition $[\JJ_2,\JJ_3]=0$ amounts to solving a system of cubic polynomial equations instead of differential equations. As a result of this simplicity we are in a position to classify \textit{all} solutions of the difference-form YBE, not just those of 8-vertex type.

\medskip

Before presenting the solutions we will comment on a certain peculiarity regarding the expansion of the $R$-matrix in terms of the Hamiltonian. We already know that we can write (restricting to difference form) 
\begin{equation}
R(u) = P\left(1 + u\,\lH + \mathcal{O}(u^2) \right)\,.
\end{equation}
The higher order corrections can by found by requiring that we reproduce the form of the conserved charges generated by the boost automorphism and satisfy the YBE. It is not hard to work out that at the next order in the expansion we have
\begin{equation}
R(u) = P\left(1 + u\,\lH + \frac{u^2}{2}\, \lH^2+\mathcal{O}(u^3) \right)\,.
\end{equation}
This is highly suggestive. We are then led to make an ansatz for the $R$-matrix of the form 
\begin{equation}
R(u) = P\left(\sum_{k=0}^\infty g_k(u) \lH^k \right)
\end{equation}
for some functions $g_k(u)$. We can then exploit the Cayley-Hamilton theorem -- since our vector space is $4$ dimensional all powers $\lH^k$ for $k\geq 4$ can be expressed in terms of $1,\,\lH,\,\lH^2,\,\lH^3$ and so our ansatz reduces to 
\begin{equation}\label{CHform}
R(u) = P\left(f_0(u)+f_1(u)\lH + f_2(u)\lH^2+ f_3(u) \lH^3 \right)
\end{equation}
and we can also use the freedom to renormalise the $R$-matrix to set $f_0(u)=1$ -- since the $R$-matrix is assumed to be regular $f_0(u)$ must be non-zero in some neighbourhood of $0$, and the other functions must be of the form 
\begin{equation}
f_k(u) = u^k \times\, \text{analytic},\quad k=1,2,3
\end{equation}
where ``analytic" refers to some function which is analytic at $0$. 

\medskip

Remarkably, despite the simplicity of our ansatz in captures a large number of models! Indeed, it can be checked that the XXX, XXZ and XYZ spin chains can all be expressed in this form. However, the coefficients are model-dependent which suggests that the $R$-matrix cannot be expressed generically as a power series in the Hamiltonian and hence something special must happen for the models where this ansatz works. 

\medskip

In order to see what is going on let's expand the Yang-Baxter equation up to third order as follows 
\begin{equation}
R(u) = P\left(1 + u\,\lH + \frac{u^2}{2}\, \lH^2+\frac{u^3}{3!}\, \left(\lH^3+G\right)+\mathcal{O}(u^4) \right)\,.
\end{equation}
for some operator $G$. If we then plug this ansatz into the YBE we find that the function $G$ must satisfy
\begin{equation}
G_{12}-G_{23} = [\lH_{12}+\lH_{23},[\lH_{12},\lH_{23}]]\,.
\end{equation}
This condition was previously obtained by Reshethikin \cite{10.1007/3-540-11190-5_8} and the function $G$ is known as Reshethikin's $G$-function. We can then repeat to fourth order and remarkably no new function is needed. We find 
\begin{equation}
R(u)=P\left(1 + u\,\lH +\frac{u^2}{2}\lH^2 + \frac{u}{3!}\left(\lH^3+G\right)+\frac{u^4}{4!}\left(\lH^4+G\lH+\lH G\right) +\mathcal{O}(u^5) \right)\,.
\end{equation}
Unfortunately at higher orders things become less clear - it is generically not possible to express the coefficient of $u^5$ in terms of $G$ and $\lH$ and a new function $\tilde{G}$ is needed which satisfies 
\begin{equation}
\tilde{G}_{12}-\tilde{G}_{23} =\, \text{polynomial in}\, \lH_{12},\, \lH_{23},\,G_{12},\, G_{23}
\end{equation}
and since $G_{12}-G_{23}$ can be expressed in terms of $\lH$ it is not clear what the simplest representation is. Furthermore it seems likely that at higher orders we need more and more functions. 

\medskip

Mysteriously, the existence of Reshethikin's $G$-function for a given $\lH$ seems to be \textit{equivalent} to the Yang-Baxter equation, despite the fact that we need at least another function $\tilde{G}$ and presumably an infinite number of others. To see this, one can consider the commutator $[\JJ_2,\JJ_3]=0$. Explicitly working it out on an infinite chain then lead to the existence of a function which we call $G$ such that
\begin{equation}
[\JJ_2,\JJ_3]=0  \quad \text{if and only if} \quad G_{12}-G_{23} =[\lH_{12}+\lH_{23},[\lH_{12},\lH_{23}]]\,.
\end{equation}
Since all evidence suggests that the condition $[\JJ_2,\JJ_3]=0$ is enough to ensure integrability and higher commutators are unnecessary this suggests that Reshethikin's $G$-function guarantees integrability. We do not have a proof of this but all evidence suggests this is the case and it has been extensively tested. The models for which the form \eqref{CHform} applies then seem to be rather degenerate since $G$ and all of the higher functions $\tilde{G}$ are expressable as polynomials in $\lH$ itself, even though this applies to a wide range of models -- certainly one does not think of the XYZ spin chain as being particularly degenerate due to the presence of a number of elliptic functions, yet remarkably the form \eqref{CHform} holds. 

\medskip

We will now present the solutions of integrability condition for difference form models. The first family of solutions are well-known 8-and-lower vertex models. Their Hamiltonians take the form
\begin{align}
\lH^{XYZ} = \begin{pmatrix}
a_1 & 0 & 0 & d_1 \\
0 & b_1 & c_1 & 0 \\
0 & c_2 & b_2 & 0 \\
d_2 & 0 & 0 & a_2
\end{pmatrix}.
\end{align}
There are eight independent generators of this type. These models are well-known in the literature but for completeness, we will list these Hamiltonians explicitly.

\paragraph{Diagonal (4 vertex)} Any diagonal Hamiltonian gives rise to an integrable system
\begin{align}
\mathcal{H}^{XYZ}_1 =
\begin{pmatrix}
a_1 & 0 & 0 & 0 \\
0 & b_1 & 0 & 0 \\
0 & 0 & b_2 & 0 \\
0 & 0 & 0 & a_2 
\end{pmatrix}.
\end{align}

\paragraph{XXZ} There are two families of XXZ type, which agrees with \cite{BFdLL2013integrable}
\begin{align}
&\mathcal{H}^{XYZ}_2 =
\begin{pmatrix}
a_1 & 0 & 0 & 0 \\
0 & b_1 & c_1 & 0 \\
0 & c_2 & b_2 & 0 \\
0 & 0 & 0 & a_1 
\end{pmatrix},
&&\mathcal{H}^{XYZ}_3 =
\begin{pmatrix}
a_1 & 0 & 0 & 0 \\
0 & b_1 & c_1 & 0 \\
0 & c_2 & b_2 & 0 \\
0 & 0 & 0 & -a_1-b_1-b_2 
\end{pmatrix}.
\end{align}
\paragraph{7--Vertex} There are two families of models which are of 7--vertex type
\begin{align}
&\mathcal{H}^{XYZ}_4 =
\begin{pmatrix}
a_1 & 0 & 0 & d_1 \\
0 & a_1+b_1 & c_1 & 0 \\
0 & -c_1 & a_1-b_1 & 0 \\
0 & 0 & 0 & a_1 
\end{pmatrix},
&&\mathcal{H}^{XYZ}_5 =
\begin{pmatrix}
a_1 & 0 & 0 & d_1 \\
0 & a_1- c_2 & c_1 & 0 \\
0 & c_2 & a_1- c_1 & 0 \\
0 & 0 & 0 & a_1-c_1-c_2 
\end{pmatrix}.
\end{align}
\paragraph{8--Vertex} Finally, there are three families of models which have all coefficients non-zero
\begin{align}
&\mathcal{H}^{XYZ}_6 =
\begin{pmatrix}
a_1 & 0 & 0 & d_1 \\
0 & b_1 & c_1 & 0 \\
0 & c_1 & b_1 & 0 \\
d_2 & 0 & 0 & a_1 
\end{pmatrix},
&\mathcal{H}^{XYZ}_7 =
\begin{pmatrix}
a_1 & 0 & 0 & d_1 \\
0 & b_1 & c_1 & 0 \\
0 & c_1 & b_1 & 0 \\
d_2 & 0 & 0 & 2b_1-a_1
\end{pmatrix}, \\
&\mathcal{H}^{XYZ}_8 =
\begin{pmatrix}
a_1 & 0 & 0 & d_1 \\
0 & a_1 & b_1 & 0 \\
0 & -b_1 & a_1 & 0 \\
d_2 & 0 & 0 & a_1 
\end{pmatrix}.
\end{align}
All corresponding $R$-matrices are listed in \cite{Vieira}.

\medskip

We now list the remaining class of models. Remarkably, the R-matrices for \text{all} of these models can be written in the form \eqref{CHform}. In some cases we will write the $R$-matrix explicitly as a $4\times 4$ matrix or in the form \eqref{CHform}, whichever is most convenient. The models are 

\paragraph{Class 1}
The generator of the next class of Hamiltonians we find takes the form
\begin{align}\label{eq:Hnilp}
\lH_1 = \begin{pmatrix}
0 & a_1 & a_2 & 0 \\
0 & a_5 & 0 & a_3 \\
0 & 0 & -a_5 & a_4 \\
0 & 0 & 0 & 0
\end{pmatrix},
\end{align} 
where $a_1 a_3-a_2 a_4 =0$. Its $R$-matrix is given by
 \begin{align}
R_1(u) =
 \begin{pmatrix}
1 & \frac{a_1 (e^{a_5 u}-1)}{a_5}&  \frac{a_2(1-e^{-a_5 u}) }{a_5} &  \frac{a_1 a_3+a_2a_4}{a_5^2}(\cosh (a_5 u) -1) \\
0 & 0 & e^{-a_5 u} & \frac{a_4(1-e^{-a_5 u}) }{a_5} \\
0 & e^{a_5 u} & 0 & \frac{a_3 (e^{a_5 u}-1)}{a_5} \\
0 & 0 & 0 & 1
\end{pmatrix}. 
\end{align} 
It is easy to check that this $R$-matrix is regular, satisfies the Yang-Baxter equation as well as braided unitarity, $R_{12}(u)R_{21}(-u)\sim 1$. 

\paragraph{Class 2}
The second class of integrable Hamiltonians is 
\begin{equation}
\lH_{2}=\left(
\begin{array}{cccc}
 0 & a_2 & a_3-a_2 & a_5 \\
 0 & a_1 & 0 & a_4 \\
 0 & 0 & -a_1 & a_3-a_4 \\
 0 & 0 & 0 & 0 \\
\end{array}
\right),
\end{equation}
which has the $R$-matrix 
\begin{align}
R_{2}(u)= u P\Big[\,\frac{a_1 }{\sinh (a_1 u)}+ \lH_{2} +\frac{ \tanh(\frac{a_1 u}{2})}{a_1}  \lH^2_{2}  \Big].
\end{align}
This $R$-matrix is regular, satisfies the Yang-Baxter equation as well as braided unitarity, $R_{12}(u)R_{21}(-u)\sim 1$.

\paragraph{Class 3}

The third family of solutions is generated by
\begin{align}
\lH_{3} = 
\begin{pmatrix}
 -a_1 & \left(2 a_1-a_2\right) a_3 & \left(2 a_1+a_2\right) a_3 & 0 \\
 0 & a_1-a_2 & 0 & 0 \\
 0 & 0 & a_1+a_2 & 0 \\
 0 & 0 & 0 & -a_1 \\
\end{pmatrix},
\end{align}
which has the following $R$-matrix 
\begin{equation}
R_{3}(u)=\left(\begin{array}{cccc}
e^{-a_1 u} & a_3 \left(e^{(a_1-a_2)u}-e^{-a_1u}\right) & a_3 \left(e^{(a_1+a_2)u}-e^{-a_1u}\right) & 0 \\
0 & 0 & e^{(a_1+a_2)u} & 0 \\
0 & e^{(a_1-a_2)u} & 0 & 0 \\
0 & 0 & 0 & e^{-a_1 u}
\end{array} \right).
\end{equation}
This Hamiltonian can be seen as a deformation of a specific case of the four-vertex model, with deformation parameter $a_3$. When we set $a_3=0$ we obtain 
\begin{align}
\lH_{12} = 
\begin{pmatrix}
 -a_1 & 0 & 0 & 0 \\
 0 & a_1-a_2 & 0 & 0 \\
 0 & 0 & a_1+a_2 & 0 \\
 0 & 0 & 0 & -a_1 
\end{pmatrix},
\end{align}
which has an $R$-matrix which appeared in the classification of \cite{Vieira}. This $R$-matrix can be expressed in terms of powers of $\lH$ as 
\begin{equation}
R_{12}(u)=P_{12}(f_0(u)+uf_1(u)\lH+u^2f_2(u)\lH^2) ,
\end{equation}
where $f_j(u)$ are easily determined functions of $u,a_1,a_2$. What is rather remarkable is that the $R$-matrix is \textit{the same function} of $\lH$ for both $a_3=0$ and $a_3\neq 0$: $a_3$ enters the $R$-matrix only through the Hamiltonian, and does not appear in the coefficient functions $f_j(u)$.

\paragraph{Class 4}
The next independent generator has a similar structure as $\lH_3$ and is
\begin{align}
\lH_4 = \begin{pmatrix}
a_1 & a_2 & a_2 & a_3 \\
0 & -a_1 & 0 & a_4 \\
0 & 0 & -a_1 & a_4 \\
0 & 0 & 0 & a_1
\end{pmatrix},
\end{align}
with $R$-matrix
\begin{equation}
R_{4}(u)=\left(
\begin{array}{cccc}
 e^{a_1 u } & \frac{a_2 \sinh (a_1 u )}{a_1 } & \frac{a_2 \sinh (a_1 u )}{a_1 } & \frac{e^{a_1 u } (a_2 a_4+ a_1 a_3 \coth (a_1 u )) \sinh ^2(a_1u )}{a_1 ^2} \\
 0 & 0 & e^{-a_1 u } & \frac{a_4 \sinh (a_1 u )}{a_1 } \\
 0 & e^{-a_1 u } & 0 & \frac{a_4 \sinh (a_1 u )}{a_1 } \\
 0 & 0 & 0 & e^{a_1 u } \\
\end{array}
\right).
\end{equation}
Braided unitarity is again satisfied.

\paragraph{Class 5}
The fifth family has a different off-diagonal structure
\begin{equation}\label{eq:weirdfishnet}
\lH_{5} = \left(\begin{array}{cccc}
a_1 & a_2 & -a_2 & 0 \\
0 & -a_1 & 2a_1 & a_3 \\
0 & 2a_1 & -a_1 & -a_3 \\
0 & 0 & 0 & a_1
\end{array}\right).
\end{equation}
The corresponding $R$-matrix is again regular and unitary
\begin{align}
R_{5} = (1-a_1 u)\left(
\begin{array}{cccc}
 2 a_1 u+1 & a_2 u & -a_2 u & a_2 a_3 u^2 \\
 0 & 2 a_1 u & 1 & -a_3 u \\
 0 & 1 & 2 a_1 u & a_3 u \\
 0 & 0 & 0 & 2 a_1 u+1 \\
\end{array}
\right).
\end{align}

\paragraph{Class 6}
The final integrable Hamiltonian is
\begin{align}
\lH_6 = \begin{pmatrix}
a_1 & a_2 & a_2 & 0 \\
0 & -a_1 & 2a_1 & -a_2 \\
0 & 2a_1 & -a_1 & -a_2 \\
0 & 0 & 0 & a_1
\end{pmatrix},
\end{align}
together with the unitary $R$-matrix
\begin{equation}
R_{6}(u)=(1-a_1 u)(1+2a_1 u)
\begin{pmatrix}
 1 & a_2 u & a_2 u & -a_2^2 u^2(2 a_1 u+1 ) \\
 0 & \frac{2 a_1 u}{2 a_1 u+1} & \frac{1}{2 a_1 u+1} & -a_2 u \\
 0 & \frac{1}{2 a_1 u+1} & \frac{2 a_1 u}{2 a_1 u+1} & -a_2 u \\
 0 & 0 & 0 & 1
\end{pmatrix}.
\end{equation}
This $R$-matrix satisfies braiding unitarity as well.

\medskip

\paragraph{Properties of the new models}

Let us briefly discuss some properties of the new classes of integrable models that we have encountered.
A feature which arises for generic choice of parameters in all of these models is non-diagonalisability of the corresponding Hamiltonians. In some cases this is more severe than in others - for example some of the Hamiltonians we find are nilpotent, i.e. they only have eigenvalue zero. A less severe case is those Hamiltonians which are non-diagonalisable but still contain different eigenvalues - in other words the conserved charges contain non-trivial Jordan blocks. While models with similar properties have been studied before, see \cite{Gainutdinov:2016pxy}, there has recently been a surge of interest in them due to their appearance in the conformal fishnet theories \cite{Caetano:2016ydc,Gromov:2017cja,Ipsen:2018fmu}. Models with non-trivial Jordan structure also appear in the context of Temperley-Lieb or Hecke type integrable models \cite{2011JSMTE..04..007M}. However, it can be checked that none of our newly formed models fall in this category. 

\paragraph{Class 1 and 2}

The conserved charges in models 1 and 2 are nilpotent. Nilpotency of the Hamiltonian is a feature of fishnet models as well \cite{Ipsen:2018fmu}.

\paragraph{Class 3, 4, 5 and 6}
 
While generically these Hamiltonians are non-diagonalisable they are actually diagonalisable for certain values of the parameters. In particular,
\begin{itemize}
\item Class 3 is diagonalizable if $a_3=0$, in which case it reduces to a simple 4 vertex model.
\item Class 4 is diagonalizable if $a_2=a_4$ and $a_1a_3=a_2a_4$.
\item Class 5 is diagonalizable if $a_2+a_3=0$.
\item Class 6 is diagonalizable if $a_2=0$.
\end{itemize}
Remarkably, all eigenvalues seem to only depend on the parameter $a_1$. Hence the eigenvalues of Hamiltonians of Classes 3 and 4 correspond to the eigenvalues of the integrable spin chain with Hamiltonian density $\mathcal{H} = S^z\otimes S^z$. The eigenvalues for the spin chains of Classes 5 and 6 correspond to a spin chain with the Hamiltonian density
$\mathcal{H} = 1-2P$.

\section{Integrable deformations of the ${\rm AdS}_2\times S^2 \times T^6$ model}

As an application of the constructed $R$-matrices we will discuss how to obtain an integrable deformation of the S-matrix governing the scattering of massive particles in the ${\rm AdS}_2\times S^2 \times T^6$ model \cite{Hoare:2014kma}. We start by reviewing the corresponding symmetry algebra and how to obtain the $R$-matrix from it and we very closely follow \cite{Hoare:2014kma}.

\paragraph{Symmetry algebra}
The symmetry algebra of this model is $\psu(1|1)_{\rm ce}$ where ${\rm ce}$ denotes ``central extension". That is we adjoin three central charges to $\psu(1|1)$ to obtain 
\begin{equation}
\psu(1|1)_{\rm ce}:=\psu(1|1)\rtimes \RR^3\,.
\end{equation}
The algebra is spanned by two fermionic generators $\fkQ_\pm$ along with three bosonic central charges $\fkP_\pm$ and $\fkC$ subject to the relations 
\begin{equation}
\begin{split}
& \{\fkQ_\pm,\fkQ_\pm \}=2 \fkP_\pm \\
& \{\fkQ_+,\fkQ_- \}=2 \fkC\\
\end{split}\,.
\end{equation}
The representation we are considering acts on a boson $\ket{\phi}$ and a fermion $\ket{\psi}$ 
and we identify 
\begin{equation}
\ket{\phi} = \left(
\begin{array}{c}
1 \\
0
\end{array}
 \right),\quad \ket{\psi} = \left(
\begin{array}{c}
0 \\
1
\end{array}
 \right)\,.
\end{equation}
In this basis the supercharges $\fkQ_\pm$ take the form 
\begin{equation}
\fkQ_+=\left(
\begin{array}{cc}
0 & b \\
a & 0
\end{array}
\right),\quad \fkQ_-=\left(
\begin{array}{cc}
0 & d \\
c & 0
\end{array}
\right)\,.
\end{equation}
We denote by $P_\pm$ and $C$ the eigenvalues of $\fkP_\pm$ and $\fkC$ respectively. Then the commutation relations imply 
\begin{equation}\label{closure}
ab = P_+,\quad cd = P_-,\quad ad+bc = 2C\,.
\end{equation}
So far we have not specified the real form we are working with. It is given by 
\begin{equation}
\fkQ_+^\dagger = \fkQ_-,\quad \fkP_+^\dagger = \fkP_-,\quad \fkC^\dagger=\fkC
\end{equation}
which further implies the constraints 
\begin{equation}
a^* = d,\quad b^*=c,\quad C^*=C,\quad P_+^* = P_-\,.
\end{equation}
Furthermore, it follows from the closure relations \eqref{closure} that
\begin{equation}
C^2 = \frac{(ad-bc)^2}{4}+P_+P_-\,.
\end{equation}
The parameter $m:=ad-bc$ is interpreted of the mass of the particle being scattered and is free to take any value. 

\paragraph{Graded coproduct}

We need to equip the above algebra with a co-algebra structure. The coalgebra structure is not the standard one $\Delta(x) = x \otimes 1 + 1 \otimes x$ and is instead constructed by twisting with an invertible (bosonic) element $\fkU$ which we will define in a moment. In terms of this the coproduct is given by
\begin{equation}
\begin{split}
& \Delta(\fkQ_\pm)= \fkQ_\pm\otimes 1 + \fkU^{\pm}\otimes \fkQ_\pm \\
& \Delta(\fkP_\pm)= \fkP_\pm\otimes 1 + \fkU^{\pm 2}\otimes \fkP_\pm\\
& \Delta(\fkC)=\fkC\otimes 1 + 1\otimes \fkC \\
& \Delta(\fkU)=\fkU\otimes \fkU\,.
\end{split}
\end{equation}
It is important to remember that since we are dealing with a superalgebra the tensor product is also graded and hence for elements $x_{1,2}$ and $y_{1,2}$ we have
\begin{equation}
(x_1\otimes y_1)(x_2 \otimes y_2)=(-1)^{[x_2][y_1]} x_1 x_2 \otimes y_1 y_2
\end{equation}
where $[x]$ denotes the grading for the element $x$, which is $0$ for bosonic elements and $1$ for fermionic ones.

\medskip

If $x$ is a central element it must cocommute with $R$, which if assumed to be invertible, implies $\Delta^{{\rm op}}(x) = \Delta(x)$. This implies strong restrictions on the relations between $\fkP_\pm$ and $\fkU$. In particular we have
\begin{equation}
\fkP_\pm = \frac{{\rm h}}{2} (1-\fkU^{\pm 2})
\end{equation}
where ${\rm h}$ is a generic number which is required to be real by our reality properties and is the interpretation of a coupling constant. 

\medskip

Denote by $U$ the eigenvalue of $\fkU$ on the representation. Our set of constraints can be conveniently solved by introducing the Zhukovski variables $x^\pm$ defined by
\begin{equation}
U^2 = \frac{x^+}{x^-},\quad 2C+m = \sfi {\rm h}(x^--x^+),\quad x^++\frac{1}{x^+}-x^- -\frac{1}{x^-}=\frac{2\sfi m}{{\rm h}}\,.
\end{equation}
In terms of these the entries $a,b,c,d$ of the supercharges $\fkQ_\pm$ take the simple form
\begin{equation}
\begin{split}
& a = e^{-\sfi \frac{\pi}{4}}\left(\frac{x^+}{x^-}\right)^{\frac{1}{4}}\sqrt{\frac{{\rm h}}{2}}\,\eta, \quad b = e^{-\sfi \frac{\pi}{4}}\left(\frac{x^-}{x^+}\right)^{\frac{1}{4}}\sqrt{\frac{{\rm h}}{2}}\frac{\eta}{x^-}\\ 
& c = e^{\sfi \frac{\pi}{4}}\left(\frac{x^+}{x^-}\right)^{\frac{1}{4}}\sqrt{\frac{{\rm h}}{2}}\frac{\eta}{x^+}, \quad d = e^{\sfi \frac{\pi}{4}}\left(\frac{x^-}{x^+}\right)^{\frac{1}{4}}\sqrt{\frac{{\rm h}}{2}}\,\eta\\ 
\end{split}
\end{equation}
where $\eta = \sqrt{\sfi(x^--x^+)}$. The $R$-matrix is now fixed in terms of these parameter by requiring that $\Delta^{\rm op}(a)R = R\Delta(a)$ together with requiring that $R$ satisfies the Yang-Baxter equation. It is important to note that the representation we have constructed in terms of $x^\pm$ can be different for each particle and so we use the labels $x^\pm_u$ for the first and $x^\pm_v$ for the second. We can solve for $x^\pm_u$ in terms of $u$ by introducing $x(u)$ with 
\begin{equation}
x(u)+\frac{1}{x(u)} = \frac{2m}{{\rm h}}u
\end{equation}
and then defining $x^\pm_u = x(u\pm\frac{i}{2})$. 

\medskip

In order to write down the ${\rm AdS}_{2}$ $R$-matrix we must introduce some notation. First, we introduce the functions
\begin{equation}
s^{\varepsilon_1\, \varepsilon_2}(u,v)=i\, \frac{x_u^{\varepsilon_1}-x_v^{\varepsilon_2}}{\gamma_u\,\gamma_v}\left(\frac{x^+_u}{x^-_u}\right)^{\frac{1}{4}(1-\varepsilon_1)}\left(\frac{x^+_v}{x^-_v}\right)^{\frac{1}{4}(1-\varepsilon_2)}\left(1+ \frac{1-\frac{1}{x^{-\varepsilon_1}_u x^{-\varepsilon_2}_v}}{x_u^{\varepsilon_1}-x_v^{\varepsilon_2}}f_{u,v} \right)
\end{equation}
where $\varepsilon_1$ and $\varepsilon_2$ are signs, $\varepsilon_1,\varepsilon_2 \in\, \{+,-\}$ and the function $f_{u,v}$ is given by 
\begin{equation}
f_{u,v}=\frac{\left(x^-_u - \frac{1}{x^+_u}\right)\sqrt{\frac{x^+_u}{x^-_u}}-\left(x^-_v - \frac{1}{x^+_v}\right)\sqrt{\frac{x^+_v}{x^-_v}}}{1-\frac{1}{x^+_u x^-_u x^+_v x^-_v}}
\end{equation}
and as usual we have 
\begin{equation}
\gamma_u =\left(\frac{x^+_u}{x^-_u} \right)^{\frac{1}{4}}\sqrt{i\left(x^-_u-x^+_u\right)}\,.
\end{equation} 
Finally, we put 
\begin{equation}
s(u,v) = \frac{f_{u,v}}{\sqrt{x^+_u x^-_u x^+_v x^-_v}}\,.
\end{equation}
In terms of these functions the ${\rm AdS}_{2}$ $R$-matrix $R^{{\rm AdS}_{2}}$ is given by 
\begin{equation}\label{AdS2Rmat}
R^{{\rm AdS}_2}(u,v)=\left( 
\begin{array}{cccc}
s^{-+} & 0 & 0 & s \\
0 & s^{++} & 1 & 0 \\
0 & 1 & s^{--} & 0 \\
s & 0 & 0 & s^{+-}
\end{array}
\right)\,.
\end{equation}

\paragraph{Embedding in 8VB}

Now that we have constructed the ${\rm AdS}_2$ $R$-matrix we will explain how to embed it in our previously obtained $R$-matrices. Obviously, it can only fit into 8VA or 8VB, and we can quickly rule out 8VA since the $(4,4)$ component is not equal to the $(1,1)$ component. 

\medskip

The main issue to be overcome is that the spectral parameters appearing in both models are different despite being denoted by the same letters $u$ and $v$. To get around this we need to transform $(u,v) \mapsto (G(u),G(v))$ in one of the $R$-matrices and we take this to be in $R^{8VB}$. We start by considering the $(1,4)$ component of both $R$-matrices which are, for $R^{8VB}$ and $R^{{\rm AdS}_2}$ respectively, 
\begin{align}
&(R^{8VB})_{14} = k\, \mathrm{sn}(G(u)-G(v))\frac{\mathrm{cn}(G(u)-G(v))}{\mathrm{dn}(G(u)-G(v))}, \\
&(R^{{\rm AdS}_2})_{14} =  \frac{1}{\sqrt{x^+_u x^-_u x^+_v x^-_v}}\, \frac{\left(x^-_u - \frac{1}{x^+_u}\right)\sqrt{\frac{x^+_u}{x^-_u}}-\left(x^-_v - \frac{1}{x^+_v}\right)\sqrt{\frac{x^+_v}{x^-_v}}}{1-\frac{1}{x^+_u x^-_u x^+_v x^-_v}}
\end{align}
Clearly, the $(1,4)$ component of $R^{\rm 8VB}$ is of difference form, that is it only depends on the difference $G(u)-G(v)$ of the spectral parameters. Let us now expand the $(1,4)$ component of the AdS2 $R$-matrix in $u$ around $v$. We find 
\begin{equation}
\frac{\left(x^+ x^-\right)'}{2\sqrt{x^-}\sqrt{x^+}(x^+x^--1)}(u-v)+\mathcal{O}\left((u-v)^2 \right)
\end{equation}
In order to be purely of difference form we must have that the coefficient of $u-v$ is a constant which we denote $A$:
\begin{equation}
\frac{\left(x^+ x^-\right)'}{2\sqrt{x^-}\sqrt{x^+}(x^+x^--1)} = A.
\end{equation}
Hence, after reinstating the $G$ dependence, we solve to obtain
\begin{equation}\label{eq:AdS2Xpm}
x^+(v) = \frac{{\rm Tanh}\left(A G(v) +\frac{c_1}{2}\right)}{x^-(v)}\,.
\end{equation}
This completely fixes $G$ in terms of $x^\pm$.

After substituting \eqref{eq:AdS2Xpm} back into the $(1,4)$ component of the AdS2 $R$-matrix we find that it reduces to simply\footnote{Working in an appropriate region such that we avoid branch cut issues}
\begin{equation}
(R^{{\rm AdS}_2})_{14} = -{\rm Tanh}\left(A(G(u)-G(v))\right)\,.
\end{equation}
A comparison with the $(1,4)$ component with the 8VB $R$-matrix then tells us that we should take the limit $k\rightarrow \infty$ in order to have this entry reduce to ${\rm Tanh}$ and furthermore the precise agreement requires that $A=-i$ and we can take $c_1=0$, and so we find that 
\begin{equation}
x^+(u) = -\frac{{\rm Tan}^2(G(u))}{x^-(u)}\,.
\end{equation}

\medskip

Next, we make the substitution $\eta(u) \rightarrow {\rm arccot}\left(k F(u)\right)$ and expand the 8VB $R$-matrix around $k \rightarrow \infty$. By subsequently expanding around $u=v$ we find that setting
\begin{equation}
F(u) = -\frac{1}{2}{\rm csc}G(u)\, {\rm sec}G(u)\, \frac{{\rm cot}G(u)\,x^-+i}{{\rm cot}G(u)\,x^--i}
\end{equation}
indeed reproduces the AdS${}_2$ $R$-matrix. 

\medskip

Since neither of the functions $F$ and $G$ depend on $k$ and the only $k$ dependence appeared in the function $\eta(u)$ and in the Jacobi elliptic functions we can simply restore the $k$-dependence and hence obtain a deformation of the ${\rm AdS}_2$ $R$-matrix which continues to satisfy the Yang-Baxter equation. We plan to return to the analysis of its physical properties in the future. 

\part{Summary and outlook}

\paragraph{Summary}

\medskip

Let us summarise the results of this work. After reviewing some fundamentals of quantum algebras, integrable systems and the separation of variables program for $\gl(2)$ spin chains we began investigating higher-rank SoV. We constructed a twist matrix \eqref{MCTcomponent} such that the structure of the $\bB$ operator simplified drastically and the Gelfand-Tsetlin generators emerged providing a direct link between SoV and Yangian representation theory. Armed with the GT basis we proved that the spectrum of $\bB$ coincided with the GT algebra. 

\medskip

One of our main results is a commutation relation \eqref{BTcomm} between $\bB$ and fused transfer matrices $\T_\lambda$ which allowed us to generate eigenvectors of $\bB$ in a way which guaranteed separation of variables. We demonstrated in various examples how the Young diagram $\lambda$ controls the excitations of the Gelfand-Tsetlin patterns labelling $\bB$ eigenvectors for rectangular $(S^A)$ representations.

\medskip

Next, we introduced the embedding morphism $\phi$ \eqref{embtwo} as a way to deal with generic representations and situations where the spectrum of $\bB$ is degenerate. The embedding morphism was an embedding of a $\gl(k)$ spin chain into a $\gl(k+1)$ spin chain which had extremely simple action on the Gelfand-Tsetlin basis and the $\bB$ operator with $\phi(\bB^{(k)}) \sim \bB^{(k+1)}$. By applying $\phi$ to the commutation relation \eqref{BTcomm} we were able to fully diagonalise $\bB$ and show that every eigenvector coincided with a GT basis vector in the auxiliary singular twist limit. 

\medskip

Our next goal was to show that the constructed basis was an SoV basis. We showed that the previously constructed basis of $\bB$ eigenvectors could be equivalently constructed by using ratios of fused transfer matrices which coincided with transfer matrices constructed from the B\"acklund flow procedure. It was then a simple application of Wronskian formulae to obtain the factorised wave functions as an ascending product of Slater determinants 
\begin{equation}\label{SoVwave3}
\Psi(\svx) =\braket{\svx|\Psi}= \displaystyle \prod_{\alpha=1}^L \prod_{k=1}^{\gn-1} \det_{1\leq i,j\leq k}\hat{\sfq}_i(\svx^\alpha_{kj})\,.
\end{equation}
The restriction of this formula to the case of symmetric power representations $(S^1)$ immediately implied that transfer matrix eigenstates could be constructed by repeated application of the $\bB$ operator
\begin{equation}
\ket{\Psi} = \prod_{j=1}^M \bB(u_j)\ket{\Omega}
\end{equation}
bypassing the nested Bethe ansatz and proving the conjecture of \cite{Gromov:2016itr} for $\gl(\gn)$ spin chains.

\medskip

We then began the task of using our new SoV wave functions to compute various quantities of interest just as scalar products of Bethe states and correlation functions. Instead of constructing the SoV measure directly we developed a functional approach to orthogonality relations generalising a construction which first appeared in the AdS/CFT context \cite{Cavaglia:2018lxi} and for non-compact spin chains \cite{Cavaglia:2019pow} to compact spin chains. 

\medskip

The developed functional scalar product linked the SoV wave functions \eqref{SoVwave3} to another set of SoV-like wave functions and so we sought out an operatorial construction of them. This involved introducing a new operator called $\bC$ which was analogous to $\bB$ for factorising left wave functions. Both of these operators corresponded to two different quantisations of classical separated variables which coincided in the classical $\hbar \rightarrow 0$ limit. 

\medskip

In order to begin the procedure of diagonalising $\bC$ and constructing the new SoV basis we introduced the $*$-map, a Yangian anti-automorphism with the property $\bC=\bB^*$. This allowed us to transfer all of the techniques developed for diagonalising $\bB$ to $\bC$, in particular the commutation relation \eqref{BTcom}. This introduced a new set of transfer matrices $\T_\lambda^*$ which we identified as transfer matrices corresponding to right-aligned (skew) Young diagrams instead of left-aligned and, similar to the relation between $\bB$ and $\bC$, constituted an alternate quantisation of classical conserved charges. We then repeated the same story as was done for $\bB$ resulting in the factorisation of left wave functions
\begin{equation}
\Psi(\svy) =\braket{\Psi|\svy}= \displaystyle \prod_{\alpha=1}^L \prod_{k=1}^{\gn-1} \det_{1\leq i,j\leq k}\hat{\sfq}^i(\svy^\alpha_{kj})
\end{equation}
as an ascending chain of Slater determinants but now featuring Hodge dual Q-functions $\hat{\sfq}^i$. 

\medskip

At this point we had extensively developed the SoV program for compact spin chains. We had access to left and right SoV wave functions as well as a highly efficient formalism for computing overlaps in this basis given by the functional orthogonality relations. We then moved on to analysing non-compact spin chains. We began by analysis the Gelfand-Tsetlin algebra for low rank $\gl(2)$ and $\gl(3)$ cases and found a generalisation of the GT branching rules for these infinite-dimensional representations. 

\medskip

Next we generalised the functional orthogonality relations to the non-compact setting. Remarkably, the only modifications were in the integration contour, which became the whole real line in contrast to the compact case which featured a circle containing finitely-many poles, and the integration measure $\mu_\alpha$ which now contained an infinite number of poles and was carefully selected to ensure convergence of the integral and functional orthogonality. 

\medskip

Afterwards, we began the task of constructing the SoV bases along the lines of what we achieved for the compact case to match with the functional approach. Diagonalisation of $\bC$ was straightforward with the only difference being the range of the separated variable spectrum. The situation was more involved for $\bB$ - we no longer had access to the lowest weight state previously used to generate $\bB$ eigenvectors from. We managed to get around this issue by demonstrating that $\bB$ could also be diagonalised starting from the highest-weight state. Doing so required application of the Yangian antipode map and allowed us to obtain a new version of the commutation relation \eqref{BTcomm} but involving fused transfer matrices built from the inverse monodromy matrix.

\medskip

After a rather abstract journey we then included numerous explicit examples of various computations using the developed techniques in $\sla(2)$ and $\sla(3)$ spin chains. We explicitly computed various overlaps and showed how they matched computations from the functional orthogonality approach and also demonstrated that the SoV measure could be computed exactly and precisely matches the functional results. 

\medskip

We finished our SoV discussion by demonstrating how our developed SoV framework involving the interplay between operator and functional approaches could be used to compute various non-trivial quantities. A key tool for this was the det-product \eqref{detproduct} which allowed us to present the obtained quantities as simple determinants in Q-functions. We successfully used it to compute overlaps between transfer matrix eigenstates corresponding to different twists, overlaps with insertions of $\bB$ and $\bC$ operators and on-shell/off-shell overlaps as well as form factors of local operators, a special case of an overlap with an insertion corresponding to a derivative of the transfer matrix with respect to some parameter.

\medskip

We then changed direction and turned our attention to the Yang-Baxter equation with the aim of developing efficient new techniques for constructing integrable spin chains. Our proposal was based on the boost automorphism which allows the conserved charges of regular models to be constructed in a systematic recursive fashion with the Hamiltonian density being the key input. By solving the integrability condition $[\JJ_2,\JJ_3]=0$ we found a plethora of new integrable systems and in all cases could reconstruct the corresponding $R$-matrix. 

\medskip

Among the new models found we obtained a complete classification of $4\times 4$ $R$-matrices which preserve fermion number in scattering processes which include the $R$-matrices of integrable superstrings on ${\rm AdS}_2 \times S^2 \times T^6$, ${\rm AdS}_3 \times S^3 \times S^3 \times S^1$ and ${\rm AdS}_3 \times S^3 \times T^4$, at least in their massive and chirality-preserving sectors, but these have recently been extended to the full models \cite{deLeeuw:2021ufg}, see also \cite{Pribytok:2021akv} for a recent overview. As an application of our classification we demonstrated how the $R$-matrix of the ${\rm AdS}_2$ model could be embedded into our classification. This provided a source of a tunable parameter $k$ hence allowing us to obtain integrable deformations of this superstring model.

\bigskip

\paragraph{Future directions}

We now summarise some future research directions, some of which are formal developments and the others are applications of the developed techniques. 

\medskip

The focus of this work has been on highest-weight bosonic spin chains. An open question is the generalisation of the discussed techniques to the case of non highest-weight representations and supersymmetric spin chains and more general classes of highest-weight representations such as those with $\sua(\mathsf{p},\mathsf{q}|\mathsf{m})$ symmetry necessary for AdS/CFT applications. For the construction of the SoV basis in a rigorous way a natural starting point is the construction of the Gelfand-Tsetlin basis and the development of the branching rules. An SoV basis was constructed in \cite{Maillet:2019ayx} for the case of the defining representation of $\gl(\gm|\gn)$ super spin chains and the Hubbard model, and it would be interesting to attempt relating the constructed basis to the $\bB$-type operator constructed in \cite{Gromov:2018cvh}, as well as generalise findings beyond the fundamental representation, as it was done here in the bosonic setting.

\medskip

One should also generalise the discussed techniques to models based on the principal series representations of $\gl(\gn)$. The SoV framework for models with principal series representations of $\gl(2)$ has been carried out in \cite{Derkachov:2001yn,Derkachov:2002tf}, with some initial progress being made for the $\gl(3)$ case in \cite{Derkachov:2018ewi}. A feature of the principal series setting is that, in contrast to the compact case, it is not necessary to introduce a boundary twist in order for the $\bB$-operator to be diagonalisable, and hence such a twist is not usually employed. However, doing so may be beneficial as the B-operator can still be related to the Gelfand-Tsetlin subalgebra with the use of the companion twist. Study of the Gelfand-Tsetlin subalgebra in the principal series setting was carried out in \cite{Valinevich:2016cwq,ValinevichGT}. The SoV framework in the principal series setting of $\gl(2)$ was recently utilised in \cite{Derkachov:2018rot} for the computation of Basso-Dixon correlators in two-dimensional fishnet CFT \cite{Gurdogan:2015csr,Kazakov:2018qez} and a set of separated variables for the case of $\mathfrak{so}(1,5)$ spin chains were constructed in \cite{Derkachov:2019tzo} and used in the computation of four-point correlation functions in \cite{Derkachov:2020zvv} which are related to the computations of \cite{Basso:2019xay}.

\medskip

It would be interesting to extend our results to other quantum integrable models not based on the rational $R$-matrices. In particular, an SoV basis for the case of $U_q(\widehat{\sla(\gn)})$ was constructed in \cite{Maillet:2018rto} and it would be interesting to check if it diagonalises the $\bB$ operator proposed in \cite{2001math.ph...9013S}. The simplest case to examine would be with the so-called spin torus model \cite{Hao:2016eeu} which is a special case of the $U_q(\widehat{\sla(\gn)})$ spin chain with a twist matrix which is a reduction of the companion matrix and so the techniques developed here are likely to be directly applicable. 

\medskip

One very promising direction of research regards the computation of correlation functions in $4d$ fishnet CFT and eventually $\lN=4$ SYM. For the case of $4d$ fishnet CFT one has access to the holographic dual \cite{Gromov:2019aku,Gromov:2019bsj,Gromov:2019jfh} which is essentially an $\sla(4)$ spin chain albeit in a non-highest-weight representation. Nevertheless the functional scalar product approach to overlaps and correlators was recently developed \cite{Cavaglia:2021mft} and would be very instructive to match it with an explicit diagonalisation of the $\bB$ operator to single out the natural basis of Q-functions. In a similar manner to the closed ${\rm AdS}_5$ fishchain mentioned above an open fishchain has also been constructed \cite{Gromov:2021ahm} to describe cusped Wilson loops in a certain ladders limit of Feynman diagrams. This warrants the further development of separation of variables for open integrable systems with high-rank symmetry which has recently been analysed in \cite{Maillet:2019hdq}.

\medskip

An open problem is the development of the SoV program for models based on algebras other than the $A_r$ series. The Q-system of $\mathfrak{so}(2r)$ integrable systems has been extensively developed recently \cite{Ferrando:2020vzk,Ekhammar:2020enr,Ekhammar:2021myw} which will hopefully lead to an analogue of the functional integral approach to scalar products developed here for $A_r$ models. 

\medskip

Another interesting problem regards the construction of the SoV basis in the Gaudin model. This was carried out by Sklyanin in \cite{sklyanin1989separation} for $\gl(2)$ highest-weight representations. Recently there has been a large amount of interest in the Gaudin model for higher-rank and non-highest-weight representations. It has been shown in \cite{Roehrig:2020kck} that the $\sla(2)$ Gaudin model describes tree-level scattering amplitudes of ambitwistor strings on ${\rm AdS}_3\times S^3$. As well as this it was demonstrated in \cite{Buric:2020dyz,Buric:2021ywo} that conformal blocks of five and higher point functions correspond to the eigenfunctions of the conserved charges in an $\mathfrak{so}(1,5)$ Gaudin model. 

\medskip

Regarding our work on the Yang-Baxter equation there are also numerous avenues of future research. The mechanism we proposed for constructing integrable spin chains produces closed spin chains. For physical applications open spin chains \cite{sklyanin1988boundary,mezincescu1991integrable} are important such as the example described above regarding cusped Wilson loops. It would be interesting to construct finite length open versions for all the new models constructed in this work. In order to do that, the first
step would be the construction of all possible integrable boundary conditions, meaning all solutions of the Boundary Yang-Baxter equation for each of the $R$-matrices introduced here.

\medskip

One natural direction involves applications to holography and the integrable systems which appear in that context. Although we did not discuss it in this work we have successfully used our construction to obtain new $R$-matrices with $\sua(2)\times \sua(2)$ symmetry \cite{deLeeuw:2020xrw}, similar to Shastry's $R$-matrix \cite{Shastry_1986} for the one-dimensional Hubbard model. Generalised Shastry-type models provide a base for a search of new types of solutions that are relevant for $ AdS_{4,5} $ integrable models. In particular it would be interesting to search for new deformations of the $ AdS_{4,5} $ $S$-matrix and establish potential contact with $q$-deformations of the underlying twisted Hopf algebra as for $ \eta $-deformed $ AdS_{5} \times S^{5} $ \cite{Beisert:2008tw,Seibold:2020ywq} or for $ \lambda $-deformed systems \cite{Appadu_2017}. We have shown that the $AdS_2$ $R$-matrix could be embedded into the $4\times4$ model 8VB and admitted a one-parameter deformation. It would be highly interesting to find a physical interpretation for this parameter and to determine the symmetry algebra of the resulting $R$-matrix and perhaps it is related to the known $\eta$-deformation \cite{Hoare:2018ngg}.

\medskip

Remarkably, all of our solutions of the Yang-Baxter equation can be characterized by the integrability condition $[\JJ_2, \JJ_3] = 0$. It is unclear to us why this is the case. Indeed, all the reverse lines in the flowchart, Figure 1, can be shown to hold. The reverse arrows that we exploit here, however, appear to be valid as well and it seems to indicate an equivalence relation. It would be very important to understand and prove these relations. There are also interesting related mathematical questions to be asked. In the case of difference form models the condition $[\JJ_2, \JJ_3] = 0$ results in a set of cubic polynomial
equations for the Hamiltonian entries which seems to be fully equivalent to the Yang-Baxter equation. It would be highly interesting to construct a proof of this claim and in doing so perhaps obtain a closed form expression for the $R$-matrix in terms of the Hamiltonian entries. In this work we have relied on a brute force approach to solving the constraint $[\JJ_2, \JJ_3] = 0$ and to a large extent have exhausted the cases where such an approach is applicable.

\medskip

In order to make more progress it could be important to make use of the extensive toolbox of algebraic geometry. Indeed, $[\JJ_2, \JJ_3] = 0$ describes an algebraic variety in projective space described by a set of coupled, cubic polynomials. For instance, in the $4 \times 4$ case the integrable models will correspond to algebraic varieties in $\CC\mathbb{P}^{16}$. It would be very interesting to exactly understand what the algebraic varieties are that describe integrable models and how exactly they can be characterized.

\appendix
\section{Appendices}
\section{Invertability of transfer matricies}\label{invertability}
Here we prove that $\T_\lambda(\theta_\alpha+\hbar\,\nu_\gn^\alpha)$ is invertible when $\lambda\subset \bar{\nu}^\alpha$, where $\bar{\nu}^\alpha$ denotes the reduced Young diagram $\bar\nu_j^\alpha=\nu_j^\alpha-\nu_\gn^\alpha$, $j=1,\dots,\gn$. We will see below that provided inhomogeneities are largely separated, that is $|\theta_\alpha-\theta_\beta|\gg 1$ for $\alpha\neq \beta$ then the transfer matricies effectively become equal to those of $L=1$. Hence, we start by considering this case. Any given transfer matrix $\T_\lambda(u)$ is a polynomial in $\theta_\alpha$ and the entries of the twist matrix $G$. Hence if we can prove the claim for a specific value of the twist then it must be true generically, {\it i.e}. away from some measure zero subset. To this end, let us make use of the fact that transfer matricies are central for $L=1$ when $G=1$ where the computation simplifies. In what follows we will omit the $\alpha$-index.

\medskip

A convenient tool to prove the claim is the quantum eigenvalues introduced in Section \ref{Qsystem}. By acting on the highest-weight state it is easy to see that $\Lambda_j(u)=(u-\theta-\hbar\, \nu_j)$. 
The transfer matrix $\T_{a,1}(u)$ can be written as a sum over quantum semi-standard Young tableaux of the form 
\begin{equation}
\ytableausetup{centertableaux}
\begin{ytableau}
i_a \\
\none[\vdots] \\
i_2 \\ 
i_1 \\
\end{ytableau}
\end{equation}
subject to the constraint $i_1<i_2<\dots <i_a$. By using the recipe to assign products of quantum eigenvalues to a tableau  we associate the factor $\prod_{k=1}^a(u-\theta-\hbar(\nu_{i_k}+a-k))$ to the above tableau. Let us now evaluate this factor at $\theta+\hbar\,\nu_\gn$. We obtain 
\begin{equation}
(-\hbar)^a(\bar\nu_{i_a})(\bar\nu_{a-1}+1)\dots(\bar\nu_{i_1}+a-1)\,.
\end{equation}
Since $\bar\nu_j\geq 0$ for all $j=1,\dots,\gn$ it follows that the above expression is non-negative. Note that if some weight $\nu_k=\nu_\gn$, it forces $\bar\nu_k=\bar\nu_{k+1}=\dots=\bar\nu_\gn=0$ and hence the indices $k,k+1,\dots,\gn$ cannot appear in the tableau as they provide vanishing contributions. Hence, in order to have a non-vanishing term we must at least have $\bar\nu_a\geq 1$ and hence $\bar\nu_1\geq \bar\nu_2\geq \dots \geq \bar\nu_a\geq 1$. Hence, $\T_{a,1}(\theta+\hbar\,\nu_\gn)$ is non-zero if 
\begin{equation}
(1^a)\subset \bar{\nu}\,.
\end{equation}
Now we consider an arbitrary Young diagram $\lambda$. $\T_\lambda(\theta+\hbar\,\nu_\gn)$ can be written as a sum over Young tableaux as before, and we will consider the factors of quantum eigenvalues associated to each column separately. The admissible indices such that a given column is non-vanishing directly effects what indices can appear in the columns to the right. Indeed, we already know the first column will always be non-negative, and we will get a non-zero contribution if 
\begin{equation}
(1^{\lambda^{\rm T}_1})\subset \bar{\nu}\,.
\end{equation}
Now we go to the second column which gives the contribution 
\begin{equation}
(-\hbar)^{\lambda^{\rm T}_2}(\bar\nu_{i_{\lambda^{\rm T}_2}}-1)(\bar\nu_{i_{\lambda^{\rm T}_2}-1}-2)\dots (\bar\nu_{i_{1}}+\lambda^{\rm T}_2-2)\,.
\end{equation}
Since the first column is non-zero, if we put some number $k$ in the top box of the second column we must have that $\bar\nu^\alpha_k> 1$ and hence the second column will be non-zero if 
\begin{equation}
\bar\nu_1\geq \bar\nu_2\geq \dots\geq \bar\nu_{\lambda^{\rm T}_2}\geq 2\,.
\end{equation}
Hence, the contribution from the first two columns will be non-zero if 
\begin{equation}
(1^{\lambda^{\rm T}_1}1^{\lambda^{\rm T}_2})\subset \bar{\nu}\,.
\end{equation}
Continuing in the same way, we find that if $\lambda\subset \bar{\nu}$ there will always be a tableau which does not vanish and the signs of the contributions of all non-vanishing tableaux are all the same and equal to the sign of $(-1)^{|\lambda|}$, where $|\lambda|$ denotes the number of boxes in the Young diagram $\lambda$. Hence for $L=1$ $\T_\lambda(\theta+\hbar\,\nu_n)$ is non-zero. 

\medskip

Now we consider $L>1$. The transfer matrix $\T_\lambda$ is obtained by taking the trace of the fused monodromy matrix $T_\lambda(u)$ which itself is a product of fused $R$-matricies $R^{\lambda,\nu^\alpha}$
\begin{equation}
\T_\lambda(u)=\displaystyle\sum_{i_1,\dots,i_L}R^{\lambda,\nu^1}_{i_1 i_2}(u-\theta_1)\otimes \dots \otimes R^{\lambda,\nu^L}_{i_L i_1}(u-\theta_L)\,,
\end{equation}
where the sum ranges over $1,2,\dots,{\rm dim }\,\lambda$. Since $R^{\lambda,\nu^\beta}(u)\sim u^{|\lambda|}$ at large $u$, with $|\lambda|$ denoting the number of boxes in the Young diagram $\lambda$, we can consider $\T_\lambda(\theta_\alpha+\hbar\,\nu^\alpha_n)$ in the limit $|\theta_\beta-\theta_\alpha|\gg 1$ for all $\beta\neq \alpha$. In this limit $\T_\lambda(\theta_\alpha+\hbar\,\nu^\alpha_n)$ coincides (up to irrelevant normalisation) with the $L=1$ transfer matrix which we know is invertible and so $\T_\lambda(\theta_\alpha+\hbar\,\nu_n^\alpha)$ is invertible for generic values of inhomogeneities, completing the proof.

\section{Action of transfer matricies -- technical details}\label{transferaction}
We need to prove that 
\be
\label{needtoprove}
\bra{\Lambda}\prod_{\alpha=1}^L\displaystyle\frac{\displaystyle \T_{F^\alpha_k+\bar{\mu}^\alpha_k}(\theta_\alpha+\hbar\,\nu^\alpha_n)}{\displaystyle \T_{F^\alpha_k}(\theta_\alpha+\hbar\,\nu^\alpha_n)}=\bra{\Lambda}\prod_{\alpha=1}^L \phi^{\gn-k-1}\left(\T_{\bar\mu_k^{\alpha}}(\theta_\alpha+\hbar\,\nu_{k+1}^{\alpha})\right)\,
\ee
if $\bra{\Lambda}\in\lV_{(k)}$. This result easily follows from the following one which we are going to prove: For a state of the form
\begin{equation}
\bra{\Lambda_I}:=\bra{\Lambda}\prod_{\gamma\in I}\phi^{\gn-k-1}\left(\T_{\bar\mu^\gamma_{k}}(\theta_\gamma+\hbar\,\nu^\gamma_{k+1})\right)\,,
\end{equation}
where $\bra{\Lambda}\in\lV_{(k)}$ and $I$ is a subset of $\{1,\ldots,L\}$, it is true that
\begin{equation}
\bra{\Lambda_I}\phi\left(\T_{\bar\mu^\alpha_{k}}(\theta_\alpha+\hbar\,\nu^\alpha_{k+1})\right)=\bra{\Lambda_I}\frac{\T_{R_{\gn-1}+\dots+R_{k-1}+\bar\mu^\alpha_k}}{\T_{R_{\gn-1}+\dots+R_{k-1}}}
\end{equation}
for $\alpha\notin I$. Here both transfer matrices on the \rhs are evaluated at $\theta_\alpha+\hbar\,\nu^\alpha_\gn$, and $R_{\gn-1}+\dots+R_{k-1}$ is a specific choice of Young diagram $F_k^{\alpha}$ to be made precise below\footnote{Recall that the ratio in the \lhs of \eqref{needtoprove} is invariant under variations of $F_k^{\alpha}$ subject to certain constraints, we are making one particular choice that simplifies computations.}.

\medskip

We will need two technical results. First, let us note that quantum minors satisfy the following commutativity property \cite{molev2007yangians}. If $\lA$ and $\lB$ are subsets of $\{1,2,\dots,\gn\}$ then
\begin{equation}\label{commutativity}
[T\left[^\lA_\lB\right](u),T_{ab}(v)]=0
\end{equation}
for all $a\in\lA$ and $b\in\lB$. Next, suppose $\bra{\Lambda}$ of $\GT_1,\dots,\GT_r$ for some $r$, for which the dual diagonal $\mu^\alpha_r$ takes its minimal allowed value $\mu^\alpha_{rj}=\nu^\alpha_{r+1}$, $j=1,\dots,r$ and $\mu^\alpha_{r+1}$ takes its maximal allowed value given the previous constraint $\mu^\alpha_{r+1,j}=\nu^\alpha_{r+1}$, $j=1,\dots,r+1$. Then we have
\begin{equation}\label{shortening}
\bra{\Lambda}T_{j,\gn-r}(\theta_\alpha+\hbar\,\mu^\alpha_{\gn-r+1,1})=0,\quad j=\gn-r-1,\dots,\gn
\end{equation}
which is simply the statement that the dual diagonal $\mu^\alpha_{r+1}$ cannot be excited further without changing $\mu^\alpha_r$ and that $\mu^\alpha_r$ cannot be lowered without first lowering $\mu^\alpha_{r+1}$. The proof of this is very similar to that of the statements (3.36-3.38) in \cite{Ryan:2018fyo} adapted to this more general setting and so we do not repeat it here. The motivation for this statement is that when we act with transfer matricies $\T_{R_{\gn-1}+\dots+R_{k-1}+\bar\mu^\alpha_k}$ the action on $\bra{\Lambda_I}$ will factorise, and each $\T_{R_j}$ factor will act as a raising operator exciting a dual diagonal to its maximal where it is equal to the next dual diagonal, allowing us to use the previous result.

\medskip

Let $\bar{\nu}^\alpha$ denote the reduced Young diagram $\bar{\nu}^\alpha_j=\nu^\alpha_j-\nu^\alpha_\gn$, $j=1,\dots,\gn-1$. 
$\bar\nu^\alpha$ splits into the rectangular regions $R_j$, $j=1,\dots,\gn-1$, where the width of $R_j$ is $\bar\nu^\alpha_j-\bar\nu^\alpha_{j+1}$ and its height is $j$. By $R_{\gn-1}+\dots+R_{k-1}$ we denote the subdiagram of $\bar\nu^\alpha$ comprising the first $\bar\nu_{k-1}^{\alpha}$ columns of $\bar\nu^\alpha$. Note that the state $\bra{\Lambda_I}$ is an admissible vector at point $\theta_\alpha+\hbar\,\nu^\alpha_\gn$ and so the action of $\T_{R_{\gn-1}+\dots+\bar\mu^\alpha_k}(\theta_\alpha+\hbar\,\nu^\alpha_\gn)$ with the MCT \eqref{MCTcomponent} coincides with that of the null twist, {\it cf.} page~\pageref{pos:admissible}.

\medskip

For simplicity of exposition, we will assume that all weights $\nu^\alpha_j$ are distinct, and will comment later on what happens when they are not. For all weights being distinct, the region $R_j$ has non-vanishing width and furthermore we have the following factorisation 
\begin{equation}\label{factorout}
\T_{R_{\gn-1}+\dots+\bar\mu^\alpha_k}(u)=\T_{R_{\gn-1}}(u)\T_{R_{\gn-2}+\dots+\bar\mu^\alpha_{k}}(u+\hbar\,\bar\nu^\alpha_{\gn-1})\,.
\end{equation}
To see this we utilise the CBR formula \eqref{CBRfla} which says that for some Young diagram $\lambda$ one has
\begin{equation}
\T_\lambda(u)=\sum_{\sigma\in S_{\gn}} \T_{\lambda_1^{\rm T}+\sigma(1)-1,1}(u+\hbar(\sigma(1)-1))\times \dots\,.
\end{equation}
When we use the null twist, all $\lambda$ are constrained to have height at most $\gn-1$, and for the case of interest to us we have $\lambda^{\rm T}_1=\gn-1$. In the above sum, if for some permutation $\sigma$ we have $\sigma(1)\neq 1$ then $\sigma(1)>1$ and so the sum contains a transfer matrix of height greater than $\gn-1$ and so must vanish. Hence, we must have that the transfer matrix factorises into $\T_{\lambda_1^{\rm T}}(u)\times \dots$ where $\dots$ refers to the transfer matrix corresponding to the Young diagram obtained from $\lambda$ by removing its first column. If the second column also has height $\gn-1$ then it also factors out and so on. Hence \eqref{factorout} follows, where now
\begin{equation}\label{Tfactor}
\T_{R_{n-1}}(u)=\T_{n-1,1}(u)\dots \T_{n-1,1}(u+\hbar(\bar\nu^\alpha_{n-1}-1))\,,
\end{equation}
and so the \rhs \eqref{Tfactor} coincides with the composite raising operator \eqref{compositeraising} for the right-most dual diagonal. Hence, evaluating at $u=\theta_\alpha+\hbar\,\nu^\alpha_\gn$ we see that acting with $\T_{R_{\gn-1}}$ takes us from $\bra{\Lambda_I}$ to the state $\bra{\Lambda'_I}$ with $\mu^\alpha_{\gn-1,j}=\mu^\alpha_{\gn-2,j}=\nu^\alpha_{\gn-1}$, $j=1,\dots,\gn-2$ and $\mu^\alpha_{\gn-1,\gn-1}=\nu^\alpha_{\gn-1}$ which satisfies {\eqref{shortening}.

\medskip

The action of $\T_{R_{n-2}+\dots+\bar\mu^\alpha_{k}}(u+\hbar\bar\nu^\alpha_{n-1})$ on $\bra{\Lambda'_I}$ is expressed as a sum over tableaux $\sum_{\lA}T\left[^\lA_{\lA+1}\right]$ where $\lA+1$ cannot contain the number $2$ by \eqref{shortening}, and so $\lA$ cannot contain $1$, forbidding us from having transfer matricies of size $\gn-1$ and so the action again factorises into 
\begin{equation}
\bra{\Lambda_I}\T_{R_{\gn-1}}\T_{R_{\gn-2}}\T_{R_{\gn-3}+\dots+\bar\mu^\alpha_{k}}(u+\hbar\bar\nu^\alpha_{\gn-2})\,.
\end{equation}
Hence when the $\T_{R_{\gn-2}}$ factor acts on $\bra{\Lambda}\T_{R_{\gn-1}}$ it will excite the dual diagonals to the configuration where $\mu^\alpha_{\gn-2,j}=\mu^\alpha_{\gn-3,j}=\nu^\alpha_{\gn-2}$, $j=1,\dots,\gn-3$ and $\mu^\alpha_{\gn-2,\gn-2}=\nu^\alpha_{\gn-2}$ and again the results of \eqref{shortening} apply, further limiting the indicies which can populate the tableaux making up the $\T_{R_{\gn-3}+\dots}$ factor. 

\medskip

The end result is that the action of $\T_{R_{\gn-1}+\dots}$ completely factorises into 
\begin{equation}
\bra{\Lambda_I}\T_{R_{\gn-1}}\T_{R_{\gn-2}}\dots \T_{R_{k-1}} \T_{\bar\mu^\alpha_k}(\theta_\alpha+\hbar\,\nu^\alpha_{k+1})\,,
\end{equation}
where we have omitted the spectral parameters of the $\T_{R_j}$ factors for brevity and $\T_{\bar\mu^\alpha_k}$ should be understood as $\sum_{\lA}T_{\bar\mu^\alpha_k}$ where $\lA$ can only be populated with indices from the set $\{\gn-k,\dots,\gn-1\}$. Then, using \eqref{commutativity} we can move this factor to the left, obtaining 
\begin{equation}
\begin{split}
& \bra{\Lambda_I}\T_{R_{\gn-1}+\dots+\bar\mu^\alpha_k}(\theta_\alpha+\hbar\,\nu^\alpha_\gn) \\ &=\bra{\Lambda_I}\phi^{\gn-k-1}\left(\T_{\bar\mu^\alpha_k}(\theta_\alpha+\hbar\,\nu^\alpha_{k+1})\right)\T_{R_{\gn-1}+\dots+R_{k-1}}(\theta_\alpha+\hbar\,\nu^\alpha_{\gn})\,.
\end{split}
\end{equation}
This completes the proof since invertiblity of the transfer matrix was proven in the previous appendix.

\medskip

Finally, let us briefly discuss the case of coinciding weights. As we have seen above, each factorisation into a rectangular region results in a reduction of the number of indices in the factors which appear to the right of it. If two weights coincide, say $\nu^\alpha_j=\nu^\alpha_{j+1}$ then the rectangle $R_j$ has vanishing width and so does not contribute to the factorisation. One could then expect that at the end the right most factor could contain more than just the indices $\gn-k,\dots,\gn-1$, ruining our conclusion. However, if two weights coincide then $\bra{\Lambda_I}$ will have extra dual diagonals  $\mu_{k+1}^{\alpha}, \mu_{k+2}^{\alpha},\ldots$ whose entries are all equal to $\nu_{k+1}^{\alpha}$. They will extend the range of indices in \eqref{shortening} which annihilate $\bra{\Lambda}$ similar to the case of rectangular representations discussed in \cite{Ryan:2018fyo}, which will further constrain the indices that can appear in the sum over tableaux. Taking this into account we find that the end conclusion is the same.

\bibliographystyle{utphys}
\bibliography{References}

\end{document}